\documentclass[12pt,oneside]{report}

\usepackage[T1]{fontenc}
\usepackage[utf8]{inputenc}
\usepackage{lmodern}   
\usepackage[letterpaper,left=1.5in,right=1in,top=1in,bottom=1in]{geometry}
\usepackage{amsmath}
\usepackage{amssymb}
\usepackage{amsfonts}
\usepackage{mathrsfs}
\usepackage{graphicx}
\usepackage{setspace}
\usepackage{indentfirst}
\usepackage{titlesec}
\usepackage{tocloft}
\usepackage{caption}
\usepackage{array}
\usepackage{longtable}
\usepackage{bm}
\usepackage{upgreek}
\usepackage[hidelinks]{hyperref}

\newlength{\figdefaultwidth}
\newcommand{\figstub}[4][\figdefaultwidth]{%
  \begin{figure}[htbp]%
    \centering
    \includegraphics[width=#1]{#3}%
    \caption{#4}%
    \label{#2}%
  \end{figure}%
}

\newcommand{\tabstub}[4][\figdefaultwidth]{%
  \begin{table}[htbp]%
    \centering
    \caption{#4}%
    \label{#2}%
    \includegraphics[width=#1]{#3}%
  \end{table}%
}

\renewcommand{\thechapter}{\arabic{chapter}}
\renewcommand{\thesection}{\thechapter.\arabic{section}}
\renewcommand{\thesubsection}{\thesection.\arabic{subsection}}
\renewcommand{\thesubsubsection}{\thesubsection.\arabic{subsubsection}}

\titleformat{\chapter}[hang]
  {\normalfont\bfseries\large}{\thechapter.}{1em}{\MakeUppercase}
\titlespacing*{\chapter}{0pt}{0pt}{20pt}

\titleformat{\section}[hang]
  {\normalfont\bfseries}{\thesection}{1em}{\MakeUppercase}
\titleformat{\subsection}[hang]
  {\normalfont\bfseries}{\thesubsection}{1em}{}
\titleformat{\subsubsection}[hang]
  {\normalfont}{\thesubsubsection}{1em}{}

\numberwithin{equation}{chapter}
\numberwithin{figure}{chapter}
\numberwithin{table}{chapter}

\begin{document}

\begin{titlepage}
\thispagestyle{empty}
\centering
\vspace*{0.7in}

{\large Electron Density Measurements from Stark Broadened Emission in a Sodium Plasma Produced by Laser Resonance Saturation}\\[1.2em]

by\\[3.0em]

{\large Mark Antony Cappelli}

\vfill
{\large Note that this is an unpublished extraction of the Ph.D. thesis of Mark Antony Cappelli, from the University of Toronto.}\\[1.2em]
\raggedright
Submitted August, 1987

\vspace{2.0cm}
\centering
\copyright\ Mark Antony Cappelli, 1987

\vspace{2.0cm}
\noindent
\begin{minipage}[t]{0.45\textwidth}
\raggedright
September, 1987
\end{minipage}%
\hfill
\begin{minipage}[t]{0.45\textwidth}
\raggedright
UTIAS Report No.\ 306\\
CN ISSN 0082-5255
\end{minipage}
\end{titlepage}

\pagenumbering{roman}
\setcounter{page}{1}

\chapter*{ACKNOWLEDGEMENTS}
\addcontentsline{toc}{chapter}{Acknowledgements}

I would like to thank Prof.\ R.M.\ Measures for his encouraging support
throughout the course of this work.  I would also like to extend my
appreciation to the other members of my committee, Profs.\ J.H.\ de Leeuw and
J.J.\ Gottlieb for their guidance and helpful suggestions, and to Prof.\ A.A.\
Haasz for his careful reading and criticism of the manuscript.

Dr.\ P.G.\ Cardinal, Mr.\ R.S.\ Kissack, and Mr.\ G.W.\ Schinn have all
assisted me in overcoming a number of hurdles encountered throughout my
earlier years spent investigating this topic and for this I am sincerely
grateful.  A special thanks is extended to Messrs.\ G.\ Bisci and A.\
Paparoni for their advice and friendship when needed most during my stay in
Toronto.

I would like to thank the staff at UTIAS, particularily Mrs.\ W.\ Ryan,
Mrs L.\ Quintero, Mrs.\ W.\ Dillon, Ms.\ P.\ Cooke and Mr.\ K.\ Bopp who have
always been willing to assist in one way or another.

The finacial assistance received from the U.\ S.\ Air Force Office of
Scientific Research and the Ontario government is gratefully acknowledged.

\chapter*{Abstract}
\addcontentsline{toc}{chapter}{Abstract}

Electron Stark broadening of the $4^2D$--$3^2P$ multiplet transition in a
sodium plasma produced by laser resonance saturation has provided a means of
undertaking the first spatial measurements of the free electron density
across and along the plasma channel created in sodium vapor of density
$10^{15}$--$5\times10^{16}$~cm$^{-3}$.  From these measurements and measurements of the neutral
sodium density within a heat sandwich oven, we are able to deduce the
corresponding electron temperature.  These temperatures compare favourably
with the electron temperature estimated from a Boltzmann analysis of line
intensities and suggest that a highly ionized ($>$10\%) plasma of electron
temperature $<$6000K can be produced within 100~ns of laser excitation from a
laser of modest irradiance ($10^6$--$10^7$~Wcm$^{-2}$).  The experimental results are in
reasonable agreement with a recently developed 3-dimensional model which
predicts a decrease in the degree of ionization along the path of the laser
beam as a result of significant depletion of laser energy.  These
experimental results demonstrate that this LIBORS code is capable of
predicting the 3-dimensional nature of this new mode of laser ionization
with reasonable accuracy, and may also explain the low electron temperatures
and free electron densities observed by other research teams.

\clearpage
\renewcommand{\contentsname}{TABLE OF CONTENTS}
\tableofcontents

\clearpage
\noindent\textbf{Tables}\par
\noindent\textbf{Figures}\par
\vspace{1em}
\noindent
\begin{tabular}{@{}p{1.05in}p{4.0in}@{}}
APPENDIX A: & EMPIRICAL FORMULA FOR COMBINED RESONANCE AND DOPPLER
              BROADENED ATOMIC ABSORPTION PROFILES IN THE IMPACT AND
              QUASI-STATIC LIMITS \\[0.4em]
APPENDIX B: & RADIATIVE TRANSFER SIMULATION CODE - SATSIM \\[0.4em]
APPENDIX C: & SODIUM ATOM DISTRIBUTION WITHIN A HEAT SANDWICH OVEN \\[0.4em]
APPENDIX D: & TWO-CHANNEL TECHNIQUE FOR STARK MEASUREMENTS OF ELECTRON
              DENSITY WITHIN A LASER-PRODUCED SODIUM PLASMA \\[0.4em]
APPENDIX E: & [see original report] \\[0.4em]
APPENDIX F: & METHOD OF LINEAR LEAST SQUARES \\[0.4em]
APPENDIX G: & EFFECTS OF INHOMOGENEITIES, OPTICAL DEPTH AND FINITE
              BANDWIDTH ON ELECTRON TEMPERATURE MEASUREMENTS IN A
              CYLINDRICAL PLASMA \\[0.4em]
APPENDIX H: & PLASMA CHANNEL FORMATION THROUGH LASER RESONANCE SATURATION \\[0.4em]
APPENDIX I: & RESONANCE AND VAN DER WAALS BROADENING OF SPECTRAL
              LINES FOR TRANSITIONS BETWEEN EXCITED STATES IN SODIUM \\
\end{tabular}

\clearpage
\pagenumbering{arabic}
\setcounter{page}{1}

\chapter{Introduction}
\label{ch:introduction}

This thesis is devoted to the study of a sodium plasma created when a
beam of moderately intense pulsed laser radiation ($10^5$-$10^7$ MW cm$^{-2}$) tuned to
the $3^2S$- $3^2P$ resonance transition passes through dense sodium vapor.  This
form of laser ionization was first experimentally demonstrated in sodium
vapor by Lucatorto and McIlrath (1976).  Since then, this Laser Ionization
Based On Resonance Saturation (LIBORS) technique has received considerable
attention and has been applied to the ionization of other members of the
alkali metals, as well as the alkaline earths.  These elements are
attractive from the point of view of the relative ease of producing and
maintaining a dense (0.1 to 10 torr) vapor in a heat pipe oven, as well as
having the lowest ionization potentials of all the elements.  Maintaining an
excited (resonance) state population density by extended laser resonance
saturation furthermore reduces the effective ionization potential by a
significant amount, as these elements also happen to possess a relatively
large resonance state energy to ionization energy ratio.

There were initial attempts to explain the physical mechanism by which
rapid and almost complete ionization occurred, shortly after the first
experimental evidence was observed (McIlrath and Lucatorto 1977, Salter
1979).  Perhaps the most satisfactory explanation however, had been
suggested nearly a decade earlier by Measures (1970).  Measures had proposed
that the large number of atoms in the resonance state (ideally saturated)
pumped by intense resonant radiation, provided a mechanism for direct and
efficient transfer of laser energy to free electrons.  The exchange would
arise via superelastic collisions between these free electrons and the large
laser maintained resonance state population.  These free electrons, rapidly
gaining energy, would subsequently collisionally excite and ionize the
neutral species.  The initial pool of free electrons necessary to get the
process going can be created through a variety of seed electron processes
(Cardinal 1986).  A qualitative description of the LIBORS process with the
dominant electron excitation and ionization rates is shown in figure
\ref{fig:1-1}.

\figstub{fig:1-1}{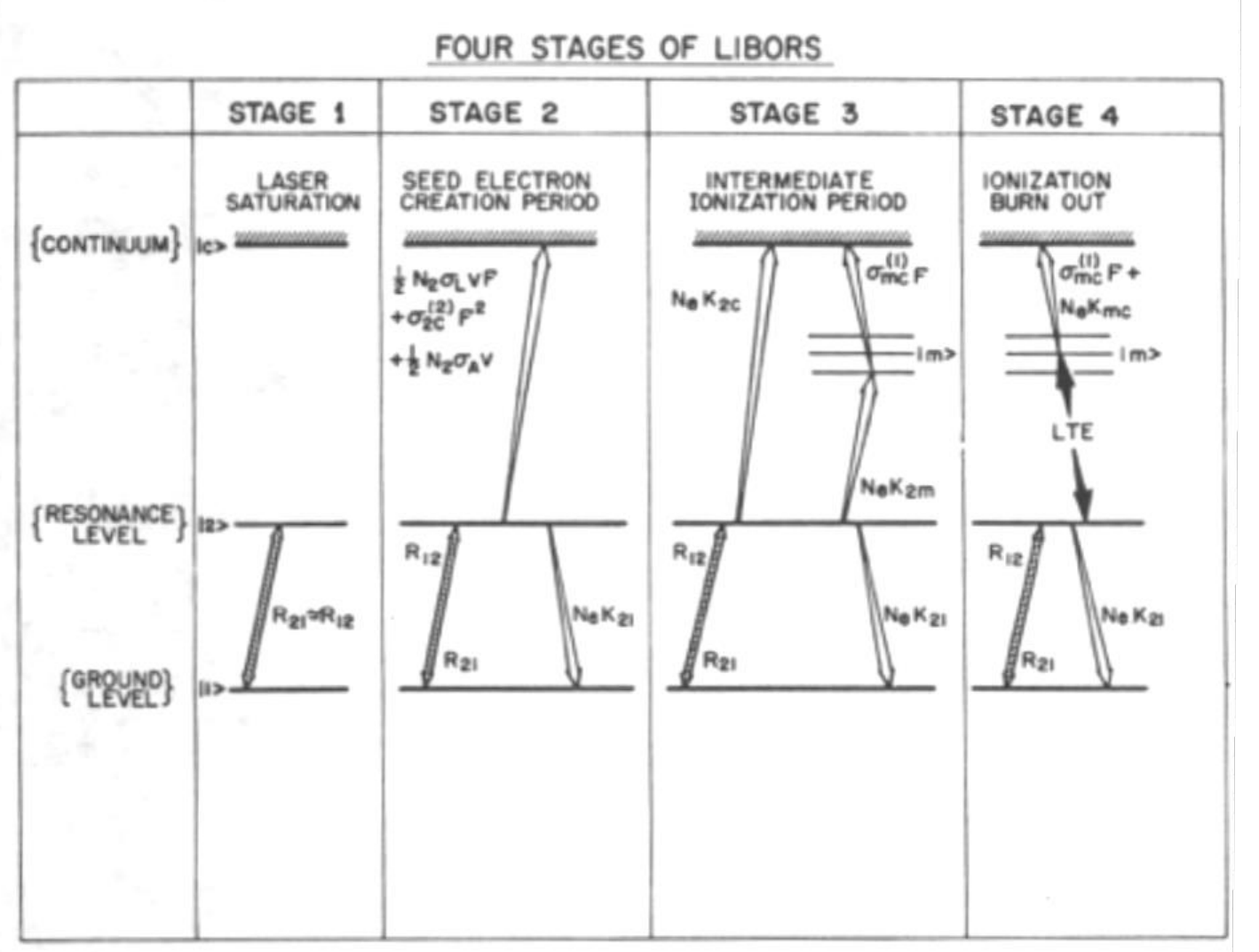}{Four stages of laser ionization based on resonance saturation. STAGE 1. Laser rapidly locks ground and resonance level populations in ratio of degeneracies. STAGE 2.(i) Rapid growth of free e1ectrons due to two -photon ionization of resonance level and laser-induced Penning ionization. Associative ionization is important for some elements. (ii) Free e1ectrons rapidly gain energy through superelastic collisions. STAGE 3. (i) Direct electron impact ionization of resonance level and single-photon ionization of collisionally populated upper levels dominate the rate of ionization. lii) Electron temperature stabilizes as the rate of superelastic heating balances rate of collisional cooling through excitation. STAGE 4.(i) Runaway collisional ionization of upper levels occurs once a critical electron density achieved. (ii) heating can no longer balance collisional cooling and electron temperature falls.
}
Since 1979, significant effort has gone into the modelling of this new
form of laser interaction (Measures, Drewell and Cardinal 1979; Measures and
Cardinal 1981; Measures, Cardinal and Schinn 1981; Measures, Wong and
Cardinal 1982) paving the road for some very sophisticated modelling to
include the effects of dimers on the seed electron creation (Cardinal 1986).
The potential application of LIBORS towards the creation of long plasma
channels to facilitate electron and light ion beam transportation in
inertial confinement fusion reactors (Yonas 1978), lithium anode plasma
sources for intense ion beam diodes (Dreike and Tisone 1986), and as well,
towards the development of short wavelength lasers (Olsen and Leeper 1982)
soon became apparent.  As a result, efforts went into modelling the three
dimensional nature of the problem (Kissack 1987) to include laser absorption
through a finite depth of sodium vapor and electron thermal conduction in
the radial domain.

As the theories became more and more elaborate, so increased the need
to provide sound and accurate experimental data with which to rigorously
test these theories.  Early quantitative measurements of the free electron
density in a sodium plasma produced by laser saturation, were made by
measuring the photoionization current across electrodes inserted directly
into the plasma (Stacewicz and Krasinski 1981).  Roussel et al.\ (1980) and
Carr\'e et al.\ (1981a, 1981b) estimated the free electron density from
measurements of the positive ion yield using a mass spectrometer arrangement
and confirmed the results of Lucatorto and McIlrath (which were based on the
measurement of the ground state atom density remaining after laser
excitation by the width of the photoabsorption spectrum of the $3^2S$-$3^2P$
transition).  These results suggested that a high degree of ionization ($>$90\%)
can be achieved.  The `Hook' technique was used quite successfully, to more
accurately measure the ground and excited state population densities in a
sodium LIBORS type plasma (Salter et al.\ 1979, Salter 1979) and in LIBORS
plasmas of the alkaline earths (Skinner 1980, Bachor and Kock 1980, 1981).

Inherent in these measurements however, is a great degree of spatial
averaging, as they rely on absorption along an optical path long enough to
overcome signal to noise or instrument resolution limitations.  The first
qualitative emission based measurements of the temporal variation in the
excited state population densities in a strontium plasma were reported by
Brehignac and Cahuzac (1982).  The observation of strong line emission from
the strontium ion well after the decay of the laser pulse, was attributed to
the dominance of superelastic electron atom collisions and collisional
excitation and ionization over radiative decay.  Once again, these early
emission based observations were spatially averaged as the signal to noise
ratio in these and earlier experiments (Wizinowich 1979), was generally
quite poor.  It was quite evident that a new direction must be taken to make
quantitative measurements (both temporally and spatially resolved) of the
free electron density so that a direct comparison with theory could be
made.

The direct perturbation of the atomic states arising from free
electrons colliding elastically and inelastically with sodium atoms in their
excited states effectively reduces the lifetime of the excited state and is
evident as a broadening of that state.  This broadening appears as a natural
and convenient indication of the free electron density in the plasma.  The
extent of the electron Stark broadening (as it is so called and referred
to in this thesis) of the excited atomic states, is directly proportional to
the free electron density and is evident as a shift and an increase in the
width of emission lines.  The measurement of the electron Stark width and
shift of emission lines, has been used as a diagnostic tool in plasmas ever
since the development of the first reliable Stark broadening theories
(Baranger 1958b, 1958c, Kolb and Griem 1958).  Krebs and Schearer (1982)
estimated the free electron density in a LIBORS sodium plasma from the Stark
shifting of emission lines arising from transitions between excited states
in sodium.  These measurements were spatially averaged across a cylindrical
plasma produced at the focus of the laser beam near the centre of a sodium
vapor cell.  Stark based measurements of electron density along the path of
a collimated laser beam (again spatially averaged across the axial symmetric
plasma cross section) were later reported by Cappelli and Measures (1984).
The natural extension of this is to make accurate measurements both along
the path of the laser beam and in the radial domain defined by the plasma
cross section.  In order to make a reasonable comparison of these results
with theory, an accurate measurement of the neutral sodium density variation
along the laser beam path is also necessary.  The objectives of this
research should then be clear and as a result, this thesis describes the
development of a facility and technique to produce and study a sodium vapor
irradiated with a laser tuned to the first resonance transition.  The first
detailed measurements of the free electron density are reported.  The plasma
that is created is analyzed in terms of the radial distribution of the free
electron density achieved (temporally resolved, shortly after the decay of
the laser pulse) at positions along the path of the laser pulse.

Spectral line broadening plays a major role in the plasma diagnostics
and in the absorption technique developed to measure the neutral sodium atom
density distribution within a heat sandwich (heat-pipe like) oven.  For this
reason, a great part of chapter two is devoted to a review of spectral line
broadening in a neutral sodium vapor, and in a sodium plasma.  The remainder
of chapter two is a review of radiative transfer theory and the inversion of
the spectral emission to obtain the radial variation in the local volume
emission.  The local volume emission coefficient is proportional to the
atomic line profile function from which we can derive the free electron
density.

Chapter three is devoted to the theory behind the vapor and plasma
diagnostics.  It is divided into four sections.  The first section reviews
the absorption based technique for measuring the neutral sodium density
distribution within the oven.  The second section is devoted to the Stark
based measurements of the radial electron density distribution.  The third
section focuses on the measurements of electron temperature from the
relative intensity of spectral lines.  The last section is a discussion of
the implications of multi-shot averaging of spectral emission from a highly
non-linear source such as a plasma channel produced by laser resonance
saturation.

An overview of the complete experimental facility is given in chapter
four, paying particular attention to the neutral sodium absorption and the
sodium plasma emission measurement sub-facilities.

The results are presented in chapter five, which is divided into three
sections.  The first section concentrates on the measurements of the radial
distribution of the free electron density across the plasma column, and how
this distribution changes with penetration depth, incident laser energy
fluence, and incident laser wavelength.  The second section deals primarily
with the electron temperature measurements, comparing the results obtained
from three somewhat different methods.  Chapter five concludes with a
presentation and discussion of measurements of the laser beam penetration,
particularily, the variation in the characteristics of the laser pulse
transmitted through the oven, with incident laser wavelength.

Theoretical predictions are compared to experiment in chapter six, for
a set of experimental results.

The conclusions are presented in chapter seven, along with a summary
and an outline of the contributions that this work has made to this
particular field of study.

\chapter{Spectral Line Broadening and Radiative Transfer}
\label{ch:broadening}

\section{Spectral Line Broadening in Neutral Cold Sodium Vapor}
\label{sec:2-1}

If absorption or emission of a photon by a sodium atom takes place
during a collision of that atom with another atom (perturber), then the
transition can be thought of as occurring between two quasimolecular states
formed as a result of the interaction between the colliding partners.  The
resulting energy shifts of the excited and ground states are a function of
the internuclear separation and depend on the type of interaction.  The
curves depicting the potential energy of these quasimolecular states as a
function of internuclear separation $r$ are referred to as the ground or
excited state potentials.  Three singlet molecular state potentials of the
sodium-sodium pair are qualitatively illustrated in figure \ref{fig:2-1}a.

\figstub{fig:2-1}{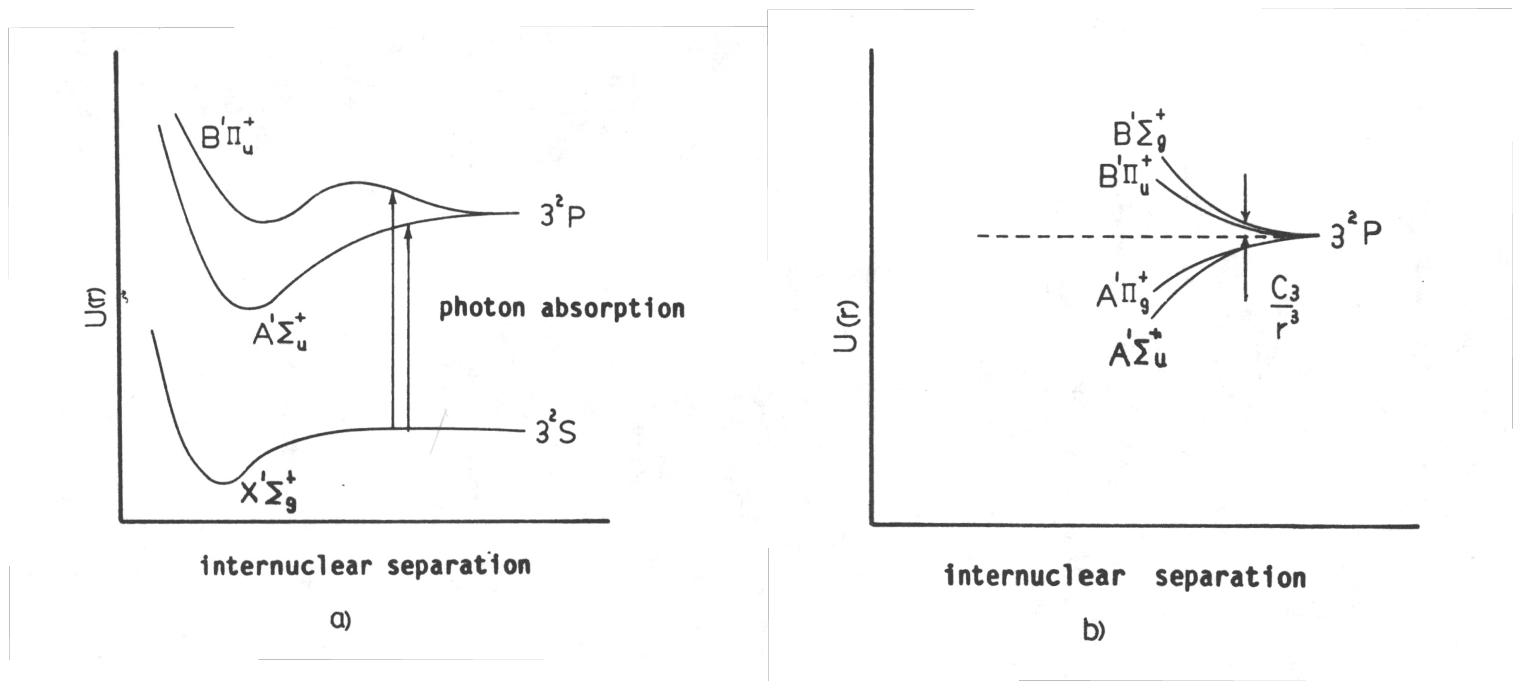}{a) a) Qualitative example of the ground and first two excited molecular state potentials in the Na-Na System (spin-spin and spin-orbit Interactions neglected). b) A qualitative description of the potentials at long range showing the symmetrical splitting about the unperturbed atomic 32P state.}

In cold neutral alkali metal vapors, the absorption transition of
general interest (at long range) occurs between the ground and first
electronic state, since the majority of the neutral species at the
temperatures of interest ($<$1000K) reside in the ground state and the
strength of the first resonance transition is considerably greater than most
others.  We have employed an absorption technique based on the $3^2S$-$3^2P$
transition, to measure the sodium atom density distribution within a heat
sandwich oven (Cappelli et al.\ 1985, and chapter 3).  Although the
discussion to follow is equally applicable to emission studies of neutral
sodium, we will refer to photon absorption as the process of interest and
will deal with the broadening of the $3^2S$-$3^2P$ transition in neutral sodium
vapor in the presence of an inert buffer gas.
\subsection{Resonance and Van der Waals Broadening of the $3^2S$-$3^2P$ Transition}
\label{sec:2-1-1}

For the case of sodium with electronic states perturbed as a result of
a collision with another sodium atom, the $3^2P$ states of the absorbing atom
split up into four quasi-molecular states with potential curves (neglecting
spin-orbit and spin-spin interactions) situated symmetrically about the
energy corresponding to the atomic $3^2P$ state at large internuclear
separations (Niemax and Pichler 1974a, 1974b, 1975, see for example the
qualitative sketch in figure \ref{fig:2-1}b).  At long range, the symmetrical
splitting of the $3^2P$ state primarily determines the shape of the atomic
absorption profile and is dominated by the long range resonant dipole-dipole
interaction (Sobelman et al.\ 1981).  The broadening of the spectral emission
which follows is generally called resonance broadening.  The magnitude of
the energy shift from the unperturbed energy of the atomic state, has the
form

\begin{equation}
\overline{h}\,\Delta\omega \;=\; \frac{C_3^{(i)}}{r^3},
\label{eq:2-1}
\end{equation}
where $C_3^{(i)}/\overline{h}$ is the resonant dipole-dipole interaction constant, $\Delta\omega = \omega-\omega_0$,
$\omega$ the frequency of the absorbed photon, $\omega_0$ that in the absence of a
perturber and $r$ is the internuclear separation.  Here the superscript (i)
refers to the i-th sub-potential.  When the duration of the collision,
defined by

\begin{equation}
\tau_d \;=\; \frac{\rho}{v}\;,
\label{eq:2-2}
\end{equation}
(here $\rho$ is the impact parameter and $v$ the relative velocity) is much less
than the time of interest $\Delta t$ (that is also typically less than the time
between collisions $\tau_c$), that is

\begin{equation}
\tau_d \;\ll\; \Delta t \;\approx\; \frac{1}{\omega-\omega_0}\;,
\label{eq:2-3}
\end{equation}
(here, $\Delta t \approx 1/(\omega-\omega_0)$, i.e.\ the correlation time between two points in
time of the amplitude of the electron treated as a semi-classical oscillator
(Chen and Takeo 1957)) then the effect of the collision can be treated as a
sudden disruption of phase in the amplitude of the oscillator.  Treated
under this ``impact'' approximation, the atomic absorption line profile
function is Lorentzian in shape,

\begin{equation}
\pounds(\omega) \;=\; \frac{\gamma}{2\pi}\;
\frac{1}{(\omega-\omega_0)^2 + \dfrac{\gamma^2}{4}}\;,
\label{eq:2-4}
\end{equation}
with fullwidth half-maximum (FWHM) $\gamma$ varying linearly with perturber
(neutral sodium atom) density $N$.  The value of $\gamma/N$ is often referred to as
the resonance broadening parameter and for resonance transitions (Sobelman
et al.\ 1981)

\begin{equation}
\frac{\gamma}{N} \;=\; \frac{2\pi^2 C_3}{\overline{h}}\;,
\label{eq:2-5}
\end{equation}
where we define the effective interaction constant (Niemax and Pichler
1974b, 1975) as

\begin{equation}
\frac{C_3}{\overline{h}} \;=\; \Bigl\{\sum_{i=1}^{2} f_i\,\frac{C_3^{(i)}}{\overline{h}}\Bigr\}
\Bigl/ \Bigl\{\sum_{i=1}^{2} f_i \Bigr\}.
\label{eq:2-6}
\end{equation}
Here, $f_i$ is the oscillator strength for the transitions to the individual
states above or below the energy of the undisturbed atomic $3^2P$ state.  We
have assumed that transitions to all sub-potentials are allowed (the
transitions do not follow molecular selection rules) and that the profiles
are symmetric.  Over the wavelength range (detuning from line centre) of
interest, this may be so for the first resonance transition in alkali metals
(Niemax and Pichler 1975), but not necessarily so for higher members of the
principle series (Niemax and Pichler 1974a).

In the other extreme, where $\tau_d \gg 1/(\omega-\omega_0)$, the field experienced by
an absorber or emitter as a result of a neighbouring perturber a distance $r_T$
away is assumed constant in time, and the atomic line profile is
proportional to the probability of finding the nearest perturber a distance
$r$ from the absorbing species.  Assuming that the nearest perturber is much
closer than the average separation between atoms (nearest neighbour or
binary approximation), yet sufficiently far apart such that the interaction
does not significantly alter the distribution of atoms about the absorbing
atom, then we can describe the probability $P(r)dr$ of the nearest perturber
being within a distance $r$ and $r + dr$ from the absorbing atom (Sobelman et
al.\ 1981),

\begin{equation}
P(r)dr \;=\; 4\pi r^2 N \exp\Bigl(-\frac{4\pi}{3}N r^3\Bigr)dr\;.
\label{eq:2-7}
\end{equation}

According to this statistical theory, the atomic absorption line profile
function arrising from absorption to the $i^{\text{th}}$ sub-potential

\begin{equation}
\pounds^{(i)}(\omega)d\omega \;\propto\; P(r)dr\;,
\label{eq:2-8}
\end{equation}
and substituting (\ref{eq:2-1}) into (\ref{eq:2-7}), we arrive at

\begin{equation}
\pounds^{(i)}(\omega)d\omega \;\propto\;
\frac{4\pi N C_3^{(i)}}{\overline{h}\,3\Delta\omega^2}\,
\exp\Bigl(-\frac{\Delta\omega_0}{\Delta\omega}\Bigr)d\omega\;,
\label{eq:2-9}
\end{equation}
where we have defined

\begin{equation}
\overline{h}\,\Delta\omega_0 \;=\; \frac{C_3^{(i)}}{r_0^{\,3}},
\label{eq:2-10}
\end{equation}
with $r_0$ representing the mean internuclear separation, that is

\begin{equation}
r_0 \;=\; \Bigl(\frac{3}{4\pi N}\Bigr)^{1/3}.
\label{eq:2-11}
\end{equation}
Applying the conditions that

\begin{equation}
\int_0^{\infty} \pounds^{(i)}(\omega)d\omega \;=\; \frac{1}{2}\;,
\label{eq:2-12}
\end{equation}
since the splitting is assumed to be in the form of $\pm C_3^{(i)}/r^3$ (Woerdman and de
Groot 1981) and that we are only considering one side of the line profile
function, and assuming that $\Delta\omega_0 \ll \Delta\omega$ (or $r \ll r_0$, i.e.\ nearest neighbour
approximation), then

\begin{equation}
\pounds(\omega) \;=\; \frac{\sum f_i\,\pounds^{(i)}(\omega)}{\sum f_i}
\;=\; \frac{2\pi}{3\overline{h}}\,N C_3 (\Delta\omega)^{-2}\;,
\label{eq:2-13}
\end{equation}
and the wings of the line profile function are also Lorentzian and
symmetrical in shape.  Of course, this ``quasi-static'' approximation is
inapplicable in describing the spectral line core as one can see that it
would violate the conditions described by equation (\ref{eq:2-6}).  In the line core,
the shape is described by equation (\ref{eq:2-4}) which conversely, is inapplicable
to the spectral line wings.

Detailed calculations of the resonance broadened line profile
describing the line wings, core, and the intermediate regions where

\begin{equation}
\tau_d \;\approx\; \frac{1}{\omega-\omega_0}\;,
\label{eq:2-14}
\end{equation}
have been performed by Srivastava and Zaidi (1975).  They have found that
the effective FWHM of the Lorentzian describing the quasi-static wings is
two-thirds that of the profile describing the line core (a comparison of
equation (\ref{eq:2-13}) with (\ref{eq:2-4}) and (\ref{eq:2-5}) would verify this),

\begin{equation}
{}^{w}\gamma \;=\; \frac{2}{3}\,{}^{c}\gamma\;,
\label{eq:2-15}
\end{equation}
where the superscripts c and w refer to the core and wings respectively.
Under the range of conditions expected for our experiments, with a
maximum sodium density $N_{\max} = 5\times10^{16}$ cm$^{-3}$, the minimum average internuclear
separation

\begin{equation}
r_0 \;=\; \Bigl(\frac{3}{4\pi N_{\max}}\Bigr)^{1/3} \;\approx\; 1.7\times10^{-6}\ \text{cm}\;.
\label{eq:2-16}
\end{equation}

The binary approximation describing the nearest perturber is valid for
$r \ll r_0$.  At an internuclear separation of $r_0/10$, it remains to check the
frequency range imposed by the criteria for applicability of the quasi-static
approximation (equation (\ref{eq:2-6})),

\begin{equation}
\Delta\omega \;\gg\; \frac{10}{r_0}\,v\;.
\label{eq:2-17}
\end{equation}
For relative velocities of approximately $10^5$ cm/s, we arrive at

\begin{equation}
\Delta\omega \;\gg\; 7\times10^{11}\ \text{rad/s},
\label{eq:2-18}
\end{equation}
which, for the sodium $3^2S$-$3^2P$ resonance transition, corresponds to

\begin{equation}
\Delta\lambda \;\gg\; 1\ \text{\AA}\;.
\label{eq:2-19}
\end{equation}

Generally, when passing the broadband radiation through dense sodium vapor
(which formed the basis for the neutral sodium density measurements -- see
section \ref{sec:3-1}), spectral absorption holes of greater than 20\,\AA\ were observed.
For the neutral sodium atom density measurements (chapter 3), the sodium
density is estimated from the absorption at approximately 10\,\AA\ from line
centre, thereby satisfying the criterion imposed by equation (\ref{eq:2-19}) allowing
us to use the quasi-static expression for the line profile in the far
wings.

Where the broadband radiation passes through regions of low density and
short optical paths (i.e.\ an outer chord of the vapor disk), the density is
estimated from the absorption at line centre of the $3^2S_{1/2}$-$3^2P_{3/2}$
transition.  The line centre of course satisfies the criteria of equation
(\ref{eq:2-3}) therby permitting the use of the impact approximation to describe the
shape of the line core.

At shorter internuclear separations ($\sim r_0/10$) one should be cautious to
check that the Van der Waals dipole-dipole interaction plays a negligible
role in the perturbation of the $3^2P$ state.  This interaction leads to an
energy shift of the state,

\begin{equation}
\overline{h}\,\Delta\omega \;=\; \frac{C_6}{r^6}\;.
\label{eq:2-20}
\end{equation}
We have estimated the $C_6$ contribution to the excited state $A'\Sigma_u^{+}$ Van der
Waals potential by fitting the numerically tabulated potentials of
Kaminsky (1977, 1980) to

\begin{equation}
\overline{h}\,\Delta\omega \;=\; -\frac{C_3}{r^3} - \frac{C_6}{r^6} - \frac{C_8}{r^8}\;.
\label{eq:2-21}
\end{equation}

In the above equation, we have also included the dipole-quadrupole
interaction term.  Using the value of $C_3$ experimentally determined by Niemax
and Pichler (1975) for the $3^2P_{3/2}$ state in sodium, that is

\begin{equation}
\frac{C_3}{\overline{h}} \;\cong\; 5\times10^{-8}\ \text{cm}^3\ \text{rad/s}\;,
\label{eq:2-22}
\end{equation}
we arrive at
\begin{equation}
\frac{C_6}{\overline{h}} \;\cong\; 5\times10^{-29}\ \text{cm}^6\text{rad/s}
\qquad\text{and}\qquad
\frac{C_8}{\overline{h}} \;\cong\; 10^{-43}\ \text{cm}^8\text{rad/s}.
\label{eq:2-23}
\end{equation}

The shift of the energy state due to the resonance and Van der Waals
interaction become comparable at internuclear separations

\begin{equation}
r \;=\; \Bigl(\frac{C_6}{C_3}\Bigr)^{1/3}\;,
\label{eq:2-24}
\end{equation}
\[
= \;10^{-7}\ \text{cm}\;.
\]
Using equations (\ref{eq:2-21}), (\ref{eq:2-22}) and (\ref{eq:2-23}) with $r=10^{-7}$ cm (10\,\AA), then

\begin{equation}
(\omega-\omega_0) \;=\; 1.1\times10^{14}\ \text{rad/sec}
\label{eq:2-25}
\end{equation}
or
\begin{equation}
(\lambda-\lambda_0) \;=\; 202\,\text{\AA}\;,
\label{eq:2-26}
\end{equation}
and it is apparent that one can not exclude the Van der Waals interaction
when calculating the shape of the extreme line wings.  If we limit our
measurements to approximately 20\,\AA\ ($\omega-\omega_0 = 1.1\times10^{13}$ rad/s) from either
resonance line centre (the sodium $3^2S$-$3^2P$ transition consists of a doublet)
then, one can estimate from equation (\ref{eq:2-21}),
\begin{equation}
r \;\approx\; 1.8\times10^{-7}\ \text{cm}
\label{eq:2-27}
\end{equation}
and at such internuclear separations, the contribution to the potential due
to the Van der Waals interaction is about 15\% that due to the resonant
dipole-dipole interaction which to first order, would approximately
translate to a 15\% change in the absorption coefficient.

For both the neutral sodium absorption measurements (section \ref{sec:3-1}) and
plasma emission simulations (section \ref{sec:2-5}), we have elected to use the
resonance broadened impact widths (FWHM) calculated by Carrington et al.\
(1973), which are in excellent agreement with the theoretical calculations
of Ali and Griem (1965), Vdovin and Dobrodeev (1969) and the experimental
results of Huennekens and Gallagher (1983), that is
\begin{equation}
{}^{c}\gamma_1 \;=\; 3.61\pi e^2 f_1 N/m_e\omega_1
\qquad (\text{for } 3^2S_{1/2}\text{-}3^2P_{1/2})\ \text{rad/s}\;,
\label{eq:2-28}
\end{equation}
\begin{equation}
{}^{c}\gamma_2 \;=\; 2.94\pi e^2 f_2 N/m_e\omega_2
\qquad (\text{for } 3^2S_{1/2}\text{-}3^2P_{3/2})\ \text{rad/s}\;.
\label{eq:2-29}
\end{equation}
Here $f_i$ and $\omega_i$ are the oscillator strength and line centre frequency for
the $i^{\text{th}}$ transition of the doublet with $m_e$ and $e$ being the mass and charge of
the electron respectively.

Since both the line core and the line wings are Lorentzian in shape,
characterized by their respective FWHM ${}^{c}\gamma_i$ and ${}^{w}\gamma_i$ $(=\frac{2}{3}\,{}^{c}\gamma_i)$, then,
ignoring the contribution to the potentials arising from Van der Waals
interactions with other sodium atoms, we can use an empirical fit to the
complete resonance broadened atomic line profile of the form
\begin{equation}
\pounds_i^{r}(\omega) \;=\; C_i\left\{
\frac{\dfrac{{}^{c}\gamma_i}{2}\,
e^{\left(-4(\omega-\omega_i)^2/{}^{c}\gamma_i^{\,2}\right)}}
{\dfrac{{}^{c}\gamma_i^{\,2}}{4} + (\omega-\omega_i)^2}
\;+\;
\frac{\dfrac{{}^{c}\gamma_i}{3}\,
\left(1-e^{\left(-4(\omega-\omega_i)^2/{}^{c}\gamma_i^{\,2}\right)}\right)}
{\dfrac{{}^{c}\gamma_i^{\,2}}{9} + (\omega-\omega_i)^2}
\right\},
\label{eq:2-30}
\end{equation}
which approaches the exact behaviour in the line core and wings ($\omega = \omega_i$ and
$(\omega-\omega_i)/{}^{c}\gamma_i \gg 1$, respectively) and resembles the numerical calculations in
the intermediate region (Srivastava and Zaidi 1975).  In equation (\ref{eq:2-30}) the
superscript r in $\pounds_i^{r}(\omega)$ refers to `resonance' and is not to be confused with
the internuclear separation.  Also, in the above equations, the subscript i
refers to the $i^{\text{th}}$ member of the doublet and $C_i$ is the normalization constant
(Appendix A),

\begin{equation}
C_i \;=\; 1.126/\pi\;,
\label{eq:2-31}
\end{equation}
which ensures that
\begin{equation}
\int_{-\infty}^{\infty} \pounds_i^{r}(\omega)d\omega \;=\; 1\;.
\label{eq:2-32}
\end{equation}

At some internuclear separation $r$, the contribution to the potential
due to the Van der Waals interaction with a foreign gas as a perturber would
be

\begin{equation}
\overline{h}\,\Delta\omega \;=\; \frac{C_6^{f}}{r^6}\;,
\label{eq:2-33}
\end{equation}
where here, $C_6^{f}/\overline{h}$ is the Van der Waals interaction constant of the foreign
gas estimated by Keilkopf (1974, see also Appendix I) for argon, to be

\begin{equation}
\frac{C_6^{f}}{\overline{h}} \;\approx\; 5\times10^{-30}\ \text{cm}^6\text{/s}\;,
\label{eq:2-34}
\end{equation}
which, for internuclear separations of approximately $1.8\times10^{-7}$ cm, leads
to

\begin{equation}
\omega-\omega_0 \;=\; 1.5\times10^{11}\ \text{rad/s}\;,
\label{eq:2-35}
\end{equation}
and is orders of magnitude less than that resulting from the resonant
dipole-dipole interaction.  At argon pressures comparable to the sodium
vapor pressure (and even in the extreme rim of the vapor disk, where the
argon pressure is as high as ten times the sodium vapor pressure, we could
safely ignore the contribution to the broadening of the atomic absorption
profile due to argon present as a buffer gas.

\subsection{Doppler Broadening}
\label{sec:2-1-2}

The frequency of the photons that are absorbed will be Doppler shifted
as a result of the atoms' thermal motion, by an amount $\omega v/c$, where $v$ now
represents the atoms' thermal speed.  If the distribution of velocity is
Maxwellian, then the Doppler broadened atomic line profile for the $i^{\text{th}}$
member of the multiplet is Gaussian in shape,

\begin{equation}
\pounds_i^{d}(\omega) \;=\; \frac{1}{\beta_i\pi^{1/2}}\exp\left(-(\omega-\omega_i)^2/\beta_i^2\right)\;,
\label{eq:2-36}
\end{equation}

with $\beta_i$, the Gaussian width, given by the relation,

\begin{equation}
\beta_i \;=\; \Bigl(\frac{2kT\omega_i^2}{mc^2}\Bigr)^{1/2}.
\label{eq:2-37}
\end{equation}

In this equation k is the Boltzmann constant and c is the speed of light, T
is the translational temperature and m is the mass of the sodium atom.  The
temperature T is related to the sodium vapor density N ($\approx N_{3^2S}$) through the
Nesmeyanov relationship (1963),
\begin{equation}
\log_{10}N \;=\; 29.84904 - 5619.4106/T - 2.04111\log_{10}T + 3.45\times10^{-6}T.
\label{eq:2-38}
\end{equation}

\section{Spectral Line Broadening in a Sodium Plasma}
\label{sec:2-2}

Most of the emission studies reported in chapter 3 are based on
analysis of the shape of the sodium $4^2D$-$3^2P$ multiplet spectra in the
vicinity of the emission line core where $\Delta\omega \approx \gamma$.  The discussion in this
section will therefore focus on the line broadening of this and other
transitions, in the ``impact'' regime (which for our laboratory conditions, is
generally applicable for the line core - see Griem 1964 for a thorough
discussion of the validity of the impact approximation).  Since this section
deals with the spectroscopy of plasmas, the discussion is extended to
include transitions between upper states which can easily be excited through
electron collisions as a result of the somewhat higher electron temperatures
expected from the superelastic electron heating process discussed in chapter
1.  Doppler broadening has already been treated in the discusion of line
broadening in neutral sodium vapor (section \ref{sec:2-1}) and will not be discussed
in this section since the expression given (equation (\ref{eq:2-36})) in that section
is generally applicable under the conditions of elevated electron
temperatures found in most laboratory plasmas.  Resonance and Van der Waals
broadening of excited state transitions that have a terminating state other
than the ground state in general are negligible in comparison to electron
Stark broadening once the fractional ionization over the electron density
range ($10^{15} - 5\times10^{16}$ cm$^{-3}$) and neutral density range ($10^{15} - 5\times10^{16}$
cm$^{-3}$) of interest, exceeds approximately 0.1\%.  For most purposes, we can
safely ignore their contributions to the linewidth, although for
completeness, we do present the results of and discuss the calculations of
the broadening arising from resonance and Van der Waals interactions in
Appendix I.  The electron impact Stark width ${}^{s}\gamma$, is linearly dependent on
electron density and a measurement of the electron Stark broadening forms
the basis for the electron density measurement technique discussed in
chapter 3.
\subsection{Electron and Ion Stark Broadening}
\label{sec:2-2-1}

The spectral lineshape of transitions between excited states of a
neutral atom immersed in a plasma is often dominated by Stark broadening as
a result of state perturbing collisions between an excited atom and
electrons and/or ions.

When treated using the impact approximation, electron collisions lead
to a Stark broadened profile that is also Lorentzian in shape (Griem 1974),

\begin{equation}
{}^{s}\pounds(\omega) \;=\; \frac{{}^{s}\gamma}{2\pi}\;
\frac{1}{(\omega-\omega_0-{}^{s}d)^2 + \dfrac{{}^{s}\gamma^2}{4}}
\label{eq:2-39}
\end{equation}
where ${}^{s}\gamma$ and ${}^{s}d$ are the electron Stark FWHM and shift respectively (both
linearly dependent on the free electron density $N_e$).  Griem (1964) has
indicated that a combination of the electron impact Stark profile with the
quadratic Stark shift associated with the quasi-static field distribution of
the ions, leads to an emission line profile of the form,

\begin{equation}
j(x,A,R) \;=\; \frac{1}{\pi}\int_0^{\infty}
\frac{H(\beta,R)d\beta}{1 + (x - A^{4/3}\beta^2)}\;,
\label{eq:2-40}
\end{equation}
where the reduced frequency variable

\begin{equation}
x \;=\; 2(\omega-\omega_0-{}^{s}d)/{}^{s}\gamma\;.
\label{eq:2-41}
\end{equation}
The reduced field strength distribution,

\begin{equation}
\beta \;=\; F/F_0\;,
\label{eq:2-42}
\end{equation}
where $F_0$ is the the Holtsmark field strength which can be thought of as the
field strength produced at one mean ion-ion radius defined in accordance
with the relation

\begin{equation}
\frac{4\pi}{3}r_0^{\,3}N_i \;=\; 1\;.
\label{eq:2-43}
\end{equation}
Here $N_i$ is the ion density.  R is the Debye shielding parameter and is given
by the ratio of the mean ion-ion radius to the Debye length, accounting for
shielding by electrons only (Griem 1974),

\begin{equation}
r_D \;=\; \bigl(\epsilon_0 k T_e/e^2 N_e\bigr)^{1/2}\;,
\label{eq:2-44}
\end{equation}
where $N_e$ and $T_e$ are the free electron density and temperature respectively.

The ion broadening parameter A relates the quadratic Stark shift due to
an ion at one mean ion-ion radius to the electron Stark halfwidth (HWHM)
${}^{s}\gamma/2$, that is

\begin{equation}
A \;=\; \bigl(C_q F_0^{\,2}/({}^{s}\gamma/2)\bigr)^{3/4}\;,
\label{eq:2-45}
\end{equation}
where $C_q$ is the quadratic Stark coefficient for the sodium transition of
interest.  Finally, $H(\beta,R)$ is the field strength distribution function of
the ions, taking into account Debye shielding and ion-ion correlations
(Griem 1974).

Griem (1974) has tabulated these profiles for a wide range of A and R
values and has also indicated that the FWHM and shifts corresponding to
these profiles are well represented by

\begin{equation}
{}^{s}\gamma^{*} \;=\; \bigl(1 + 1.75A(1-0.75R)\bigr)\,{}^{s}\gamma\;,
\label{eq:2-46}
\end{equation}

and
\begin{equation}
{}^{s}d^{*} \;=\; {}^{s}d \pm 2A(1-0.75R)\,{}^{s}\gamma\;,
\label{eq:2-47}
\end{equation}
for $A \leqslant 0.5$ and $R \leqslant 0.8$.  The sign in the shift equation is the same as that
of the low temperature limit of ${}^{s}d$ (Griem 1964, 1974).  It is found
convenient to introduce the modified reduced frequency variable,

\begin{equation}
x^{*} \;=\; 2(\omega-\omega_0-{}^{s}d^{*})/{}^{s}\gamma^{*}\;,
\label{eq:2-48}
\end{equation}
so that the peak of the profile appears at approximately $x^{*} = 0$ and the
halfwidth (HWHM) equals unity in terms of the new scale.

We have found for our earlier work (Cappelli and Measures 1984) that it
was possible to approximate the tabulated profiles of Griem (1974) by an
empirical profile of the form
\begin{equation}
j(x^{*}) \;=\; \frac{b_1}{\pi(1+x^{*2})} + \frac{b_2}{\pi^{1/2}}\exp(-x^{*2}/\mathbf{a_0}^{\,2})\;,
\label{eq:2-49}
\end{equation}
where here $b_1$ and $b_2$ are fitting parameters determined by the method of
linear least squares and $\mathbf{a_0}$ as a third parameter chosen to minimize the
error in fitting the tabulated points to the line profile wing (Cappelli
1983).

In the present work, we have assumed that the quadratic Stark shift due
to ions is much less than the electron Stark shift, that is

\begin{equation}
C_q F_0^{\,2} \;\ll\; {}^{s}d\;,
\label{eq:2-50}
\end{equation}
which will allow us to express the profile as a Lorentzian thereby saving
considerable computational effort.  For the electron densities ($10^{15} - 10^{16}$
cm$^{-3}$) and temperatures ($\sim$5000K) of interest, $C_q F_0^{\,2}/{}^{s}d$ varies from 0.1 - 0.24
and in the high electron density range, equation (\ref{eq:2-50}) may not be
satisfied.  As a result, in this region, there may be an overprediction in
the free electron density by as much as 20\%.

Most recent Stark broadening theories are formulated around the
original quantum mechanical treatment of pressure or collision broadening of
Baranger (1958a) and Kolb and Griem (1958).  At some point in the derivation
of the expression for the shift and width, the electron is assumed to act as
a point charge travelling in a straight path (classical path assumption) for
collisions with neutral emitters.  This is the case for the semi-classical
theories of Griem (1974) and Sahal-Br\'echot (1969) and is justified providing
that the electron energy is much greater than the separation between the
initial or final states (i,f) and their nearest perturbing level (j,j'),

\begin{equation}
kT_e \;\gg\; E_{ij},\ E_{fj'}\;.
\label{eq:2-51}
\end{equation}

The concept of perturbing levels is perhaps most easily exemplified in
the expression for the fullwidth of an ``isolated line'' in the treatment of
Sahal-Br\'echot (1969) which, unlike that of Griem (1974), separates the
elastic and inelastic terms,

\begin{equation}
{}^{s}\gamma \;=\; N_e \int_v v f(v)dv
\Bigl\{\sum_{j\neq i}\sigma_{ij}(v) + \sum_{j'\neq f}\sigma_{fj'}(v) + \sigma_{e\ell}(v)\Bigr\},
\label{eq:2-52}
\end{equation}
here, $\sigma_{ij}$ and $\sigma_{fj'}$ are the inelastic electron collisional excitation
cross-sections for the initial and final states involved in the transition of
interest, $\sigma_{e\ell}$ is the elastic scattering cross-section.  The cross-sections
are averaged over a velocity distribution $f(v)$ which is assumed to be
Maxwellian.  The summation extends over the nearest states (perturbing
levels) that are connected through optically allowed dipole transitions.
Griem (1974) extends the summation over the nearest five perturbing levels
for the upper (initial) state of the radiating atom and the nearest three
perturbing levels for the lower (terminating) state.  A line is considered
isolated if

\begin{equation}
{}^{s}\gamma \;\ll\; E_{ij}/\overline{h},\ E_{fj'}/\overline{h}\;,
\label{eq:2-53}
\end{equation}
that is, there is little or no interference from perturbing levels.  The two
criteria specified by equations (\ref{eq:2-51}) and (\ref{eq:2-53}) (as well as satisfying the
criteria for use of the impact approximation for electrons and quasi-static
approximation for ions, most often such is the case -- see Griem 1974) must be
met before one can justifiably use the tabulated shifts and widths along
with the expressions given as equations (\ref{eq:2-39}) and (\ref{eq:2-52}).

Extensive calculations of ${}^{s}d$ and ${}^{s}\gamma$ have been performed for many
transitions in neutral sodium (Dimitrijevi\'c and Sahal-Br\'echot 1985, Griem
1974).  Table \ref{tab:2-1} compares the results of these calculations for six
transitions of interest.  Also listed in the table are the critical electron
densities $N_e^{*}$ above which the inequality in equation (\ref{eq:2-53}) breaks down and
the line can no longer be considered isolated.  Clearly, for most of the
transitions, the results of Griem (1974) exceed those of Dimitrijevi\'c and
Sahal-Br\'echot (1985) over the 5000K - 10000K temperature range predicted by
the LIBORS theory.  As will be shown in chapter 3, the free electron density
is primarily derived from the electron Stark broadened line profile of the
$4^2D - 3^2P$ transition.  Although the results quoted in chapter 5 use the
tabulated values of Griem (1974), use of the widths and shifts tabulated by
Dimitrijevi\'c and Sahal-Br\'echot (1985) would systematically increase our
measured electron densities by roughly 20\%.

\tabstub{tab:2-1}{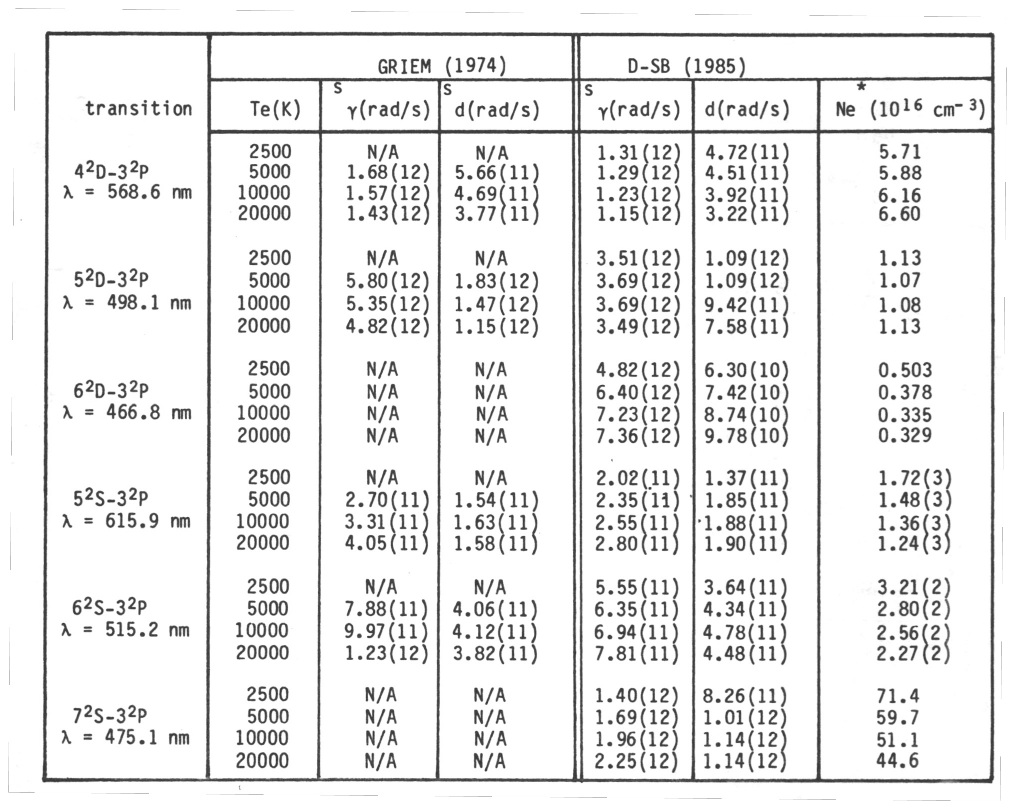}{Comparison of the calculated Stark widths and shifts for
six transitions of interest, together with the critical electron densities
$N_e^{*}$.}

In our modelling of the plasma emission (section \ref{sec:2-5}), it was found
convenient to approximate the electron temperature dependence of the Stark
shift and widths of Griem (1974) by

\begin{equation}
{}^{s}\gamma \;=\; \left\{\frac{{}^{s}\tilde{\gamma}\,N_e}{10^{16}}\right\}
\left(\frac{5000}{T_e}\right)^{\alpha}\;,
\label{eq:2-54a}
\tag{2.54a}
\end{equation}

and

\begin{equation}
{}^{s}d \;=\; \left\{\frac{{}^{s}\tilde{d}\,N_e}{10^{16}}\right\}
\left(\frac{5000}{T_e}\right)^{\beta}\;,
\label{eq:2-54b}
\tag{2.54b}
\end{equation}

which fit the tabulated values generally to within 5\% over the 5000K-10000K
range.  ${}^{s}\tilde{\gamma}$ and ${}^{s}\tilde{d}$ are the widths and shifts for an electron density of
$10^{16}$ cm$^{-3}$ and electron temperature of 5000K.  Table (\ref{tab:2-2}) lists the fitting
parameters $\alpha$ and $\beta$ (arrived at by fitting equations (2.54) exactly at $T_e =$
5000K and 20000K) for the six sodium transitions of interest.

\tabstub{tab:2-2}{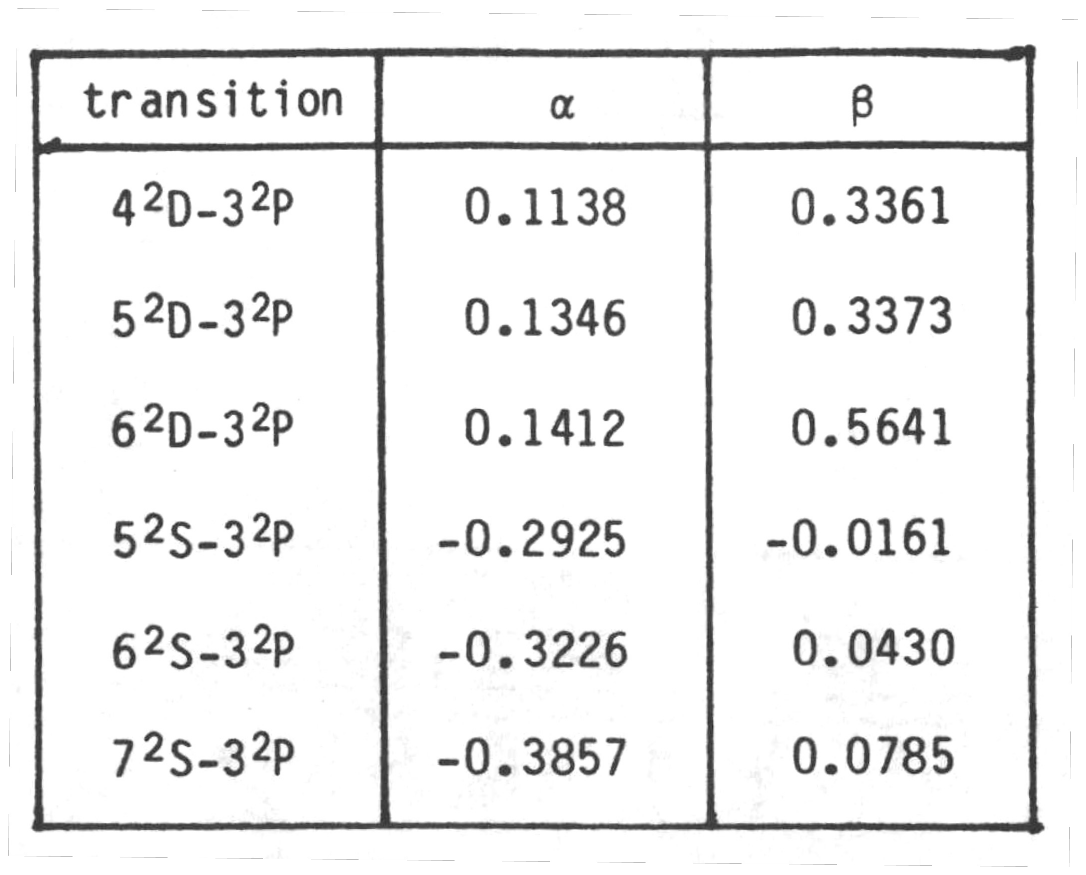}{Fitting parameters $\alpha$ and $\beta$ for the six sodium
transitions of interest.}

\section{AC Stark Effect and Virtual State Transitions}
\label{sec:2-3}

Optical pumping of the $3^2S$-$3^2P$ resonance transition can cause the
atomic energy levels involved to be split as a result of the direct
perturbation of the laser field on the atomic levels.  This splitting is a
result of the well known AC Stark effect (Cohen-Tanoudji 1974) and for a two
level system, the splitting of the upper (u) and lower ($\ell$) states are
asymmetric, with the separation between the split states represented by the
generalized Rabi frequency (see figure \ref{fig:2-2}),

\begin{equation}
\Omega' \;=\; \sqrt{(\Omega^2 + \Delta\omega_L^{\,2})}\;,
\label{eq:2-55}
\end{equation}

\figstub{fig:2-2}{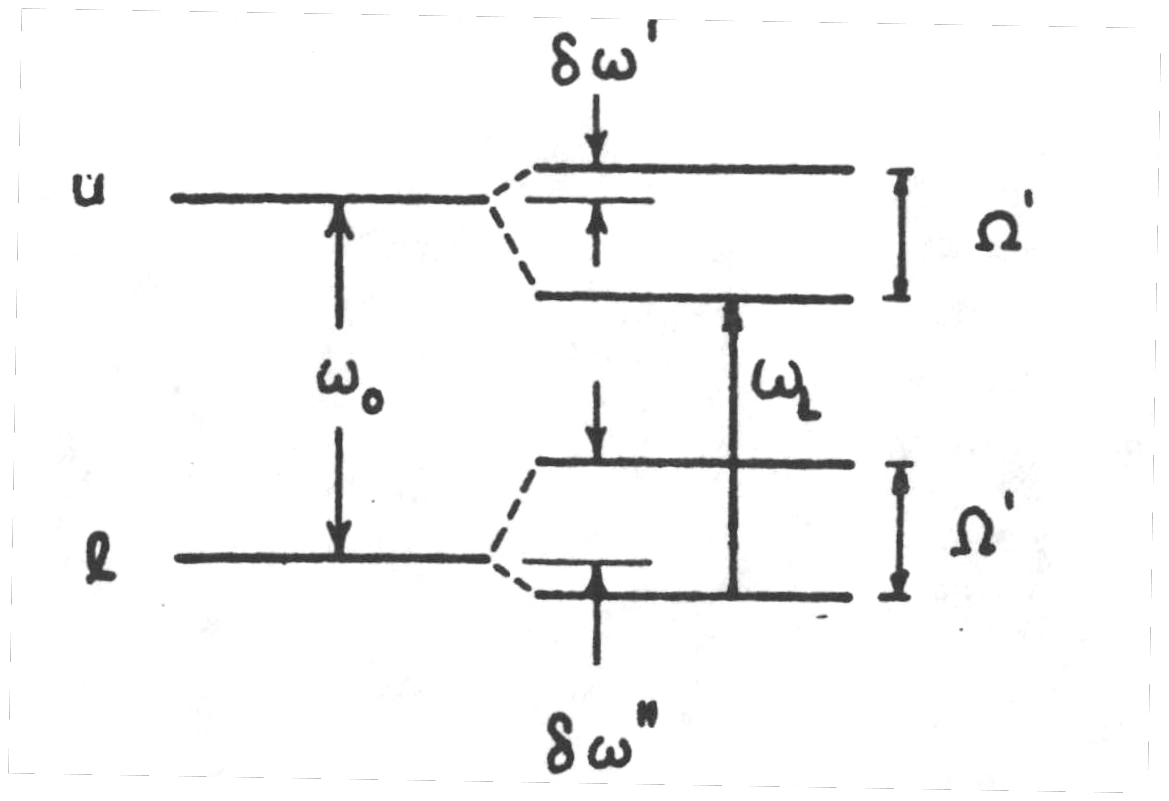}{Splitting of the upper (u) and lower ($\ell$) states of a two
level system by the AC Stark effect, showing the generalized Rabi frequency.}

with $\Delta\omega_L = \omega_0-\omega_L$ and $\Omega$, the Rabi frequency given by

\begin{equation}
\Omega \;=\; \frac{\mu E_0}{\overline{h}}\;.
\label{eq:2-56}
\end{equation}

Here, $\mu$ is the electric dipole matrix element for the transition, $E_0$ is the
amplitude of the electric field disturbance and $\omega_L$ and $\omega_0$ are the laser
frequency and frequency separation between the unperturbed states
respectively.  The shifting of the states from the unperturbed levels are
expressed in terms of the generalized Rabi frequency (Bjorkholm and Liao,
1974) and for $\omega_0 > \omega_L$,

\begin{align}
\delta\omega'_u &= -(\Omega'-\Delta\omega_L)/2\;, &
\delta\omega'_{\ell} &= (\Omega'-\Delta\omega_L)/2\;,
\tag{2.57a}\label{eq:2-57a}\\[0.4em]
\delta\omega''_u &= (\Omega'+\Delta\omega_L)/2\;, &
\delta\omega''_{\ell} &= -(\Omega'+\Delta\omega_L)/2\;.
\tag{2.57b}\label{eq:2-57b}
\end{align}

In the limit, with $\Omega/\Delta\omega_L \gg 1$, equations (2.57) become,

\begin{align}
\delta\omega'_u &= -\Omega/2\;, & \delta\omega'_{\ell} &= \Omega/2\;,
\tag{2.58a}\label{eq:2-58a}\\[0.4em]
\delta\omega''_u &= \Omega/2\;, & \delta\omega''_{\ell} &= -\Omega/2\;,
\tag{2.58b}\label{eq:2-58b}
\end{align}

and the splitting of the states is linear with the electric field and
independent of the laser detuning $\Delta\omega_L$.  In the opposite limit, with $\Omega/\Delta\omega_L$
$\ll 1$, we can write,

\begin{align}
\delta\omega'_u &= -\Omega^2/4\Delta\omega_L\;, & \delta\omega'_{\ell} &= \Omega^2/4\Delta\omega_L\;,
\tag{2.59a}\label{eq:2-59a}\\[0.4em]
\delta\omega''_u &= \Delta\omega_L\;, & \delta\omega''_{\ell} &= -\Delta\omega_L\;.
\tag{2.59b}\label{eq:2-59b}
\end{align}

Equation (\ref{eq:2-59b}) indicates that there will be a component that is linear
with the detuning ($\delta\omega''_{u,\ell}$), which could have been deduced from an energy
conservation argument and the creation of a virtual state (Bjorkholm and
Liao 1974).  During the period of laser pumping, one of these virtual states
can act as a terminal state for transitions from an excited upper state u*
(see figure \ref{fig:2-3}).  For the case where $\omega_L < \omega_0$ (the laser wavelength is to the
red of the unperturbed resonance line wavelength), then one would expect to
see a component of emission that is blue shifted relative to the unperturbed
emission line centre wavelength.  The second component (arising from $\delta\omega'_u$),
would be virtually indistinguishable from the spectrally broadened
unperturbed emission at $\omega_0^{*}$, for $\Omega \ll \gamma$, $\gamma$ being the collisionally broadened
FWHM of the atomic emission profile.

\figstub{fig:2-3}{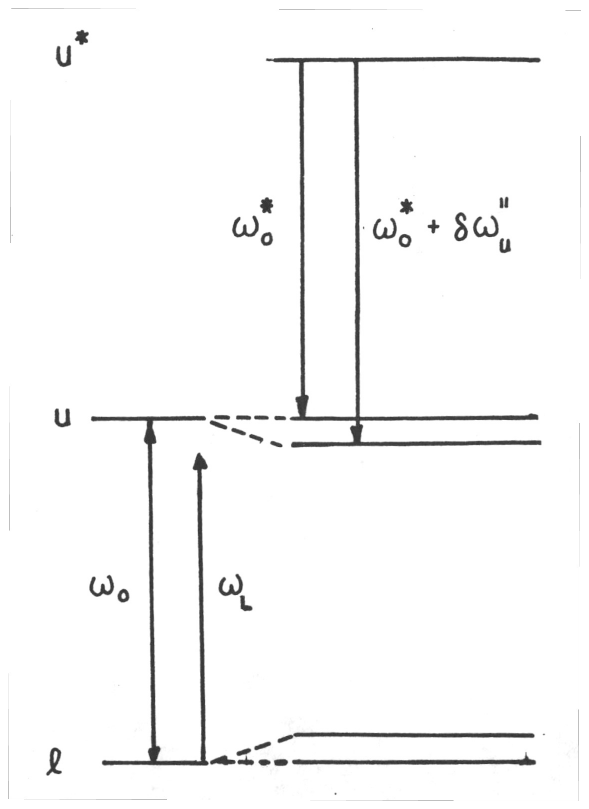}{Virtual state acting as a terminal state for transitions
from an excited upper state u* during the period of laser pumping.}

This effect has been observed in the $4^2D$-$3^2P$ emission spectrum in
sodium, for times less than the laser pulse duration (chapter 5).  Although
it is a relatively small effect at large laser detunings, it is difficult to
ascertain the importance of this AC Stark effect for $\Delta\omega_L \sim 0$.  To avoid any
distortion in the spectral emission resulting from the creation of virtual
states, we limited our experimental measurements to times greater than the
duration of the laser pulse.  As we shall see however in section \ref{sec:2-5},
radiation trapping of emission appears to have a much greater effect on the
spectral emission than the AC Stark effect, and is the major factor in
determining the time during which a measurement can be made.

\section{Atomic Line Profile}
\label{sec:2-4}

Providing that the broadening mechanisms are a result of statistically
independent events (Sobelman et al.\ 1981), the atomic line profile $\pounds(\Delta\omega)$
(for convenience, it is usually written as $\pounds(\omega)$ but $\pounds(\Delta\omega)$ is implied)
resulting from the combined effects which individually contribute a line
profile ${}^{i}\pounds(\Delta\omega)$, can be written as a convolution or ``folding'' of the
individual line profiles (Griem 1964),

\begin{equation}
\pounds(\Delta\omega) \;=\; {}^{1}\pounds(\Delta\omega) * {}^{2}\pounds(\Delta\omega)\;\ldots\ldots*\;{}^{i}\pounds(\Delta\omega)\;.
\label{eq:2-60}
\end{equation}

Here $*$ denotes the convolution operation,

\begin{equation}
{}^{1}\pounds(\Delta\omega) * {}^{2}\pounds(\Delta\omega) \;=\;
\int_{-\infty}^{\infty} {}^{1}\pounds(\Delta\omega-\Delta\omega')\,{}^{2}\pounds(\Delta\omega')d\Delta\omega'\;,
\label{eq:2-61}
\end{equation}

with $\Delta\omega = \omega-\omega_0$ and $\Delta\omega' = \omega'-\omega_0$, the detunings from line centre.

\subsection{Neutral Sodium Vapor}
\label{sec:2-4-1}

In a cold sodium vapor, ($<$1000K), both Doppler and Resonance broadening
have to be considered as the broadening mechanisms for the resonance $3^2S$ -
$3^2P$ absorption transition.  A simple ``back of the envelope'' calculation
using equations (\ref{eq:2-28}), (\ref{eq:2-37}) and (\ref{eq:2-38}) for T = 1000K would give

\begin{equation}
{}^{c}\gamma_1 \;=\; 7.7\times10^{11}\ \text{rad/s}\;,
\tag{2.62a}\label{eq:2-62a}
\end{equation}

and

\begin{equation}
\beta \;=\; 1.8\times10^{11}\ \text{rad/s}\;.
\tag{2.62b}\label{eq:2-62b}
\end{equation}

Under these circumstances, the resultant atomic line profile function takes
on the convolution of the corresponding profiles given by equations (\ref{eq:2-30})
and (\ref{eq:2-36}), that is

\begin{equation}
\pounds_i(\omega) \;=\; \frac{1}{\beta_i\pi^{1/2}}\int_{-\infty}^{\infty}
e^{-(\Delta\omega_i-\Delta\omega_i^{*})^2/\beta_i^2}\,\pounds_i^{r}(\Delta\omega_i^{*})\;d\Delta\omega_i^{*}\;.
\label{eq:2-63}
\end{equation}

In the above equations, it is important to remember that the subscripts i
refer to the $i^{\text{th}}$ member of the sodium $3^2S$ - $3^2P$ transition doublet and $\Delta\omega_i$
refers to the frequency detuning from the line centre of that member.

Upon substitution of equation (\ref{eq:2-30}) into equation (\ref{eq:2-63}), one obtains
three integrals (see Appendix A) which all involve some form of the Voigt
integral,

\begin{equation}
W(u,\eta) \;=\; \frac{1}{\eta\pi^{3/2}}\int_{-\infty}^{\infty}
\frac{e^{-y^2/\eta^2}}{1+(u-y)^2}\,dy\;.
\label{eq:2-64}
\end{equation}

We have elected to use an expression for the Voigt integral based on the
approximation of Whiting (1969),

\begin{multline}
W(u,\eta) \;=\; \left[\Bigl(1-\frac{2}{w_v}\Bigr)\exp\Bigl(-2.772\,\frac{u^2}{w_v^{\,2}}\Bigr)
+ \frac{2}{w_v}\;\frac{1}{1+\dfrac{4u^2}{w_v^{\,2}}}\right]\\[0.6em]
\times\;\left\{\frac{w_v\sqrt{\pi}}{2}
\left[\Bigl(1-\frac{2}{w_v}\Bigr)\Big/\sqrt{\ln 2} + \frac{2\sqrt{\pi}}{w_v}\right]\right\}^{-1},
\label{eq:2-65}
\end{multline}

with

\begin{equation}
w_v \;=\; 1 + (1 + 4\eta^2\ln 2)^{1/2}\;,
\label{eq:2-66}
\end{equation}

and we have normalized the function $W(u,\eta)$ such that

\begin{equation}
\int_{-\infty}^{\infty} W(u,\eta)du \;=\; 1\;.
\label{eq:2-67}
\end{equation}

\subsection{Sodium Plasma}
\label{sec:2-4-2}

Treating the resonance and Van der Waals broadened emission profiles in
the core as Lorentzian in shape (Appendix I), then providing we can ignore
the contribution to the Stark broadening resulting from ions, the Stark
broadening due to the electrons is also described by a Lorentzian and the
combined convoluted profile takes on a Lorentzian shape,

\begin{equation}
\pounds(\omega) \;=\; \frac{\gamma}{2\pi}\;
\frac{1}{(\omega-\omega_0-d)^2 + \dfrac{\gamma^2}{4}}
\label{eq:2-68}
\end{equation}

where now,

\begin{equation}
d \;=\; {}^{s}d + {}^{v}d\;,
\tag{2.69a}\label{eq:2-69a}
\end{equation}

and

\begin{equation}
\gamma \;=\; {}^{s}\gamma + {}^{r}\gamma + {}^{v}\gamma\;,
\tag{2.69b}\label{eq:2-69b}
\end{equation}

represent the total shift and width (FWHM) of the line profile.  Here ${}^{r}\gamma$, ${}^{v}\gamma$
and ${}^{v}d$ represent the contribution to the width and shift arising from
resonance and Van der Waals broadening.  The Doppler broadening can be
incorporated by convolving equation (\ref{eq:2-68}) with equation (\ref{eq:2-36}).  The
resulting line profile is also represented by the Voigt function and we can
use the same approximation given by equation (\ref{eq:2-65}) (see also Appendix A).

The emission lines of interest for our work, namely the $n^2S$ - $3^2P$ (n =
5-7) and the $n^2D$ - $3^2P$ (n = 4-6), are actually multiplet spectra:

\begin{align}
& n^2S_{1/2}\text{-}\,3^2P_{1/2,\,3/2}\;, \tag{2.70a}\label{eq:2-70a}\\
& n^2D_{3/2}\text{-}\,3^2P_{1/2,\,3/2}\;, \tag{2.70b}\label{eq:2-70b}\\
& n^2D_{5/2}\text{-}\,3^2P_{3/2}\;, \tag{2.70c}\label{eq:2-70c}
\end{align}

and the atomic line profile function describing the multiplet can be
expressed as

\begin{equation}
\pounds(\omega) \;=\; \sum_{JJ'}\xi(J,J')\,\pounds(\omega-\omega_{JJ'})\;,
\label{eq:2-71}
\end{equation}

where $\omega_{JJ'}$ represents the line centre frequency of the $J\rightarrow J'$ transition (J
and J' are the upper and lower total angular momentum quantum numbers
respectively), $\xi(J,J')$ represents the relative strength of this line within
the multiplet (Allen 1963) and the sum over J and J' takes into account the
allowed transitions that constitute the multiplet.

\section{One Dimensional Radiative Transfer}
\label{sec:2-5}

The plasma which has been investigated throughout the course of this
research, has been produced by the passage of a collimated laser beam
through sodium vapor.  Providing the laser is axially symmetric and its
energy and spatial profile do not vary significantly along its path over the
region of observation, then we can also assume cylindrical symmetry.

The steady state solution to the one-dimensional radiative transfer
equation enables the spectral radiance $J_{nm}(\nu,x,y)$ arising from the n to m
transition emitted in the x direction from a cylindrically symmetric plasma
column of radius R at some height y from the column axis (see figure \ref{fig:2-4}) to
be written in the form

\begin{equation}
J_{nm}(\nu,y) \;=\; \int_{-\sqrt{(R^2-y^2)}}^{\sqrt{(R^2-y^2)}}
\varepsilon_{nm}(\nu,x,y)\exp\left\{-\int_{x}^{\sqrt{(R^2-y^2)}}
\frac{\varepsilon_{nm}(\nu,x^{*},y)\,dx^{*}}{P(\nu_{nm},x^{*},y)}\right\}dx\;,
\label{eq:2-72}
\end{equation}

\figstub{fig:2-4}{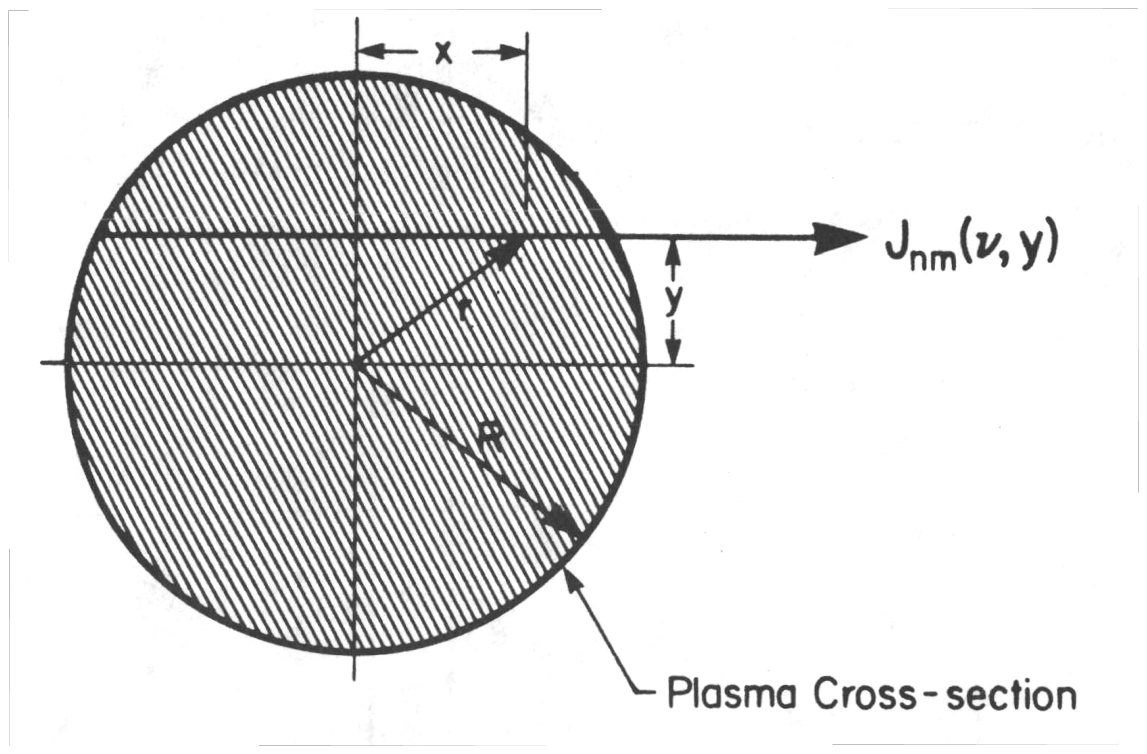}{Geometry for the emission of spectral radiance $J_{nm}(\nu,x,y)$
in the x direction from a cylindrically symmetric plasma column of radius R
at height y from the column axis.}

where

\begin{equation}
\varepsilon_{nm}(\nu,x,y) \;=\; \frac{h\nu_{nm}}{4\pi}N_n(x,y)A_{nm}\pounds_{nm}(\nu,x,y)\;,
\label{eq:2-73}
\end{equation}

represents the volume emission coefficient, $N_n(x,y)$ the upper state
population density at position (x,y), $A_{nm}$ the Einstein spontaneous emission
coefficient and $\pounds_{nm}(\nu,x,y)$ the atomic line profile for the n to m
transition.  For convenience, we have now expressed the frequency dependence
in terms of $\nu(\text{s}^{-1}) = \omega(\text{rad/s})/2\pi$.

In general, the source function

\begin{equation}
P(\nu_{nm},x,y) \;=\; \frac{\varepsilon_{nm}(\nu,x,y)}{\kappa_{nm}(\nu,x,y)}\;,
\label{eq:2-74}
\end{equation}

where

\begin{equation}
\kappa_{nm}(\nu,x,y) \;=\; \frac{h\nu_{nm}^{3}}{4\pi}
\bigl(N_m(x,y)B_{mn} - N_n(x,y)B_{nm}\bigr)\pounds_{nm}(\nu,x,y)\;,
\label{eq:2-75}
\end{equation}

represents the volume absorption coefficient, $N_m$ the lower state population
density and $B_{mn}$ and $B_{nm}$ the Milne stimulated absorption and emission
coefficients respectively.  The source function is independent of frequency
as we are assuming complete frequency redistribution.

If the plasma is in local thermodynamic equilibrium (LTE), then the
source function

\begin{equation}
P(\nu_{nm},x,y) \;=\; \frac{2h\nu_{nm}^{3}}{c^2}
\left[\exp\left\{\frac{h\nu_{nm}}{kT_e(x,y)}\right\}-1\right]^{-1},
\label{eq:2-76}
\end{equation}

which is the Planck function, $h\nu_{nm}$ being the n to m transition photon
energy, $T_e(x,y)$ the free electron temperature at position (x,y) and h,
Planck's constant.

The earlier sections in this chapter have been devoted to determining
the functional form of the atomic line profile $\pounds_{nm}(\nu)$, which of course,
depends on the various broadening mechanisms.  In order to test the accuracy
of the computational routines, we have found it useful to simulate the
plasma emission by direct numerical integration of equation (\ref{eq:2-72}).  This
would also assist us in comparison of the theoretical prediction with
experiment as well as to understand the effects arising from multi-shot
averaging (section \ref{sec:3-4}).  In any event, it is necessary to formulate a model
describing the local state of the plasma so that the shape of the atomic
line profile, electron temperature, electron density and excited state
population densities can be evaluated.

\subsection{LTE Model}
\label{sec:2-5-1}

We have applied the above computation to the $n^2D$ - $3^2P$ (n = 4,5,6) and
$n^2S$ - $3^2P$ (n = 5,6,7) spectral series of a cylindrically symmetric sodium
plasma, with a preassigned radial variation in either the free electron
density $N_e(r)$ or electron temperature $T_e(r)$.

We begin with the assumption of continuity,

\begin{equation}
N_e + \sum_{n=1}^{n_{\max}} N_n \;=\; N\;,
\label{eq:2-77}
\end{equation}

where we have omitted the explicit dependence of the population densities on
position (x,y) and the summation includes only those levels below the
reduced ionization limit (Griem 1964).  If we further assume that LTE
applies down to the ground state, then

\begin{equation}
N_n \;=\; \frac{g_n}{g_1}N_1\exp\left\{-\frac{E_{n1}}{kT_e}\right\},
\label{eq:2-78}
\end{equation}

where $E_{n1}$ represents the energy separation between the excited and ground
states and $g_n$ the degeneracy of state n.  Combining equations (\ref{eq:2-77}) and
(\ref{eq:2-78}), we can write

\begin{equation}
\frac{N_1}{g_1} \;=\; \frac{\{N-N_e\}}{Z(T_e)}\;,
\label{eq:2-79}
\end{equation}

where $Z(T_e)$ represents the partition function,

\begin{equation}
Z(T_e) \;=\; \sum_{n=1}^{n_{\max}} g_n\exp\left\{-\frac{E_{n1}}{kT_e}\right\}.
\label{eq:2-80}
\end{equation}

For purpose of this investigation, we have used the tabulated data taken
from Drawin and Felenbok (1965) assuming an ionization energy depression of
$\Delta E_{c1} = 0.1$ eV.  This ionization energy reduction is a result of the
interaction between the bound electron and an ion or free electron at one
Debye radius.  For a plasma of maximum electron density $N_e = 2\times10^{16}$ cm$^{-3}$
and minimum electron temperature $T_e = 2500$K, the ionization reduction is at
most 0.1 eV.  Equations (\ref{eq:2-79}) and (\ref{eq:2-80}) together with one form of the Saha
relationship,

\begin{equation}
\frac{N_e^{\,2}}{N_1} \;=\; \frac{2\,g_1^{+}}{g_1}
\left\{\frac{m_e kT_e}{2\pi\overline{h}^{2}}\right\}^{3/2}
\exp\left\{-\frac{E_{c1}-\Delta E_{c1}}{kT_e}\right\},
\label{eq:2-81}
\end{equation}

represent three equations in four unknowns, given a known neutral sodium
atom density N.  In equation (\ref{eq:2-81}), $g_1^{+}$ is the degeneracy of the ground
state of the ion and we have assumed that the plasma is singly ionized.
Combining equations (\ref{eq:2-79}), (\ref{eq:2-80}) and (\ref{eq:2-81}), we arrive at

\begin{equation}
\frac{N_e^{\,2}}{N-N_e} \;=\; \frac{2\,g_1^{+}}{Z(T_e)}
\left\{\frac{m_e kT_e}{2\pi\overline{h}^{2}}\right\}^{3/2}
\exp\left\{-\frac{E_{c1}-\Delta E_{c1}}{kT_e}\right\},
\label{eq:2-82}
\end{equation}

which is a transcendental equation for the electron temperature and a
quadratic equation for the electron density.  For a given electron
temperature radial distribution and neutral atom density, the electron
density can be solved for directly (or conversely, given $N_e(r)$ and N, $T_e(r)$
can be solved for iteratively) allowing for the calculation of $N_1(r)$ and
finally, the radial variation of the upper state population density $N_n(r)$.
All plasma parameters are now available for us to evaluate the width of the
atomic line profile for the transition of interest, and numerically
integrate equation (\ref{eq:2-72}).  The variation of the plasma along the line of
sight at height y from the axis can be obtained through $r^2 = x^2+y^2$,
allowing us to calculate $\varepsilon_{nm}(\nu,x,y)$ and $P(x,y)$.  The plasma along the line
of sight is divided into 200 elemental slabs of equal thickness.  A set of
rational splines (Kissack, 1984 - see also Appendix B, subroutine RASPCOEF)
is used to describe the smooth variation of $\varepsilon_{nm}$ and P between elements for
each frequency point, and integration over x is performed by the subroutine
RASPEVAL which interpolates between nodes of the rational spline from the
spline coeficients determined in RASPCOEF.

Figure \ref{fig:2-5}a shows the simulated $4^2D$ - $3^2P$ spectrum that one would
expect to observe along x through y = 0 mm, for a plasma of constant radial
temperature $T_e$ = 5000K, N = $2\times10^{16}$ cm$^{-3}$ and a column radius R = 0.5 cm.
Also illustrated in the figure is the spectrum that one would observe under
ideal ``optically thin'' conditions, that is, with

\begin{equation}
\int_{x}^{\sqrt{(R^2-y^2)}}\frac{\varepsilon_{nm}(\nu,x,y)}{P(\nu_{nm},x,y)}\,dx \;\ll\; 1\;.
\label{eq:2-83}
\end{equation}

\figstub{fig:2-5}{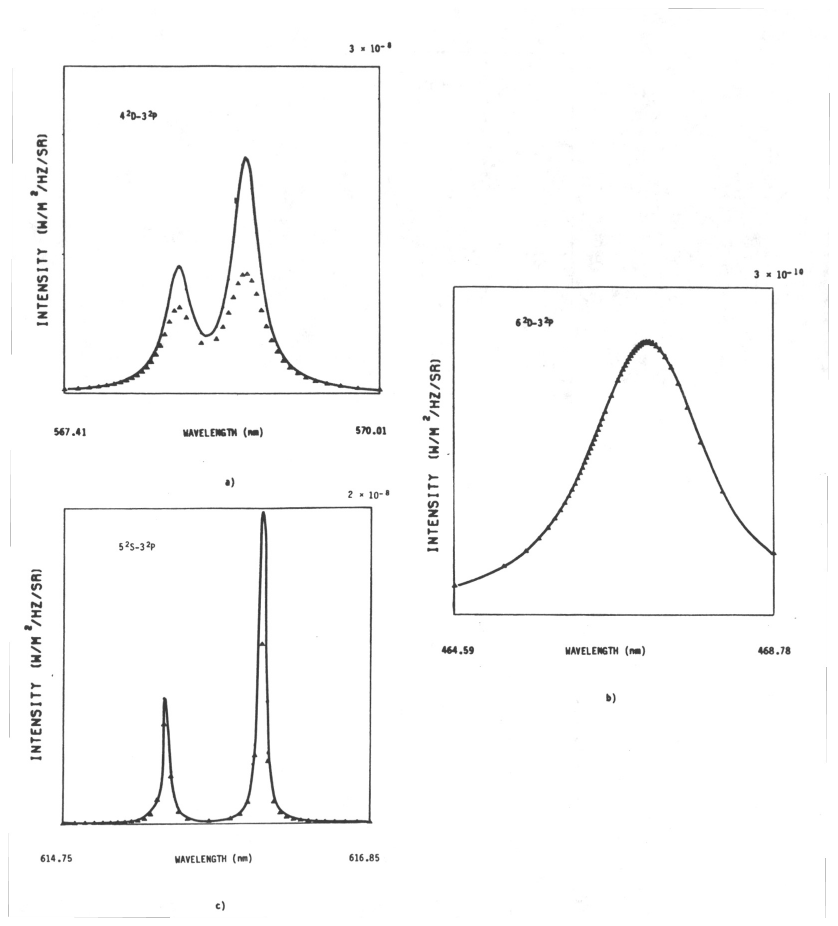}{Simulated spectra for a plasma of constant radial
temperature $T_e$ = 5000K, N = $2\times10^{16}$ cm$^{-3}$ and column radius R = 0.5 cm:
a) $4^2D$ - $3^2P$; b) $6^2D$ - $3^2P$; c) $5^2S$ - $3^2P$.}

This comparison clearly illustrates the effect of self-absorption of
radiation under the selected plasma conditions, and allows one to estimate
the effects of self-absorption on the particular measurements being made
(chapter 3).  For comparison, the spectrum of the $6^2D$ - $3^2P$ and $5^2S$ - $3^2P$
multiplet transitions have been computed for the same plasma conditions and
are illustrated in figures \ref{fig:2-5}b and \ref{fig:2-5}c respectively.  Clearly, the effects
of self-absorption for these transitions are less pronounced, as one might
expect, seeing that the absorption oscillator strengths for these
transitions are less than that for the $4^2D$ - $3^2P$ transition.  The lower
oscillator strength coupled with the lower population density of the upper
state (in the Boltzmann ratio with $N_{4D} = 3.32\times10^{12}$ cm$^{-3}$) results in an
overall decrease in the radiance defined as

\begin{equation}
J_{nm}(y) \;=\; \int_{-\infty}^{\infty} J_{nm}(\Delta\nu,y)\,d\Delta\nu\;.
\label{eq:2-84}
\end{equation}

As well, the emission spectra suffer less broadening for the $5^2S$ - $3^2P$
transition and greater broadening for the $6^2D$ - $3^2P$ transition, primarily a
result of the difference in the electron Stark widths (Table \ref{tab:2-1}) at the
Saha equilibrium electron density of $8.83\times10^{15}$ cm$^{-3}$.

At electron temperatures of 10000K, the degree of ionization is
significantly higher ($N_e = 1.99\times10^{16}$ cm$^{-3}$) and the excited state
population densities are less than that at 5000K.  The result is an overall
decrease in the radiance and an increase in the broadening of the computed
spectra, as illustrated for the $4^2D$-$3^2P$ transition in figure \ref{fig:2-6}.
Interesting to note is the decrease in the self-absorption as a result of
the decrease in the $3^2P$ state population density ($1.67\times10^{13}$ cm$^{-3}$ as
opposed to $3.14\times10^{14}$ cm$^{-3}$).

\figstub{fig:2-6}{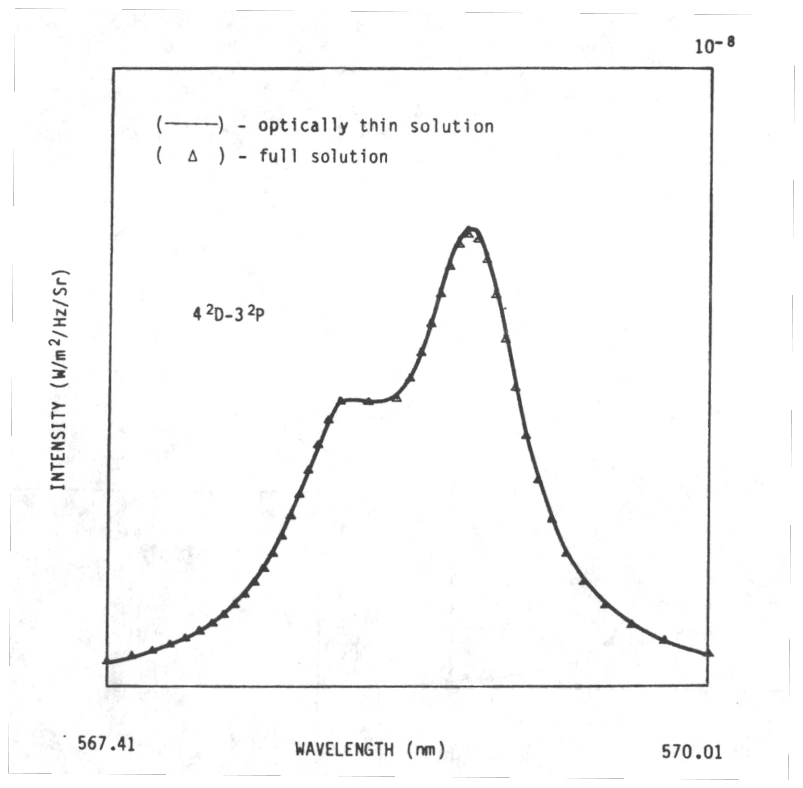}{Computed $4^2D$-$3^2P$ spectrum at an electron temperature of
10000K.}

Perhaps a more interesting case is one in which the electron
temperature has a radial variation of the form

\begin{equation}
T_e(r) \;=\; 2000 + 8000\exp(-r^2/r_0^{\,2})\quad\text{K}\;,
\label{eq:2-85}
\end{equation}

with $r_0$ = 2.5 mm.  For a core (r = 0) temperature of 10000K and N = $2\times10^{16}$
cm$^{-3}$, one would then expect the neutral species to be highly ``burned out''
in the core as a result of the high degree of ionization.  Figure \ref{fig:2-7}
depicts the $4^2D$ - $3^2P$ state population density versus radius and clearly
illustrates this point, suggesting that situations can arise where maximum
radiances can be observed at y values other than y = 0 mm.  The emission
spectrum of the $4^2D$ - $3^2P$ transition at y = 0 mm (figure \ref{fig:2-8}a) and at y =
2.5 mm (figure \ref{fig:2-8}b) show significant differences in both overall radiance
and broadening.  Similar results are obtained for the $5^2S$ - $3^2P$ and $6^2D$ -
$3^2P$ transitions (Figures \ref{fig:2-9} and \ref{fig:2-10}).  The narrower, more resolved spectra
at y = 2.5 mm reflects the lower electron density along the line of sight.
This clearly suggests that measurements of this nature (i.e., spectral
variance for various y values) can be used for the evaluation of the spatial
distribution of the free electron density, and as a result, the y variation
of the spectral radiance of the $4^2D$ - $3^2P$ transition forms the basis for the
radial free electron density measurements discussed in sections \ref{sec:3-2} and
\ref{sec:5-1}.

\figstub{fig:2-7}{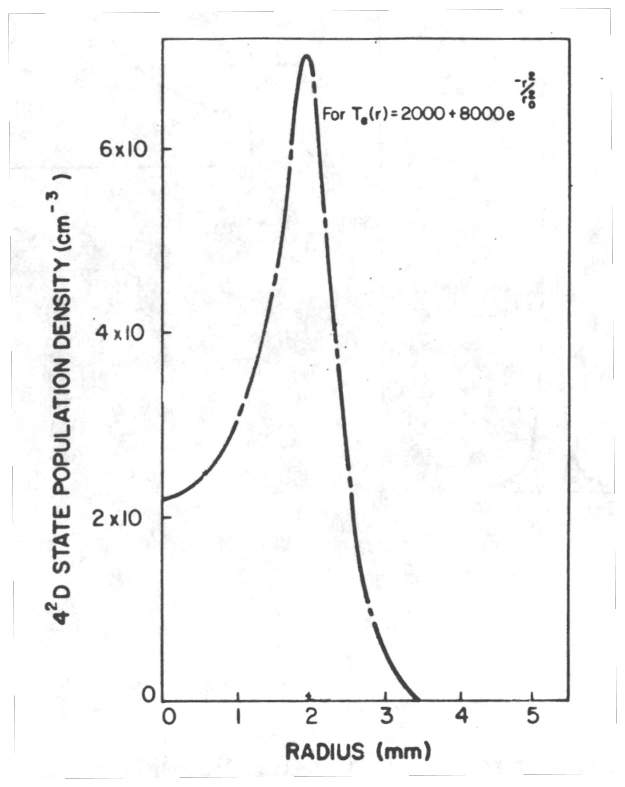}{$4^2D$ - $3^2P$ state population density versus radius for the
radially varying electron temperature of equation (\ref{eq:2-85}).}

\figstub{fig:2-8}{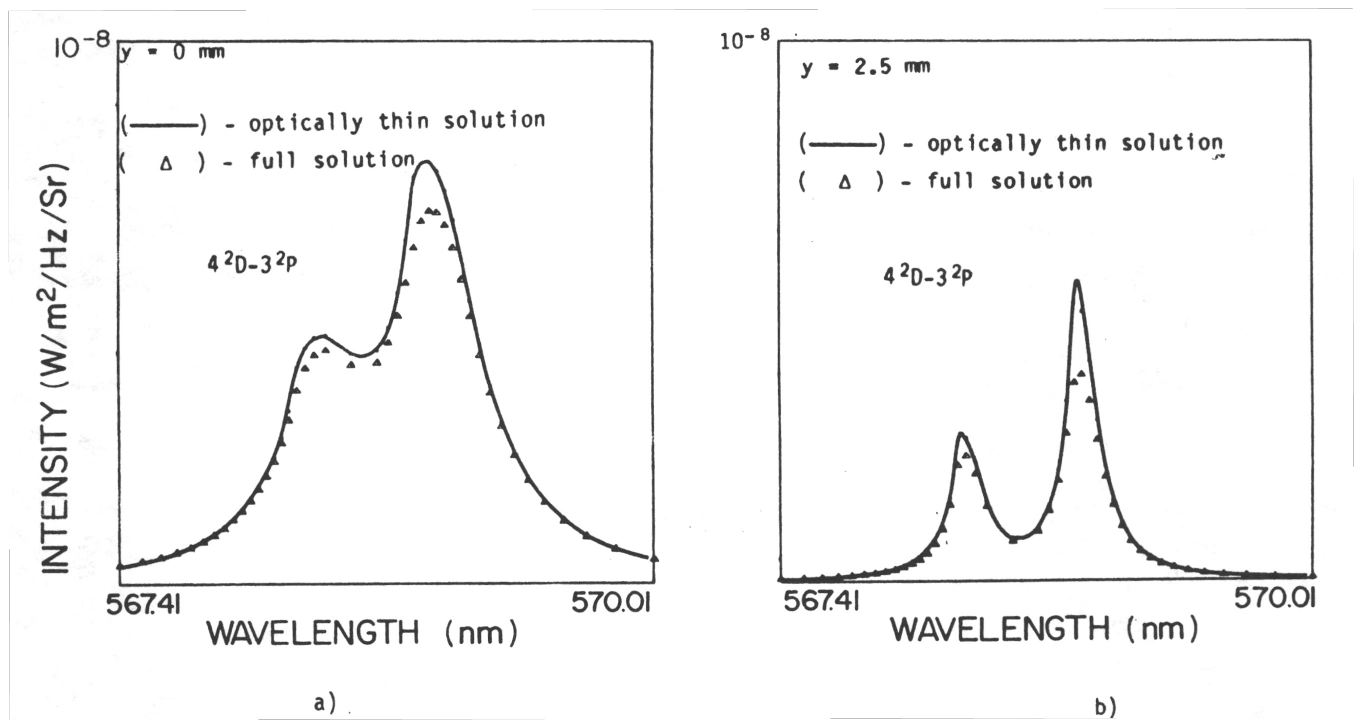}{Emission spectrum of the $4^2D$ - $3^2P$ transition at
a) y = 0 mm and b) y = 2.5 mm.}

\figstub{fig:2-9}{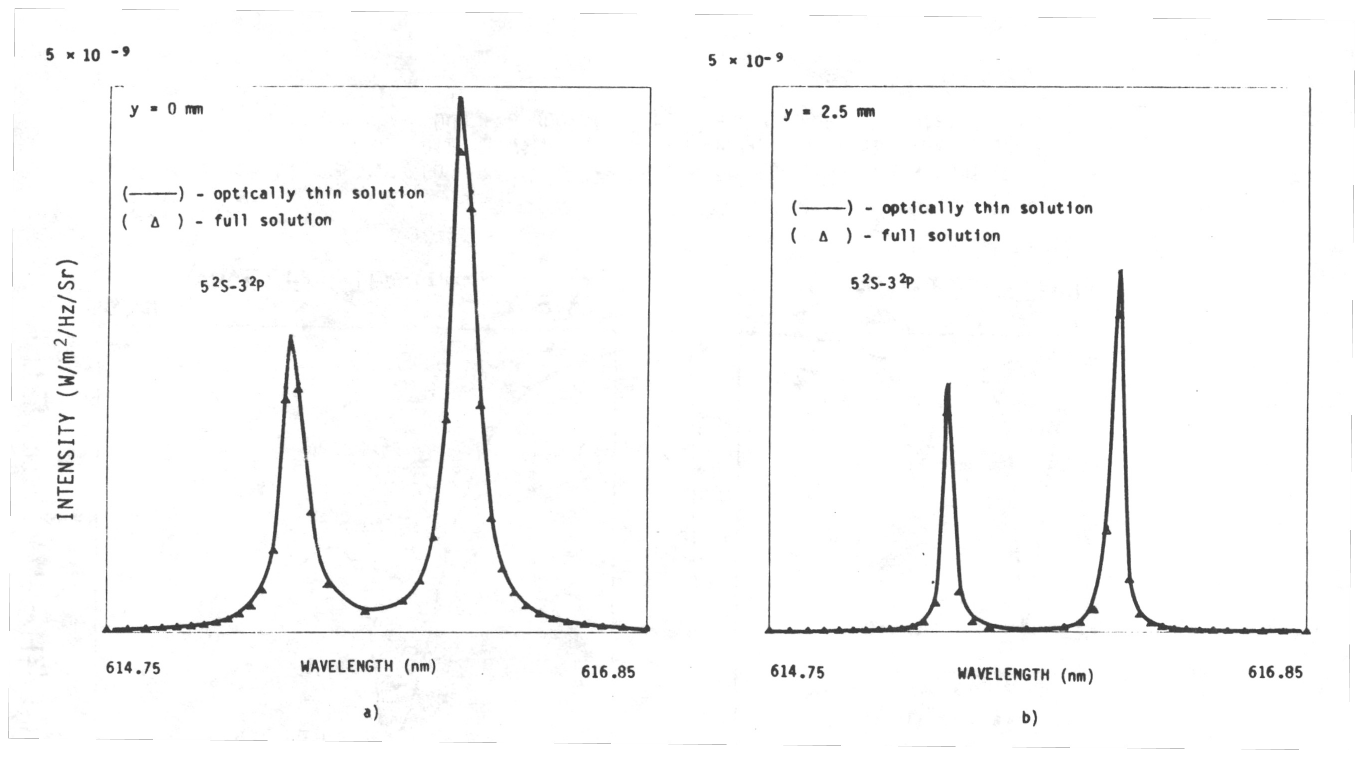}{Emission spectrum of the $5^2S$ - $3^2P$ transition at
y = 0 mm and y = 2.5 mm.}

\figstub{fig:2-10}{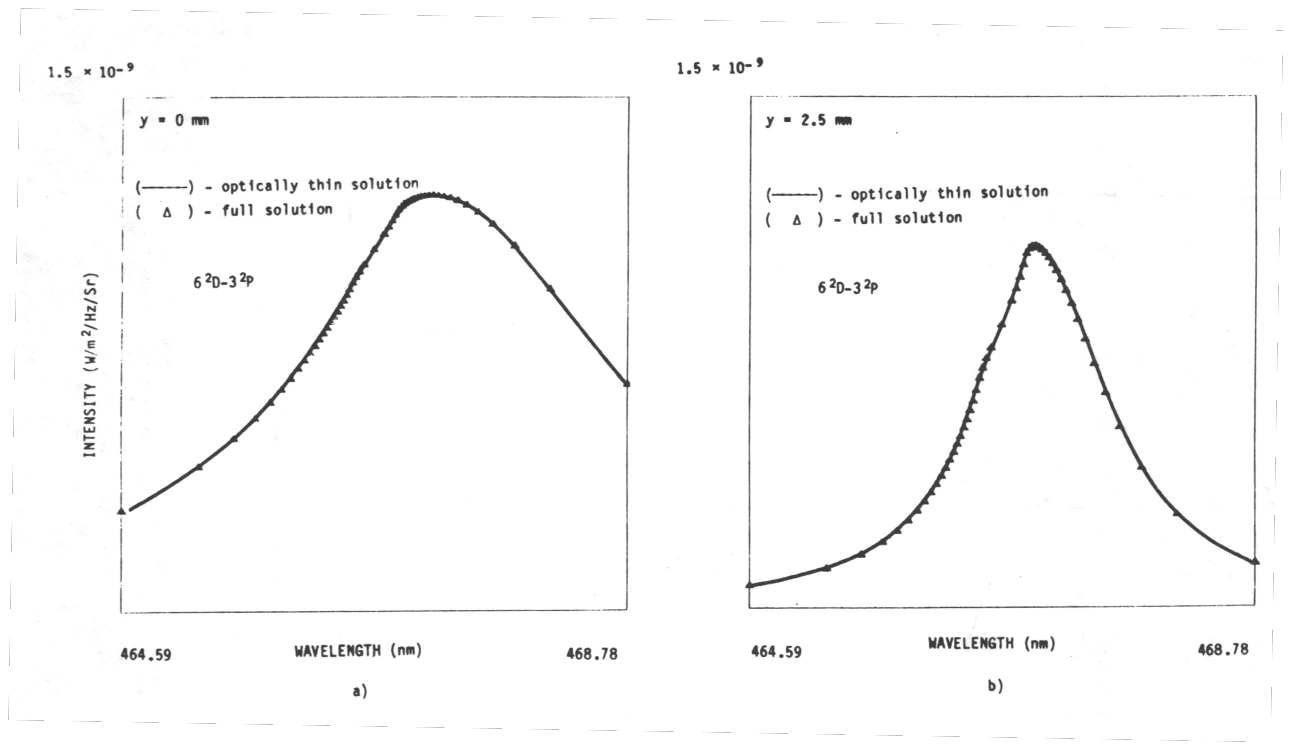}{Emission spectrum of the $6^2D$ - $3^2P$ transition at
y = 0 mm and y = 2.5 mm.}

Figure \ref{fig:2-11} shows the computed y variation of the radiance for the six
transitions.  As expected, there appears a maximum (at approximately 1.5
mm) at a radial position other than r=0.  One should note that although
self-absorption under these circumstances appears to have secondary effects,
a Boltzmann plot of the radiances for y = 0 mm (used for homogeneous
plasmas) does not necessarily have to exhibit a linear relationship
(Cappelli and Measures 1987a) and even if it did, the resulting slope would
tend to reflect the temperature arising from the region of greatest emission
(section \ref{sec:3-3} and Appendix G).

\figstub{fig:2-11}{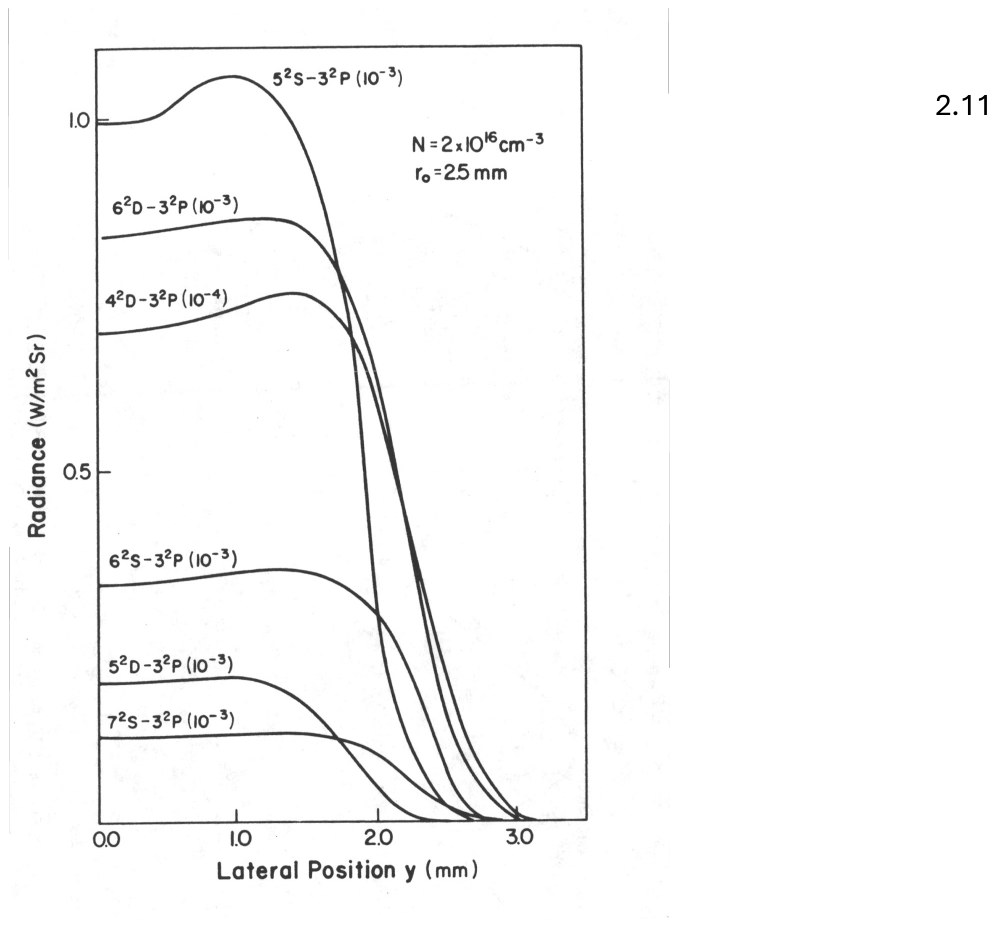}{Computed y variation of the radiance for the six
transitions.}

\subsection{Collisional-Radiative Equilibrium Model}
\label{sec:2-5-2}

In the above model, we have assumed that LTE extends down to the sodium
ground state.  A collisional-radiative model has been formulated to include
stimulated absorption and emission of resonance radiation as a mechanism of
populating or depopulating the resonance state.  The resonance radiation can
arise either directly from the laser field, or as a result of the strong
imprisonement of the spontaneous emission (Holstein 1949, 1951).  LTE is
assumed to extend down to the resonance state, thereby allowing us to
express the upper state population densities (n $>$ 2) as

\begin{equation}
N_n \;=\; \frac{g_n}{g_2}N_2\exp\Bigl(-\frac{E_{n2}}{kT_e}\Bigr)\;.
\label{eq:2-86}
\end{equation}

Continuity now takes on the form,

\begin{equation}
N_e + N_1 + N_2 + \sum_{n>3}^{n_{\max}} N_n \;=\; N\;,
\label{eq:2-87}
\end{equation}

and applying Saha equilibrium to the resonance level, we have

\begin{equation}
\frac{N_e^{\,2}}{N_2} \;=\; \frac{2\,g_1^{+}}{g_2}
\Bigl(\frac{m_e kT_e}{2\pi\overline{h}^{2}}\Bigr)^{1/2}
\exp\Bigl(-\frac{E_{c2}-\Delta E_{c1}}{kT_e}\Bigr)\;.
\label{eq:2-88}
\end{equation}

If we further assume that $N_1$ and $N_2$ are principally determined by stimulated
and spontaneous decay of the resonance state atoms, stimulated absorption
and emission of resonance radiation by atoms in the ground and resonance
state respectively, and by electron excitation and de-excitation collisions
between atoms in the ground and resonance states, then in a steady state
analysis, we can express their ratio,

\begin{equation}
\frac{N_2}{N_1} \;=\; \frac{R_{12}+N_e K_{12}}{R_{21}+N_e K_{21}+A_{21}}\;,
\label{eq:2-89}
\end{equation}

where $R_{12}$ and $R_{21}$ are the respective rates of stimulated absorption and
emission.  $K_{12}$ and $K_{21}$ represent the electron excitation and de-excitation
collisional rate coefficients, respectively (Seaton 1962).  With the
assumption of detailed balance,

\begin{equation}
K_{12} \;=\; \frac{g_2}{g_1}K_{21}\exp\Bigl(-\frac{E_{21}}{KT_e}\Bigr)\;,
\label{eq:2-90}
\end{equation}

and we can also write

\begin{equation}
R_{12} \;=\; \frac{g_2}{g_1}R_{21}\;.
\label{eq:2-91}
\end{equation}

Equation (\ref{eq:2-89}) becomes, with $g = g_2/g_1$,

\begin{equation}
\frac{N_2}{N_1} \;=\; \frac{gR_{21}+gN_e K_{21}\exp(-E_{21}/kT_e)}
{R_{21}+N_e K_{21}+A_{21}}\;.
\label{eq:2-92}
\end{equation}

This can be expressed in terms of the ratio of stimulated to spontaneous
emission rates $\Lambda = R_{21}/A_{21}$,

\begin{equation}
\frac{N_2}{N_1} \;=\; \frac{g\Lambda A_{21}+gN_e K_{21}\exp(-E_{21}/KT_e)}
{A_{21}(1+\Lambda)+N_e K_{21}}
\label{eq:2-93}
\end{equation}
\[
= \;\Phi(N_e,\,T_e,\,\Lambda)\;.
\]

Equation (\ref{eq:2-87}) can now be rewritten using equation (\ref{eq:2-86}), (\ref{eq:2-88}) and
(\ref{eq:2-93}),

\begin{equation}
N_2 \;=\; \Gamma(T_e,\,N_e,\,\Lambda)(N-N_e)\;,
\label{eq:2-94}
\end{equation}

where

\begin{equation}
\Gamma(T_e,\,N_e,\,\Lambda) \;=\;
\frac{f(T_e,\,N_e,\,\Lambda)}
{1 + f(T_e,\,N_e,\,\Lambda)\displaystyle\sum_{n=3}^{n_{\max}}
\frac{g_n}{g_2}\exp\Bigl(-\frac{E_{n2}}{kT_e}\Bigr)}\;,
\label{eq:2-95}
\end{equation}

and we have set

\begin{equation}
f(T_e,\,N_e,\,\Lambda) \;=\; \frac{\Phi(T_e,\,N_e,\,\Lambda)}{1+\Phi(T_e,\,N_e,\,\Lambda)}\;.
\label{eq:2-96}
\end{equation}

We can express $\Gamma$ in terms of the partition function,

\begin{equation}
\Gamma(T_e,\,N_e,\,\Lambda) \;=\;
\frac{f(T_e,\,N_e,\,\Lambda)}
{1 + f(T_e,\,N_e,\,\Lambda)\Bigl\{\dfrac{Z(T_e)}{g_1}\alpha(T_e)-\alpha(T_e)-1\Bigr\}}\;,
\label{eq:2-97}
\end{equation}

where we define

\begin{equation}
\alpha(T_e) \;=\; \frac{g_1}{g_2}\exp\Bigl(\frac{E_{21}}{kT_e}\Bigr)\;.
\label{eq:2-98}
\end{equation}

Using the Saha relationship (equation (\ref{eq:2-88})) and equation (\ref{eq:2-94}), we can
now derive a transcendental equation for $T_e$ and $N_e$,

\begin{equation}
\frac{N_e^{\,2}}{\{N-N_e\}} \;=\; \Gamma(T_e,\,N_e,\Lambda)\,S(T_e)\;.
\label{eq:2-99}
\end{equation}

In this equation, $S(T_e)$ represents the right hand side (RHS) of equation
(\ref{eq:2-88}).  To illustrate the numerical procedure used to solve equation
(\ref{eq:2-99}), we define

\begin{equation}
h(T_e,\,N_e,\,\Lambda) \;=\; \frac{N_e^{\,2}}{(N-N_e)\Gamma S}\;,
\label{eq:2-100}
\end{equation}

and for a given $T_e$ and $\Lambda$, we can evaluate h over a range of electron
densities $N_e$.  A cubic spline (ICSCCU-IMSL subroutine library) is used to
represent the data, from which an interpolation in $N_e$ for h = 1 is
performed.

Figure \ref{fig:2-12} illustrates the difference obtained in the computed $4^2D$ -
$3^2P$ spectrum with $\Lambda$ = 0 and $\Lambda$ = 0.1, using the radial temperature
distribution $T_e(r) = 2000 + 8000\exp(-r^2/2.5^2\text{mm})$ described in the previous
section.  Clearly evident is the increased self-absorption which now appears
as a self-reversal in the spectrum.  This figure indicates that the
inclusion of a stimulated term increases the population of the $3^2P$ state, as
one may expect, however, the dominance of this stimulated term is only in
the plasma ``halo'' where the electron density is too low for superelastic
collisional quenching to take effect.  Figure \ref{fig:2-13} shows the computed radial
variation in the ratio of electron density and $3^2P$ state population density
achieved when $\Lambda$ = 0 and $\Lambda$ = 0.1 for the temperature distribution indicated
by equation (\ref{eq:2-85}).  The electron density ratio increases as one moves from
the core but more importantly (for self-absorption consideration), the
resonance state population varies as the square in the electron density.

\figstub{fig:2-12}{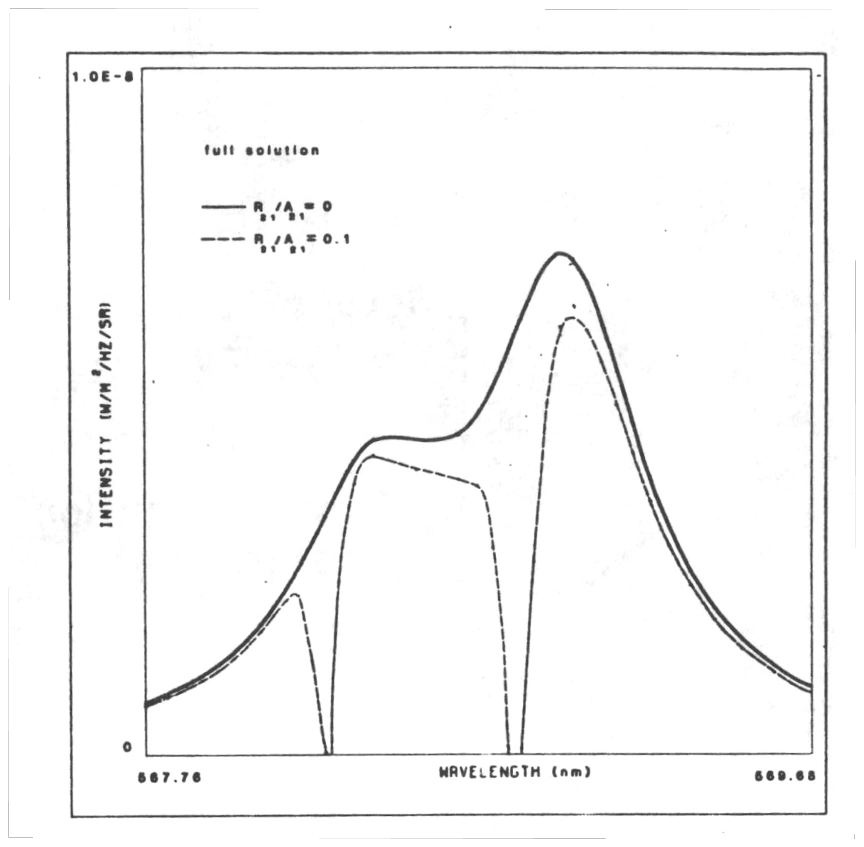}{Computed $4^2D$ - $3^2P$ spectrum with $\Lambda$ = 0 and $\Lambda$ = 0.1.}

\figstub{fig:2-13}{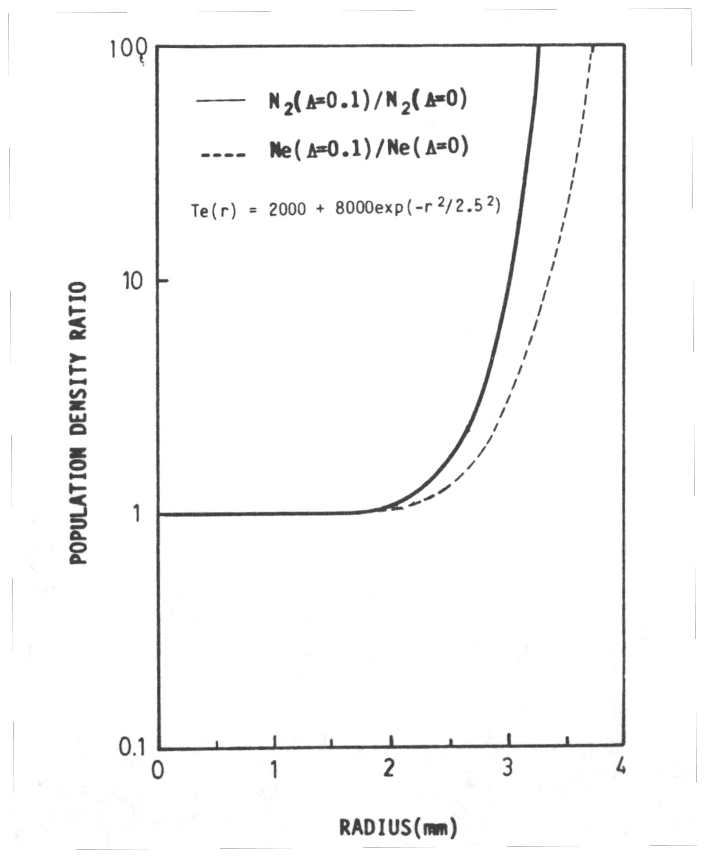}{Computed radial variation in the ratio of electron density
and $3^2P$ state population density achieved when $\Lambda$ = 0 and $\Lambda$ = 0.1.}

Self-reversal in the emission spectrum has been observed (chapter 6) at
early times in the plasma evolution (t $\leqslant \tau_L$ = duration of the laser pulse-
FWHM), suggesting that the tail of the laser pulse (or, radiation trapped
spontaneous emission) plays a role in the creation of an extended halo of
excited state atoms surrounding the plasma core.  This of course suggests
that measurements of $T_e$ and $N_e$ based on the observed optically thin radiance
or spectral radiance, have to be limited to times at least greater than the
sum of the laser pulse duration and the resonance state spontaneous decay
time, in order that any excess resonance state population has a chance to
decay through superelastic collisional de-excitation or spontaneous
emission.

\subsection{Collisional-Radiative LIBORS Model}
\label{sec:2-5-3}

The three-dimensional (five-level) LIBORS code (Kissack 1987) that has
been used in this thesis (Chapter 6) to model the plasma development
resulting from the interaction of laser radiation tuned to the $3^2S$ - $3^2P$
resonance transition in sodium, is an extension of the work undertaken in
this laboratory over the past two decades (Measures 1970; Measures, Drewell
and Cardinal 1979; Measures and Cardinal 1981; Measures, Cardinal and Schinn
1981; Measures, Wong and Cardinal 1982; Wong 1985, Cardinal 1986).  All of
the LIBORS computer codes are based on solutions of the energy and
population rate equations (the justification of this can be found in a
comprehensive review by Cardinal 1986), to predict the temporal history of
the ground and excited state population densities, the free electron
density, the free-electron energy density and the sodium ion and atom energy
densities.  Unlike the early twenty-level model (Measures, Drewell and
Cardinal 1979), which represents the atom by twenty electronic levels and
only solved for the temporal variation, the more recent LIBORS computer code
uses a much simpler five level model but takes account of the variation of
the laser pulse as it propagates through a non-uniform atom density
distribution.  This is achieved by dividing the vapor into a series of slabs
and solving the rate equations for each slab, allowing for attenuation
within each slab by solving the one dimensional radiative transfer equation
for the laser irradiance (Wong 1985).

With a predetermined incident radial variation of the laser flux
density, and the variation of sodium density along the direction of beam
propagation, the five-level model can compute both the radial and temporal
evolution in the free electron density and temperature, from which we can
estimate the population density of any excited state through the Saha
equation,

\begin{equation}
N_n(r,t) \;=\; \frac{N_e^{\,2}(r,t)}{S_n(T_e(r,t))}\;,
\label{eq:2-101}
\end{equation}

where now, we define

\begin{equation}
S_n(T_e) \;=\; \frac{2\,g_1^{+}}{g_n}
\Bigl(\frac{m_e kT_e}{2\pi\overline{h}^{2}}\Bigr)^{3/2}
\exp\Bigl(-\frac{E_{cn}-\Delta E_{c1}}{kT_e}\Bigr)\;.
\label{eq:2-102}
\end{equation}

Calculating $N_2(r,t)$ in this manner permits equation (\ref{eq:2-72}) to be directly
integrated, with the source function $P(\nu_{n2},r,t)$ now taking on the form

\begin{equation}
P(\nu_{n2},r,t) \;=\; \frac{2h\nu_{n2}^{3}}{c^2}
\left\{\frac{1}{\dfrac{g_n N_2(r,t)}{g_2 N_n(r,t)}-1}\right\}\;.
\label{eq:2-103}
\end{equation}

Computation of the spectral radiance predicted by the five-level model
has allowed us to understand the line emission that is observed especially
at earlier times during the peak of the laser pulse where saturation of the
resonance transition is evident.  In addition, such calculations have
allowed us to simulate the effects of noise on the radial laser profile and
its implication on multi-shot averaging (section \ref{sec:3-4}) of emission signals
from a plasma based on a highly non-linear interaction.

\section{Abel Transformation}
\label{sec:2-6}

The Abel transformation is a special case of the more general Radon
transformation (Clough and Barrett 1983) and is applicable where axial or
cylindrical symmetry can be assumed.  In emission spectroscopy, it is most
often used to transform the optically thin lateral radiance $J(y)$ into the
radial variation of the local volume emission coefficient $\varepsilon(r)$.  In the case
of the optically thin nm-transition,

\begin{equation}
J_{nm}(\nu,y) \;=\; \int_{-\sqrt{(R^2-y^2)}}^{\sqrt{(R^2-y^2)}}
\varepsilon_{nm}(\nu,y,r)\,dx\;.
\label{eq:2-104}
\end{equation}

A simple coordinate transformation $r^2 = x^2+y^2$ allows us to write

\begin{equation}
J_{nm}(\nu,y) \;=\; 2\int_{y}^{R}\frac{\varepsilon_{nm}(\nu,r)\,r\,dr}{\sqrt{(r^2-y^2)}}\;.
\label{eq:2-105}
\end{equation}

The radial variation of the local volume emission is then obtained from the
Abel transform of $J_{nm}(\nu,y)$,

\begin{equation}
\varepsilon_{nm}(\nu,r) \;=\; -\frac{1}{\pi}\int_{r}^{R}
\frac{\{dJ_{nm}(\nu,y)/dy\}\,dy}{\sqrt{(y^2-r^2)}}\;.
\label{eq:2-106}
\end{equation}

We can non-dimensionalize the coordinate reference frame by setting

\[
y^{*} \;=\; y/R\;,
\]
\[
r^{*} \;=\; r/R\;,
\]
\begin{equation}
J_{nm}(\nu,y^{*}) \;=\; J_{nm}(\nu,y)\,dy/dy^{*}\;,
\label{eq:2-107}
\end{equation}

and

\[
\varepsilon(\nu,r^{*}) \;=\; \varepsilon(\nu,r)\,dr/dr^{*}\;.
\]

Equation (\ref{eq:2-106}), in non-dimensionalized variables $r^{*}$ and $y^{*}$, becomes

\begin{equation}
\varepsilon_{nm}(\nu,r^{*}) \;=\; -\frac{1}{\pi}\int_{r^{*}}^{1}
\frac{\{dJ_{nm}(\nu,y^{*})/dy\}\,dy^{*}}{\sqrt{(y^{*2}-r^{*2})}}\;.
\label{eq:2-108}
\end{equation}

The above integral which is the basis for the Abel transformation, is
solved using the procedure proposed by Deutsch (1983).  The lateral
distribution of the line of sight spectral radiance from a plasma of
non-dimensional radius unity, can be approximated by the product of a polynomial
and an exponential function (Choi and Kim 1982, Deutsch 1983), viz.,

\begin{equation}
J_{nm}(\nu,y) \;=\; \Bigl\{\sum_{k=0}^{n} a_k y^{2k}\Bigr\}\exp(-\alpha y^2)\;.
\label{eq:2-109}
\end{equation}

In the above equation, we revert back to y and r for clarity, yet maintain
the non-dimensional interpretation, that is, $y = y^{*}$, $r = r^{*}$, and frequency
dependence $\nu$ is implied but not indicated explicitly.

Under these conditions, direct integration of equation (\ref{eq:2-108}) enables
us to write

\begin{equation}
\varepsilon_{nm}(r) \;=\; -\frac{1}{\pi}\sum_{k=0}^{n} b_k f_k(r)
\label{eq:2-110}
\end{equation}

where

\begin{align}
b_k &= 2(k+1)a_{k+1} - 2\alpha a_k\;, && \text{for } 0 < k \leqslant n-1\;,
\label{eq:2-111}\\
&= -2\alpha a_k\;, && \text{for } k = n\;,\nonumber
\end{align}

and

\begin{multline}
f_k(r) \;=\; \sum_{p=0}^{k}\frac{k!}{(k-p)!\,p!}\,r^{2p}
\Bigl[(2\alpha)^{p-k-1}(\alpha\pi)^{1/2}\exp(-\alpha r^2)\,
\mathrm{erf}\{\alpha(1-r^2)\}^{1/2}\\[0.4em]
-\;\exp(-\alpha)\sum_{s=1}^{k-p}(2\alpha)^{s}(1-r^2)^{k-p-s+1/2}\Bigr]\;.
\label{eq:2-112}
\end{multline}

It should be noted that the last sum in equation (\ref{eq:2-112}) vanishes for p = k.
Here erf denotes the error function,

\begin{equation}
\mathrm{erf}(y) \;=\; \frac{2}{\sqrt{\pi}}\int_{0}^{y}\exp\{-t^2\}dt\;.
\label{eq:2-113}
\end{equation}

Although equation (\ref{eq:2-112}) is fairly complicated, it is not too difficult to
evaluate numerically since in most practical situations, four or five terms
are adequate (Deutsh 1983).  Indeed, in our analysis (Chapter 3), we have
found that we can represent the lateral emission profiles by a function of
the form

\begin{equation}
J(y) \;=\; \{a_0 + a_1 y^2 + a_2 y^4\}\exp(-\alpha y^2)\;,
\label{eq:2-114}
\end{equation}

where the fitting parameters $a_0$, $a_1$, $a_2$ and $\alpha$ are chosen to minimize the
residual sum of the squares of the deviation from the actual data.

We can check the results of the above formulation on the simple
expression

\begin{equation}
J(y) \;=\; a_0\exp(-\alpha y^2)\;,
\label{eq:2-115}
\end{equation}

that is, equation (\ref{eq:2-109}) with n=0.  This of course is a Gaussian function,
which with the substitution $u^2 = (y^2-r^2)$, can be inserted directly into
equation (\ref{eq:2-106}) giving

\begin{equation}
\varepsilon(r) \;=\; \frac{2\alpha a_0}{\pi}\exp\{-\alpha r^2\}
\int_{0}^{\sqrt{(1-r^2)}}\exp(-\alpha u^2)\,du\;.
\label{eq:2-116}
\end{equation}

Using the definition of the error function, this expression can be directly
integrated, resulting in

\begin{equation}
\varepsilon(r) \;=\; \alpha^{1/2}a_0\pi^{-1/2}\exp(-\alpha r^2)\,
\mathrm{erf}\bigl(\sqrt{\alpha(1-r^2)}\bigr)\;.
\label{eq:2-117}
\end{equation}

Equations (\ref{eq:2-111}) and (\ref{eq:2-112}), for n=0, become

\begin{equation}
b_0 \;=\; -2\alpha a_0\;,
\label{eq:2-118}
\end{equation}

\begin{equation}
f_0 \;=\; 2^{-1}\alpha^{-1/2}\pi^{1/2}\exp(-\alpha r^2)\,
\mathrm{erf}\bigl(\sqrt{\alpha(1-r^2)}\bigr)\;,
\label{eq:2-119}
\end{equation}

and upon substitution into equation (\ref{eq:2-110}), lead to

\begin{equation}
\varepsilon(r) \;=\; \alpha^{1/2}a_0\pi^{-1/2}\exp(-\alpha r^2)\,
\mathrm{erf}\bigl(\sqrt{\alpha(1-r^2)}\bigr)
\label{eq:2-120}
\end{equation}

which is exactly the expression given by equation (\ref{eq:2-117}) above.

\chapter{Sodium Vapor and Sodium Plasma Characterization}
\label{ch:characterization}

\section{Neutral Sodium Density Characterization}
\label{sec:3-1}

Alkali metal vapors are highly reactive and in the absence of adequate
precautions, can quickly spoil the windows of any containment vessel built
to optically study them.  One method of minimizing this degradation is to
interpose a cooled inert buffer gas between the hot vapor and the windows.
Vidal and Cooper (1969) first demonstrated that a heat pipe concept with a
buffer gas not only protected the windows but led to the formation of a well
defined column of vapor.

The limited optical access of the traditional heat pipe was overcome by
Boyd et al.\ (1980) and by Boyd and Harter (1980), who proposed a
sandwich-like oven in which the vapor was trapped between two disk-shaped wire mesh
wicks that were heated at their centre and were cooled at their periphery.
This design leads to the formation of a disk-shaped vapor region within the
oven and to $360^{\circ}$ optical access in the plane of the disk.

A specially designed heat sandwich oven was constructed (Herchen 1982,
Cappelli 1983, Cappelli et al.\ 1985, Cardinal 1986) for undertaking
experiments involving laser ionization of sodium vapor based on resonance
saturation.  The oven design is similar to that of Boyd et al.\ (1980).
Details of the design and its operation is the subject of section \ref{sec:4-1}.

Although thermocouple sensors were employed to estimate the vapor
temperature within the oven, it was felt that they could at best provide
only a rough estimate of the sodium atom density using the empirical fit of
Nesmeyanov (equation (\ref{eq:2-38})).  Temperature readings from the thermocouples
positioned outside of the oven may not necessarily reflect that of the
liquid sodium within the oven wick, which actually determines the vapor
pressure.  For that reason, we decided to employ the absorption measurement
technique described below in order to measure the sodium atom density
distribution within the oven.

\subsection{Sodium Atom Density Measurement Technique}
\label{sec:3-1-1}

The spectral irradiance $I(\nu,z_0+\ell)$ of a well collimated beam of
radiation of frequency $\nu$ propagating in the z-direction through a vapor of
length $\ell$ can be related to the incident spectral irradiance $I(\nu,z_0)$ by the
Beer-Lambert law,

\begin{equation}
I(\nu,z_0+\ell) \;=\; I(\nu,z_0)\exp\left\{-\int_{z_0}^{z_0+\ell}\kappa(\nu,z)dz\right\},
\label{eq:3-1}
\end{equation}

where $\kappa(\nu,z)$ represents the volume absorption coefficient at position z
along the path.  In the event that $\nu$ is close to the resonance doublet of
sodium, we can write (Cappelli et al.\ 1985)

\begin{equation}
\kappa(\nu,z) \;=\; \frac{\lambda_0^{\,2}}{8\pi}N_1(J',z)\sum_{i=1}^{2}A_{J_iJ'}\pounds_i(\nu,z)\;,
\label{eq:3-2}
\end{equation}

where $N_1(J',z)$ represents the population density of the lower level (at a
depth z along the path) having a total angular momentum quantum number $J'$
($J' = 1/2$) for sodium, $A_{J_iJ'}$ represents the Einstein transition probability
for the $J_i$ to $J'$ transition of the doublet (where $J_1 = 1/2$ and $J_2 = 3/2$),
$\pounds_i(\nu,z)$ is the corresponding line profile function given by equation (\ref{eq:2-79})
and $\lambda_0$ represents the mean wavelength of the doublet.

If we treat the absorption as arising from a single line with an
``effective'' line profile function $\pounds(\nu,z)$, we can write

\begin{equation}
\kappa(\nu,z) \;=\; \frac{\lambda_0^{\,2}}{8\pi}N_1(J',z)\{A_{J_1J'}+A_{J_2J'}\}\pounds(\nu,z)\;,
\label{eq:3-3}
\end{equation}

where

\begin{equation}
\pounds(\nu,z) \;=\; \zeta(J_1,J')\pounds_1(\nu,z) + \zeta(J_2,J')\pounds_2(\nu,z)\;.
\label{eq:3-4}
\end{equation}

In this instance $\zeta(J_i,J')$ and $\pounds_i(\nu,z)$ represent the fractional line strength
and line profile function of the $J_i$ to $J'$ transition of the doublet.

It should be recognized that use of equation (\ref{eq:3-1}) implies that the
radiation is well below the saturated value (Herchen and Measures 1982) and
that scattering is negligible.  The atomic line profile $\pounds(\nu,z)$ has explicit
z dependence in that its shape is predominantly determined by resonance
broadening, that is, broadened through collisions with other sodium atoms
(section \ref{sec:2-1}) whose density depends on position z along the path.

\subsection{Correction for Finite Instrument Resolution}
\label{sec:3-1-2}

In the facility for neutral sodium absorption studies (section \ref{sec:4-1}), a
Heath scanning monochromator was employed to determine the spectral
characteristics of the transmitted radiation (emitted from a broadband
xenon arc source).  For this kind of monochromator, where the entrance and
exit slits are of comparable width, the spectral transmission function
$T(\nu_T-\nu)$ (normalized) is well represented by a Gaussian profile (section
\ref{sec:4-1}), and we can write

\setcounter{equation}{6}
\begin{equation}
T(\nu_T-\nu) \;=\; \frac{1}{\gamma_T\pi^{1/2}}\exp\{-(\nu_T-\nu)^2/\gamma_T^{2}\}\;,
\label{eq:3-7}
\end{equation}

where $\nu_T$ is the frequency of peak transmission (determined by rotation of
the grating) and

\begin{equation}
\gamma_T \;=\; \delta_T/2(\ln 2)^{1/2}\;.
\label{eq:3-8}
\end{equation}

Here $\delta_T$ is the instrument's FWHM bandwidth corresponding to a given slit
setting and can be expressed in terms of the slope $b_T$, and intercept $a_T$, of
the linear relationship between the observed FWHM bandwidth of an extremely
narrow spectral line and the monochromator slit width $s_T$,

\begin{equation}
\delta_T \;=\; a_T + b_T s_T\;.
\label{eq:3-9}
\end{equation}

The observed spectral irradiance at frequency $\nu$ can thus be written in
the form

\begin{equation}
\psi(\nu,z_0+\ell) \;=\; \int_{-\infty}^{\infty} I(\nu',z_0+\ell)\,T(\nu-\nu')\,d\nu'\;,
\label{eq:3-10}
\end{equation}

or, using equation (\ref{eq:3-1}) and (\ref{eq:3-7}) and assuming broadband incident radiation
so that $I(\nu,z_0) \approx I(z_0)$, then the observed spectral profile,

\begin{equation}
\Phi(\nu,z_0+\ell) \;=\; \frac{\psi(\nu,z_0+\ell)}{I(z_0)}
\;=\; \frac{1}{\gamma_T\pi^{1/2}}\int_{-\infty}^{\infty}
\exp\left\{\left[-\int_{z_0}^{z_0+\ell}\kappa(\nu',z)dz\right]-(\nu-\nu')^2/\gamma_T^{2}\right\}d\nu'
\label{eq:3-11}
\end{equation}

can be computed using equations (\ref{eq:2-63}), (\ref{eq:3-3}) and (\ref{eq:3-4}) with a numerical
quadrature routine (Herchen 1982).  For most of the work presented in this
thesis, the slit width $s_T$ was 150 $\mu$m and the experimentally determined
values for $a_T$ and $b_T$ are provided with other parameters of relevance in Table
\ref{tab:3-1}.

\tabstub{tab:3-1}{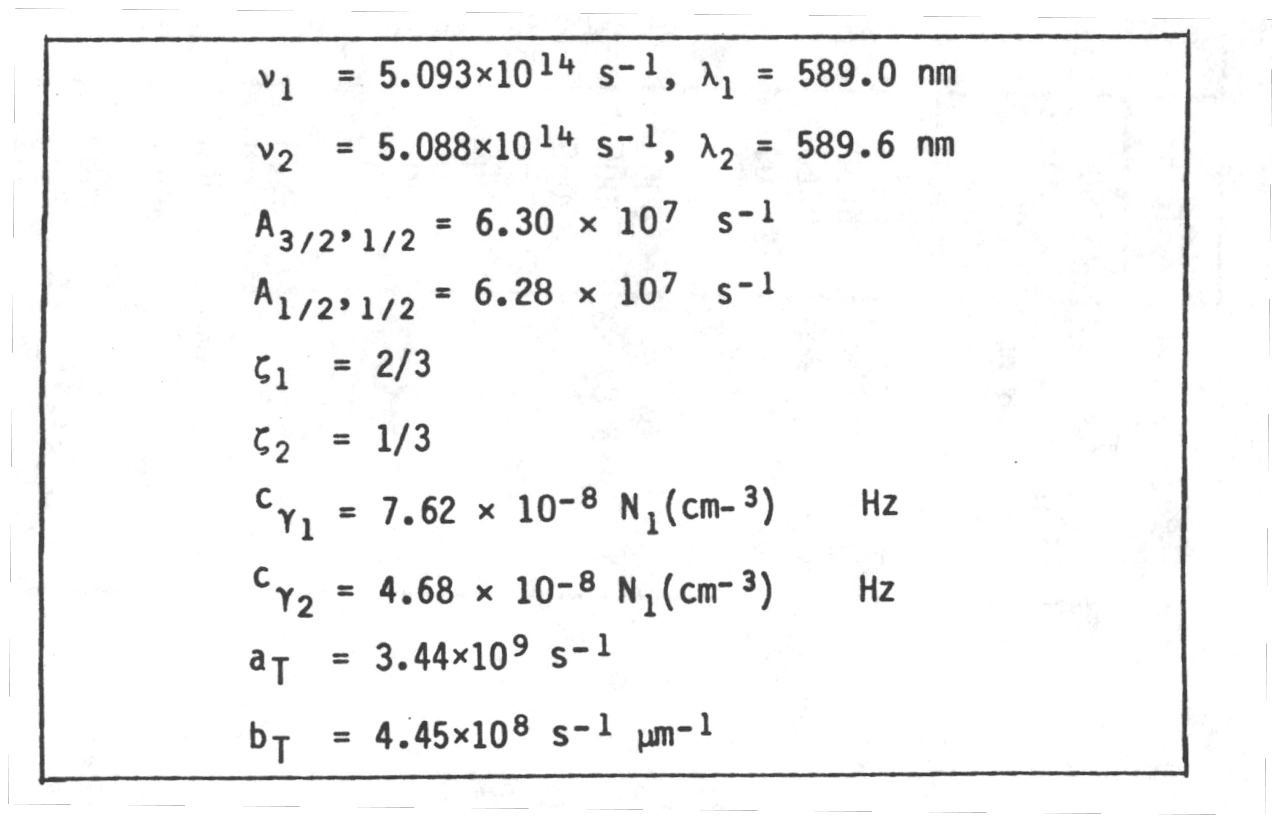}{Experimentally determined values for $a_T$ and $b_T$, together
with other parameters of relevance.}

To evaluate the atom density distribution, it is necessary to undertake
a series of measurements.  The sodium vapor disk is assumed to be divided
into a number of equally spaced rings of constant density (see figure \ref{fig:3-1})
and the absorption is evaluated for a series of chords which intercept an
increasing number of these rings.  Reference to figure \ref{fig:3-1}, reveals that the
element of path length

\figstub{fig:3-1}{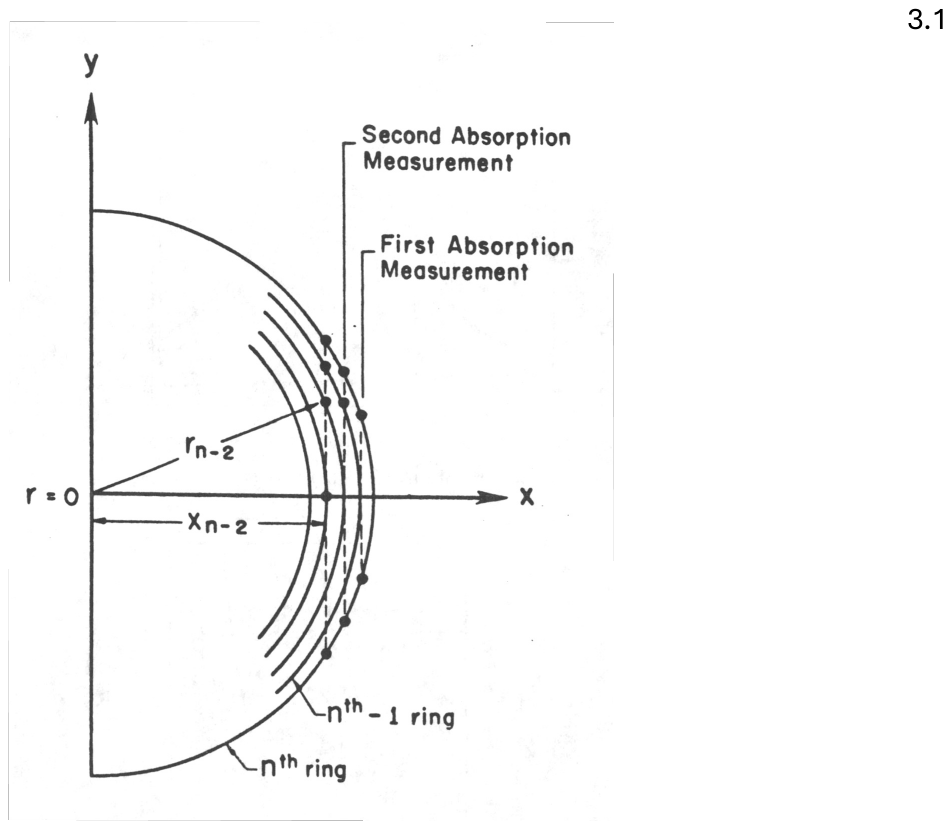}{The sodium vapor disk divided into a number of equally
spaced rings of constant density, showing the series of measurement chords.}

\begin{equation}
\ell_{mn} \;=\; \{r_n^{\,2}-x_m^{\,2}\}^{1/2} - \{r_{n-1}^{\,2}-x_m^{\,2}\}^{1/2}\;,
\label{eq:3-12}
\end{equation}

where $x_m$ is the x-displacement of the $m^{\text{th}}$-measurement chord and $r_n$ the
radius of the $n^{\text{th}}$-density ring.  The total optical depth for the $m^{\text{th}}$-chord

\begin{equation}
\tau(\nu,x_m) \;=\; 2\sum_{k=n_{\min}}^{k=n}\kappa_k(\nu)\,\ell_{mk}\;,
\label{eq:3-13}
\end{equation}

where

\begin{equation}
\kappa_k(\nu) \;=\; \frac{\lambda_0^{\,2}}{8\pi}N_1(r_k)\{A_1+A_2\}\pounds^{k}(\nu)\;,
\label{eq:3-14}
\end{equation}

represents the absorption coefficient for the $k^{\text{th}}$-ring of atom density
(approximately the ground state density) $N_1(r_k)$ and $\pounds^{k}(\nu)$, the relevant
total absorption profile for the sodium D-doublet, as given by equation
(\ref{eq:3-4}).  In the above equations, we have simplified the notation by setting
$A_i = A_{J_iJ'}$ and $N_1 = N_1(J')$.  In order to evaluate the Doppler contribution
to $\pounds^{k}(\nu)$ for each ring, the temperature is assumed to be related to the atom
density $N_1(r_k)$ of that ring through the equilibrium relation given earlier
as equation (\ref{eq:2-38}).

The density for the $n^{\text{th}}$-ring is first guessed at and the appropriate
absorption spectrum is calculated.  This spectrum is then compared to that
measured for that particular chord.  The density is then adjusted
accordingly until an acceptable agreement is obtained.  The average density
of the $n^{\text{th}}$-ring is then known and is used in conjunction with a slightly
higher (or possibly lower) density for the two segments of the next ring
required to compute the absorption spectrum for the second chord.  The
density for this second ring is adjusted until reasonable agreement
(discussed below) is obtained between the calculated and the measured
absorption spectrum.  This series of calculations and comparison with
measurements is continued towards the centre of the heat sandwich oven
leading to an effective radial density profile for the sodium vapor.

Reasonable agreement is obtained when the fullwidth half-maximum of the
predicted absorption spectrum closely matches that measured well within the
limit of instrument resolution (see top inset in figure \ref{fig:3-2}).  When the
sodium doublet is clearly resolvable, usually for x $\geqslant$ 5 cm, we used the
absorption at line centre of the $3^2P_{3/2}$ - $3^2S_{1/2}$ transition as our matching
condition (see lower inset in figure \ref{fig:3-2}).  It should be noted that best
results were obtained when a preliminary smoothing of the data was
performed.  The FWHM of the absorption spectrum was fit to a function in
position x of the form

\begin{equation}
P(x) \;=\; \{\alpha_1 + \alpha_2 x^2 + \alpha_3 x^4\}\exp(-\alpha_4 x^2)\;,
\label{eq:3-15}
\end{equation}

where the $\alpha_i$ (i = 1 to 4) represents the various fitting parameters.  The
variation of the transmission at line centre of the $D_2$ line with position x
was approximated by a straight line.  Data from these two curves were then
directly used in the density reconstruction routine.

An example of the variation of the FWHM and $D_2$ line centre transmission
with position x is presented in figure \ref{fig:3-2}.  The two insets display two
representative absorption spectra, one corresponding to a path close to the
centre of the oven, the other in the low density rim of the vapor disk.  The
corresponding radial density distribution of sodium atoms within the heat
sandwich oven derived from these data, is presented in figure \ref{fig:3-3}.  The
maximum temperature registered by the thermocouple near the oven centre was
938K, which would suggest a sodium density considerably higher than that
which is estimated for the oven centre.  Mechanical equilibrium between the
argon buffer gas and sodium vapor was ensured by continuously monitoring and
adjusting the buffer gas pressure to match the sodium vapor pressure at the
centre of the oven (measured according to the technique described).  The
density measurements are estimated to be accurate to within $\pm$20\%.  This is
primarily determined by the uncertainty in the halfwidth of the resonance
broadened atomic absorption profile.  The theoretical values by Carrington
et al.(1973) used here agree fairly well with the experimental measurements
of Heunenekens and Gallagher (1973) which were estimated to have an error of
no more than $\pm$20\%.  We have found that over the density range of interest,
there exists a direct relationship between the percentage change in the
theoretical halfwidth of the absorption profile used for the density
evaluation and the percentage change in the peak atom density computed.

\figstub{fig:3-2}{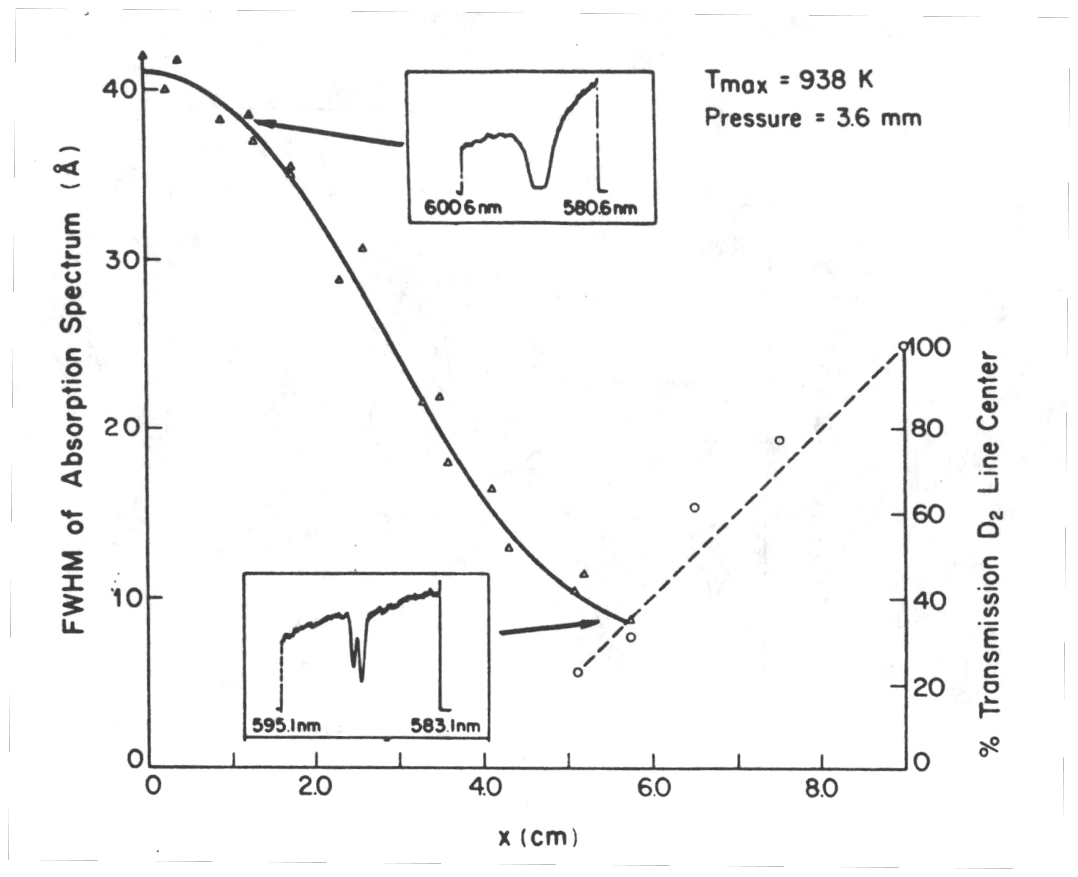}{Variation of the FWHM and $D_2$ line centre transmission with
position x.  The insets display two representative absorption spectra.}

\figstub{fig:3-3}{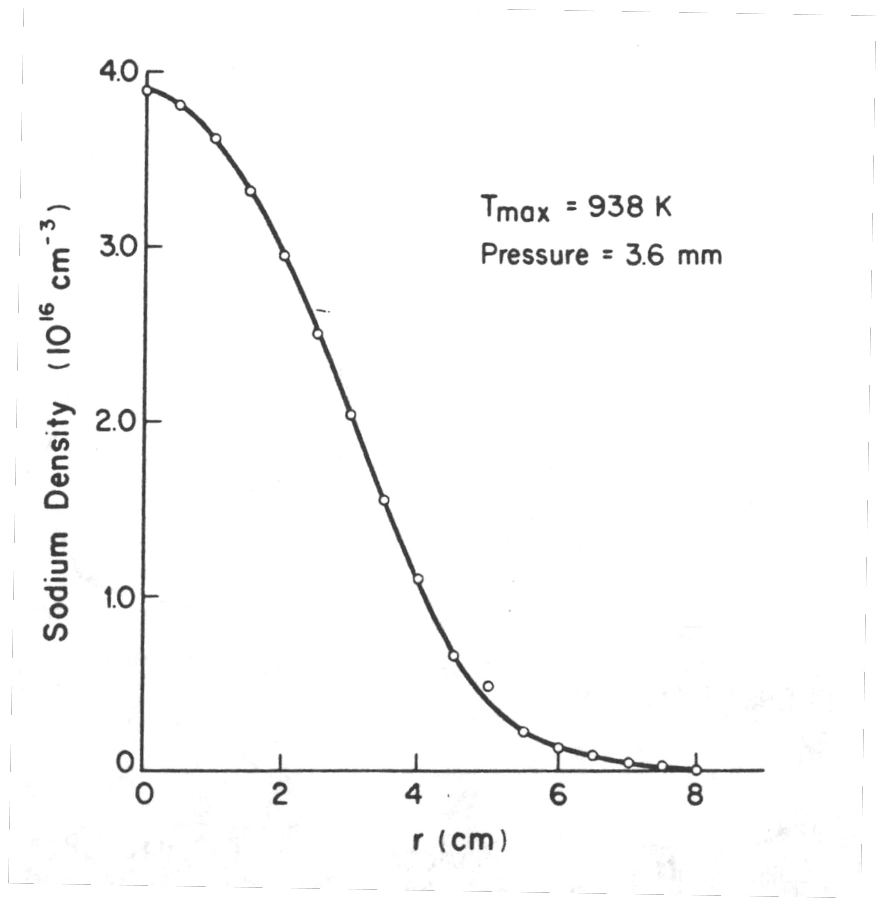}{Radial density distribution of sodium atoms within the heat
sandwich oven.}

Secondary sources of error are contributions to the atomic line
profile arising from the Van der Waals interactions, as well as fluctuations
in the oven temperature during the series of measurements.  There was no
noticeable change in the output of the xenon arc lamp, which provided the
broadband continuous wave source, between each scan.

\section{Electron Density Measurements from Stark Broadened Emission Lines}
\label{sec:3-2}

Electron density measurements from the Stark broadening of plasma
emission lines has developed into a useful and common diagnostic tool since
the formulation of reliable Stark broadening theories (Griem 1964,
Sahal-Brechot 1969, Griem 1974).  The most attractive feature of this technique is
that it is noninvasive and has been widely used in applications where plasma
ionization probes either directly interfere with the processes studied or
are easily contaminated by highly reactive constituents of the plasma.  The
viability of the technique was clearly demonstrated by Agnew and Reichelt
(1969) in determining the free electron density in a cesium diode, and
afterwards, in wall stabilized arcs (Grumberg et al.\ 1976, Helbig et al.\
1976, Waszink and Flinsenberg 1978, Kelleher 1981, Goldbach et al.\ 1982)
where cylindrical symmetry and plasma stability permitted the Abel inversion
of lateral intensity measurements to obtain the local volume emission
coefficient, and hence, the radial variation in the free electron density.
Stark broadening has also been widely investigated and employed as a
diagnostic tool in transient plasmas produced in shock tubes (Neiger and
Griem 1976, Baur and Cooper 1977, Chiang et al.\ 1977, Vaessen et al.\ 1985),
laser ablation of solid targets (Hashimoto and Yamaguchi 1983, Lee et al.\
1984), low induction vacuum sparks (Datla and Griem 1978, 1979) and in
plasmas produced by laser resonance saturation (Krebs and Schearer 1982,
Cappelli and Measures 1984 and Appendix 4, Landen et al.\ 1985).  Most of
these measurements were spatially and/or temporally averaged, depending on
the nature or reproducibility of the plasma source.  In some cases,
multi-shot averaging has been performed (Krebs and Schearer 1982, Landen et
al.\ 1985).  The question of multi-shot averaging in highly non-linear
systems, such as a plasma column produced by laser resonance pumping, will
be addressed in a numerical simulation of the spectral emission of a sodium
plasma produced by resonance saturation (section \ref{sec:3-4}).

In this section, the measurement of the radial variation in the free
electron density in a sodium plasma column produced by laser resonance
saturation is described.  The neutral sodium density range of interest is
$10^{15} - 10^{17}$ cm$^{-3}$.  At comparable electron densities, lines of the $n^2D$-$3^2P$
series in sodium suffer significant electron Stark broadening (section
\ref{sec:2-2-1}).  We have elected to use the Stark broadening of the $4^2D$-$3^2P$
transition in sodium to measure the free electron density.  The broadening
of this transition, and its equivalent in potassium has been studied
extensively (Grunberg et al.\ 1976, Oettinger and Cooper 1969, Hohimer 1984,
1985, Konjevic 1985) providing us with confidence in using the tabulated
values for the electron Stark shifts and widths of Griem (1974) and
Dimitrijevi\'c and Sahal-Br\'echot (1985).  The electron densities are derived
from the local volume emission coefficients which in turn are obtained from
the Abel inversion of the lateral spectral radiances observed perpendicular
to the direction of laser propagation into the sodium vapor.

In the case of sodium, the visible wavelength spectral lines of
greatest intensity either terminate on the ground or resonance state.
Unfortunately, lines which terminate on the ground state will always suffer
severe radiation trapping, and usually are of minimal use for spectroscopic
measurements (an exception to this is the use of self-reversed Stark
broadened lines to determine the electron density in exploding lithium wire
plasmas - Ya'akobi 1969a, 1969b, 1969c, 1971).  Furthermore, under
conditions of laser resonance saturation, the lines terminating on either of
the $3^2P$ levels can be influenced by the intense radiation field of the laser
and suffer radiation trapping due to the large population of atoms which are
excited into the resonance level.  This appears to preclude measurements of
the free electron density during the period of laser radiation.  However,
this limitation might be avoided if infrared spectral lines terminating on
the $4^2S$ level are employed.

Lines that terminate on the $3^2P$ level can provide meaningful data after
the laser field has significantly diminished and the excess $3^2P$ resonance
state population density has had adequate time to decay by means of
spontaneous emission and electron collision quenching.  Since the electron
density and temperature will vary considerably across the plasma column
created by the laser pulse, it is necessary to undertake an elaborate series
of measurements across the plasma column, then perform an appropriate
inversion in order to ascertain the radial variation of $N_e$.  This inversion
was made much simpler by the assumption of cylindrical symmetry.

\subsection{Theory of the Measurement}
\label{sec:3-2-1}

The spectral radiance at frequency $\nu$, arising from the n to m
transition and observed in the x direction at some height y above the axis
of a cylindrically symmetric plasma column of radius R, under optically thin
conditions (equations (\ref{eq:2-72}) and (\ref{eq:2-83})) can be expresed in the form

\begin{equation}
J_{nm}(\nu,y) \;=\; \int_{-\sqrt{(R^2-y^2)}}^{\sqrt{(R^2-y^2)}}\varepsilon_{nm}(\nu,x,y)\,dx\;.
\label{eq:3-16}
\end{equation}

We can write

\begin{equation}
J_{nm}(\nu,y) \;=\; 2\int_{y}^{R}\frac{\varepsilon_{nm}(\nu,r)\,r}{\sqrt{(r^2-y^2)}}\,dr
\label{eq:3-17}
\end{equation}

by introducing $r^2 = x^2 + y^2$.  If the plasma emission is imaged onto the
entrance slit of a monochromator with the slit aligned in the direction
parallel to the cylinder's axis, then we can express the output current
signal (Amps) of a photomultiplier tube positioned at the exit slit in the
form

\begin{equation}
I_{nm}(\nu,y) \;=\; S_{nm}\int_{y-\Delta y/2}^{y+\Delta y/2}\int_{z-\Delta z/2}^{z+\Delta z/2}
\int_{-\infty}^{\infty} J_{nm}(\nu',y)\,T(\nu'-\nu)\,d\nu'\,dz'\,dy'\;,
\label{eq:3-18}
\end{equation}

where, for long focal lengths (f $\gg \Delta y$) and 1:1 imaging, $\Delta y$ represents the
entrance slit width and $\Delta z$ the slit height over which the plasma is assumed
to not vary appreciably.  $S_{nm}$ is the photomultiplier sensitivity
(Amps/Watt) in the vicinity of the frequency of the n to m transition (a
slowly varying function of $\nu$ over the range of the line spectra) and $T(\nu'-\nu)$
is the transmission function of the monochromator and input optics, and as
in equation \ref{eq:3-7} is described by a Gaussian function,

\begin{equation}
T(\nu'-\nu) \;=\; \frac{K}{\gamma_T}\Bigl\{\frac{\ln 2}{\pi}\Bigr\}^{1/2}
\exp\left[-\Bigl(\frac{\nu'-\nu}{\gamma_T}\Bigr)^2\ln 2\right],
\label{eq:3-19}
\end{equation}

with $\gamma_T$ the instrument halfwidth half-maximum (HWHM) spectral width and K
the filter function defined by the relation,

\begin{equation}
K \;=\; \int_{-\infty}^{\infty} T(\nu'-\nu)\,d\nu'\;.
\label{eq:3-20}
\end{equation}

Since $T(\nu'-\nu)$ is independent of r, then equation (\ref{eq:3-18}) can be expressed as

\begin{equation}
I_{nm}(\nu,y) \;=\; S_{nm}\int_{\Delta y}\int_{\Delta z}
\left\{2\int_{y}^{R}\frac{\varepsilon_{nm}(\nu,r)*T(\nu)\,r\,dr}{\sqrt{(r^2-y^2)}}\right\}dz\,dy\;,
\label{eq:3-21}
\end{equation}

where $*$ denotes the convolution operation, that is

\begin{equation}
\varepsilon_{nm}(\nu,r)*T(\nu) \;=\; \int_{-\infty}^{\infty}\varepsilon_{nm}(\nu',r)\,T(\nu'-\nu)\,d\nu\;.
\label{eq:3-22}
\end{equation}

The Abel transformation (section \ref{sec:2-6}) can then be used to provide the
corresponding radial distribution of the volume emission coefficient
convoluted with $T(\nu)$,

\begin{equation}
\xi(\nu,r) \;=\; \varepsilon_{nm}(\nu,r)*T(\nu) \;=\;
\frac{-1}{S_{nm}\Delta y\,\Delta z}\Bigl(\frac{1}{\pi}\Bigr)
\int_{r}^{R}\left\{\frac{dI_{nm}(\nu,y)}{dy}\right\}\frac{dy}{\sqrt{(y^2-r^2)}}\;.
\label{eq:3-23}
\end{equation}

Equation (\ref{eq:2-73}) implies that $\varepsilon_{nm}(\nu,r)$ is directly proportional to the
atomic line profile $\pounds_{nm}(\nu,r)$.  For our experimental conditions, Doppler,
resonance and Van der Waals broadening of the $4^2D$ - $3^2P$ multiplet transition
were negligible in comparison to electron Stark broadening.  The
contribution to the broadening of the lines by the quasi-static fields of
the ions is evident as a red shift which can lead under our conditions, at
most, to a 25\% increase in the linewidth (Griem 1974).  If we ignore the
effect of the ions, the Stark profile is Lorentzian in shape with a
fullwidth half-maximum of ${}^{s}\gamma$ and shift ${}^{s}d$, both linearly dependent on the
free electron density $N_e$.  $N_e(r)$ can be evaluated by suitably fitting a
Lorentzian function convolved with a Gaussian [to represent the instrument
response function in accordance with equation (\ref{eq:3-19})] to the Abel inverted
radial emission coefficient $\varepsilon_{nm}(\nu,r)$.  The results of the semi-classical
calculations of Griem (1974) and Dimitrijevi\'c and Sahal-Br\'echot (1985) for
${}^{s}\gamma$ and ${}^{s}d$ were presented and compared in the previous chapter (Table (\ref{tab:2-1}))
for a wide range of electron temperatures.

The Abel inversion is performed using the procedure proposed by Deutsch
(1983) described in some detail in section \ref{sec:2-6}.  The measured lateral
distribution of the line of sight spectral emission from the plasma is fit
to the function described by equation (\ref{eq:2-109}),

\begin{equation}
I_{nm}(\nu,y) \;=\; \Bigl\{\sum_{k=0}^{2} a_k(\nu)\,y^{2k}\Bigr\}\exp\{-\alpha(\nu)y^2\}\;.
\label{eq:3-24}
\end{equation}

In this equation, we have explicitly written the frequency dependence
of the fitting parameters.  $I_{nm}(\nu,y)$ can then be inverted at each
predescribed frequency value using the procedure described in section \ref{sec:2-6}.
The radial distribution of $N_e$ can then be obtained from the fitting of a
Voigt profile to the inverted spectral profiles $\xi(\nu,r)$.

The linear coefficients in equation (\ref{eq:3-24}) are obtained by the method
of linear least squares (Appendix F) and $\alpha$ is selected to minimize the least
squares residual.  This approach allows us to calculate the coefficients $a_k$
and the variance in the coefficients $\Delta a_k^{2}$ in terms of the basis
functions,

\begin{equation}
\phi_k \;=\; y^{2k}\exp(-\alpha y^2)\;,
\label{eq:3-25}
\end{equation}

and the variance in the experimental data,

\begin{equation}
\Delta a_k^{2} \;\propto\; \sigma^2\;.
\label{eq:3-26}
\end{equation}

Here, $\sigma$ describes the standard deviation of the spread in the experimental
data values $\{I^{*}(y_i)\}$ at particular lateral positions and is approximately
the same for all lateral positions $y_i$.

We can express the elemental change in $\xi(\nu,r)$ as

\begin{equation}
\Delta\xi(\nu,r) \;=\; \sum_{k=0}^{2}\frac{\partial\xi(\nu,r)}{\partial a_k}\Delta a_k
+ \frac{\partial\xi(\nu,r)}{\partial\alpha}\Delta\alpha\;,
\label{eq:3-27}
\end{equation}

and assuming that the functional form of $\xi(\nu,r)$ is relatively insensitive to
small changes in $\alpha$ as compared to changes in the linear coefficients, then
using equation (\ref{eq:3-27}), we can write

\begin{equation}
\Delta\xi(\nu,r) \;=\; \frac{\partial\xi(\nu,r)}{\partial a_0}\Delta a_0
+ \frac{\partial\xi(\nu,r)}{\partial a_1}\Delta a_1
+ \frac{\partial\xi(\nu,r)}{\partial a_2}\Delta a_2\;,
\label{eq:3-28}
\end{equation}

which using equations (\ref{eq:2-110}) through (\ref{eq:2-113}) (only we have $\varepsilon_{nm}*T$ as opposed
to just $\varepsilon_{nm}$ on the left hand side) and that $\partial f_i/\partial a_j = 0$ for all i,j, reduces
to

\begin{equation}
\Delta\xi(\nu,r) \;=\; -\frac{1}{\pi}\bigl\{-2f_0\Delta a_0 + 2(f_0-\alpha f_1)\Delta a_1
+ 2(2f_1-\alpha f_2)\Delta a_2\bigr\},
\label{eq:3-29}
\end{equation}

with $f_0(r)$, $f_1(r)$ and $f_2(r)$ given by equation (\ref{eq:2-112}).

The value $\Delta\xi(\nu,r)$ represents the degree of uncertainty in the inverted
spectrum at frequency $\nu$ and position r arising from the experimental
uncertainty in the value of $I(\nu,y)$ which may sometimes reflect the scatter
in the experimental spectra.  It is difficult to estimate the degree of
uncertainty arising from how well the chosen function does or does not fit
the experimental data.  In order to get a feel for this, the next section
describes a simulated experiment whereby the spectral radiances are
generated by the model described in section \ref{sec:2-5-1} and the inversion is
carried through on data which has superimposed on it, random error which is
comparable to that observed in experiment.  The plasma conditions have been
suitably chosen to represent that expected in our experiments.

\subsection{Numerical Simulation}
\label{sec:3-2-2}

Two somewhat different cases of electron density and temperature
distributions have been investigated.  The first deals with the conditions
given in section (\ref{sec:2-5}) with the electron temperature distribution described
by equation (\ref{eq:2-85}).  Three simulated spectral radiances at y=0, 1.5 and 2.5
mm are shown as the data in figure \ref{fig:3-4}.  The y variation of the data at four
selected frequency values are illustrated in figure \ref{fig:3-5}, along with the
corresponding least squares polynomial fit.  A random scatter (with $\sigma \approx$ 2\%
of peak signal) was superimposed on the simulated y-data, to represent that
expected from experiment.  The inverted radial distributions are presented
in figure \ref{fig:3-6}.  As expected, the radial variations peak at approximately 1.9
mm, where the maximum population density was observed for this set of
conditions.  The error bars illustrated in figure \ref{fig:3-6} are computed on the
basis of the superimposed scatter.  The reconstructed spectra at several
radial values are illustrated in figure \ref{fig:3-7}.  The large scatter (greater
than the error bars designated in figure \ref{fig:3-6}) for small r reflects the
sensitivity of the inversion to the accuracy of the function fit near r=0
(discussed below).  In a typical analysis, the spectra would be fitted to a
Voigt profile, describing the Lorentzian shaped emission convoluted with a
Gaussian shaped instrument response function (solid lines in figure \ref{fig:3-7}).
In this case however, the data represent emission that has not passed
through a monochromator, and therefore, simple Lorentzian profiles were
utilized to represent the inverted profiles (solid lines).  The electron
densities and temperatures used for the generation of the solid lines in the
figure are listed.  The spectral profiles are assumed to be principally
electron Stark broadened.

\figstub{fig:3-4}{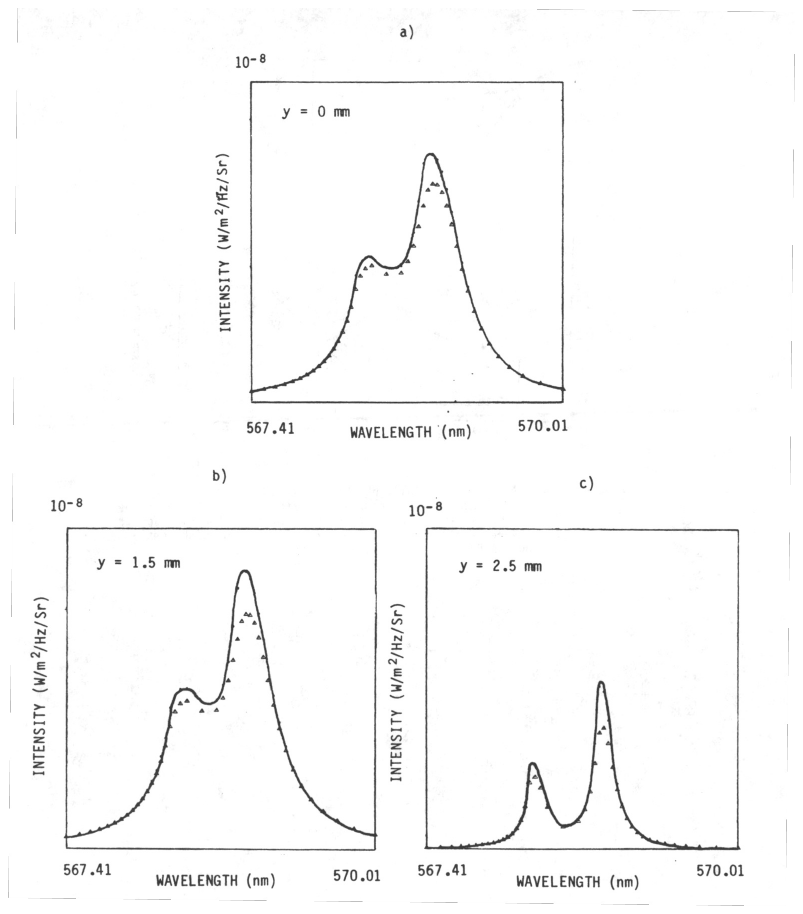}{Three simulated spectral radiances at y=0, 1.5 and 2.5 mm.}

\figstub{fig:3-5}{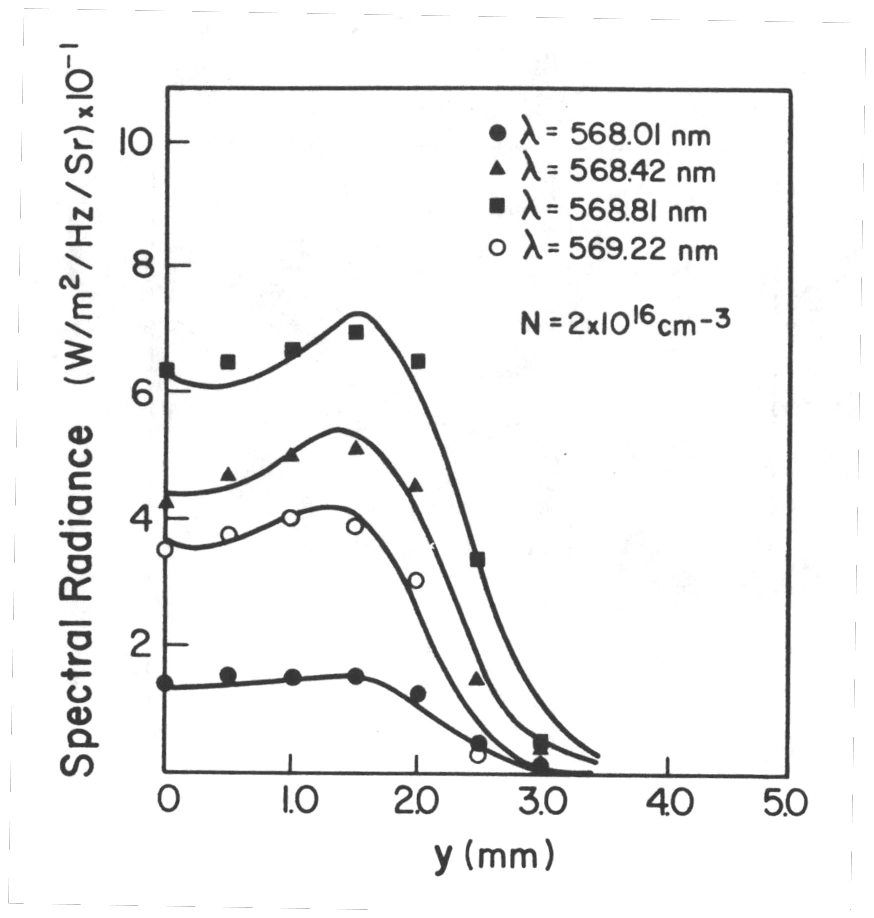}{The y variation of the data at four selected frequency
values, along with the corresponding least squares polynomial fit.}

\figstub{fig:3-6}{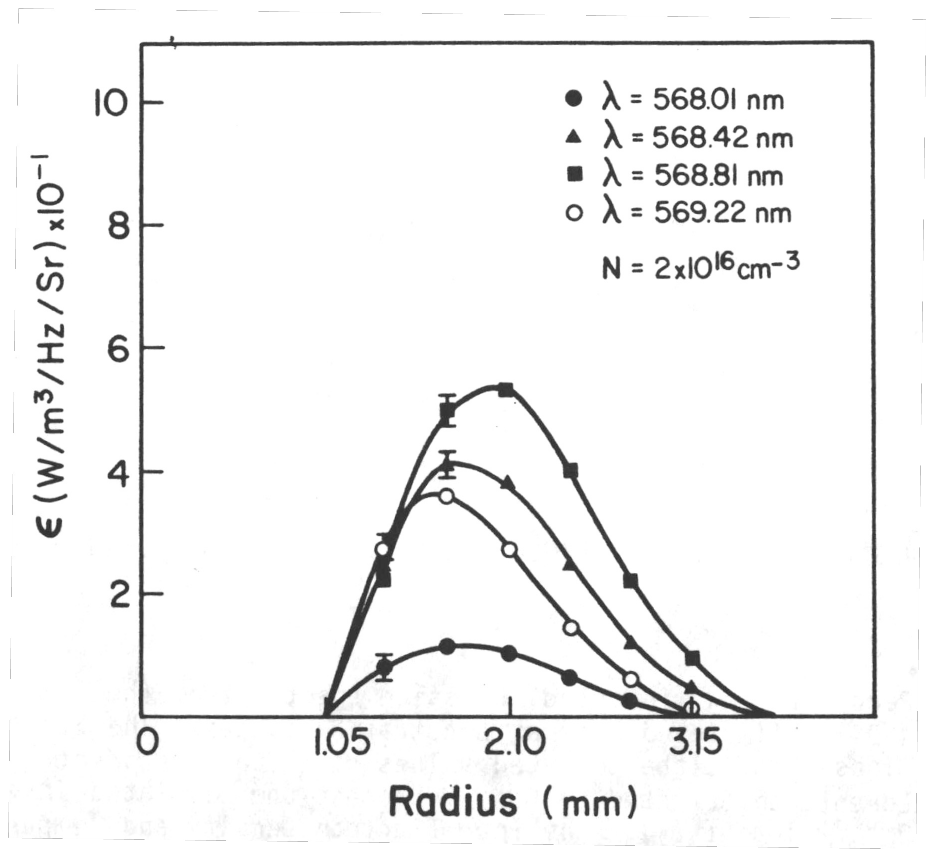}{Inverted radial distributions, with error bars computed on
the basis of the superimposed scatter.}

\figstub{fig:3-7}{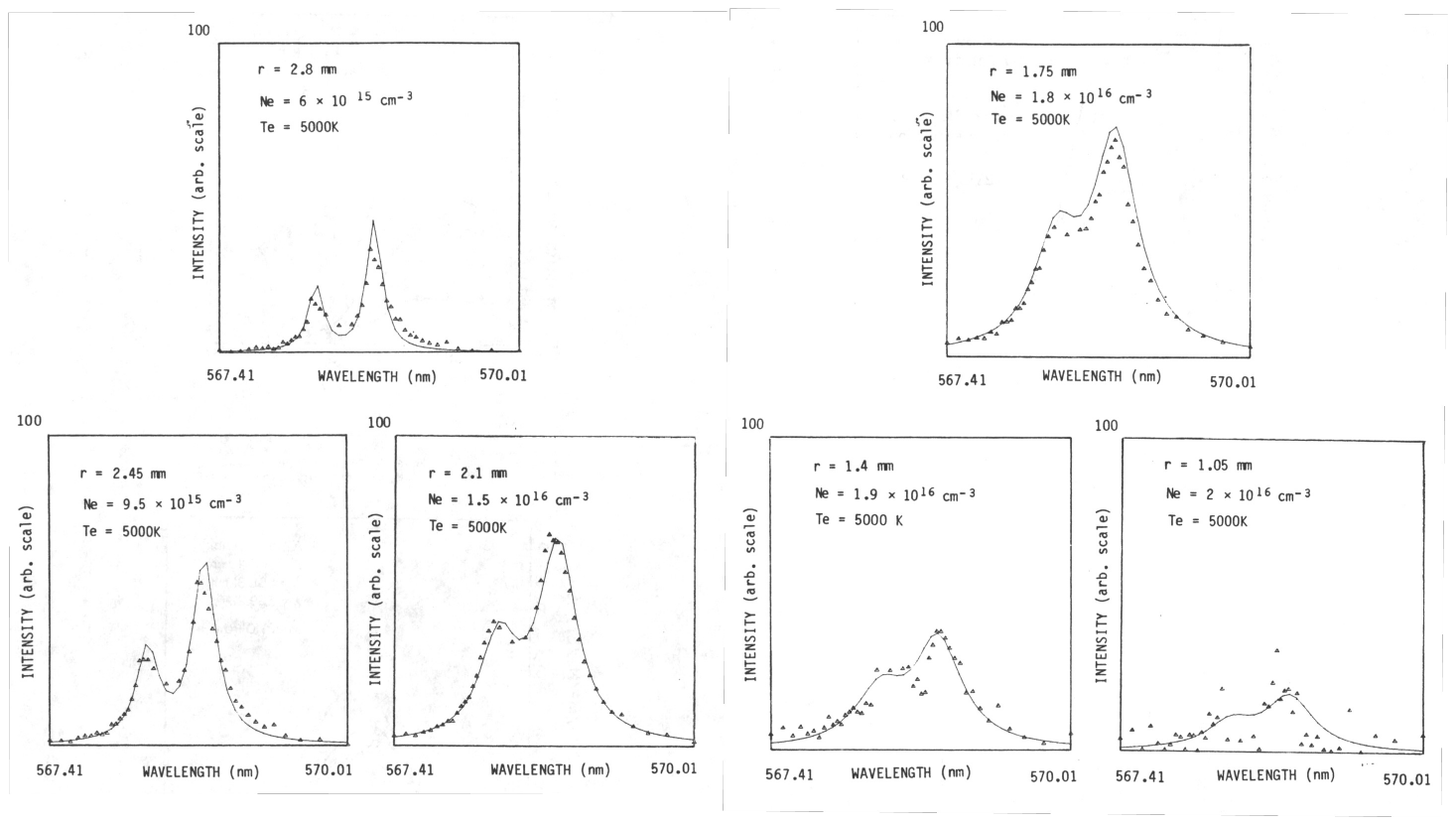}{Reconstructed spectra at several radial values.}

The increased scatter in the data as one approaches r=0 can be
explained with reference to figure \ref{fig:3-5}.  The form of the Abel transformation
suggests that it is sensitive to intensity gradients (the integrand is
proprtional to $dI(y)/dy$).  For the data presented in figure \ref{fig:3-5}, the scatter
band is negligible and the deviation of the points from the polynomial line
is simply a result of the ability of the chosen polynomial form to fit the
experimental data.  This deviation increases and becomes comparable to the
increase in the actual data as one approaches the centre of the column.  For
a situation such as this which clearly suggests that minimal emission arises
from the plasma core, it is difficult to extract any useful information
about the electron density and temperature in the core other than that the
neutral species population density is highly depleted in this region.  The
useful data lies in the 1.4-2.8 mm region.  The electron density distribution
derived from the inverted profiles is shown in figure \ref{fig:3-8}, along with the
actual distribution used to generate the data.  Although we cannot extract
information about the plasma core, the agreement is very good elsewhere and
suggests that the original electron density can be retrieved, especially in
regions where there is a strong electron density gradient.

\figstub{fig:3-8}{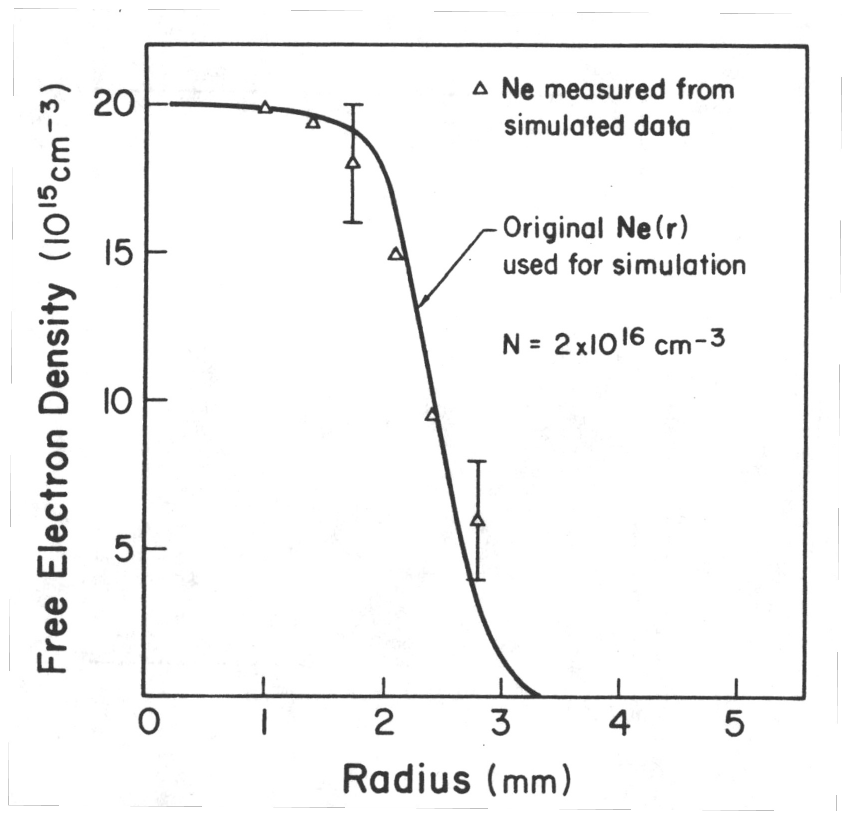}{Electron density distribution derived from the inverted
profiles, along with the actual distribution used to generate the data.}

The second case investigated was one where the electron temperature
distribution takes on the form

\begin{equation}
T_e(r) \;=\; 2000 + 3000\exp(-r/r_0^{\,2})\;,
\label{eq:3-30}
\end{equation}

with N = $2\times10^{16}$ cm$^{-3}$ and $r_0$=2.5 mm as in the previous example.  Here, the
excited state population densities peak at r=0 (see figure \ref{fig:3-9}), suggesting
that a more meaningful measurement of the electron density can be made of
the plasma core.  The y-variation in the spectral radiance at four selected
frequency values, are illustrated in figure \ref{fig:3-10}.  Here, the fitting
polynomial represents the actual data, somewhat more accurately.  The data
in figure \ref{fig:3-11} show the corresponding inverted functions.  As expected, the
volume emission coefficients peak in the plasma core for this temperature
distribution.  A reconstruction of the spectra at selected radial positions
(figure \ref{fig:3-12}) gives rise to an acceptable scatter at r=0 and the radial
variation in the free electron density derived from these spectra retrieves
the original electron density distribution fairly well, as illustrated by
the results in figure \ref{fig:3-13}.

\figstub{fig:3-9}{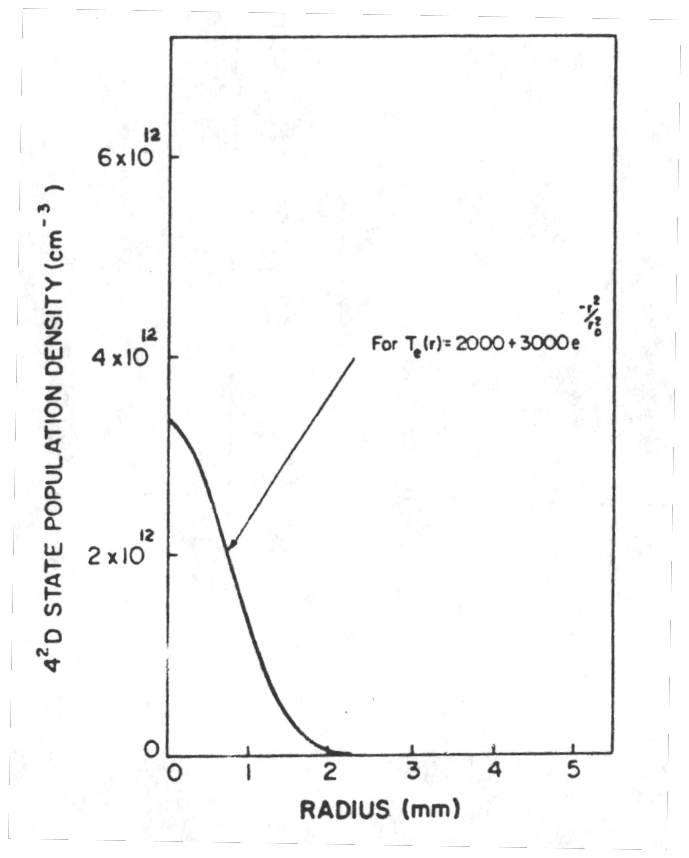}{Excited state population densities for the electron
temperature distribution of equation (\ref{eq:3-30}).}

\figstub{fig:3-10}{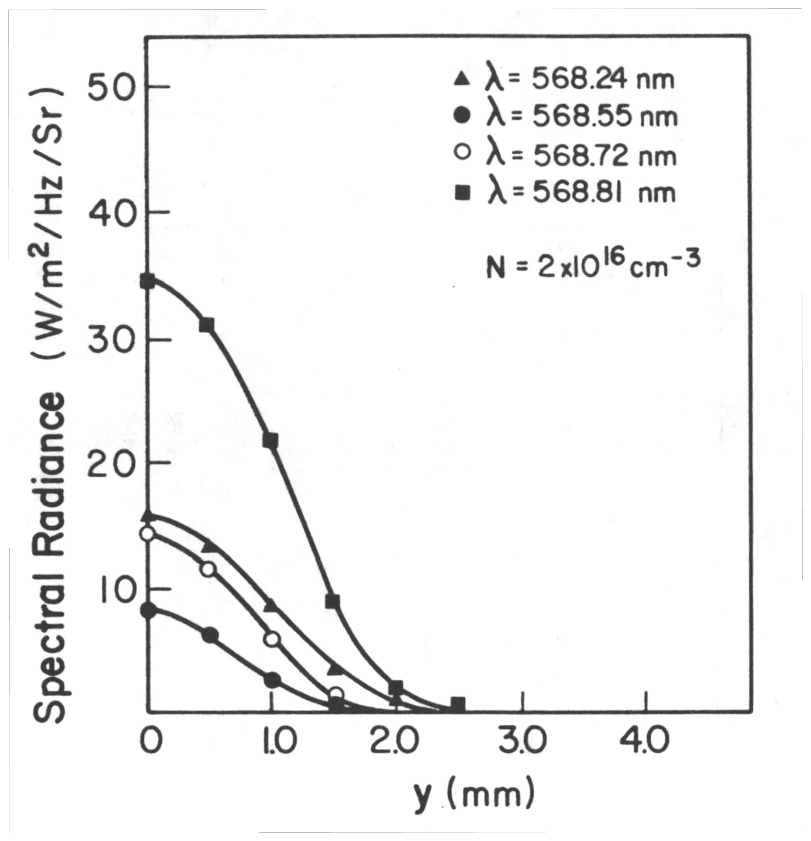}{The y-variation in the spectral radiance at four selected
frequency values.}

\figstub{fig:3-11}{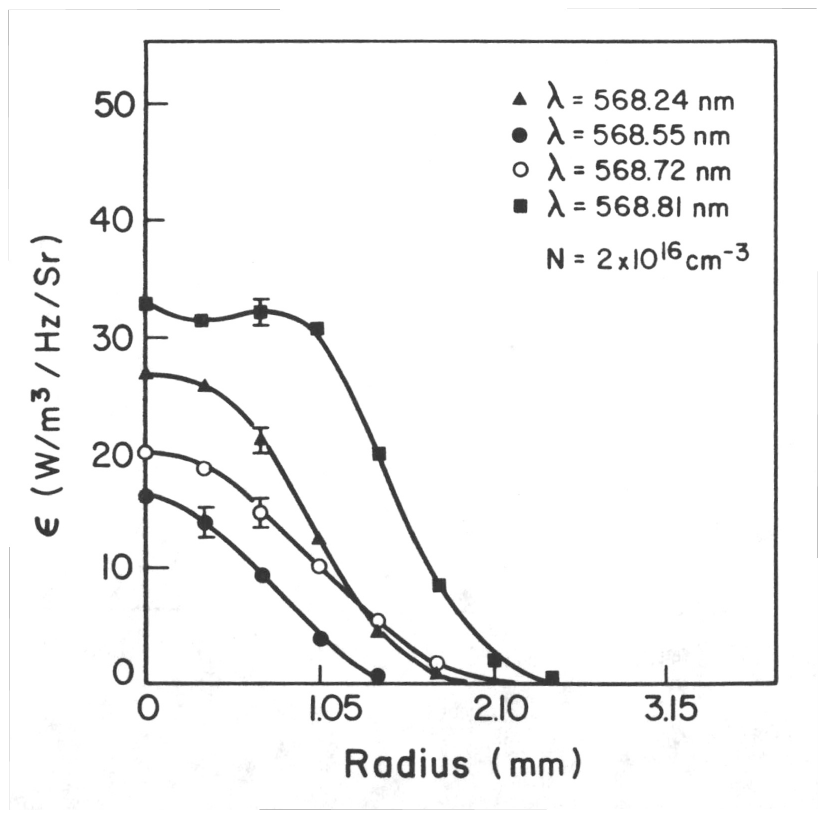}{Corresponding inverted functions.}

\figstub{fig:3-12}{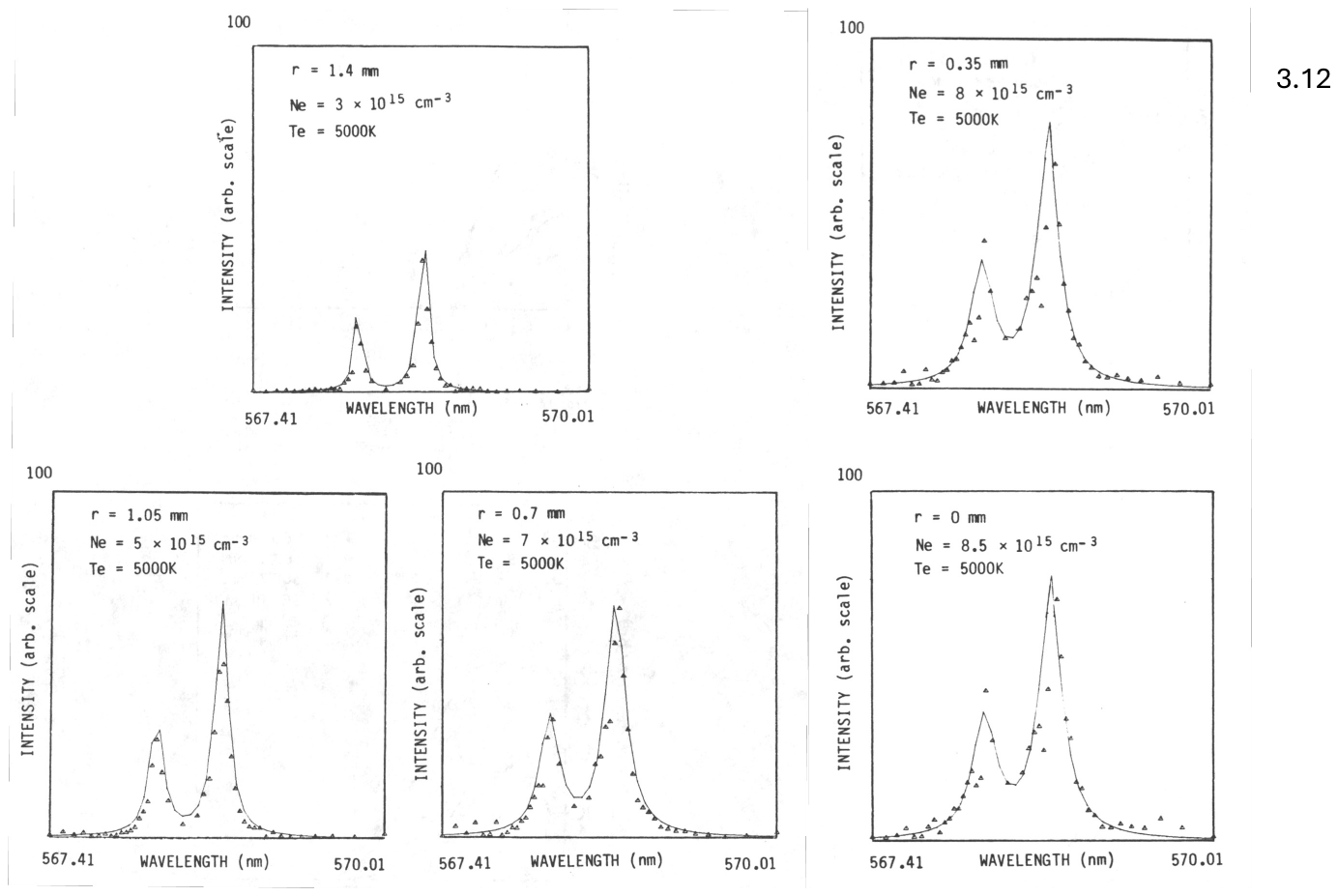}{Reconstruction of the spectra at selected radial
positions.}

\figstub{fig:3-13}{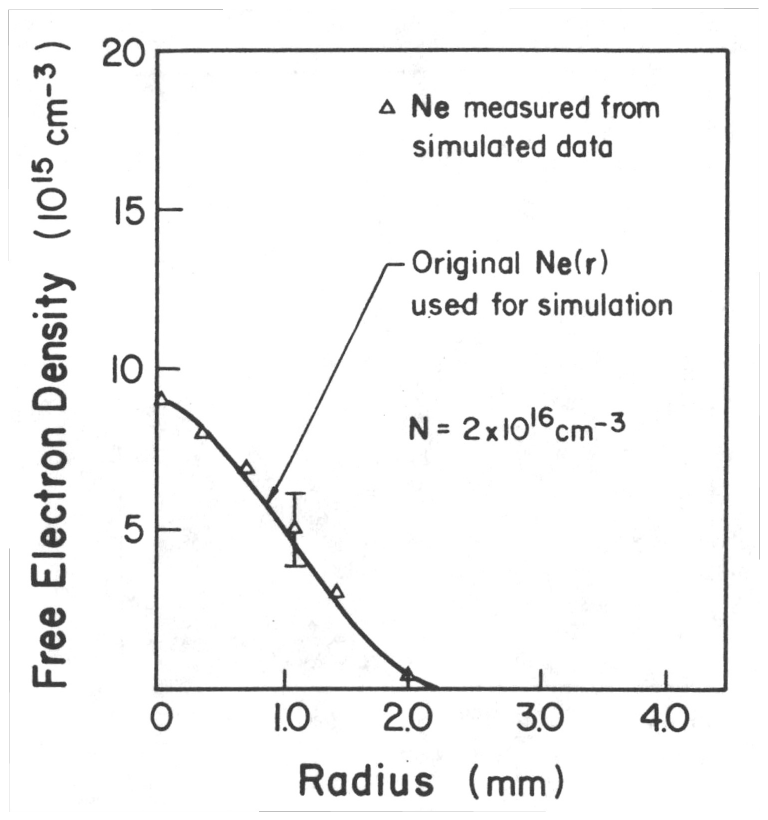}{Radial variation in the free electron density derived from
these spectra, compared with the original electron density distribution.}

\section{Line Emission Based Measurements of the Electron Temperature}
\label{sec:3-3}

Line emission based measurements of the electron temperature can be
classified under four major catagories: (i) absolute line intensity
measurements, (ii) relative intensity of emission lines from the same
ionization stage, (iii) relative measurements of emission lines from
different ionization stages and (iv) measurements of the shift to width
ratios of the electron Stark broadened emission line profiles (Burgess
and Cooper 1965).  When working on the nanosecond time scales, signal to
noise considerations usually demand the use of relatively strong emission
lines.  In alkali metals, the second ionization stage has a relatively large
energy gap between the ground and first excited state.  This generally
results in weak emission from the ion, the third method is therefore a
relatively unlikely choice.  Shift to width ratios are generally not
sensitive to electron temperature changes expected in our work (2000-7000K)
and along with the first method, would require expensive and accurately
calibrated detection equipment.

Electron temperature measurements from the relative intensity of
emission lines from the same ionization stage appear as the only supplement
to the electron temperatures derived from the assumption of Saha
equilibrium (having measured $N_e$ and the neutral population density N).
There are potentially three major drawbacks to this technique, that one
should be clearly aware of before drawing any conclusions from the results
derived.  The first of course, the assumption that LTE exists between the
upper states of each transitions involved, may not be valid.  Second,
the plasma may not be optically thin to these transitions.  Finally, unless
$kT_e \ll E_{nq}$, (where $E_{nq}$ represents the energy seperation between the upper
states of the n to m and q to p-transitions), the accuracy in the measured
electron temperature is at best equal to the sum of the accuracies in the
relative intensity and ratio of the oscillator strength of the emission
lines considered.  The effects of plasma inhomogeneities and optical depth
on electron temperatures derived from Boltzmann plots of line emission have
been discussed in a recent article by Cappelli and Measures (1987a, see also
Appendix G).

\subsection{Electron Temperatures from Homogeneous Plasma Sources}
\label{sec:3-3-1}

In accordance with equation (\ref{eq:3-16}) and the formulation derived in
section \ref{sec:3-2}, we can express the spectral radiance emitted from a homogeneous
plasma of length $\ell$, as

\begin{equation}
J_{nm}(\nu) \;=\; \varepsilon_{nm}(\nu)\,\ell\;,
\label{eq:3-31}
\end{equation}

where we are assuming that the source is invariant to y and z displacements
perpendicular to the direction of observation x.  Experimentally, the
radiance $J_{nm}$ is measured by collecting the spectral emission over a finite
bandwidth $\Delta\nu_I$ centred about the unperturbed line centre frequency $\nu_{nm}$.
Using equation (\ref{eq:3-31}) above yields the radiance, that is

\begin{equation}
J_{nm} \;=\; \frac{h\nu_{nm}}{4\pi}N_n A_{nm}\ell
\int_{\nu_{nm}-\Delta\nu_I/2}^{\nu_{nm}+\Delta\nu_I/2}\pounds_{nm}(\nu)\,d\nu\;.
\label{eq:3-32}
\end{equation}

The bandwidth $\Delta\nu_I$ is determined by the size of the exit slit and the
dispersion of the monochromator grating.  As introduced in section \ref{sec:3-2}, we
can express the output current from a photomultiplier tube positioned at the
exit slit as

\begin{equation}
I_{nm} \;=\; S_{nm}J_{nm}\Delta y\,\Delta z\;,
\label{eq:3-33}
\end{equation}

with $\Delta y$ and $\Delta z$ being the entrance slit dimensions for 1:1 imaging.
Substitution of equation (\ref{eq:3-32}) into (\ref{eq:3-33}), leaves,

\begin{equation}
I_{nm} \;=\; \frac{h\nu_{nm}S_{nm}A_{nm}N_n\ell}{4\pi}\,\Theta_{nm}(\Delta\nu_I)\;,
\label{eq:3-34}
\end{equation}

where we have defined

\begin{equation}
\Theta_{nm}(\Delta\nu_I) \;=\;
\int_{\nu_{nm}-\Delta\nu_I/2}^{\nu_{nm}+\Delta\nu_I/2}\pounds_{nm}(\nu)\,d\nu\;.
\label{eq:3-35}
\end{equation}

In most cases, $\Delta\nu_I$ is chosen such that $\Delta\nu_I \gg \gamma$, $\gamma$ being the FWHM of the
atomic emission profiles, in which case, $\Theta_{nm}\approx 1$.

If we assume that the upper state n is in LTE with the $3^2P$ resonance
state, then the upper state population densities are in a Boltzmann
distribution with $N_2$ and, using equations (\ref{eq:3-32}) to (\ref{eq:3-35}), it is relatively
easy to show that

\begin{equation}
\ln B \;=\; C - \frac{E_{n2}}{kT_e}\;,
\label{eq:3-36}
\end{equation}

with

\begin{equation}
B \;=\; \frac{I_{nm}}{(S_{nm}/S_{qp})\,\nu_{nm}A_{nm}g_n\Theta_{nm}}\;,
\label{eq:3-37}
\end{equation}

and

\begin{equation}
C \;=\; \ln\left\{\frac{S_{qp}\,\Delta y\,\Delta z\,\ell N_2}{4\pi g_2}\right\}.
\label{eq:3-38}
\end{equation}

We have intentionally expressed the photomultiplier response relative to the
response at a reference transition (in this case q$\rightarrow$p), $S_{qp}$.  Plotting the
left side of equation (\ref{eq:3-36}) against $E_{n2}$ for the spectral series $n^2D$-$3^2P$ (n
= 4,5,6) and $n^2S$-$3^2P$ (n= 5,6,7) should result in a straight line with slope
($-1/kT_e$) and intercept C, as C is a constant, independent of the transition.
A measurement of C can lead to an estimate of $N_2$, providing that an absolute
calibration of the photomultiplier tube and monochromator is performed.  The
assumption that $\Theta_{nm}$= 1 can lead to significant deviations from the expected
straight line (Cappelli and Measures 1987a) which one should be aware of,
especially when the lines used suffer significant electron Stark
broadening.

\subsection{Electron Temperature Measurements of a Cylindrically Symmetric Plasma}
\label{sec:3-3-2}

\subsubsection{Optically Thin Source}
\label{sec:3-3-2-1}

We can begin by expressing $I_{nm}(\nu,y)$ as in equation (\ref{eq:3-18}) using
equation (\ref{eq:3-16}) and the definition of $\varepsilon_{nm}(\nu,x,y)$ (equation (\ref{eq:2-73})), that is

\begin{equation}
I_{nm}(\nu,y) \;=\; \frac{h\nu_{nm}S_{nm}\Delta y\,\Delta z A_{nm}}{4\pi}
\int_{-\sqrt{(R^2-y^2)}}^{\sqrt{(R^2-y^2)}}\pounds_{nm}(\nu,x,y)N_n(x)\,dx\;.
\label{eq:3-39}
\end{equation}

In accordance with the previous section, we can spectrally integrate
equation (\ref{eq:3-39}) over the interval $\Delta\nu_I$ centred about $\nu_{nm}$ and write

\begin{align}
I_{nm}(y,\Delta\nu_I) &= \int_{\nu_{nm}-\Delta\nu_I/2}^{\nu_{nm}+\Delta\nu_I/2} I_{nm}(\nu,y)\,d\nu
\nonumber\\[0.5em]
&= \frac{h\nu_{nm}S_{nm}\Delta y\,\Delta z A_{nm}}{4\pi}
\int_{-\sqrt{(R^2-y^2)}}^{\sqrt{(R^2-y^2)}} N_n(x)\Theta_{nm}(x)\,dx\;,
\label{eq:3-40}
\end{align}

and in the limit $\Delta\nu_I \gg \gamma$ (for all x), then $\Theta_{nm}(x)\approx 1$, and equation (\ref{eq:3-40})
reduces to

\begin{equation}
I_{nm}(y) \;=\; \frac{h\nu_{nm}S_{nm}\Delta y\,\Delta z A_{nm}}{4\pi}
\int_{-\sqrt{(R^2-y^2)}}^{\sqrt{(R^2-y^2)}} N_n(x)\,dx\;.
\label{eq:3-41}
\end{equation}

The functional form of $\Theta_{nm}(x)$ is of course difficult to ascertain,
therefore, it is important that the exit slit width is selected accordingly,
such that $\Theta_{nm}$ is approximately unity and equation (\ref{eq:3-41}) applies.  If we now
perform an Abel transformation on $I_{nm}(y)$, we arrive at

\begin{equation}
N_n(r) \;=\; \frac{4\pi}{h\nu_{nm}S_{nm}\Delta y\,\Delta z A_{nm}}
\left\{-\frac{1}{\pi}\int_{r}^{R}\left[\frac{dI_{nm}(y)}{dy}\right]
\frac{dy}{\sqrt{(y^2-r^2)}}\right\}.
\label{eq:3-42}
\end{equation}

If the plasma is in LTE with $N_n$ in a Boltzmann ratio with $N_2$, then the
radial distribution in the electron temperature $T_e(r)$, can be described as

\begin{equation}
\frac{E_{nq}}{kT_e(r)} \;=\; f(r) \;=\; -\ln
\left\{\frac{g_q\nu_{qp}A_{qp}S_{qp}}{g_n\nu_{nm}A_{nm}S_{nm}}\;
\frac{\displaystyle\int_{r}^{R}\frac{I'_{nm}\,dy}{\sqrt{(y^2-r^2)}}}
{\displaystyle\int_{r}^{R}\frac{I'_{qp}\,dy}{\sqrt{(y^2-r^2)}}}\right\}.
\label{eq:3-43}
\end{equation}

Here, $I'_{nm}$ denotes differentiation with respect to lateral position y.  The
elemental change in $f(r)$ can be expressed as

\begin{equation}
\Delta f(r) \;=\; \Bigl|\frac{\partial f}{\partial S}\Bigr|\Delta S
+ \Bigl|\frac{\partial f}{\partial A}\Bigr|\Delta A
+ \Bigl|\frac{\partial f}{\partial I}\Bigr|\Delta I\;,
\label{eq:3-44}
\end{equation}

with

\begin{equation}
S \;=\; S_{qp}/S_{nm},
\tag{3.45a}\label{eq:3-45a}
\end{equation}
\begin{equation}
A \;=\; A_{qp}/A_{nm}\;,
\tag{3.45b}\label{eq:3-45b}
\end{equation}

and

\begin{equation}
I(r) \;=\; I_{nm}(r)/I_{qp}(r)\;.
\tag{3.45c}\label{eq:3-45c}
\end{equation}

Equations (\ref{eq:3-43}) through (3.45) can be combined, giving

\setcounter{equation}{45}
\begin{equation}
\left\{\frac{\Delta T_e(r)}{T_e(r)}\right\} \;=\;
\left\{\frac{kT_e(r)}{E_{nq}}\right\}
\left[\frac{\Delta S}{S}+\frac{\Delta A}{A}+\frac{\Delta I}{I}\right],
\label{eq:3-46}
\end{equation}

which explicitly shows that the effects of uncertainty in S, A and I(r) on
the error in $T_e(r)$, can be minimized if the upper state energy separations
$E_{nq}$ are as high as possible.

\subsubsection*{Numerical Simulation}
\addcontentsline{toc}{subsubsection}{\quad Numerical Simulation}

The LTE model described in section \ref{sec:2-5}, in conjunction with the
radiative transfer code, was used to simulate the spectral radiance versus
lateral position y, under conditions of neutral density N = $10^{16}$ cm$^{-3}$,
R = 3.25 mm and $\Lambda$ = 0.005.  The radial temperature distribution used in the
simulation is Gaussian in shape ($T_e(r) = T_e(0)\exp(-r^2/r_0^{\,2})$ with $T_e(0)$ =
5140K and $r_0$ = 3.25 mm).  In accordance with the above formulation, we have
solved the radiative transfer equation assuming an optically thin source
(equation (\ref{eq:3-16})) and extended the spectral integration such that $\Theta$ = 1.
The inverted radial distribution of the volume emission coefficient derived
from the lateral radiances of the $4^2D$-$3^2P$ and $6^2D$-$3^2P$ transitions are shown
as data points in figure \ref{fig:3-14}.  The ratio of these curves is solely
dependent on the electron temperature and when the ratio of the Abel
inverted simulated lateral radiances is computed, the original temperature
distribution used for the simulation is to a great extent retrieved.  Figure
\ref{fig:3-15} illustrates the radial variation in $T_e$ originally used for the
simulation to generate the lateral radiances, along with the electron
temperature derived using the Abel inversion on these lateral radiances.
The deviation in the radial wing is taken to arise from the accuracy of the
polynomial fit to the lateral radiance.  Although the test has not been
performed on data with scatter introduced, this exercise does indicate that
the inversion routine is behaving and that the method can lead to a useful
measurement of $T_e(r)$.

\figstub{fig:3-14}{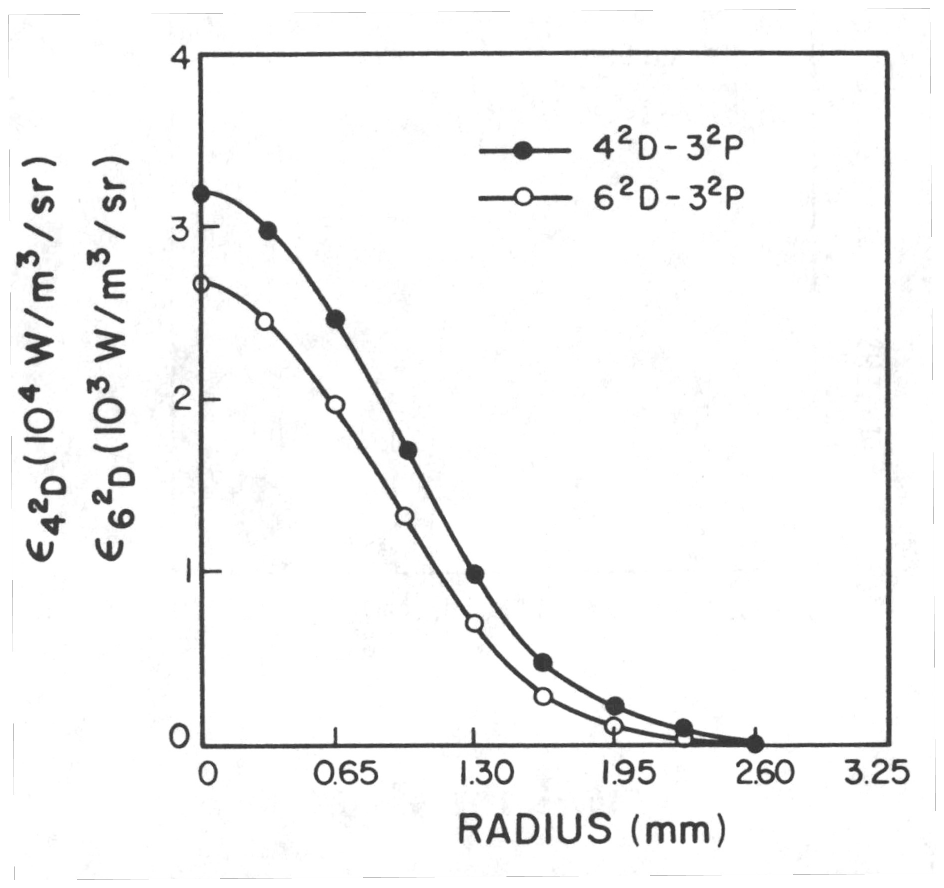}{Inverted radial distribution of the volume emission
coefficient derived from the lateral radiances of the $4^2D$-$3^2P$ and $6^2D$-$3^2P$
transitions.}

\figstub{fig:3-15}{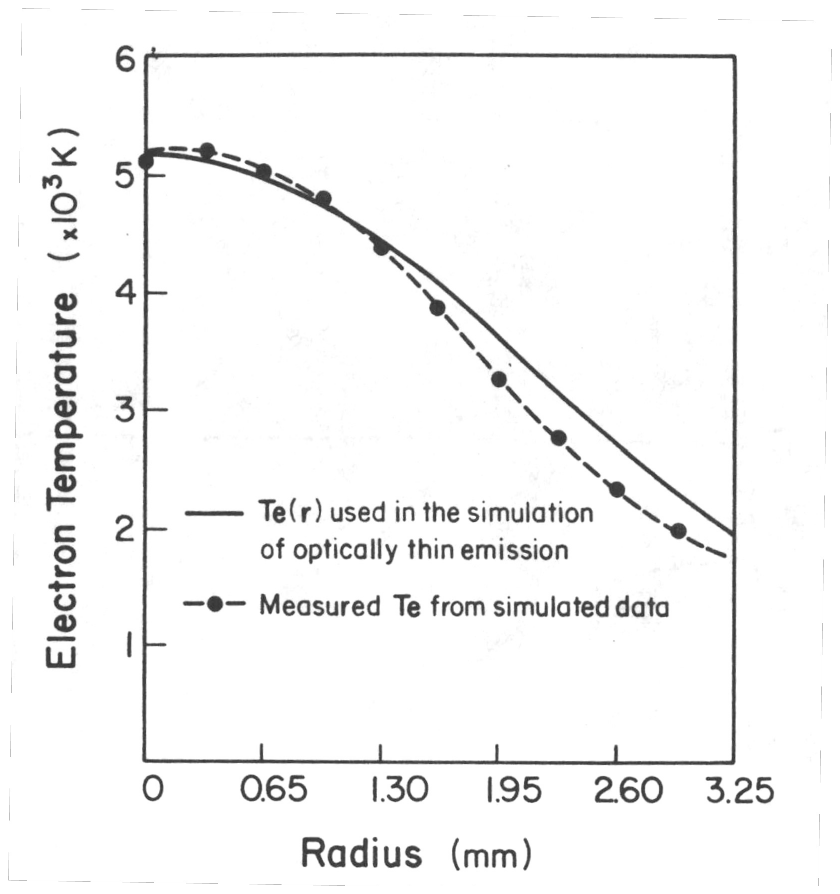}{Radial variation in $T_e$ originally used for the simulation,
along with the electron temperature derived using the Abel inversion.}

In practice, the measured data may be affected by self-absorption.  To
illustrate this effect, figure \ref{fig:3-16} shows the electron temperature
distribution that would be obtained if the full solution of the equation of
radiative transfer were used to generate the lateral radiances.  In
comparison with that obtained using the optically thin source, there is an
overprediction of as much as 25\% in the radial wing.  This arises from the
fact that emission from the $4^2D$-$3^2P$ transition would be self-absorbed to a
greater extent than that from the $6^2D$-$3^2P$ transition.  This would be
interpreted as an underprediction in the $4^2D$ state population density,
leading to higher overall electron temperatures.

\figstub{fig:3-16}{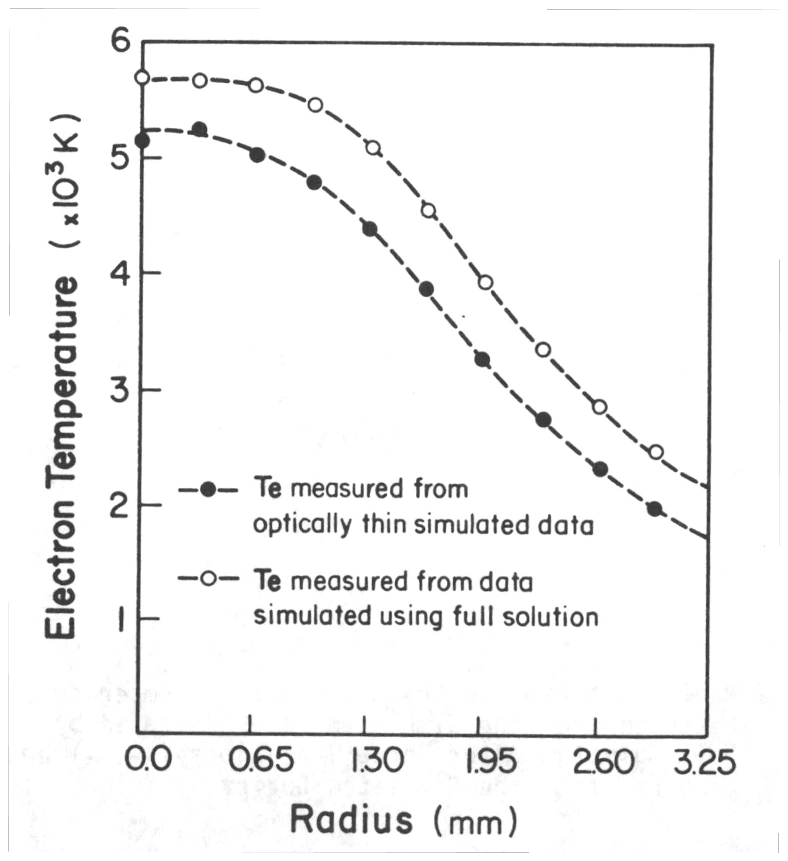}{Electron temperature distribution obtained if the full
solution of the equation of radiative transfer were used to generate the
lateral radiances.}

The section to follow describes a novel iterative scheme which can be
utilized to derive the radial electron temperature distribution from data
suffering from second order self-absorption losses.

\subsubsection{Correction for Finite Optical Depth}
\label{sec:3-3-2-2}

Techniques for measuring the radial distribution of $T_e$ in an
inhomogeneous, cylindrically symmetric, optically thick source by correcting
for self-absorption, in conjunction with the Abel transformation, has been
used previously by Bober and Tanken (1969).  Iterative schemes have been
shown to converge onto a solution (Elder et al.\ 1965, Young 1981), however,
controversy exists wether the technique of Elder et al.\ (1965) converges
onto the correct solution (Young 1981).

Presented in this section, is an iterative algorithm based on the two
line ratio technique described in the previous section.  Subsequent steps in
the iteration require using the $T_e(r)$ estimated from the previous step into
the full solution of the radiative transfer equation.  The ratio of the
radiance generated using the full solution and the optically thin solution,
is used as a correction factor for the subsequent iteration.  As shall be
shown for an example used in the previous section, convergence can be rapid
and retrieves the electron temperature distribution that is generated from
the use of the optically thin solution with the original distribution.

The $i^{\text{th}}$ iteration of the measured radiance $J_{nm}(y)$, can be writen as

\begin{equation}
{}^{i}J_{nm}(y) \;=\; {}^{i-1}J_{nm}(y)
\left\{\frac{{}^{i-1}j_n^{*}(y)}{{}^{i-1}j_n(y)}\right\},
\label{eq:3-47}
\end{equation}

where ${}^{i-1}j_n^{*}(y)$ and ${}^{i-1}j_n(y)$ are the respective optically thick and thin
solutions of the radiative transfer equation computed on the basis of the
previously evaluated electron temperature distribution ${}^{i-1}T_e(r)$, viz.,

\begin{equation}
{}^{i-1}j_{nm}^{*}(y) \;=\; \int_{-\infty}^{\infty}\!\!d\nu
\int_{-\sqrt{(R^2-y^2)}}^{\sqrt{(R^2-y^2)}}
\varepsilon_{nm}({}^{i-1}T_e)\exp\left\{-\int_{x}^{\sqrt{(R^2-y^2)}}
\frac{\varepsilon_{nm}({}^{i-1}T_e)\,dx^{*}}{P({}^{i-1}T_e)}\right\}dx
\tag{3.48a}\label{eq:3-48a}
\end{equation}

\begin{equation}
{}^{i-1}j_{nm}(y) \;=\; \int_{-\infty}^{\infty}\!\!d\nu
\int_{-\sqrt{(R^2-y^2)}}^{\sqrt{(R^2-y^2)}}\varepsilon_{nm}({}^{i-1}T_e)\,dx\;.
\tag{3.49b}\label{eq:3-49b}
\end{equation}

In the above equations, ${}^{i-1}T_e$ is the position dependent electron temperature
obtained using (in analogy with equation (\ref{eq:3-43})),

\setcounter{equation}{49}
\begin{equation}
\frac{E_{nq}}{k\,{}^{i-1}T_e} \;=\; -\ln
\left\{\frac{g_q\nu_{qp}S_{qp}A_{qp}}{g_n\nu_{nm}S_{nm}A_{nm}}\right\}
\frac{\displaystyle\int_{r}^{R}\frac{{}^{i-1}J_{nm}'(y)\,dy}{\sqrt{(y^2-r^2)}}}
{\displaystyle\int_{r}^{R}\frac{{}^{i-1}J_{qp}'(y)\,dy}{\sqrt{(y^2-r^2)}}}\;,
\label{eq:3-50}
\end{equation}

with ${}^{0}T_e$ being the position dependent electron temperature derived by
assuming that the measured lateral emission profiles are optically thin.

This iteration procedure is performed on the results in figure \ref{fig:3-17}.
The temperature resulting from the third iteration is compared to the
temperature distribution obtained from the use of the optically thin data,
in figure \ref{fig:3-17}.  The difference between this distribution and the third
iteration is well within the error that we can expect from the experiment.
Although no general test for convergence can be performed, this iteration is
expected to raise the measured electron temperature and converge rather
quickly in situations where the optical depth is not too severe.

\figstub{fig:3-17}{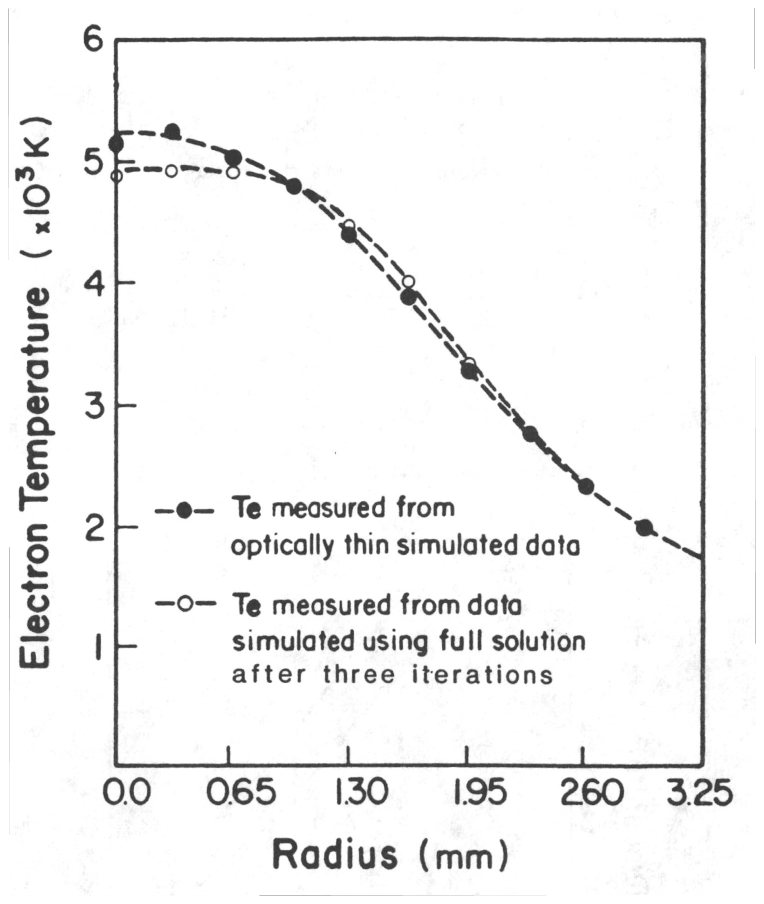}{Temperature resulting from the third iteration compared to
the temperature distribution obtained from the use of the optically thin
data.}

\section{Discussion on Multishot Averaging of Spectral Emission}
\label{sec:3-4}

The experimental measurements of spatial distributions of quantities
such as electron density and temperature in a plasma are frequently derived
from an analysis of the spatially integrated spectral radiances arising from
high lying atomic or ionic transitions.  In many pulsed experiments, the
spectral distribution of the emission is determined from a series of
measurements at different wavelengths.  If the shot to shot reproducibility
is poor, or the signal is weak, then the experiment is arranged to record
emission in each wavelength interval and lateral position as an average over
several shots.  When the results of such experiments are to be compared with
those predicted by a theoretical model, difficulties can arise if the
interaction responsible for the observed emission is highly nonlinear.

This situation is well illustrated in the case of plasmas created by
laser resonance saturation (Kissack 1987).  The strong attenuation of the
laser pulse and the highly nonlinear nature of the interaction may not have
been fully appreciated by some experimenters (Krebs and Schearer 1982,
Landen et al.\ 1985) and can lead to error in the electron densities derived
from the electron Stark broadened emission lines.

In this section, an attempt is made to understand the situation by
reference to comparison of simulated experiments and theoretical modelling.
In essence, a laser pulse is fired through sodium vapor (confined within a
heat sandwich oven).  The wavelength of the laser is tuned to closely
coincide with the 589.6 nm sodium D-resonance line and its energy fluence is
sufficient to saturate and subsequently ionize the vapor along its path.
The laser pulse is assumed to be axisymmetric and is assigned some radial
(noisy) structure.

In order to measure the electron density and temperature distributions,
within the resulting plasma column, as described in the previous section,
the lateral spectral emission from the $4^2D$-$3^2P$ transition is monitored as a
function of lateral (or y) displacement for a number of axial (or z)
locations (see figure \ref{fig:3-18}).  The emission recorded in each spectral
interval is averaged over a number of shots and the resultant signal
variation with y for a given wavelength is inverted using the Abel
transformation, to yield the emission spectrum of the multiplets
corresponding to different radial positions.  The free electron density is
then evaluated from these spectra using a knowledge of the electron Stark
broadening of this multiplet.

\figstub{fig:3-18}{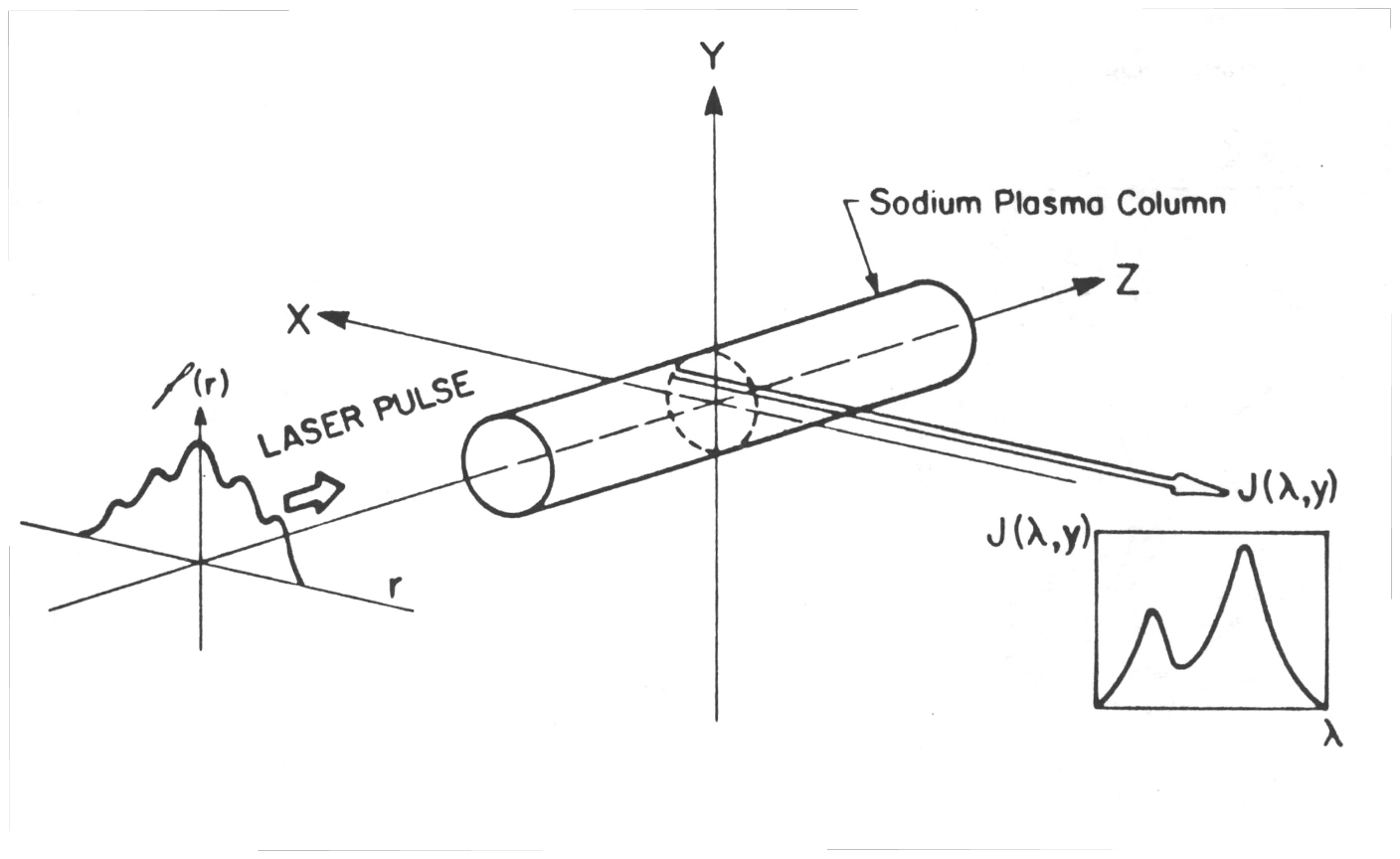}{Arrangement for monitoring the lateral spectral emission
from the $4^2D$-$3^2P$ transition as a function of lateral (y) displacement for a
number of axial (z) locations.}

Kissack (1987) developed a collisional-radiative computer code for
modelling this interaction (section \ref{sec:2-5-2}).  His five-level model of the
sodium atom predicts the temporal variation of the free electron density,
electron temperature and other dynamic variables along the path of the laser
pulse for a given incident laser pulse intensity and sodium atom
distribution, taking into account absorption of the laser radiation.  An
example of the predicted variation of these parameters with z, for a laser
pulse of energy fluence of 65 mJ cm$^{-2}$ is displayed as figure \ref{fig:3-19}.  The
experimentally based sodium atom distribution N(z) employed in this
calculation is also indicated in the figure.

\figstub{fig:3-19}{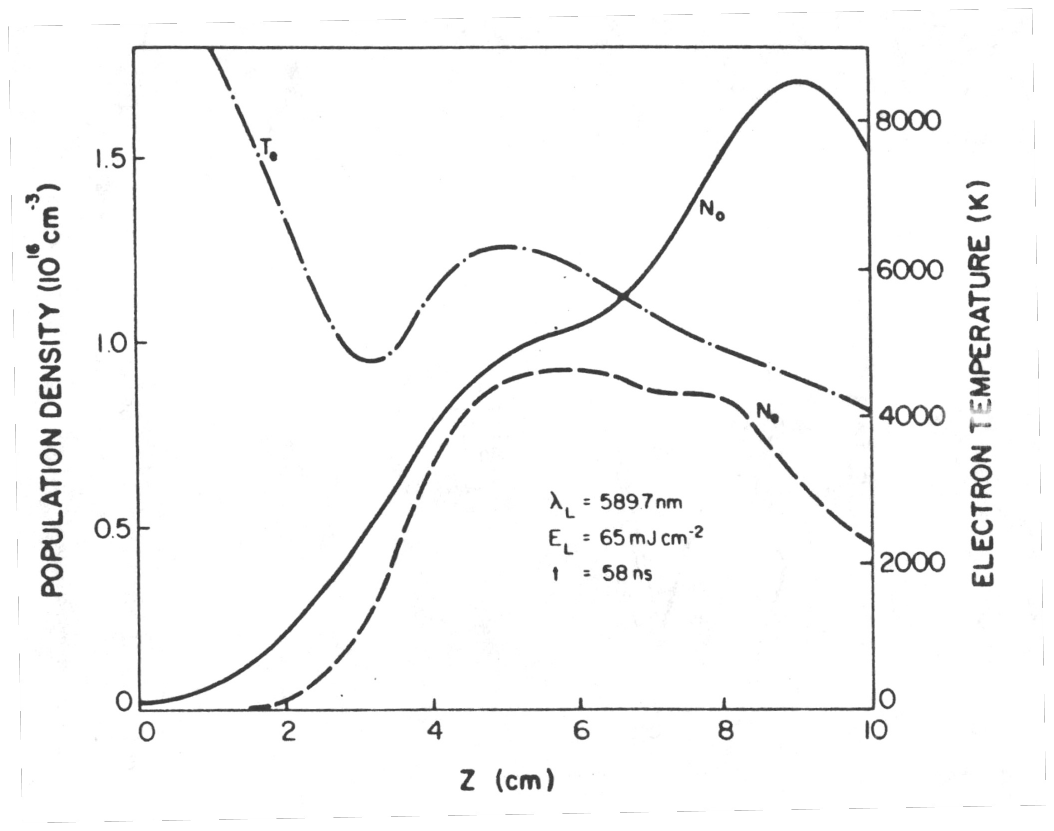}{Predicted variation of the plasma parameters with z, for a
laser pulse of energy fluence of 65 mJ cm$^{-2}$, together with the
experimentally based sodium atom distribution N(z).}

To illustrate the nonlinear nature of the interaction, the data
presented in figure \ref{fig:3-20} show the predicted variation of the free electron
density, temperature, and the $4^2D$ level population density $N_{4D}$ with incident
laser energy fluence at the centre of the heat sandwich oven (z=0) where the
sodium density is approximately $1.7\times10^{16}$ cm$^{-3}$, at 58 ns from the start of
the laser pulse having a 40 ns duration (FWHM, nearly Gaussian in temporal
shape).  These computational results clearly indicate that full ionization
should be attained at this location, provided the incident laser energy
fluence is greater than 100 mJ cm$^{-2}$.  Both the free electron density and
temperature are predicted to fall rapidly as the energy fluence drops below
100 mJ cm$^{-2}$.  This suggests that a variation of about 10\% about a mean
energy fluence of 90 mJ cm$^{-2}$, can lead to much greater fluctuations in the
state of the plasma at this location.

\figstub{fig:3-20}{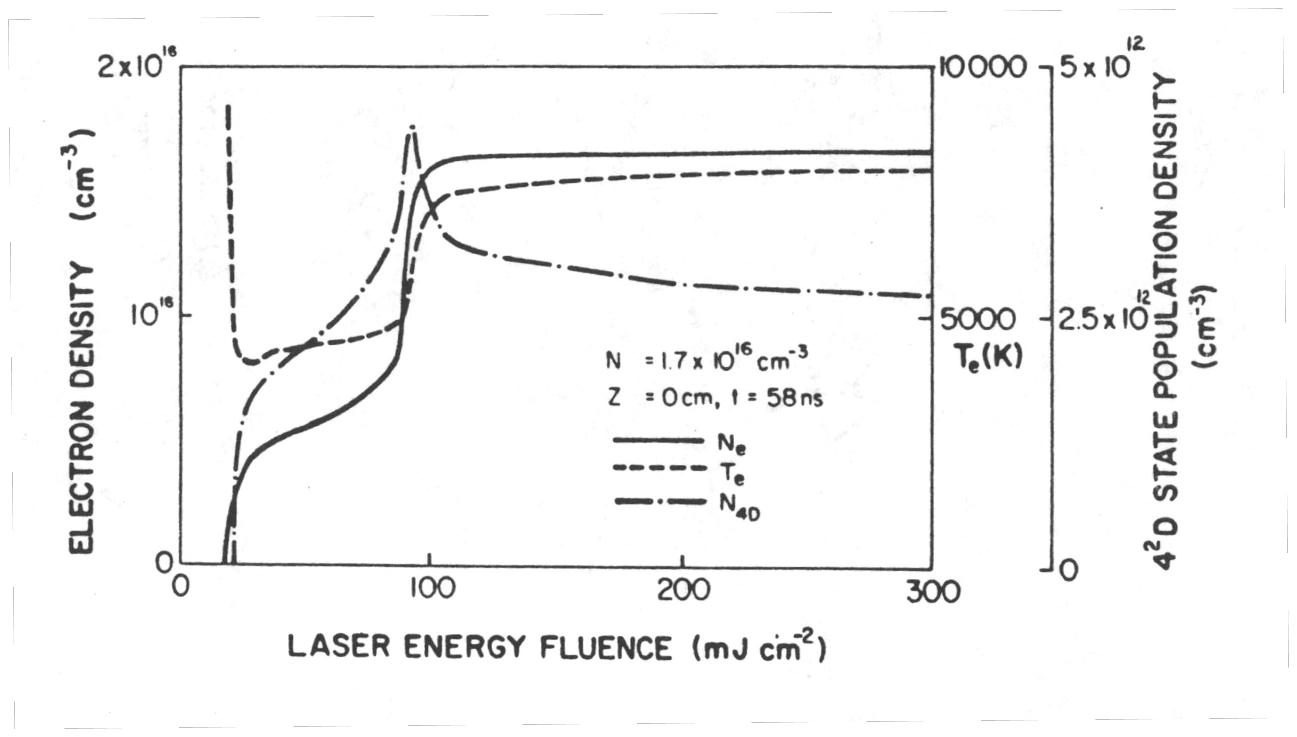}{Predicted variation of the free electron density,
temperature, and the $4^2D$ level population density $N_{4D}$ with incident laser
energy fluence at z=0.}

The real (experimental) situation is further complicated by the radial
variation of the incident laser pulse.  To model the three dimensional
nature of the interaction, the data from a series of computer runs at
various energy fluences are used to synthesize the radial variation of the
plasma properties at the z position of interest.  For example, the curves in
figure \ref{fig:3-21}a display the predicted electron density and temperature radial
profiles (t=58 ns, z=0) corresponding to the incident laser radial profile
shown.  The radial variation of this laser pulse is an extreme
representation of the actual fluctation observed in the experimental laser
pulse.  Presented in figure \ref{fig:3-21}b, is the radial variation of $N_{4D}$
corresponding to the same incident laser pulse.

\figstub{fig:3-21}{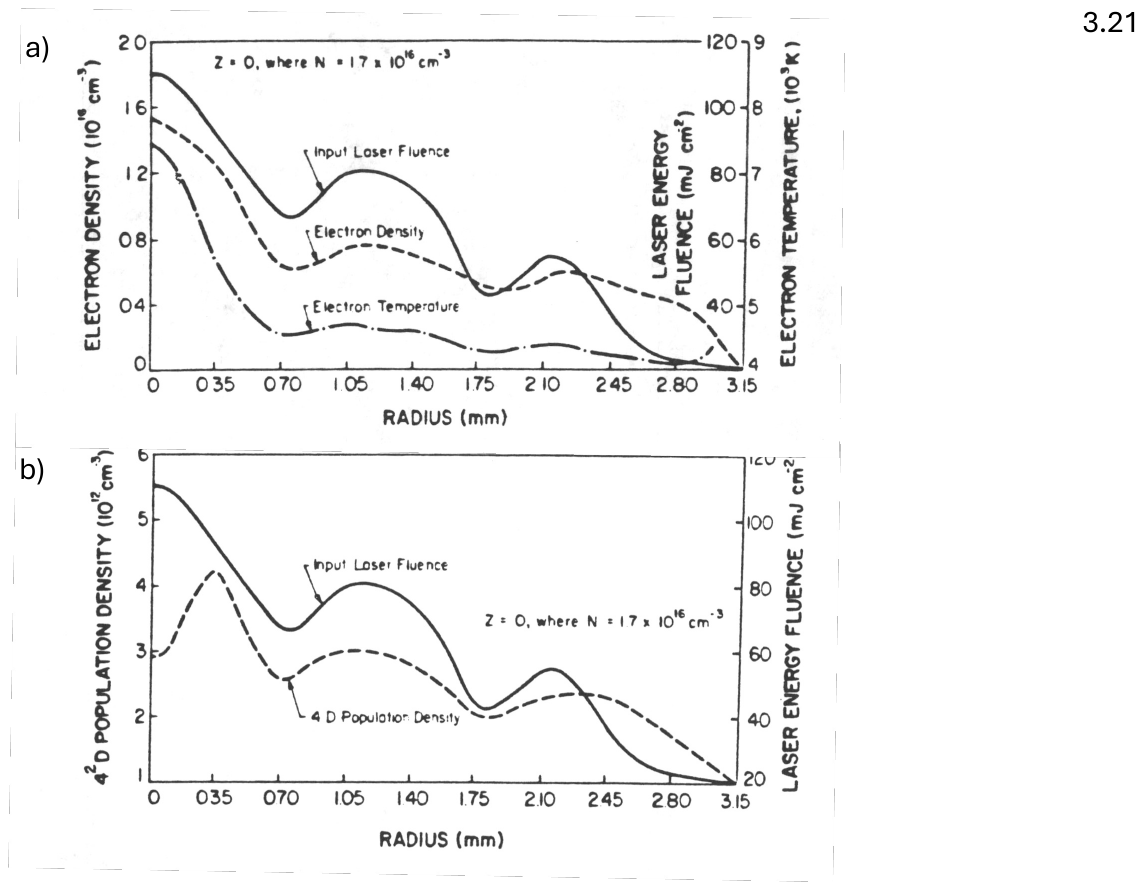}{a) Predicted electron density and temperature radial
profiles (t=58 ns, z=0) corresponding to the incident laser radial profile
shown.  b) Radial variation of $N_{4D}$ corresponding to the same incident laser
pulse.}

In the experiments, these variations in the incident laser pulse from
shot to shot, led to a shot to shot fluctuation in the emission and
necessitated the averaging of the emission detected in each spectral
interval of a scan to ascertain the multiplet emission spectrum.

\subsection{Multishot Average Spectrum Definitions}
\label{sec:3-4-1}

Generally speaking, the incident laser pulse is characterized by a
given distribution and a number of defining parameters (e.g., peak fluence
$E_p$, characteristic radius $r_c$, horizontal and vertical displacements of the
axis of symmetry of the pulse $\delta_x$ and $\delta_y$, spectral bandwidth $\beta_{\ell}$, etc\ldots)
which may vary from shot to shot.  That is, the incident spectral radiance
may be written as

\begin{equation}
I^{\ell} \;=\; I^{\ell}(\lambda,x,y,t,c_1,c_2,\ldots)\;,
\label{eq:3-51}
\end{equation}

where the $c_j$ are the characteristic parameters mentioned above, and in the
context of this analysis, are random variables.  From a mathematical
viewpoint, the corresponding spatially integrated spectral radiance
$J(\lambda,y,z,t)$, generated by the nonlinear interaction of this pulse with the
sodium vapor, can be understood as the action of a nonlinear operator L
acting on this input pulse (Kissack 1987) viz.,

\begin{multline}
J^{(i)}(\lambda,y,z,t)\\
= L(\lambda,x,y,z,t,c_1^{(i)},c_2^{(i)},\ldots)
\bigl[I^{\ell}(\lambda,x,y,t,c_1^{(i)},c_2^{(i)},\ldots)\bigr]\;.
\label{eq:3-52}
\end{multline}

It is convenient to introduce the shot to shot average of a random variable
x over n shots as

\begin{equation}
\langle x\rangle_n \;=\; \frac{1}{n}\sum_{i=1}^{n} x^{(i)}\;,
\label{eq:3-53}
\end{equation}

and over an infinite number of shots as

\begin{equation}
\overline{x} \;=\; \langle x\rangle_{\infty}\;.
\label{eq:3-54}
\end{equation}

With this in mind we can define three types of ``multishot averaged spectra'',
each discussed below.

\subsubsection*{Type 1 Average Spectrum}
\addcontentsline{toc}{subsubsection}{\quad Type 1 Average Spectrum}

The characteristic parameters of the incident laser pulse assume their
mean values.  The multishot averaged ($4^2D$-$3^2P$) multiplet spectrum is then
generated from the radiative transfer code presented in section \ref{sec:2-5}, which
solves the one-dimensional radiative transfer equation using the radial
distributions of the parameters computed from the LIBORS five-level code
(Kissack 1987) for the average incident laser pulse described above.
Mathematically, this can be written as

\begin{equation}
J(\lambda,y,z,t) \;=\; L\bigl[I^{\ell}(\lambda,x,y,t,\overline{c}_1,\overline{c}_2,\ldots)\bigr]\;.
\label{eq:3-55}
\end{equation}

For purposes of this computational investigation, the spatial profile
of the laser energy fluence is assumed to be Gaussian with mean peak value
$E_p$ = 90 mJ cm$^{-2}$, $r_c$ = 2.5 mm, $\delta_x = \delta_y$ = 0, $\beta_{\ell}$ = 0.015 nm.  The line centre of
the laser is assumed to be 589.7 nm.

\subsubsection*{Type 2 Average Spectrum}
\addcontentsline{toc}{subsubsection}{\quad Type 2 Average Spectrum}

The incident laser pulse for the LIBORS code is the arithmetic mean of
a finite number of shots.  The corresponding multishot average spectrum is
constructed as above.  Mathematically, this average is

\begin{equation}
J(\lambda,y,z,t) \;=\; L\bigl[\langle I^{\ell}(\lambda,x,y,t,c_1,c_2,\ldots)\rangle_n\bigr]\;.
\label{eq:3-56}
\end{equation}

In this computation, simplifying assumptions must be made.  Those
parameters which characterize the temporal and spectral characteristics are
assumed to be constant from shot to shot.  $\delta_x$ and $\delta_y$ are assumed to be zero
and the radial variation in the laser energy fluence fluctuates about some
Gaussian mean, with characteristic radius $r_c$ = 2.5 mm and a peak (r=0) laser
energy fluence $E_p$ = 90 mJ cm$^{-2}$.  The laser pulse is assumed to be symmetric
about r=0.

\subsubsection*{Type 3 Average Spectrum}
\addcontentsline{toc}{subsubsection}{\quad Type 3 Average Spectrum}

Each noisy laser pulse (with a random displacement of its centre
relative to the z axis) is used to determine the radial distribution of
excited states from which the lateral emission in the various spectral
intervals is computed.  The average value (over a given number of shots) of
this spectral radiance is then computed and used to generate the emission
spectrum.  Mathematically, this average is

\begin{equation}
J(\lambda,y,z,t) \;=\; \bigl\langle L\bigl[I^{\ell}(\lambda,x,y,t,c_1,c_2,\ldots)\bigr]\bigr\rangle_n\;.
\label{eq:3-57}
\end{equation}

To rigorously test this simulation, the range of incident laser energy
fluence was deliberately chosen to encompass the greatest rate of change of
the free electron density.

To conclude this section, it must be noted that neither all of the
characteristic spatial parameters which describe the incident laser pulse
nor their associated probability distributions are accurately known.  This,
coupled with the fact that any given laser pulse has a somewhat noisy
spatial distribution experimentally, suggests that the radial profile for
each computational shot needs to be constructed in a special manner.  The
following section outlines a method for doing this.

\subsection{Comparison of Computational Average Spectra}
\label{sec:3-4-2}

The ``noiseless'' incident radial energy fluence function,

\begin{equation}
E(r) \;=\; E_p\exp\left[-\frac{r^2}{r_c^{\,2}}\right],
\label{eq:3-58}
\end{equation}

is assumed to be Gaussian with a peak value $E_p$ and a characteristic radius
$r_c$.  Laser pulses with a noisy radial distribution are produced in the
following manner: the energy fluence at the $r_j$ (j = 1,2,\ldots.10) radial
position, $E(r_j)$, is regarded as a random variable with an associated
Gaussian probability distribution characterized by the mean energy fluence,

\begin{equation}
\overline{E}_j \;=\; E_p\exp\left[-\frac{r_j^{\,2}}{r_c^{\,2}}\right]
\label{eq:3-59}
\end{equation}

and by a standard deviation $\sigma_j$.  In order to produce an appreciable noise
level, we set $\sigma_j = 0.25\,\overline{E}_j$.  $E(r_j)$ is selected via a random number generator
and weighted according to the above distribution.  To allow for the possible
jumping around of the laser pulse with respect to the z axis, we set the
standard deviation in this displacement to be 20\% of $r_c$.  In addition, the
computed spectrum corresponded to that which would have been observed at y =
1.5 mm, in order to accentuate the sensitivity of the measurement to lateral
displacement of the laser beam axis relative to the field of view.

In the numerical simulation, we have chosen to model the $4^2D$-$3^2P$
multiplet spectrum created at z=0 for a sodium atom distribution that is
symmetrical about z=0 and has its maximum value of $1.7\times10^{16}$ cm$^{-3}$.  The
time of observation is 58 ns after the start of the laser pulse (40 ns FWHM)
thereby matching some of the experimental conditions (see chapter 5).  The
mean value of the peak energy fluence at r=0, was chosen as 90 mJ cm$^{-2}$ so
that much of the energy fluence would lie in the range where the most rapid
nonlinear change in $N_e$ and $T_e$ occurs.  The exponential radius $r_c$ corresponds
to that in an actual experiment (2.5 mm).

The comparison of type 1 and type 2 average spectra can be gauged by
reference to figures \ref{fig:3-22}a and \ref{fig:3-22}b.  The type 2 average spectra is derived
from 4 shots in the former case and 16 shots in the latter.  It is quite
apparent that a considerable difference exists when only four shots are
considered.  A 20\% difference in the free electron density from the
broadening of the spectra, would be predicted on the basis of this
comparison.  In contrast, the type 2 average spectrum derived from 16 shots
appears to agree very closely with the type 1 average spectrum which is the
norm used in theoretical modelling and assumes the incident laser pulse to
have a Gaussian radial distribution.  This seems to indicate that little
difference would be expected if we used either a Gaussian radial profile or
the actual averaged radial profile provided we averaged over at least 16
shots.

A more interesting comparison is that of the type 2 and type 3 average
spectra for 4 and 16 shots respectively (see figure \ref{fig:3-22}c and \ref{fig:3-22}d).  It is
again clear that while there is a significant difference between the spectra
based on the 4 shot average, the spectra based on the 16 shot average still
show a slight difference.  This suggests that although 16 shots may suffice,
the interpreted value of the free electron density can differ from that
predicted using a type 2 analysis.  The difference however is well below the
experimental accuracy that is expected from the current facility albeit for
the conditions chosen.  Results of this nature should shed some light,
however, onto the amount of care and understanding that is necessary of the
domain over which the measurements encompass.

\figstub{fig:3-22}{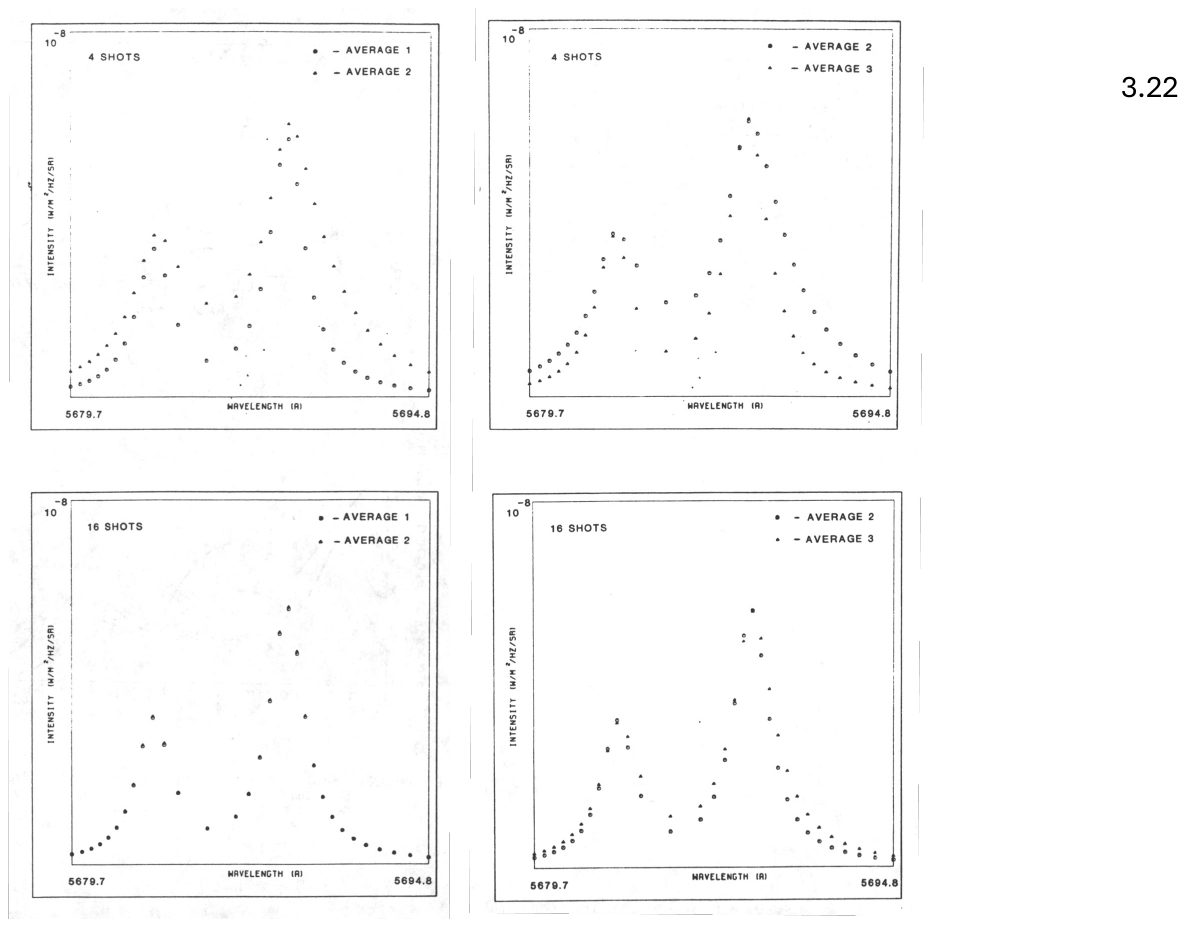}{Comparison of computational average spectra: a) type 1 and
type 2 (4 shots); b) type 1 and type 2 (16 shots); c) type 2 and type 3
(4 shots); d) type 2 and type 3 (16 shots).}

\chapter{Experimental Facility}
\label{ch:facility}

The complete LIBORS facility (less the Nd:YAG and Dye Laser) was
erected on a granite table approximately $1.8\times2.5$ m in dimensions.  The
facility will be discussed in five separate parts, each serving a specific
role.

The sodium vapor is housed in a specially designed heat sandwich oven.
Absorption spectroscopy is performed, to estimate the sodium density
distribution within the oven.  A Xenon lamp acts as a broadband radiation
source and a Heath scanning monochromator is used for wavelength
discrimination, with a photomultiplier at the exit slit (section \ref{sec:4-1}).  The
laser facility used to generate tunable laser radiation in the vicinity of
the sodium D-lines comprises a JK/Lumonics HY750 Nd:YAG laser with a
frequency doubling KD*P crystal, in conjunction with a Quanta-Ray dye Laser
(section \ref{sec:4-2}).  Plasma emission is observed with a Spex scanning
monochromator and a thermionically cooled photomultiplier at its exit slit
(section \ref{sec:4-3}).  A set of photodiodes is used to monitor the temporal shape
of the laser pulse prior to and after passing through the heat sandwich oven
(section \ref{sec:4-4}).  Data aquisition is performed with an EG\&G boxcar averager
and gated sampled integrators (section \ref{sec:4-5}).  A schematic of the complete
facility is given in figure \ref{fig:4-1}.  The overhead photograph of the facility
shown in figure \ref{fig:4-2} gives the reader some perspective of the relative size
and position of the components.

\figstub{fig:4-1}{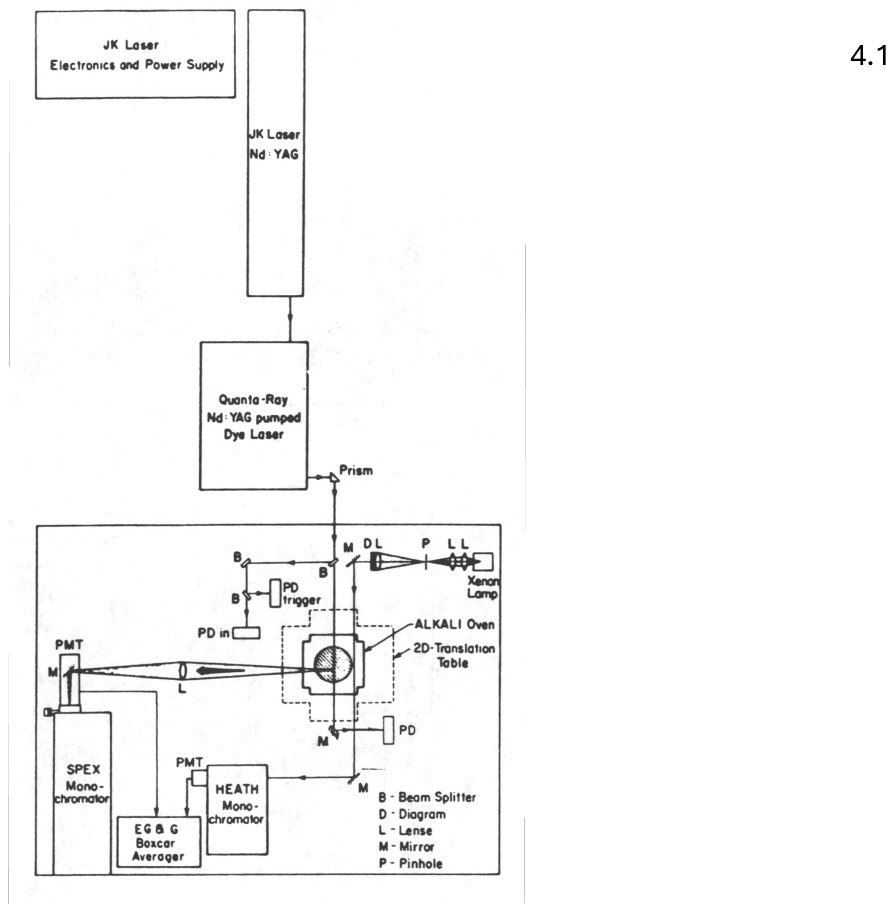}{Schematic of the complete facility.}

\figstub{fig:4-2}{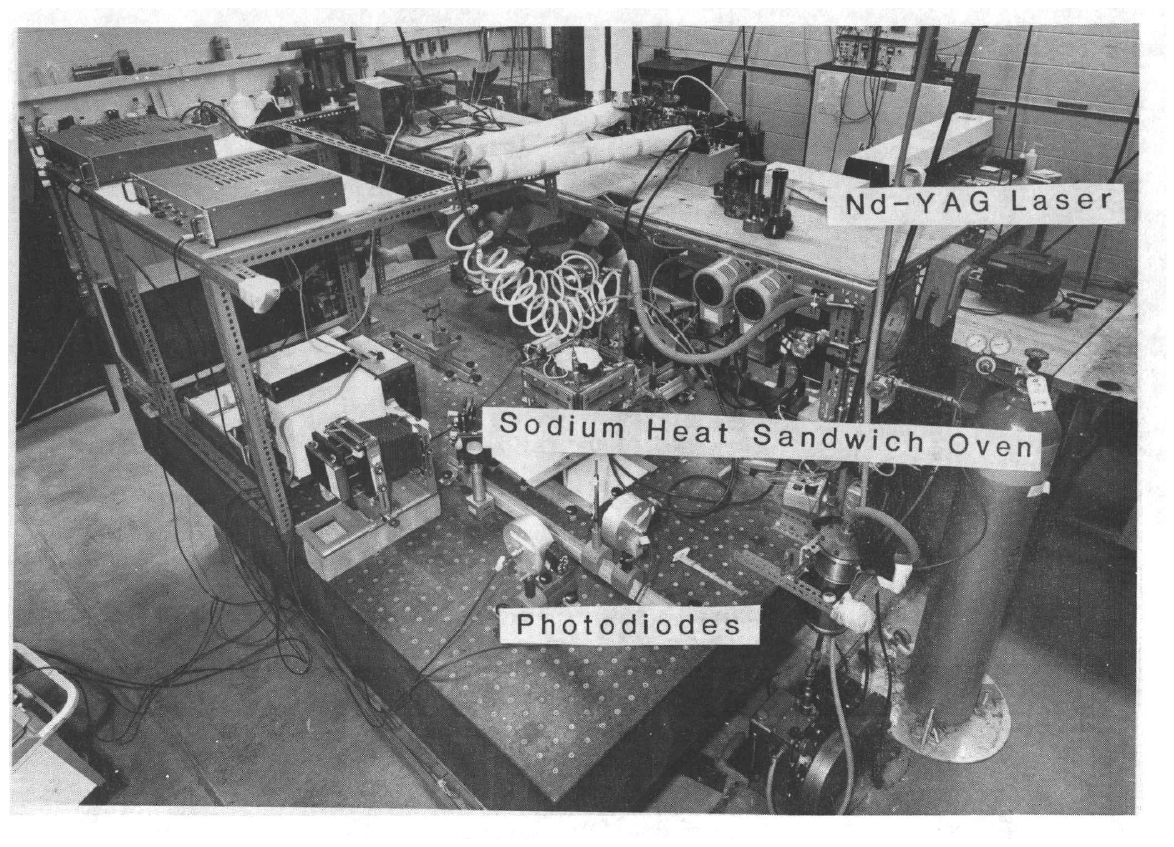}{Overhead photograph of the facility.}

\section{Heat Sandwich Oven and Neutral Sodium Atom Density Measurement Facility}
\label{sec:4-1}

The original design of the heat sandwich oven was started by P.\
Wizinowich (1979) and completed by a summer student, J.\ Zapfe, in the summer
of 1980.  The oven's design was based on the design of Boyd et al.\ (1980).
Our original design differs from the design of Boyd et al (1980) in two
ways: we use a rectangular frame which permits us to view the laser
produced plasma channels through optically flat windows, and we have
introduced a thermal expansion buckle to relieve the radial stress that can
potentially develop (Boyd and Harter 1980) in each of the heated plates.  A
cross sectional view of one half of this heat sandwich oven is presented in
figure \ref{fig:4-3}.  The axis of symmetry (A-A) corresponds to the extreme right of
figure \ref{fig:4-3}.  A photograph of the heat sandwich oven illuminated by the
ionizing laser pulse is shown in figure \ref{fig:4-4}.  The narrow sodium plasma
channel created by the laser is clearly visible through the front window.

\figstub{fig:4-3}{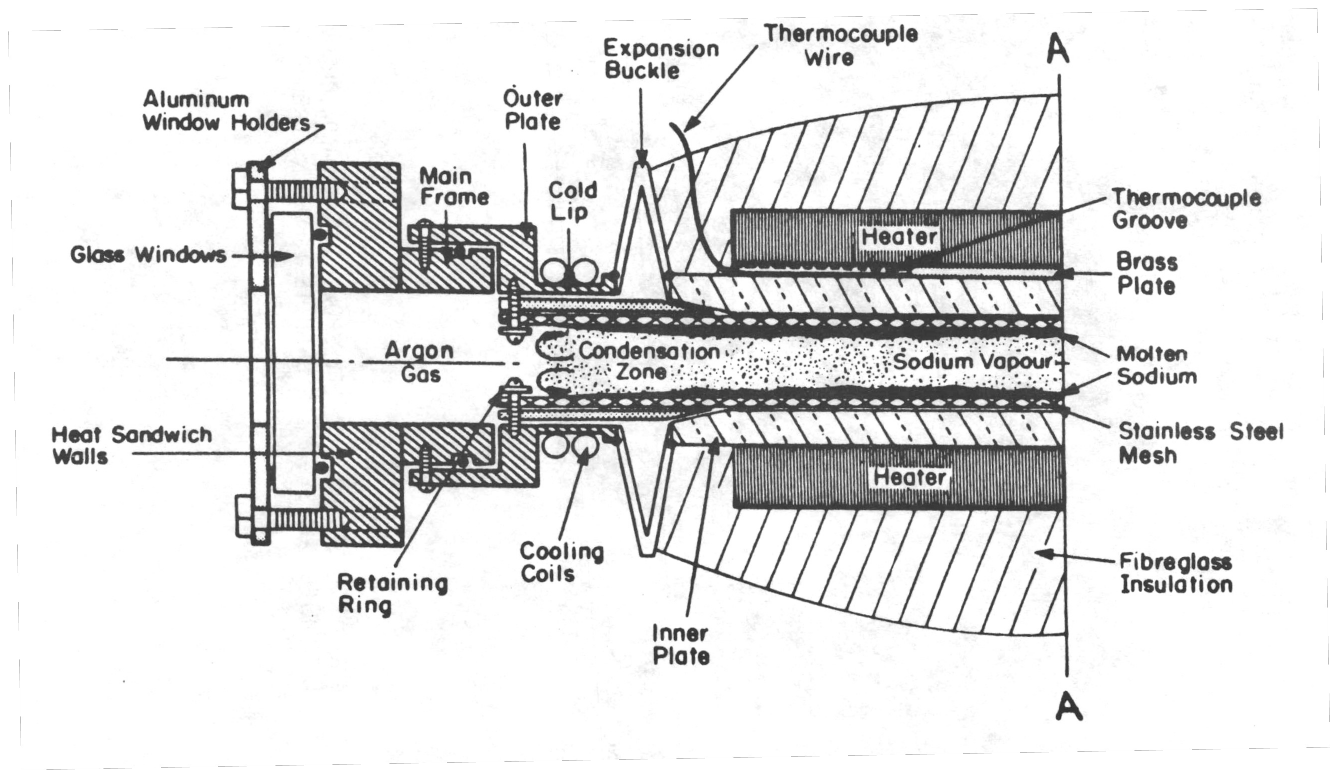}{Cross sectional view of one half of the heat sandwich oven.
The axis of symmetry (A-A) corresponds to the extreme right of the figure.}

\figstub{fig:4-4}{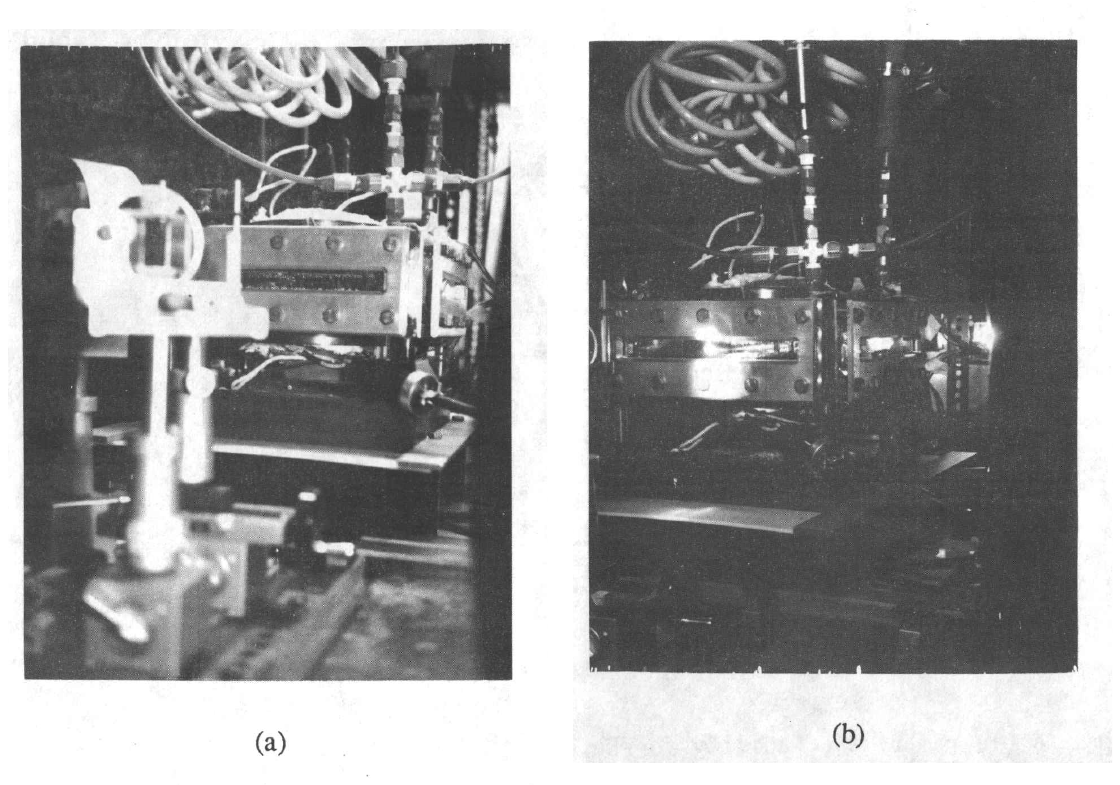}{Photograph of the heat sandwich oven illuminated by the
ionizing laser pulse.}

The basic rectangular frame is constructed of stainless steel (type
304) and is fastened to the circular shaped stainless steel (type 304) outer
plates.  The outer plates are held close to room temperature by circulating
water through the copper cooling coils indicated in figure \ref{fig:4-3}.  The outer
plates are welded to the inner plates via a specially designed expansion
buckle.  The upper and lower wicks (required for heat pipe action) are made
of three layers of stainless steel (type 60) mesh (0.152 mm wires) and are
secured to the inner plates (high grade stainless steel-type 316) by a large
number of small screws and to the stainless steel cold lip (which extends
into the heated zone and prevents condensation in the buckle) by a stainless
steel retaining ring (fastened to the mainframe just outside the cooled
region).  The rectangular viewports are each 20 cm $\times$ 3 cm in size and the
heat sandwich walls have O-ring grooves machined into them in order to
provide a vacuum seal for the windows.  Cylindrical ceramic heaters
(Electro-applications Inc.\ - 11.4 cm dia.\ and 1.6 cm thick) rated for up to
$2210^{\circ}$F @ 115V, are positioned just outside of the inner plates.  A more
detailed description of the oven and modifications made to it can be
obtained from Cappelli (1983) and Cardinal (1986).

Operation is commenced by first introducing a buffer gas followed by
the pumping down of the pressure until it is just below the desired sodium
partial pressure required for a given experiment.  Two pressure gauges
(Wallace \& Tiernan 0-20 torr; McLeod 0.01-10 torr) were connected to a
common port on the oven mainframe allowing the operator to continuously
monitor the buffer gas presure.  The sodium is then heated until it melts,
whereupon its vapor pressure increases and the resulting radial flow of the
sodium atoms effectively pumps the buffer gas into the cool outer region of
the oven.  An essential feature of this heat pipe action is the radially
inward flow (from cold to hot) of the liquid sodium due to capillary forces
in the wicks.  It should be noted that an overpressure of the buffer gas
leads to an observed drop in the vapor density of the sodium, for under
these conditions, the buffer gas is able to permeate the hot core of the
oven and thereby transfer energy from the hot vapor to the cool boundary
region.

Although thermocouple temperature sensors (chromel-alumel
thermocouples attatched to a C-A thermoelectric gauge) were employed, it was
felt that they could at best only provide a rough estimate of the sodium
atom density using the empirical fit of Nesmeyanov (equation (\ref{eq:2-38})).  As
one can see from figure \ref{fig:4-3}, the thermocouple wires were placed just beneath
the heaters in a series of radial grooves machined into a circular brass
plate 1/8" in thickness.  The temperature readings obtained from the
thermocouples reflect the temperature on the heater (outer) surface of the
inner plate and not necessarily the temperature of the molten sodium in the
stainless steel wick.  It was suggested and shown previously (Cappelli
1983), that a substantial temperature gradient can exist across this inner
plate.  It was for this reason that the sodium atom density measurement
technique described in section \ref{sec:3-1} was derived.

To position the heat sandwich oven and accommodate the sodium atom
density measurements, four stainless steel legs were welded onto the main
frame and were rested on a translatable mount (Hanratty 1987).  This allowed
us to irradiate any region of the sodium vapor disc either with the
radiation from the broadband source such as the Xenon lamp or the laser
itself.

Three minor modifications to the oven were necessary for operation at
sodium densities greater than $10^{16}$ cm$^{-3}$.  The first problem encountered was
the need for increased thermal power deposited into the inner plates from
the heaters.  Since we were already at the current limit of the heater
power supplies (rectified AC current sources), we decided to remove the
brass plates and thermocouples thereby relying only on the absorption
measurements to provide us with an estimate of the sodium density
distribution.  The increased sodium density (or vapor pressure) resulted in
an increased fogging of the windows (a result of the slow deposition of an
alkali metal film).  This problem was alleviated by the insertion of a 2.5
cm aluminum spacer between each window and oven mainframe.  The last problem
encountered was the formation of large sodium droplets on the upper wick.
These droplets directly interfered with the passage of the laser beam
through the oven and sometimes obstructed the observation of the plasma
emission at right angles to the direction of beam propagation.  It was
therefore necessary to increase slightly the separation of the inner plates
by the insertion of a 0.5 cm stainless steel spacing ring between each outer
plate and the oven mainframe.

The facility for measuring the sodium atom density distribution
was comprised of a broadband continuous wave xenon light source, an aperture
and collimating optics, the heat sandwich oven, a Heath model EU-700
scanning grating monochromator (1180 lines/mm, f/6.8, 35 cm focal length)
and an RCA type 1P28 photomultiplier with associated housing.  A schematic
of the optical arrangement is presented as figure \ref{fig:4-5}.  The diameter of the
probe beam of broadband radiation was set to about 0.3 cm by the stop.

\figstub{fig:4-5}{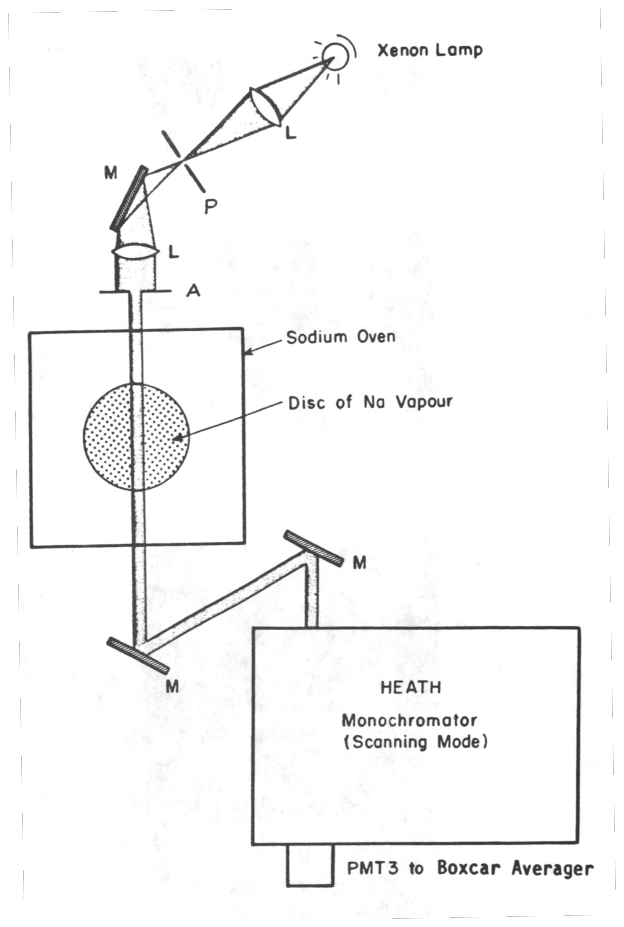}{Schematic of the optical arrangement for measuring the
sodium atom density distribution.}

The signal from the photomultiplier was processed by an EG\&G Boxcar
Averager and Gated Integrator (described in section \ref{sec:4-5}).  The trigger for
the EG\&G was provided by the output gate of an oscilliscope.  To maximize
the signal handling capability, ideally, one would want the gate width (the
integration time for the sample integrator)

\begin{equation}
\tau_g \;=\; \frac{1}{f_t}
\label{eq:4-1}
\end{equation}

where $f_t$ represents the trigger frequency (in this case determined by the
time range on the oscilliscope).  The gate width or trigger frequency is
therefore determined by the desired signal to noise ratio.

As described in section \ref{sec:3-1-2}, the transmission function of the system
is generally dependent on the entrance and exit slit height and width, as
well as the grating ruling.  With the entrance and exit slit widths
comparable, one would expect a symmetric transmission function that is
triangular to Gaussian in shape.  Diffuse scattering of a He-Ne laser beam
at the entrance slit of the Heath monochromator acted as a narrow line
source at 632.8 nm.  The profile measured can be attributed to the response
of the system and its shape represents the transmission function.  The
output of the spectral scans shown in figure \ref{fig:4-6} for two slit widths ($\delta_{sw}$ =
150 and 300 $\mu$m respectively) were recorded on a Hewlett Packard X-Y plotter.
The broken line in each figure represents the Gaussian function with FWHM
$\delta_T$ given in the figure.  The curve of $\delta_T$ verses $\delta_{sw}$ can be represented by a
straight line for $\delta_{sw} >$ 100 $\mu$m,

\[
\delta_T \;=\; 0.08 + 0.023\,\delta_{sw}\ \ (\text{\AA})
\]

as can be seen from the graph in figure \ref{fig:4-7}.  The deviation from linearity
at lower slit widths is a result of both the gradual wear of the spring
mechanisms involved in controlling the slit width, as well as the fact that
there exists a safety feature which prevents the slits from closing thereby
preventing permanent dammage to the slit edges.

\figstub{fig:4-6}{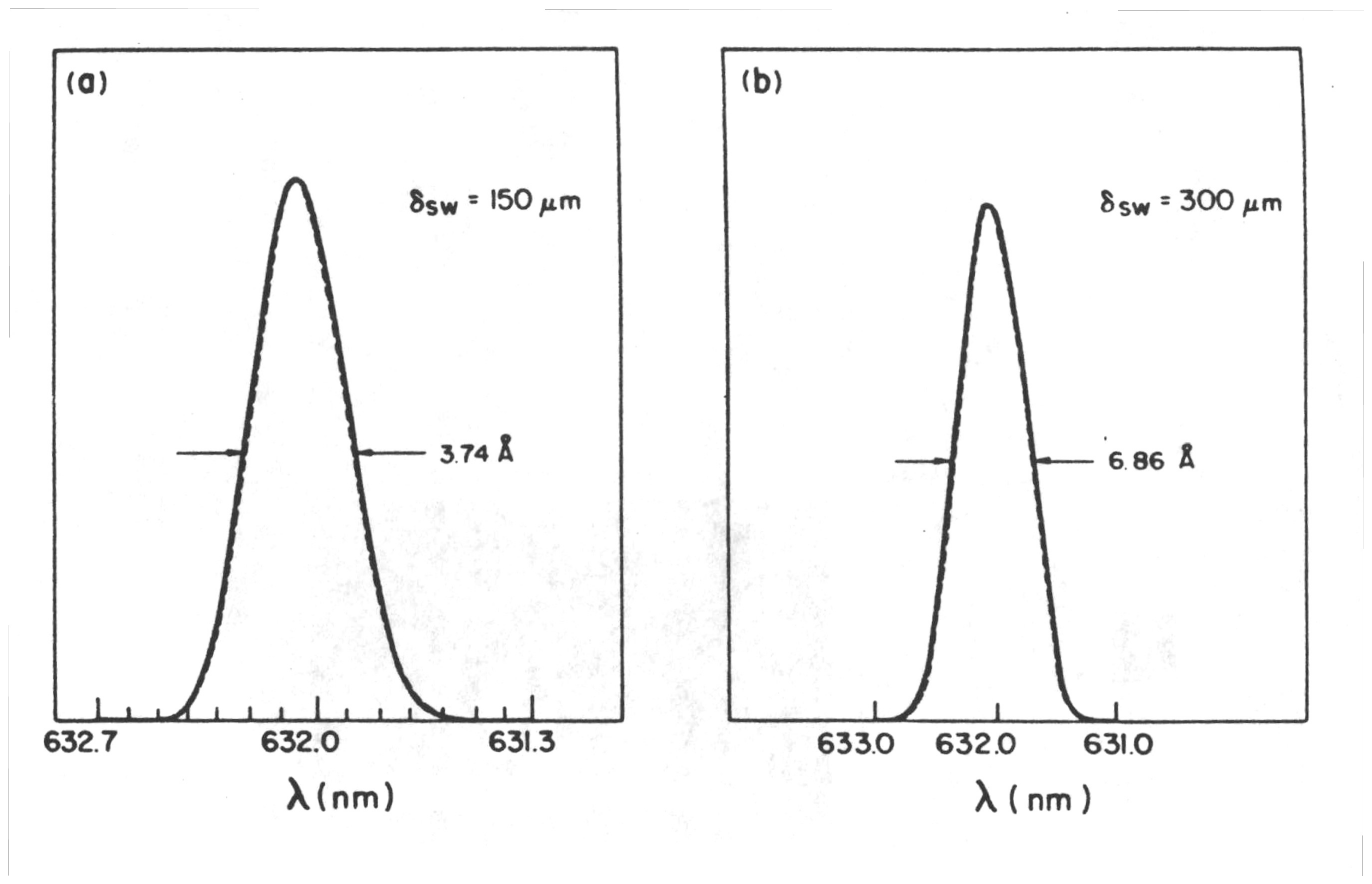}{Output of the spectral scans for two slit widths
($\delta_{sw}$ = 150 and 300 $\mu$m).  The broken line in each figure represents the
Gaussian function with FWHM $\delta_T$ given in the figure.}

\figstub{fig:4-7}{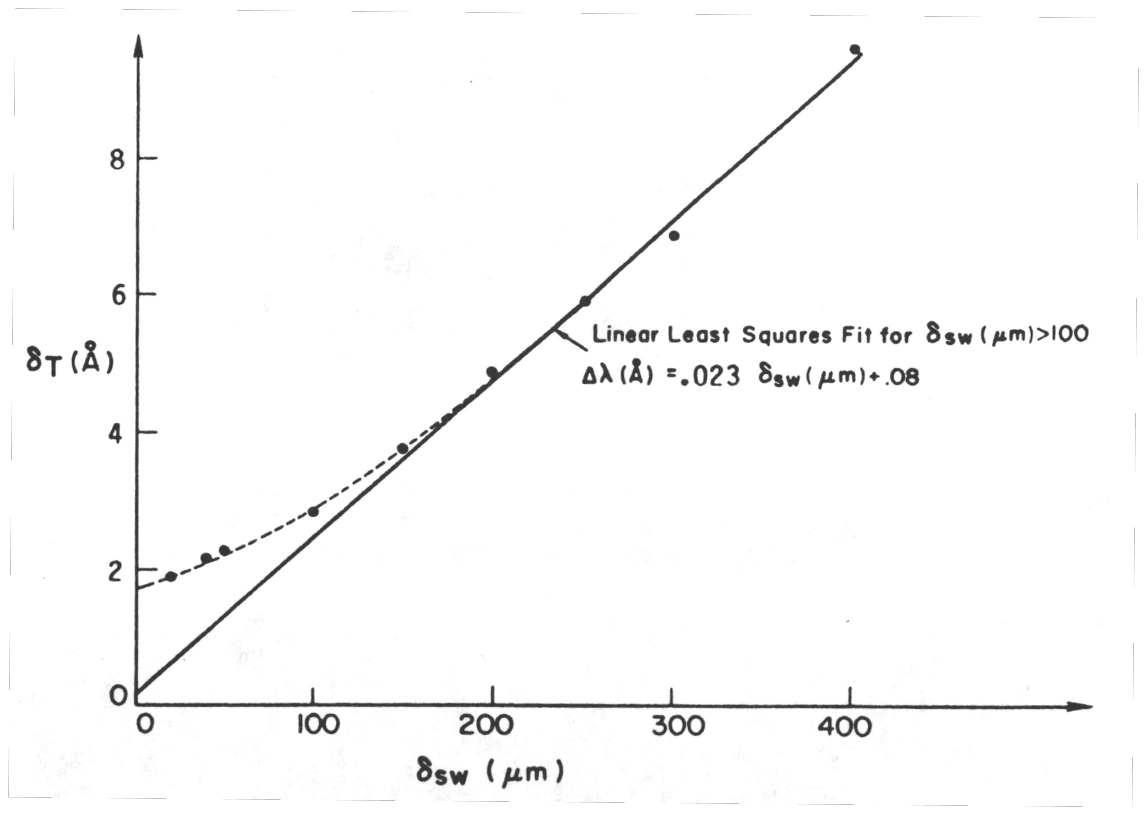}{Graph of $\delta_T$ verses $\delta_{sw}$.}

\section{Nd:YAG Pumped Dye Laser}
\label{sec:4-2}

A Nd:YAG HY750 oscillator/amplifier laser (JK Lasers distributed by
Lumonics in North America) represents the main component in the facility.
The 1.06 $\mu$m output can be frequency doubled with a KD*P doubling crystal
giving rise to approximately 300 mJ of energy at 532 nm (at optimized
lasing) with a first harmonic energy of 850 mJ.  The temporal pulse width
(FWHM) is specified as 20 ns and 17 ns at 1.06 $\mu$m and 532 nm respectively.
Operation was at 10 Hz although it is capable of operating at 20 Hz with
minor realignment and a sacrifice in the flashlamp life expectancy.

The frequency doubled output was used to pump a Quanta-Ray
(oscillator/amplifier arrangement) PDL-1 dye laser with Kiton Red (Exciton)
dye disolved in methanol as the lasing medium.  Concentrations of 130 mg/l
for the oscillator and 16 mg/l for the amplifier were typical for optimized
lasing near the sodium D-lines.  With a maximum pump energy of 300 mJ and a
somewhat optimistic conversion factor of 24\% near 589 nm, one can expect a
maximum dye laser energy output of 70 mJ.  Table \ref{tab:4-1} summarizes the typical
characteristics of the dye laser output pulse.

\tabstub{tab:4-1}{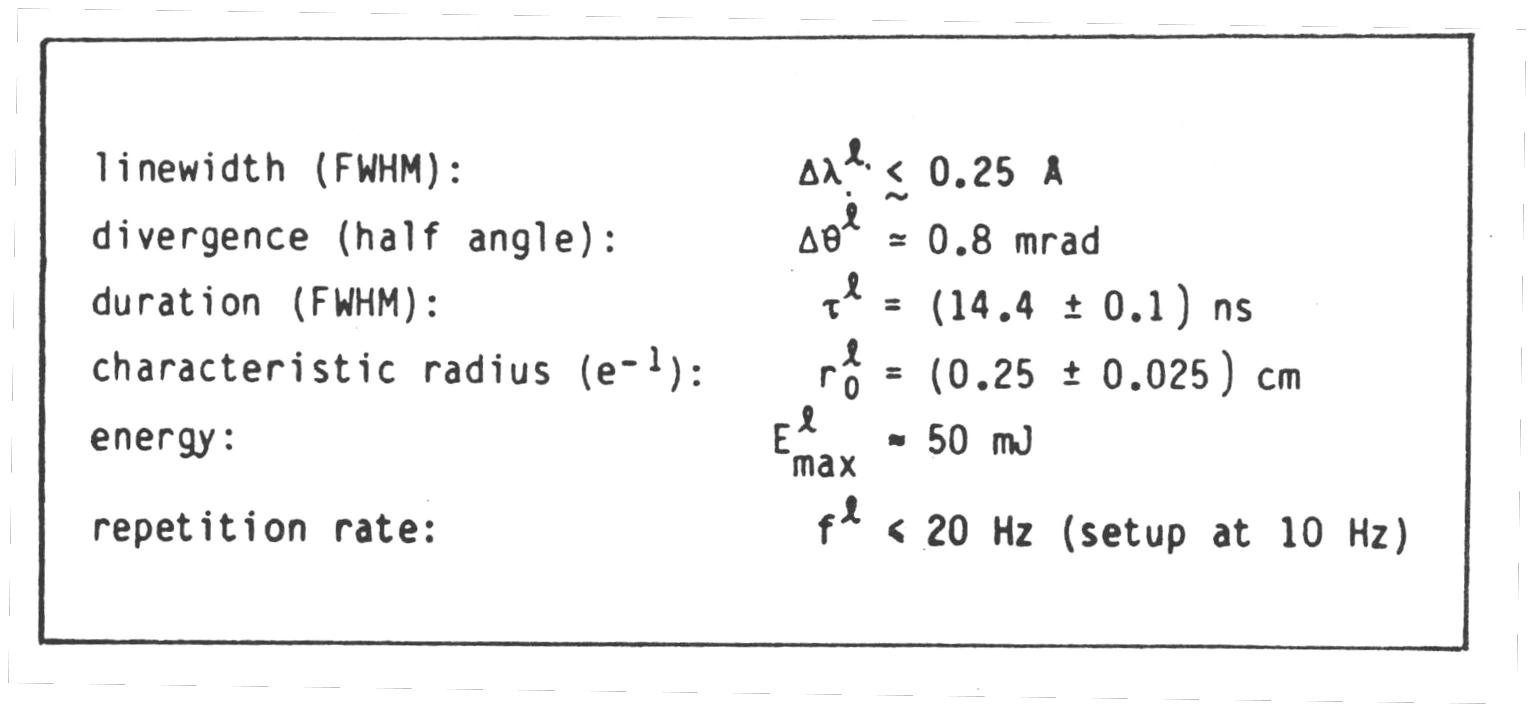}{Typical characteristics of the dye laser output pulse.}

A saturated ``burn spot'' on photographic film (Figure \ref{fig:4-8}a) as well as
a linear photodiode array trace across the midsection of the beam
cross-section (Figure \ref{fig:4-8}b) reveals that the cross-section is not quite circular
but can be described to have an effective characteristic radius of
approximately 2.0 mm.  For some of the experiments, the laser beam is
apertured giving rise to a cross-section that is significantly smaller and
perhaps more uniform across its diameter.  The results of aperturing with a
1 mm diameter aperture are shown via the saturated burn spot and laser
photodiode array trace in Figure \ref{fig:4-9}a and \ref{fig:4-9}b.

\figstub{fig:4-8}{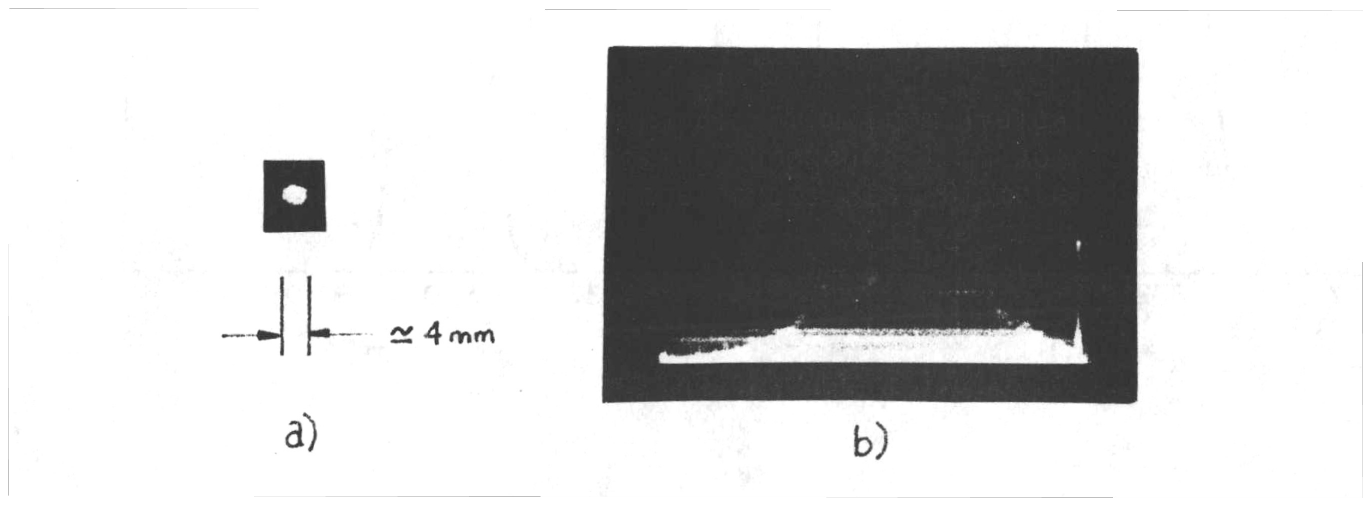}{a) Saturated ``burn spot'' on photographic film.  b) Linear
photodiode array trace across the midsection of the beam cross-section.}

\figstub{fig:4-9}{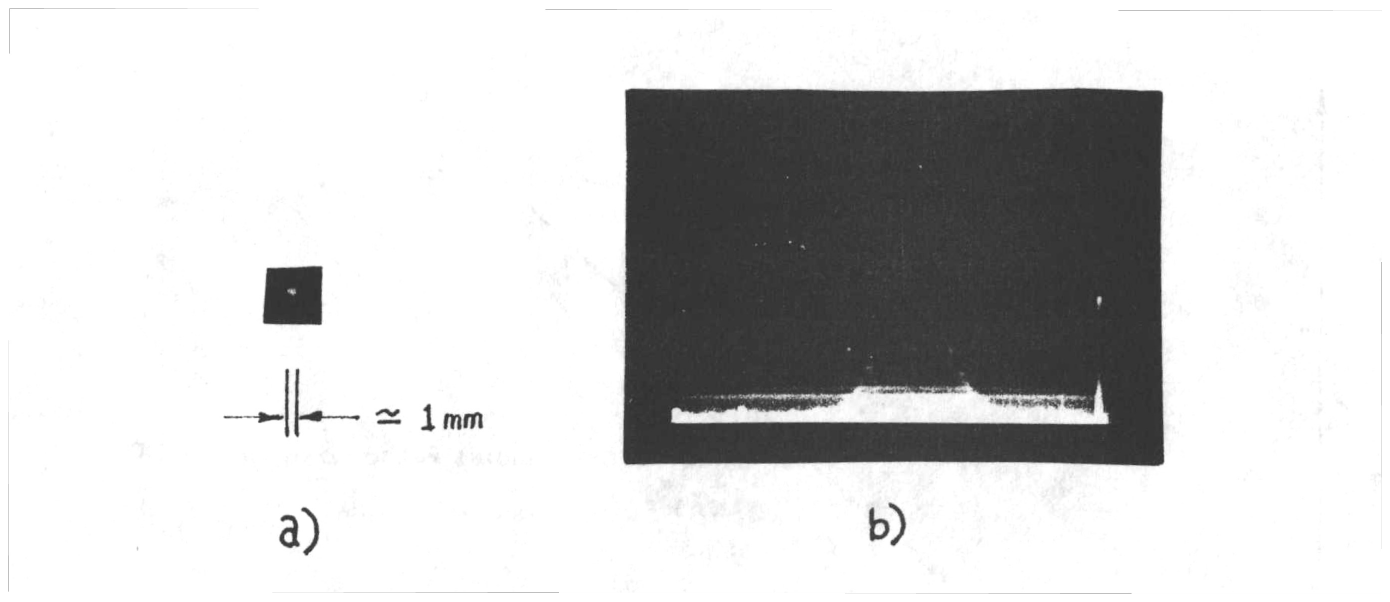}{a) Saturated burn spot and b) laser photodiode array trace
for the beam apertured with a 1 mm diameter aperture.}

For comparison of theory with experimental results, it is convenient
to express the output irradiance of the dye laser by a function of the form
(Cappelli 1983, Wong 1985, Cardinal 1986)

\begin{equation}
I^{\ell}(t) \;=\; I^{\ell}_{\text{peak}}\,a\,t^{b}e^{-ct^2}\;,
\label{eq:4-2}
\end{equation}

where a, b and c are fitting constants that can be obtained by a least
squares fitting procedure and $I^{\ell}_{\text{peak}}$ represents the peak laser irradiance
defined by

\begin{equation}
I^{\ell}_{\text{peak}} \;=\; \frac{E_0}
{\pi\bigl(r_0^{\ell}\bigr)^{2}\displaystyle\int_{0}^{\infty}a\,t^{b}e^{-ct^2}}\;,
\label{eq:4-3}
\end{equation}

with $E_0$ as the incident laser energy and $\pi\bigl(r_0^{\ell}\bigr)^{2}$ the effective
cross-sectional area of the laser beam as determined by either the burn spot or
photodiode array trace.  With the time t expressed in nanoseconds, an
excellent fit to the experimental laser pulse was obtained using (Cardinal
1986)

\begin{align}
a &= 3.352\times10^{-3}\ (\text{ns})^{-b}\;,\nonumber\\
b &= 2.642\;,
\label{eq:4-4}\\
\intertext{and}
c &= 6.50\times10^{-3}\ (\text{ns})^{-2}\;.\nonumber
\end{align}

The data in figure \ref{fig:4-10} illustrates the excellent agreement between the
empirical function given by equation (\ref{eq:4-2}) and that obtained by directing
the laser beam onto the face of a PIN photodiode (section \ref{sec:4-4}) whose output
current has been digitized using the EG\&G Boxcar Averager (section \ref{sec:4-5}).

\figstub{fig:4-10}{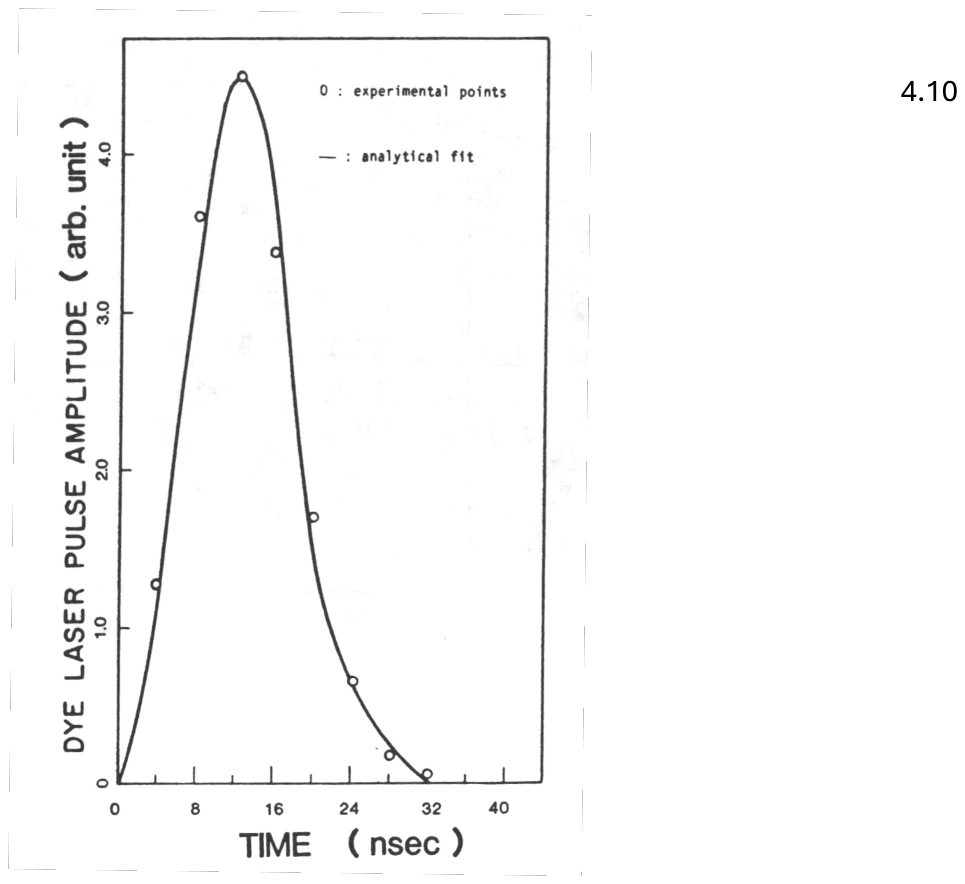}{Agreement between the empirical function given by equation
(\ref{eq:4-2}) and that obtained by directing the laser beam onto the face of a PIN
photodiode.}

\section{Facility for Plasma Emission Studies}
\label{sec:4-3}

The electron density and temperature measurements were undertaken
using a sideways mounted SPEX model 1700 II monochromator.  This enabled the
entrance slit of the monochromator to sample a thin horizontal slab of the
plasma emission at some height y above the axis of the laser beam.  The
image height was scanned by vertical displacement of the imaging lens which
had a 1m focal length.  The magnification of the imaging optics was 1:1 with
the entrance slit set at 1cm $\times$ 40$\mu$m.  With a wavelength dispersion of
approximately 1.0 nm per mm over the visible spectral range, and with the
exit slits also set at the same dimensions, the transmission function of the
monochromator was well described by a Gaussian with FWHM of approximately
0.05 nm.

An RCA C31034 photomultiplier tube (peak QE: 20\% @ 360 nm; 10\% relative
QE @ 230 and 840 nm) with a GaAs photocathode, was mounted at the exit slit
for most of the experimental runs.  This particular photomultiplier tube has
extended sensitivity in the red region of the visible spectrum and requires
cooling well below room temperature.  A Thermionically cooled housing
(Products For Research model TE 104) was used to minimize the dark noise.  A
prewired socket was obtained from Hamamatsu Corporation.  The socket is
designed to have a rise-time of less than 2.5 ns with an average output
current of 100 nA.  A schematic wiring diagram of the C31034 socket used for
the experiments is shown in Figure \ref{fig:4-11}.

\figstub{fig:4-11}{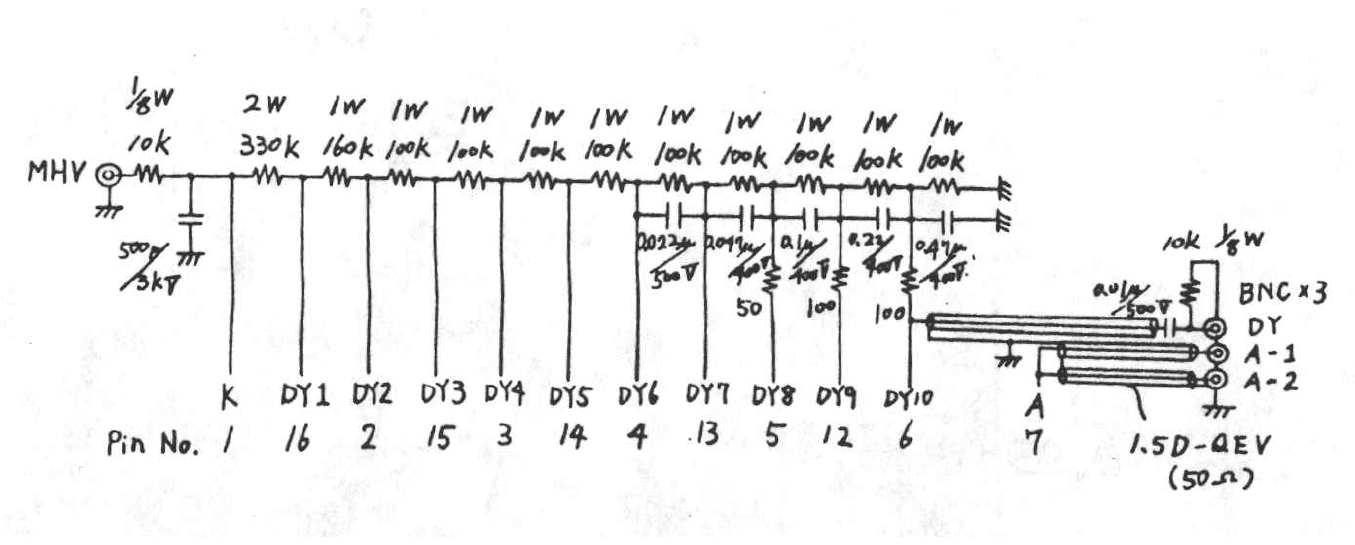}{Schematic wiring diagram of the C31034 socket used for the
experiments.}

A calibrated tungsten blackbody source was used to estimate the
systems' relative spectral response.  The slit widths were set at the values
expected to be used during the experiments.  The systems relative spectral
response $S(\lambda)/S(\lambda_r)$ (where $\lambda_r$ is the reference wavelength, which we have
taken as 400 nm) can be estimated from the measured intensity ratio
$I(\lambda)/I(\lambda_r)$ and that predicted by the lamps calibration (assumed to be a
Blackbody function) $B(\lambda)/B(\lambda_r)$, that is

\begin{equation}
\frac{S(\lambda)}{S(\lambda_r)} \;=\; \frac{B(\lambda)}{B(\lambda_r)}\times\frac{I(\lambda)}{I(\lambda_r)}\;.
\label{eq:4-5}
\end{equation}

The relative spectral response curve for the SPEX monochromator with
the C31034 is shown in figure \ref{fig:4-12}.  As one may expect, the response is
reasonably flat from 400 to 550 nm and falls off gradually for $\lambda >$ 550 nm.
There is virtually no red response for $\lambda >$ 850 nm.

\figstub{fig:4-12}{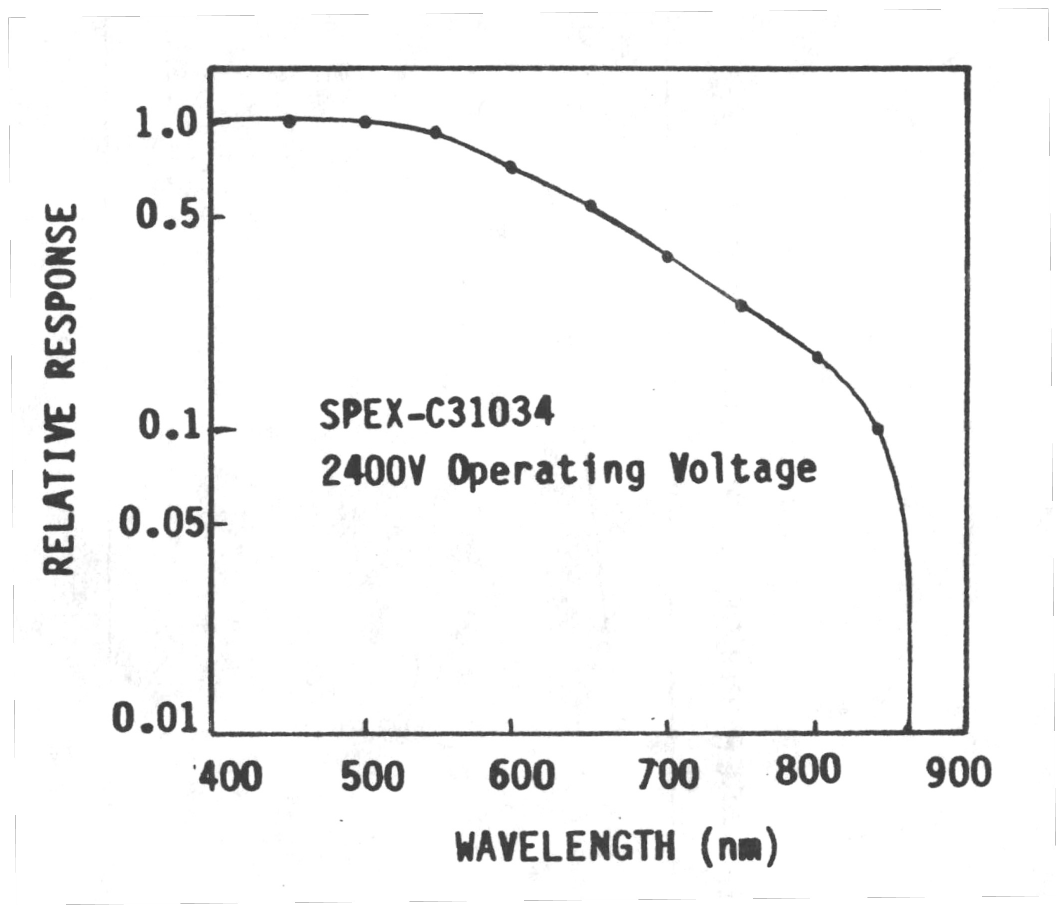}{Relative spectral response curve for the SPEX
monochromator with the C31034.}

For some of the experimental runs, an RCA 7265 photomultiplier tube
(peak QE: 19 \% @ 410 nm; 10 \% relative QE @ 290 and 710 nm) was employed.
The relative spectral response for this tube is shown in figure \ref{fig:4-13}.  The
RCA 7265 having an NaKCsSb (multialkali) photocathode is designed for
efficient blue response and as one can see from the figure, falls off
towards the red much sooner than the C31034.

\figstub{fig:4-13}{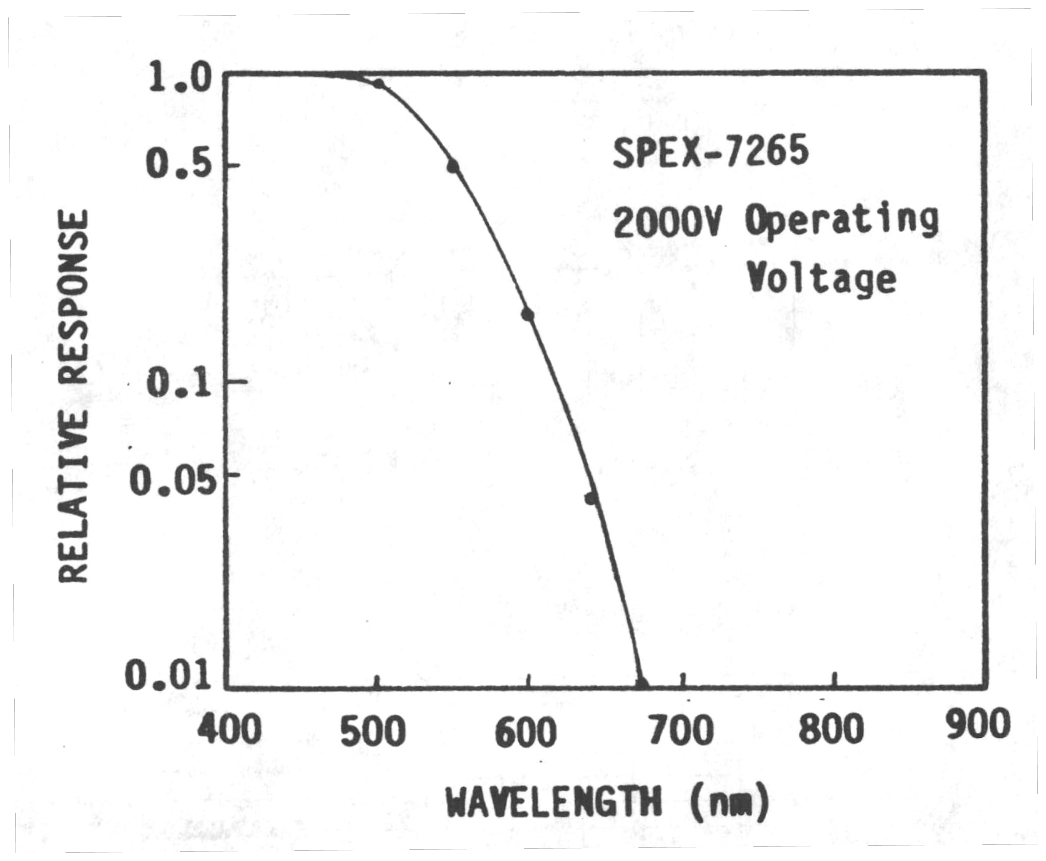}{Relative spectral response for the RCA 7265
photomultiplier tube.}

\section{Photodiodes}
\label{sec:4-4}

Three photodiodes were used in the facility (see figure \ref{fig:4-1}) to
monitor the characteristics of the laser pulse.  Each of the photodiodes
(Hewlett Packard type 5082-4220) are installed in a high speed mount similar
to the original design of McCall (1972).  The total risetime of these mounts
is estimated to be less than 2 ns (Drewell 1979).  The design allowed for a
50$\Omega$ output impedance.  Low noise co-axial cable (Beldon type 9223) was used
to connect the photodiode assemblies to the data aquisition system (see
section \ref{sec:4-5}).

Prior to entering the heat sandwich oven, a fraction of the incident
laser beam (typically 8\%) was deflected via a thin microscope slide onto the
input photodiode which when necessary was calibrated against a Hadron (model
TRG 100) thermopile connected to an energy meter (Hadron model 102A).
Tracing paper across the face of the photodiode mount served to attenuate
and spatially average the directly scattered laser beam.  Providing the
temporal shape of the incident laser pulse does not change with incident
laser energy (this can be ensured by using neutral density filters rather
than the applied voltage on the laser flashlamp as a means of energy
control), then the incident laser energy $E_0$ is directly proportional to the
peak current from the photodiode.  Absolute calibration of the output
photodiode is not as easy since the temporal profile of the outgoing laser
pulse can take on various shapes depending on the dynamics of the plasma
that is created (Cardinal et al 1982, Wong 1985).  When a measurement of the
energy transmitted was required, the outgoing laser pulse was monitored with
the thermopile and energy meter (Hanratty 1977).

A fraction of the beam incident on the input photodiode was used as a
trigger for the data aquisition.  It was discovered that the triggering of
the data acquisition electronics had an internal delay of approximately 40
ns.  As a result, all signals other than that from the trigger photodiode
itself were sent through a co-axial delay cable approximately 13 m in length
to create a 45 ns delay between the arrival of the trigger pulse and the
signal that was to be analyzed.

\section{Data Acquisition Facility}
\label{sec:4-5}

All photomultiplier and photodiode signals were digitized and
processed by an EG\&G (model 4420) boxcar averager equiped with an EG\&G
(model 4422) gated integrator as an amplifier sampling head.  The sampling
head had a risetime of $<$1 ns with a variable sensitivity and a choice of 50$\Omega$
or 1M$\Omega$ as imput impedences.  The minimum sampling gate width was
approximately 2 ns.

The boxcar averager was driven by an Apple IIe microcomputer
interfaced to the boxcar averager via an RS-232 serial interface.  The
driving program (written in Applesoft Basic), has provision for two modes of
operation.  The first one allows the user to sample and integrate an
analogue current pulse at equidistant time spacings after appropriate
triggering of the boxcar averager thereby digitizing the time history of the
current pulse.  The basic options to this mode allow the user to vary the
time window of interest, the number of triggers to be averaged at each
position in time, the number of positions in time, the gate width (period
over which the current is averaged for a single time point) and of course,
the scale sensitivity.  The reader should refer to the operating manual for
the more sophisticated options such as zoom delays and zoom time which allow
the user to zoom in to a particular time window.

The second mode of operation allows the user to sample and report on
the status of a selected position in time.  As an example, this was used to
monitor the shot to shot fluctuations in the signal near the peak of the
temporal history of the laser pulse, as well as to reconstruct the spectral
fluorescence emission at this predescribed position in time of the
fluorescence signal.  This was achieved by continuously scanning the SPEX
monochromator while aquiring data at a fixed position in time.  In this
application, the time scale is converted to frequency scale provided that
the scanning rate is reasonably uniform.

\section{Data Reduction and Analysis}
\label{sec:4-6}

The digitized spectral radiances for the radial free electron density
measurements are stored on mini-floppy diskettes in Applesoft format.  Each
spectrum represent an array of length j for a given $m^{\text{th}}$ lateral position at
$y_m$.  A simple basic program is used to rearrange the data into a single file
containing an array of dimension (i $\times$ j) so that the data can be ordered as
$I_n(y_m)$ with i representing the total number of frequency positions $\nu_n$ and j
represents the total number of lateral positions.  The data are converted
into CP/M format in accordance with the requirements of the file transfer
protocol CROSSTALK.  Using CAPTURE/S available on the UTIAS Perkin-Elmer
mini-computer, the data is transfered to the mainfraim from the Apple
through a dedicated line, the handshaking performed with the CROSSTALK
protocol.

All further data analysis is performed on the UTIAS Perkin-Elmer
mini-computer.  A fortran program has been written to perform the
least-squares fitting of the $I_n(y_m)$ data to the polynomial function described by
equation (\ref{eq:3-24}).  The non-linear fitting parameter $\alpha$, is selected by
inspection of the least squares residual for a range of values of $\alpha$ spanning
three decades (0.01 to 10).  Once the linear co-efficients and $\alpha$ have been
obtained, a second program performs the Abel inversion outlined in sections
(\ref{sec:2-6}) and (\ref{sec:3-2}), thereby generating the inverted spectra for various radial
positions.  A third fortran program prepares the inverted data for plotting.
A fortran program based on the theoretical profiles discussed in chapter 2
generates the Voigt profiles for a trial electron density and electron
temperature, to evaluate the most appropriate values of $N_e$ and $T_e$ for
comparison with the experimental spectra.

\chapter{Experimental Results}
\label{ch:results}

The influence of the laser field on the perturbation of the resonance
state (section \ref{sec:2-3}) and the effect of radiation trapping arising from the
elevated and extended $3^2P$ population density (section \ref{sec:2-5}) suggested that
measurements of the free electron density and temperature are limited to
times later than the laser pulse duration.  Although earlier theoretical
predictions based on a step-wise laser excitation (Measures and Cardinal
1981) indicated ionization times (time where runaway ionization burnout
occurs) well within 50 ns for sodium densities N $\sim 10^{16}$ cm$^{-3}$ and laser
irradiances $I^{\ell}\sim 10^{7}$ Wcm$^{-2}$, more realistic modelling to include the actual
temporal shape of the laser pulse and attenuation resulting from the
propagation through a finite length of sodium vapor (Wong 1985, Kissack
1987) indicates that the free electron density can sustain a positive growth
well after the peak of the laser pulse.  This of course arises from the
energy stored in the large pool of atoms that have been resonantly excited.

Measurements of the $4^2D$-$3^2P$ emission spectra from a sodium plasma
column produced by resonance excitation for times before ($\sim$10 ns), during
($\sim$20 ns) and shortly after ($\sim$30 ns) the peak of the laser pulse reveal the
distortion arising from self-absorption as a result of the heavily populated
$3^2P$ state due to direct laser pumping (see figure \ref{fig:5-1}).  As time progresses,
the reversal near line centre disappears (figure \ref{fig:5-2}) and the spectrum takes
on the expected electron Stark broadened multiplet shape.

\figstub{fig:5-1}{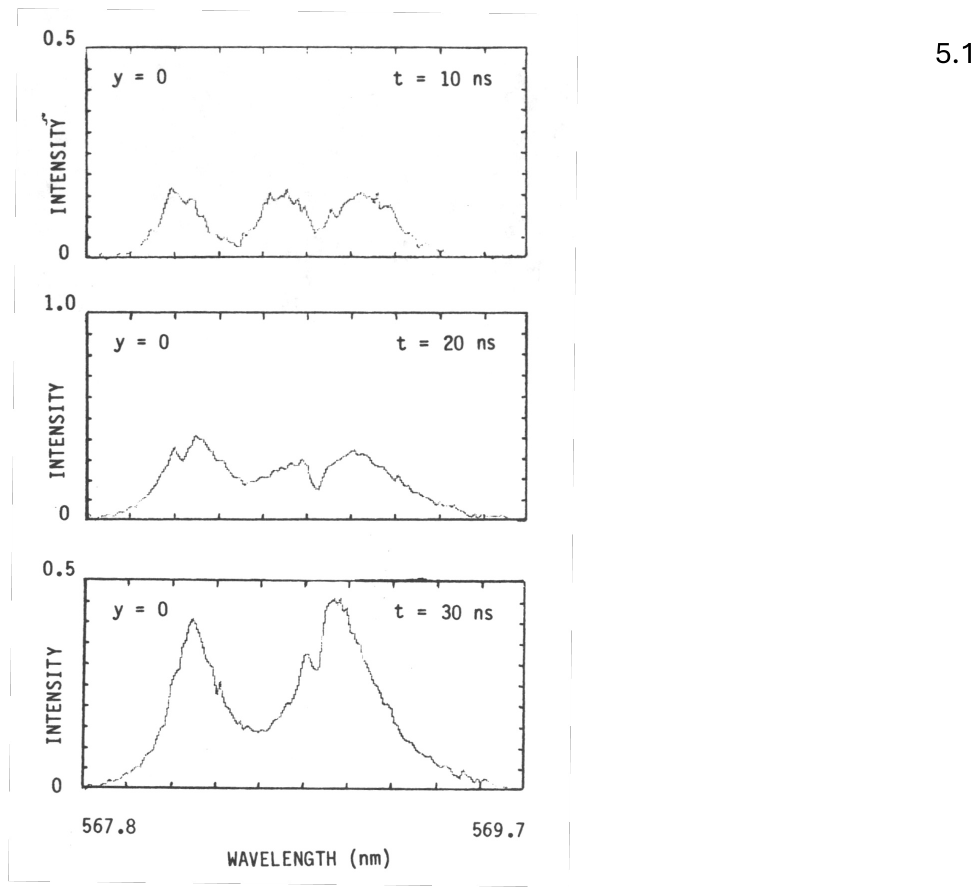}{Measurements of the $4^2D$-$3^2P$ emission spectra for times
before ($\sim$10 ns), during ($\sim$20 ns) and shortly after ($\sim$30 ns) the peak of
the laser pulse, revealing the distortion arising from self-absorption.}

\figstub{fig:5-2}{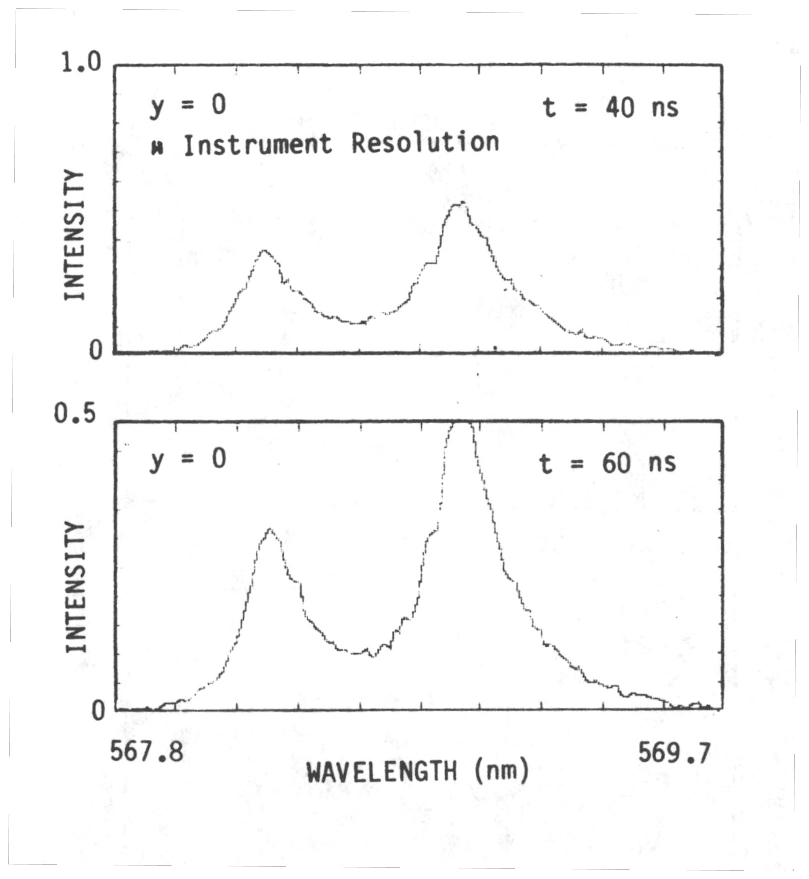}{Disappearance of the reversal near line centre as time
progresses.}

Although the direct perturbation of the laser field due to the AC
Stark effect on the $3^2P$ state is expected to be small in comparison to the
distortion from self-absorption, the creation of a virtual state at the
laser detuning is none the less visible.  Spectra recorded at a time of
15 ns after the onset of the laser pulse (near the peak) for various laser
detunings are shown in figure \ref{fig:5-3} and reveal that a small but noticeable
peak is visible at wavelengths which correspond to the radiative decay from
the $4^2D_{5/2}$ state down to this virtual state.  The correlation between laser
detuning and the position of this peak is illustrated in figure \ref{fig:5-4}.  The
relative magnitude of the peak in comparison to that of the unperturbed
spectra is dependent on a number of variables such as laser cross-section
uniformity, observation geometry and of course, strength of the interaction,
which depends on laser detuning from the $3^2P$ state.

\figstub{fig:5-3}{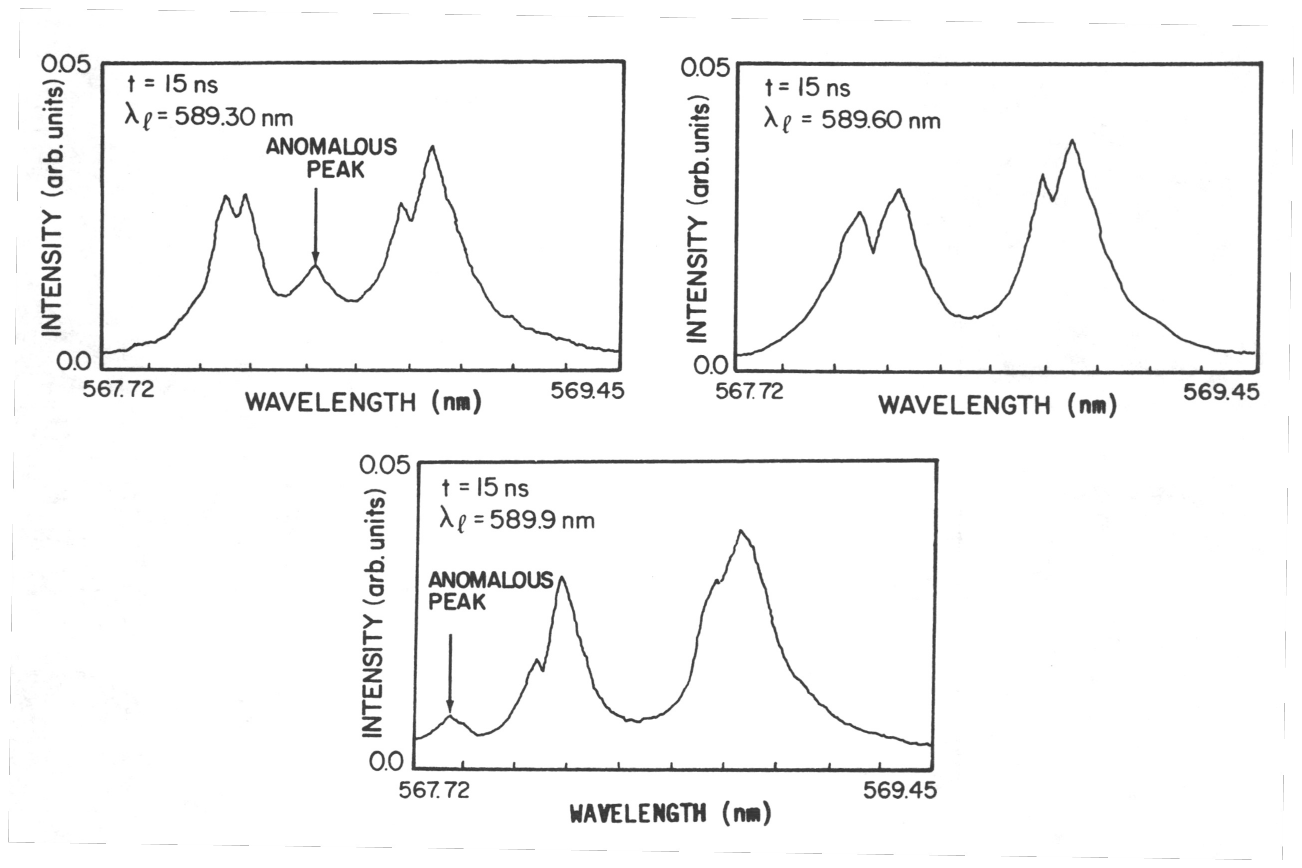}{Spectra recorded 15 ns after the onset of the laser pulse
for various laser detunings.}

\figstub{fig:5-4}{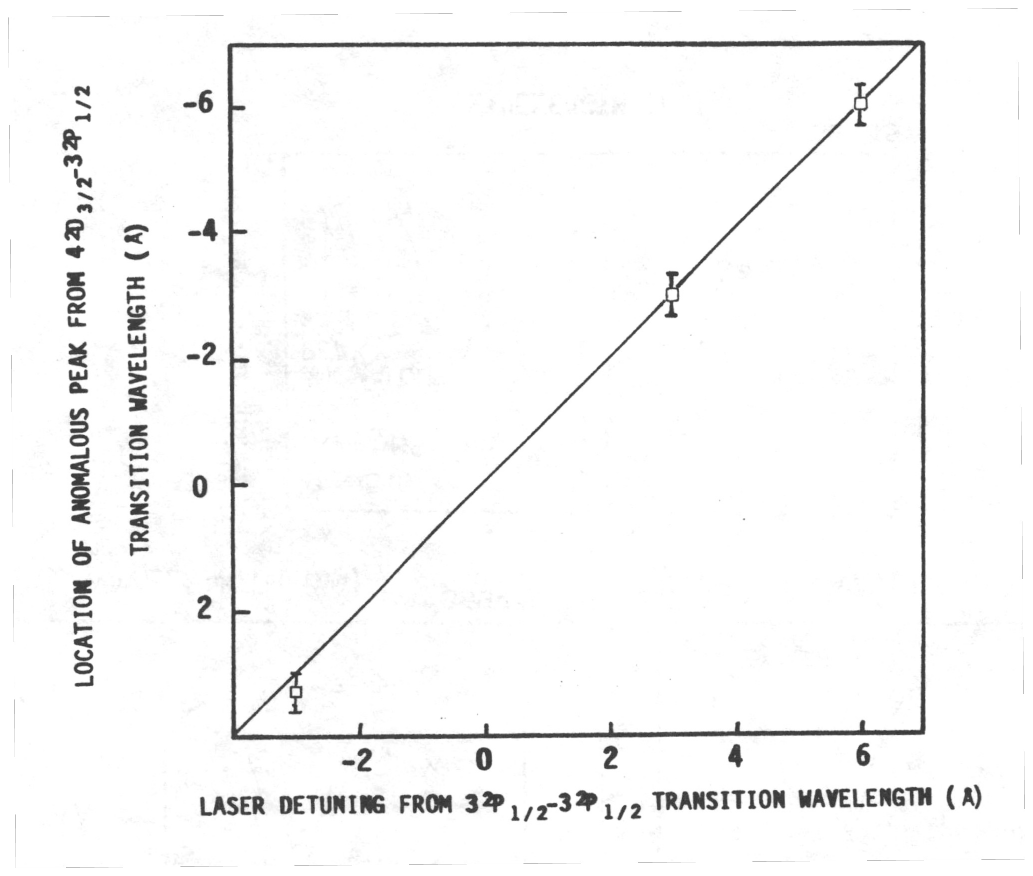}{Correlation between laser detuning and the position of the
virtual state peak.}

In the following sections, we shall report on the results of a number
of experiments based on the observation of the spectral broadening and
relative intensity of the $4^2D$-$3^2P$ transition in sodium.  The times for which
the useful (for quantitative measurements) spectra are recorded are limited
to times greater than 45 ns after the onset of the laser pulse.  Section \ref{sec:5-1}
deals primarily with the measurement of the radial variation in the free
electron density.  The measurements of electron temperature are presented in
section \ref{sec:5-2}.  The electron temperature distributions are derived from the
relative intensity of emission from the $4^2D$-$3^2P$ and $6^2D$-$3^2P$ transitions.
Electron temperatures are also estimated from the Saha relationship and the
measured electron and neutral sodium atom densities.  It should be
remembered that the neutral sodium density can have a systematic error of as
much as 25\% associated with it (section \ref{sec:3-1}), which may not have a
significant effect on the relative temperature radial distribution, but
certainly on its magnitude.  Section \ref{sec:5-3} is devoted to the study of the
laser beam penetration through the dense sodium vapor, particularly to the
analysis of the transmitted laser beam with attention to the implication
that laser attenuation may have to the creation of long plasma channels.

\section{Electron Density Measurements}
\label{sec:5-1}

The first measurements of the radial variation in the free electron
density were made in a plasma channel created in sodium vapor irradiated
with 37 mJ of laser energy over a beam diameter of approximately 5 mm.  The
plasma was observed at the oven centre where the sodium vapor density was
estimated at $1.7\times10^{16}$ cm$^{-3}$.  The laser was tuned to the 589 nm ($3^2S_{1/2}$-
$3^2P_{3/2}$) transition and the electronics was triggered and gated to sample
data over 2 ns, 58 ns after the onset of the laser pulse incident on the
sodium vapor.  Spectral scans of the $4^2D$-$3^2P$ emission multiplet were taken
for 16 lateral (y) positions across the plasma column at right angles to the
direction of laser propagation.  The spectral resolution was estimated to be
0.05 nm (FWHM) from the width of the emission lines observed at large y
values.  Sixteen laser shots were averaged to constitute one intensity
measurement at a given frequency, and 128 frequency values constitute one
spectral scan.  With the laser operating at 10 Hz, the complete scan takes
205 seconds.

Wavelength calibration was performed by the use of the relative
position and relative separation of the members in the multiplet observed at
large y values.  An approximate wavelength calibration can be obtained from
the monochromator setting indicator and scanning motor controller which is
used to rotate the grating (thereby changing the frequency at the exit slit)
at a predescribed rate.  Two representative spectral scans are presented in
figure \ref{fig:5-5}.  The emission originating from the central core of the plasma
(y=0.25 mm) is seen to suffer a much greater electron Stark broadening than
the emission arising from the more weakly ionized rim of the plasma column
(y=2.25 mm).  Figure \ref{fig:5-6} illustrates the variation in the emission with
lateral position y, for four wavelength values.  Along with the 16 data
points for each curve, we have plotted the corresponding polynomial fit
(solid line) generated by the method of least squares.  The generated
functions are quite versatile and represent the data reasonably well.  When
inverted using the Abel transformation, one obtains the corresponding
spectral emission versus radial coordinate r (see figure \ref{fig:5-7}).  All of the
functions show a remarkably different variation, and at first glance there
appears no recognizable pattern.  The error bars (arising from the estimated
scatter in the signal) of course increase for smaller r.  Reconstruction of
the spectra at various radial positions allow us to make a clearer
interpretation of these data (see figure \ref{fig:5-8}).  The solid lines in figure
\ref{fig:5-8} are Voigt functions which are generated by convoluting electron Stark
broadened Lorentzian functions (using the electron density and temperature
indicated in the figure) with a Gaussian function to represent the
instrument resolution.  Whiting's approximation (1968) for the Voigt
integral was used to simplify the computation.  A FWHM of 0.05 nm was used
as the instrument resolution for r $\geqslant$ 1.2 mm.  At smaller radii, smoothing of
the experimental data was necessary and 5-point moving averages were
performed on the inverted spectra.  This in effect introduces a further
instrumental contribution to the experimental data and to compensate for
this, the theoretical profiles incorporate an instrument resolution that can
be approximated by,

\figstub{fig:5-5}{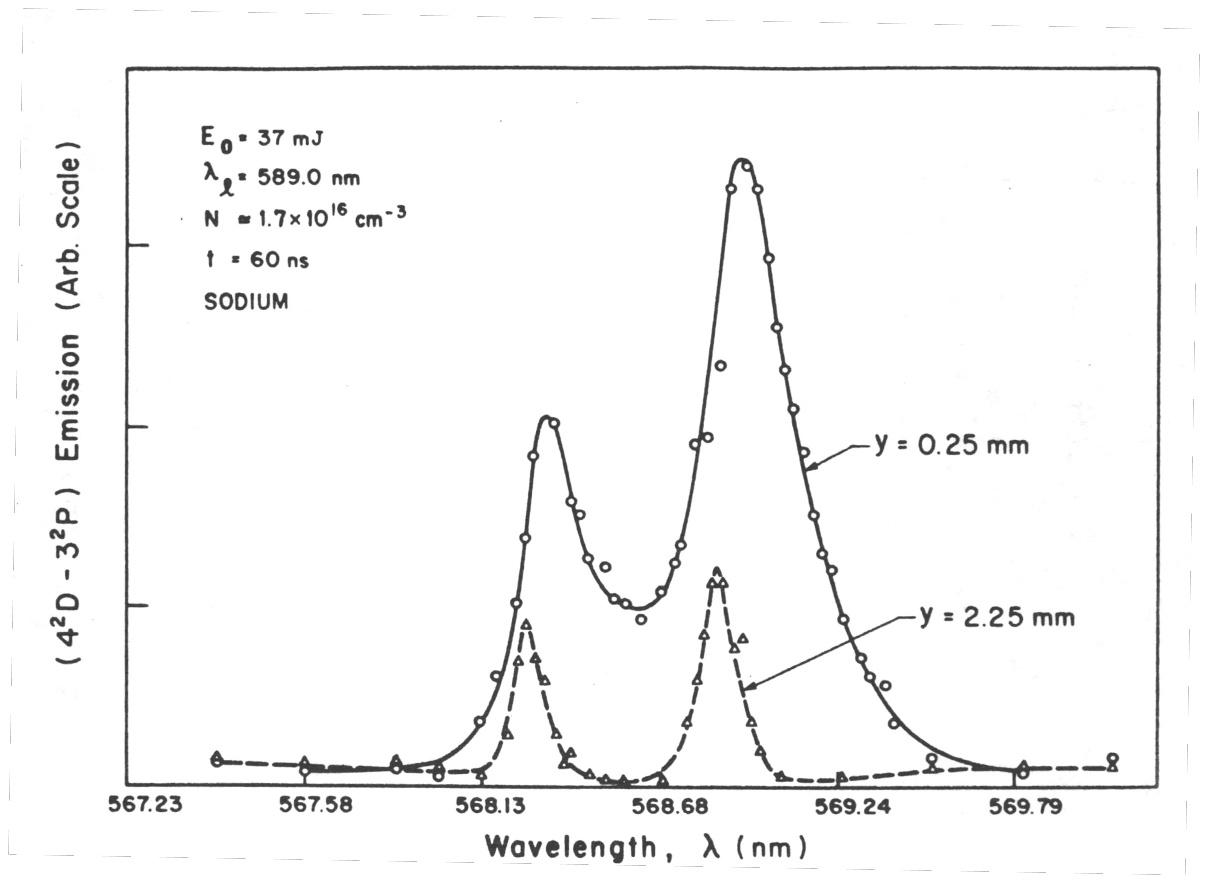}{Two representative spectral scans, at y=0.25 mm and
y=2.25 mm.}

\figstub{fig:5-6}{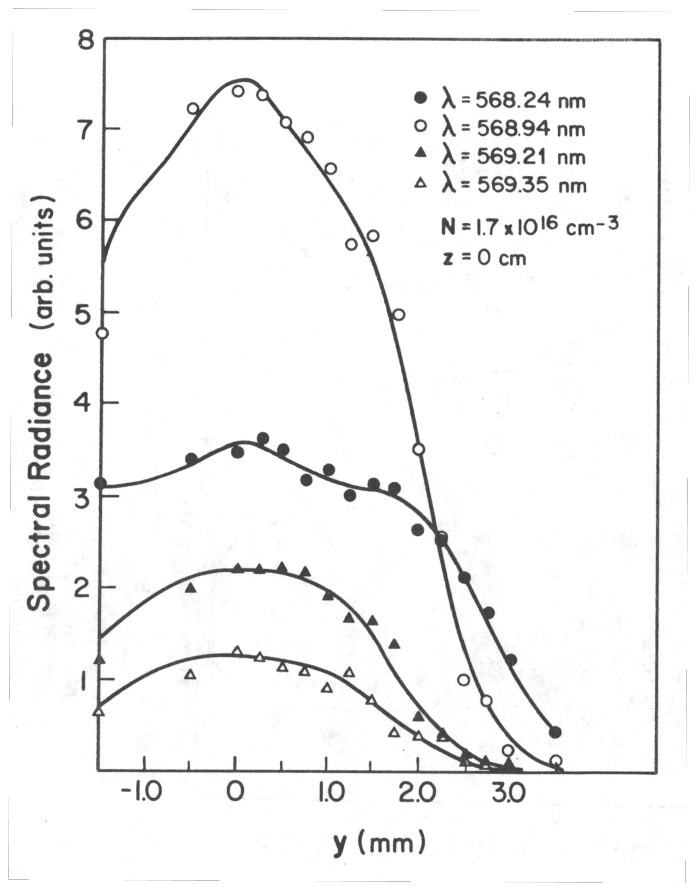}{Variation in the emission with lateral position y, for four
wavelength values, with the corresponding polynomial fit (solid line).}

\figstub{fig:5-7}{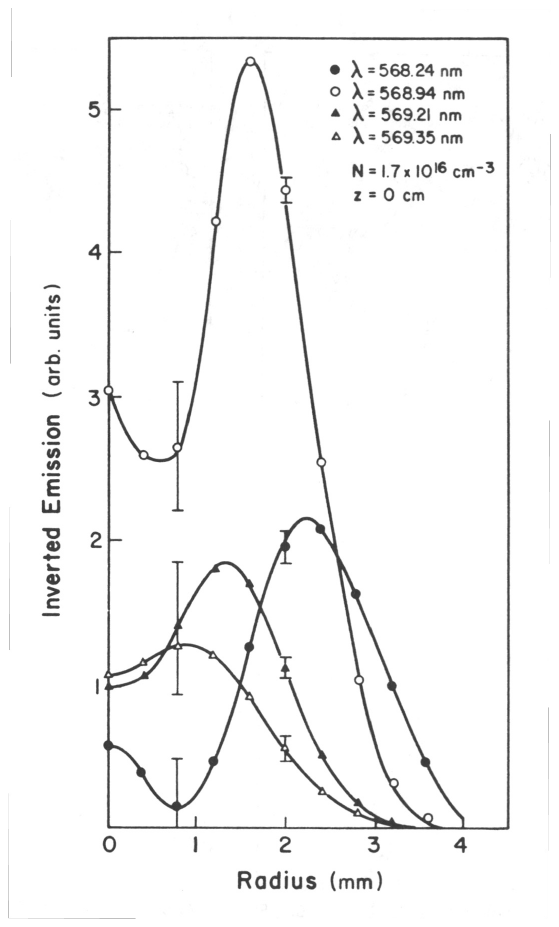}{Corresponding spectral emission versus radial coordinate r.}

\[
\Delta\lambda_I \;=\; \sqrt{(0.05^2 + Dn^2)}
\]

where D represents the monochromator dispersion (nm per experimental point)
and n is the number of experimental points used in the moving average.  $\Delta\lambda_I$
for the inverted spectra is indicated in the figure.

As expected, the scatter in the spectra increases towards the plasma
core and reasonably accurate measurements of the electron density for this
particular set of results, are restricted to r $\geqslant$ 1.2 mm.  The results
illustrated in figure \ref{fig:5-7} do suggest that significant ionization may exist
in the plasma core as the radial variation in the local volume emission
(spectrally integrated inverted emission) takes a noticeable drop for r $<$
1.2 mm.  The radial variation in the free electron density (inferred from
figure \ref{fig:5-8}) is illustrated in figure \ref{fig:5-9}.  Although it is difficult to
determine the degree of ionization in the plasma core, it is apparent that
less than 50\% is achieved for r $\geqslant$ 2.0 mm.  The gradual fall-off in laser
power as one approaches 2.5 mm (recall figure \ref{fig:4-8} which shows an example of
the laser intensity cross-section) is assumed to be the cause of the
relatively shallow electron density gradient observed.

\figstub{fig:5-8}{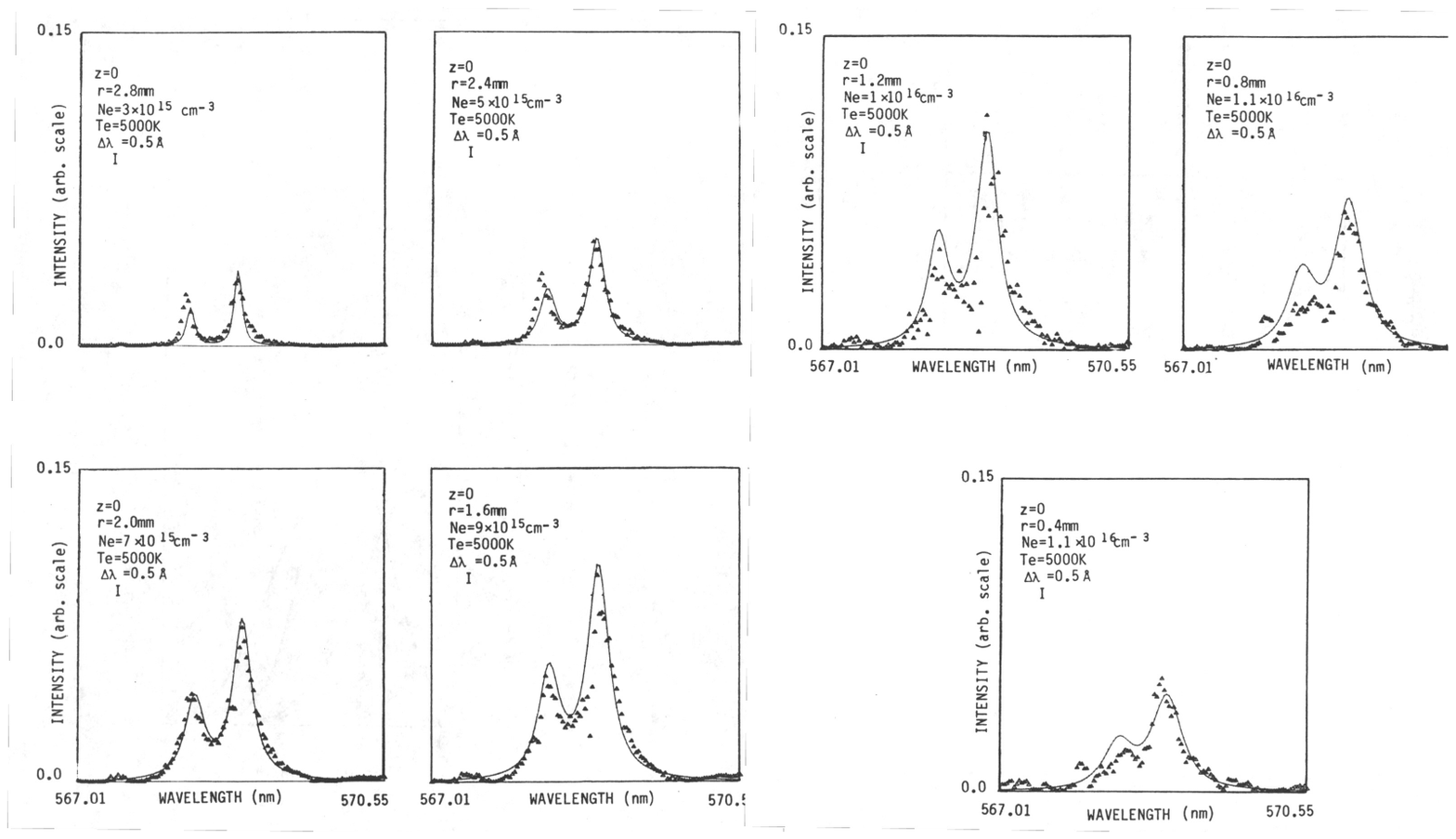}{Reconstruction of the spectra at various radial positions.
The solid lines are Voigt functions generated by convoluting electron Stark
broadened Lorentzian functions with a Gaussian instrument function.}

\figstub{fig:5-9}{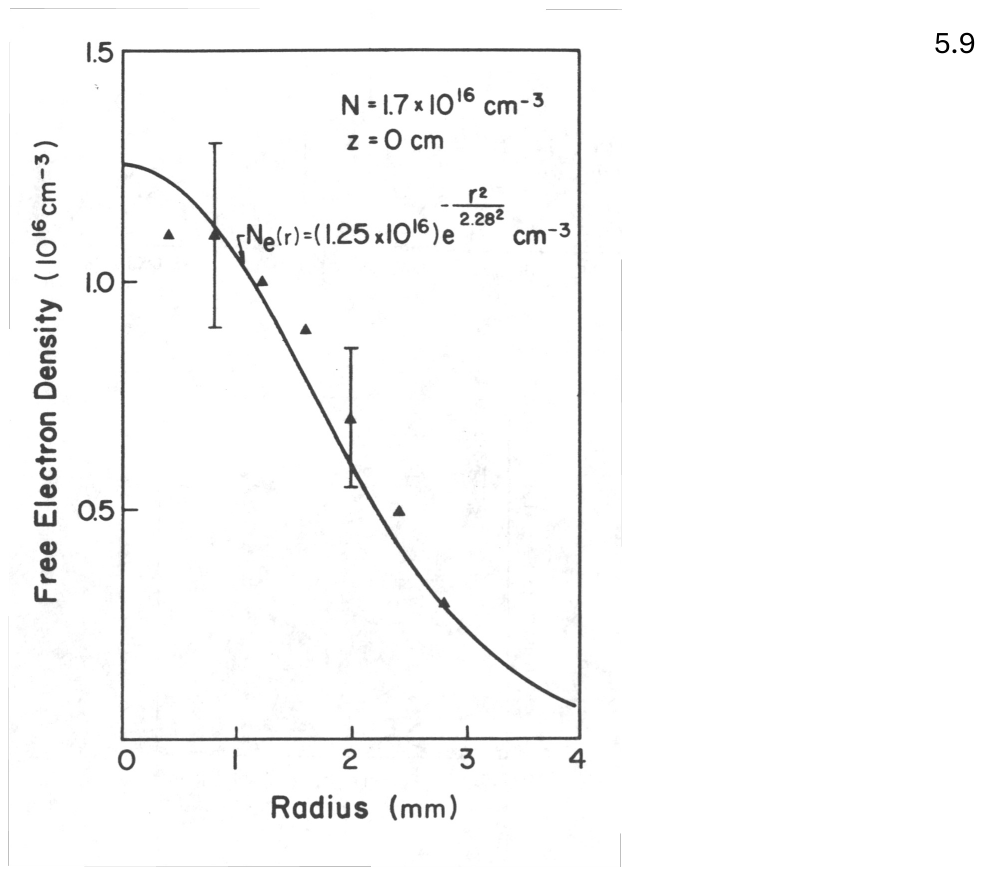}{Radial variation in the free electron density inferred from
figure \ref{fig:5-8}.}

\subsection{Variation of $N_e(r)$ with Beam Penetration Depth}
\label{sec:5-1-1}

The following section reports on the results of an experiment which
measured the difference in the radial electron density distribution between
two positions along the direction of laser beam propagation (Cappelli and
Measures 1987b).  With the laser beam passing directly through the oven
centre, the positions of observation were selected such that they lie an
equidistance (2.0 cm) from the oven centre defined by z = 0 cm.  The
variation of the neutral sodium atom density along the laser path is
illustrated in figure \ref{fig:5-10}.  The neutral density at z = -2 cm and z = 2 cm
is approximately $10^{16}$ cm$^{-3}$.  Providing that there is axial symmetry about
the oven centre (which for this experiment corresponds to z = 0 cm), then
the difference in the electron density observed is strictly a result of the
laser energy depletion as it penetrates through the extra 4 cm of sodium
vapor.  In this experiment, the laser beam cross-section is similar to that
of the previous experiment described, and the laser energy incident onto the
sodium vapor was measured to be approximately 25 mJ.  The laser wavelength
was such that it was tuned to the $3^2S_{1/2}$-$3^2P_{1/2}$ transition at 589.6 nm and
the electronics was gated to sample 2 ns, 65 ns after the onset of the laser
pulse.

\figstub{fig:5-10}{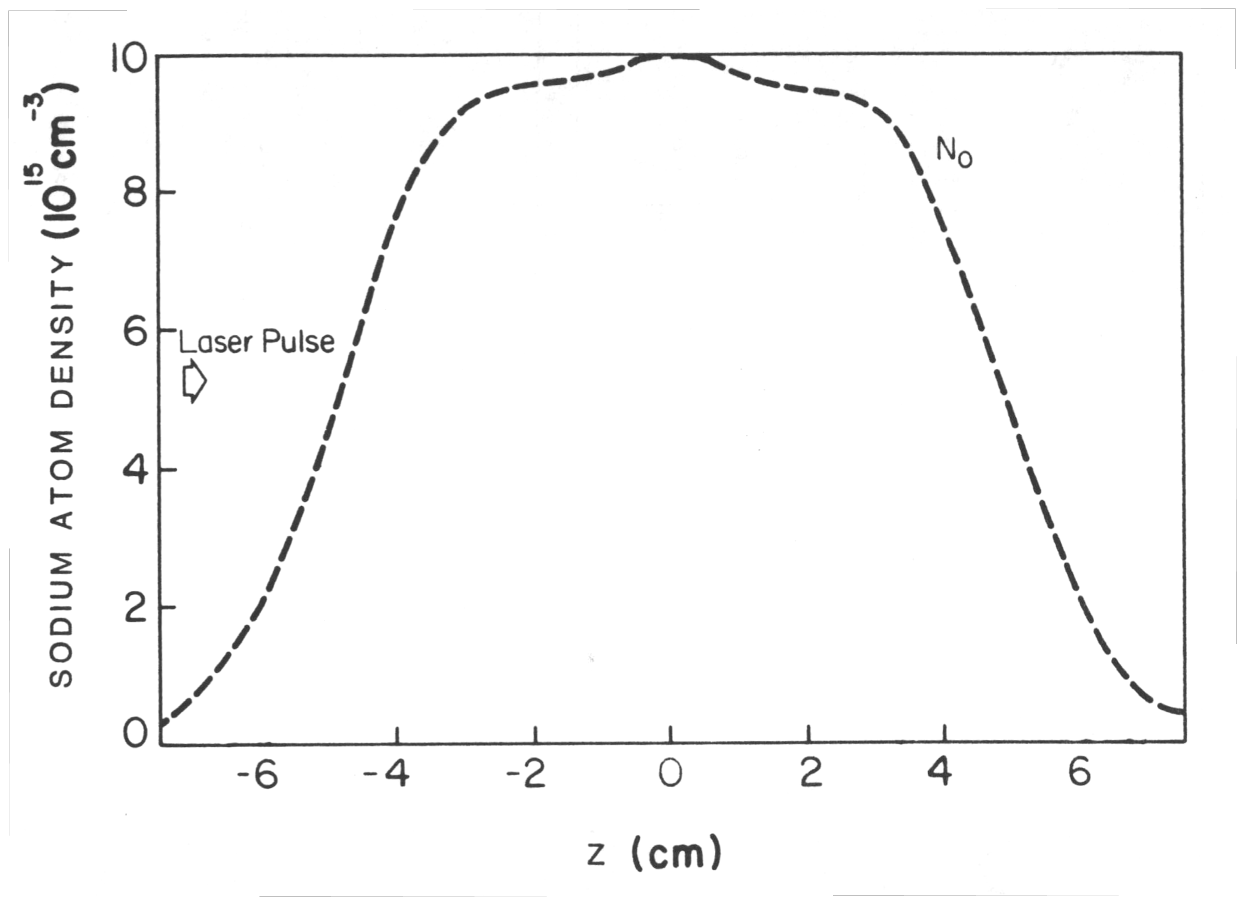}{Variation of the neutral sodium atom density along the
laser path.}

A set of representative spectral scans of the $4^2D$-$3^2P$ multiplet
recorded at z= -2 cm and z= 2 cm (the laser pulse enters at z= $-\infty$) are
presented as figure \ref{fig:5-11}.  Again, the emission arising from small lateral
displacements is seen to suffer a much greater Stark broadening and red
shift than the emission originating from the more weakly ionized rim of the
plasma column.  The difference in broadening between z= -2 cm and z = 2 cm
clearly indicates that the electron density is decreasing along the path of
the laser pulse.

\figstub{fig:5-11}{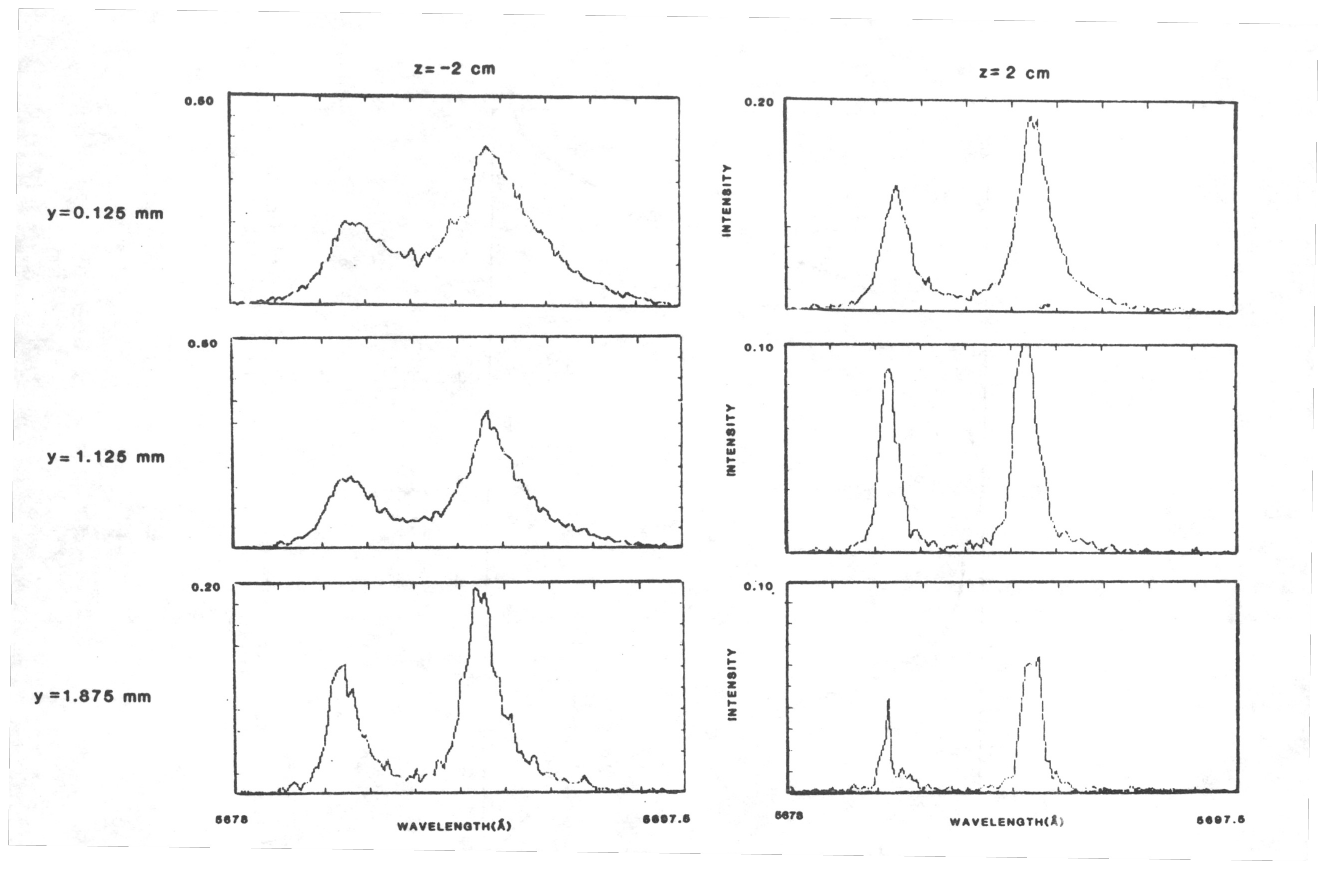}{Representative spectral scans of the $4^2D$-$3^2P$ multiplet
recorded at z = -2 cm and z = 2 cm.}

Two representative inverted ($4^2D$-$3^2P$) multiplet spectra for z = -2 cm
and r = 1.2 mm, and z = 2 cm and r = 0.3 mm are displayed as ``o'' data points
in figure \ref{fig:5-12}.  Each of these inverted multiplet spectra were fitted by
three theoretically computed Stark electron impact broadened profiles that
have been convoluted with a Gaussian function having a 0.05 nm FWHM (curves
in the figure).  This permits us to ascertain both the most likely value for
the free electron density and the spread in this value for each radial
position.  For the curves depicted in figure \ref{fig:5-12} (as well as for all the
results presented), we recall that we have used the Stark widths and shifts
of Griem (1974) and the temperature of the free electrons was assumed to be
5000K.  Fortunately, these calculations are quite insensitive to electron
temperature (see Table \ref{tab:2-1}) for the expected range of conditions.

\figstub{fig:5-12}{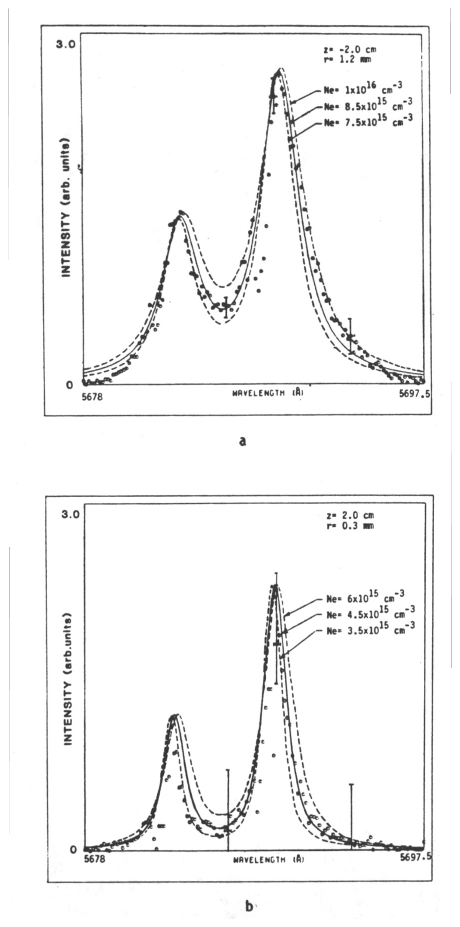}{Two representative inverted ($4^2D$-$3^2P$) multiplet spectra
for z = -2 cm, r = 1.2 mm and z = 2 cm, r = 0.3 mm (``o'' data points), with
fitted theoretical profiles.}

In figures \ref{fig:5-13} and \ref{fig:5-14}, we present examples of the inverted
multiplet spectra with the theoretically computed Stark broadened profiles
for the most likely free electron density.  These figures reaffirm the
qualitative interpretation of the raw spectral data presented in figure
\ref{fig:5-11}, that is to say, the electron density and radius of the plasma is
decreasing along the direction of laser beam propagation.  This deduction is
even more graphically illustrated in figure \ref{fig:5-15} where we have evaluated the
free electron density radial profiles at z = -2 cm ($\Delta$ data) and z = 2 cm (o
data).  Also shown are two empirical curves for $N_e(r)$ that represent this
data quite well.

\figstub{fig:5-13}{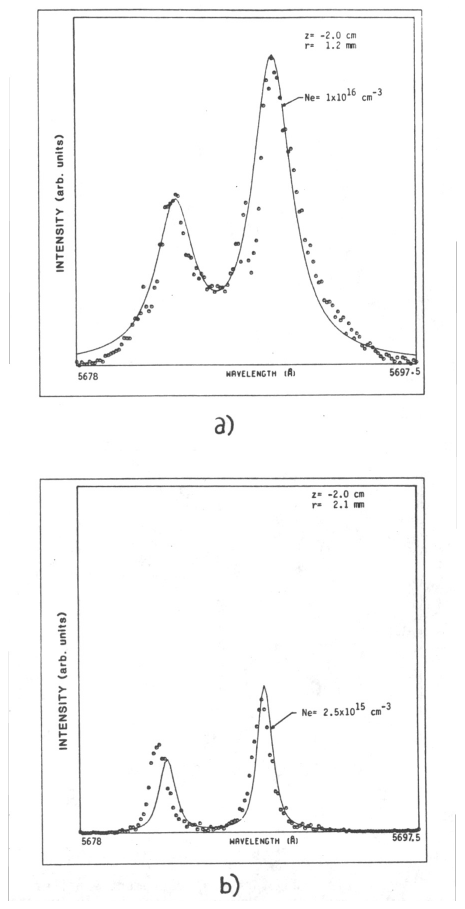}{Inverted multiplet spectra with the theoretically computed
Stark broadened profiles for the most likely free electron density,
z = -2 cm.}

\figstub{fig:5-14}{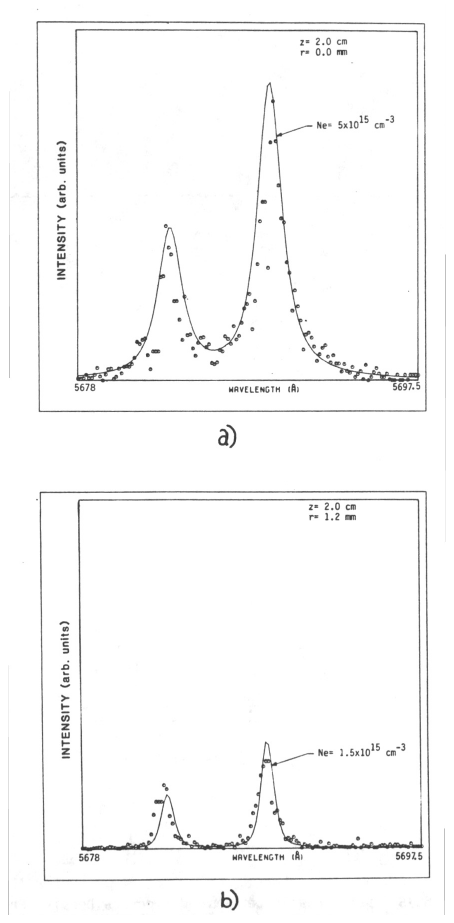}{Inverted multiplet spectra with the theoretically computed
Stark broadened profiles for the most likely free electron density,
z = 2 cm.}

\figstub{fig:5-15}{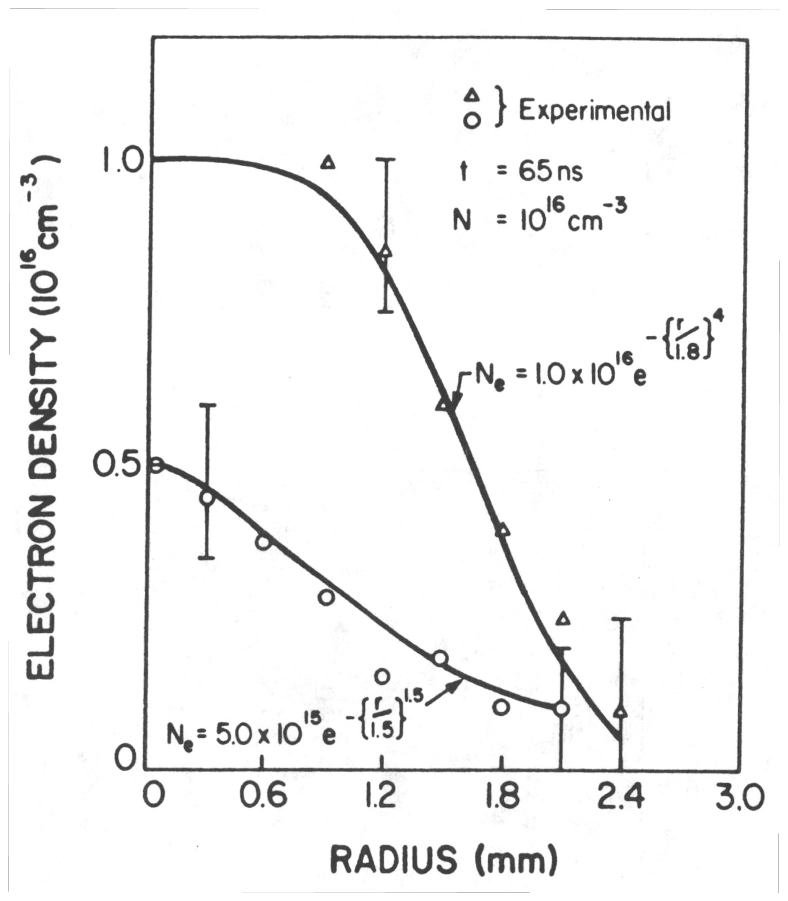}{Free electron density radial profiles at z = -2 cm
($\Delta$ data) and z = 2 cm (o data), with two empirical curves for $N_e(r)$.}

The results indicate that although the sodium atom density is roughly
the same at these two axial positions, the radius of the electron density
distribution, and the peak density of the free electrons is dramatically
different.  This can be understood in terms of the absorption of the laser
pulse as it propagates through the sodium vapor.  A comparison between these
experimentally evaluated free electron density radial profiles and those
predicted on the basis of a computational model of LIBORS is presented in
Chapter 6.

As a check on the consistency of our measurements, we have computed
the optically thin solution (equation (\ref{eq:2-104})) and full solution (equation
(\ref{eq:2-72})) generating the multiplet spectra that would be observed at z = -2 cm
and y = 1.125 mm, using the free electron density radial profile (empirical
fit) measured at z = -2 cm.  These computed spectra are then compared to the
actual measured spectra at z = -2 cm and y = 1.125 mm.

The computed multiplet spectra were obtained by first deriving the
radial profile of the free electron temperature using the Saha equation, the
empirical fit to the free electron density profile and the initial sodium
atom density.  This temperature distribution, combined with the assumption
of LTE allowed us to ascertain the radial profiles of the $4^2D$ and $3^2P$
population densities.  The multiplet spectrum observed at any given lateral
(y) position was then determined by solving the radiative transfer equation
assuming that either the $3^2P$ population was zero everywhere (optically thin
solution) or as determined by the full solution under LTE conditions.

This comparison is presented as figure \ref{fig:5-16} and suggests that the
distortion to the profiles due to optical depth is within experimental
uncertainty and can be neglected.  Both of the computed spectra are in good
agreement with the experimental data, implying that optical depth plays a
minor role in the interpretation of the Stark broadened spectra.

\figstub{fig:5-16}{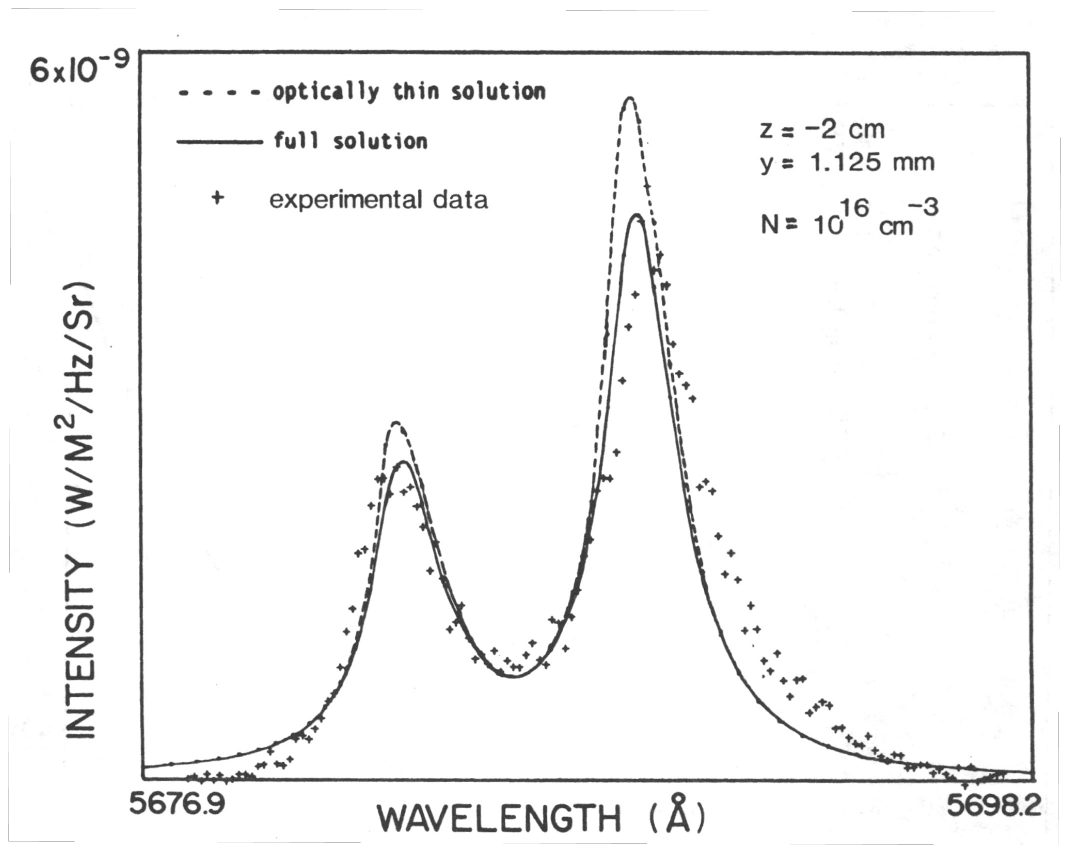}{Comparison of the computed optically thin and full
solutions with the actual measured spectra at z = -2 cm and y = 1.125 mm.}

As a further check on these measurements, we compare the computed $4^2D$
population density radial profiles at z = -2 cm and z = 2 cm (curves) with
the measured $4^2D$ population density (normalized at r = 0 cm to the computed
density radial profile at z = -2 cm) based on the Abel inversion of the
spectrally integrated lateral emission profiles.  Of course, this comparison
is only valid if the lateral emission can be taken as optically thin.  This
comparison is presented as figure \ref{fig:5-17} and indicates a general, but not
exact agreement.

\figstub{fig:5-17}{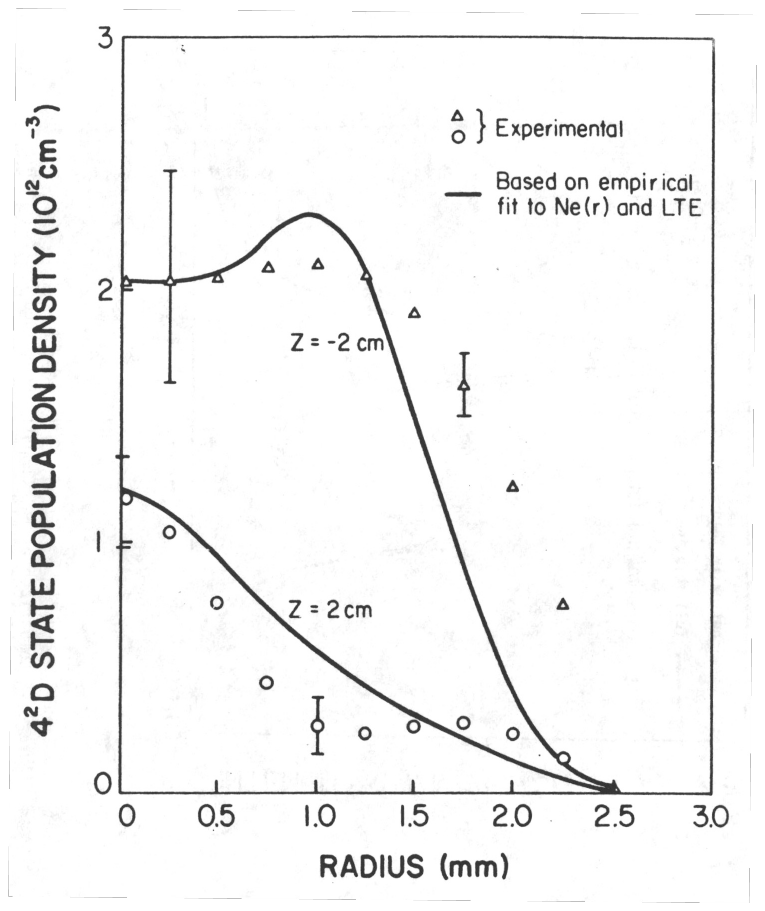}{Computed $4^2D$ population density radial profiles at
z = -2 cm and z = 2 cm (curves) compared with the measured $4^2D$ population
density.}

In addition to the radial measurements of the free electron density,
the $4^2D$-$3^2P$ multiplet has been recorded for y = 0, along five axial
locations (z = -4, -2, 0, 2, 4 cm).  These spectra can be used to give us a
fairly reasonable indication of the variation of the average electron
density along the path of the laser beam (of course, the electron density
evaluated from the spectra is not an average, but rather, the emission is
weighted to represent the region of greatest $4^2D$ population density).  This
is accomplished by matching the spectra at each z location to that evaluated
from the full solution of the one-dimensional radiative transfer equation
for a uniform plasma of radius $r_0$ and electron density $N_e$.  In effect, $N_e$
and $r_0$ are used as fitting parameters, after estimating $r_0$ and the scaling
factor for the absolute intensity, from the actual radial electron density
masurements at z = -2 cm.  One can see from figure \ref{fig:5-15} that $r_0$(z=-2cm) $\approx$
2.0 mm.  Figure \ref{fig:5-18} presents this set of spectra.  The computed spectra
(principly Stark broadened) represented by the smooth curve, are seen to
approximate the experimental spectra in regards to: the ratio of peak
heights, the peak to minimum ratio and the widths of the
profiles.  Confirmation that these spectra (which are integrated along the
line of sight at y = 0) can be used to estimate the core electron density,
is obtained by comparing the predictions at z = -2 cm and z = 2 cm from
figure \ref{fig:5-18} to the actual radial profiles presented in figure \ref{fig:5-15}.

\figstub{fig:5-18}{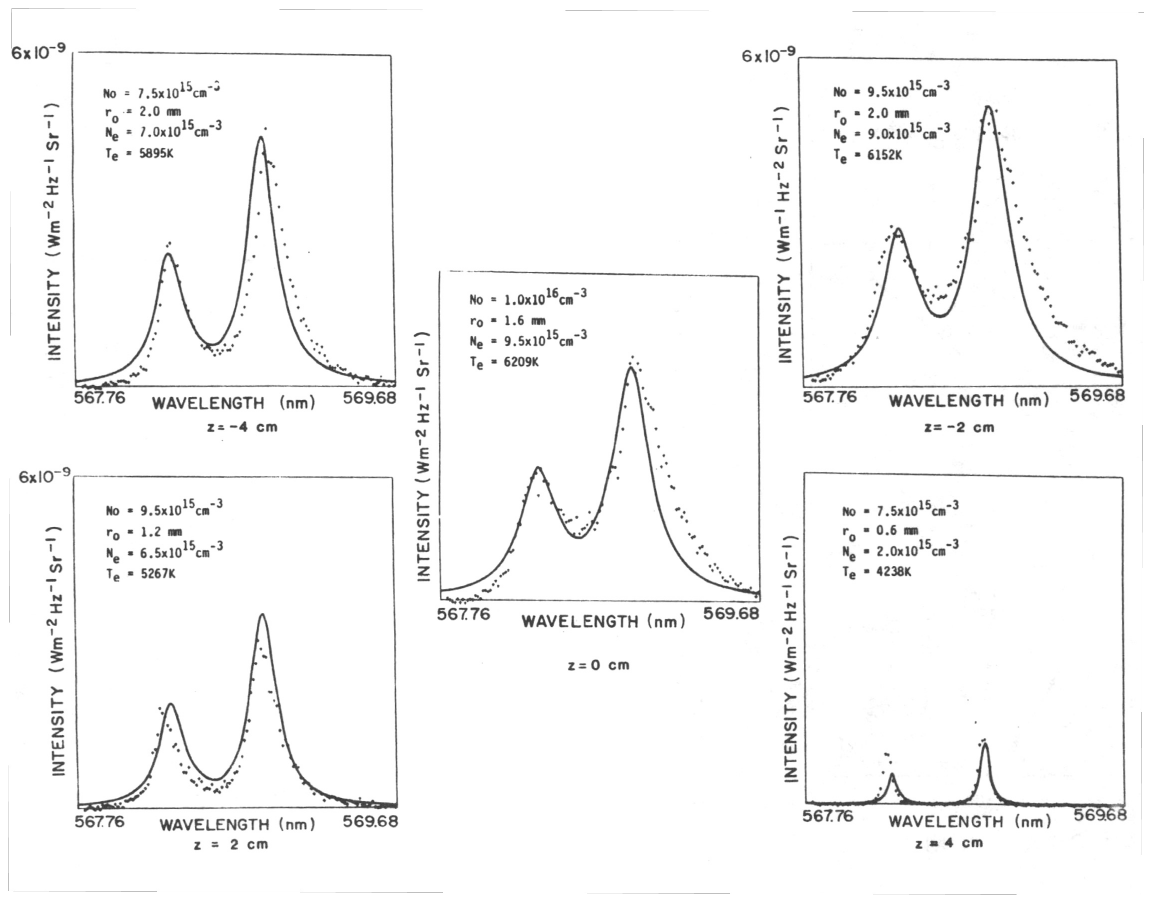}{Measured lateral spectra for the five axial positions,
along with the spectra computed by the solution of the radiative transfer
equation on the basis of a uniform plasma.}

The variation of $N_e$ with z at these five locations is illustrated in
figure \ref{fig:5-19}.  Also drawn in the figure is the variation of the electron
temperature obtained from the assumption of Saha equilibrium and knowledge
of the neutral sodium atom density variation along z.

\figstub{fig:5-19}{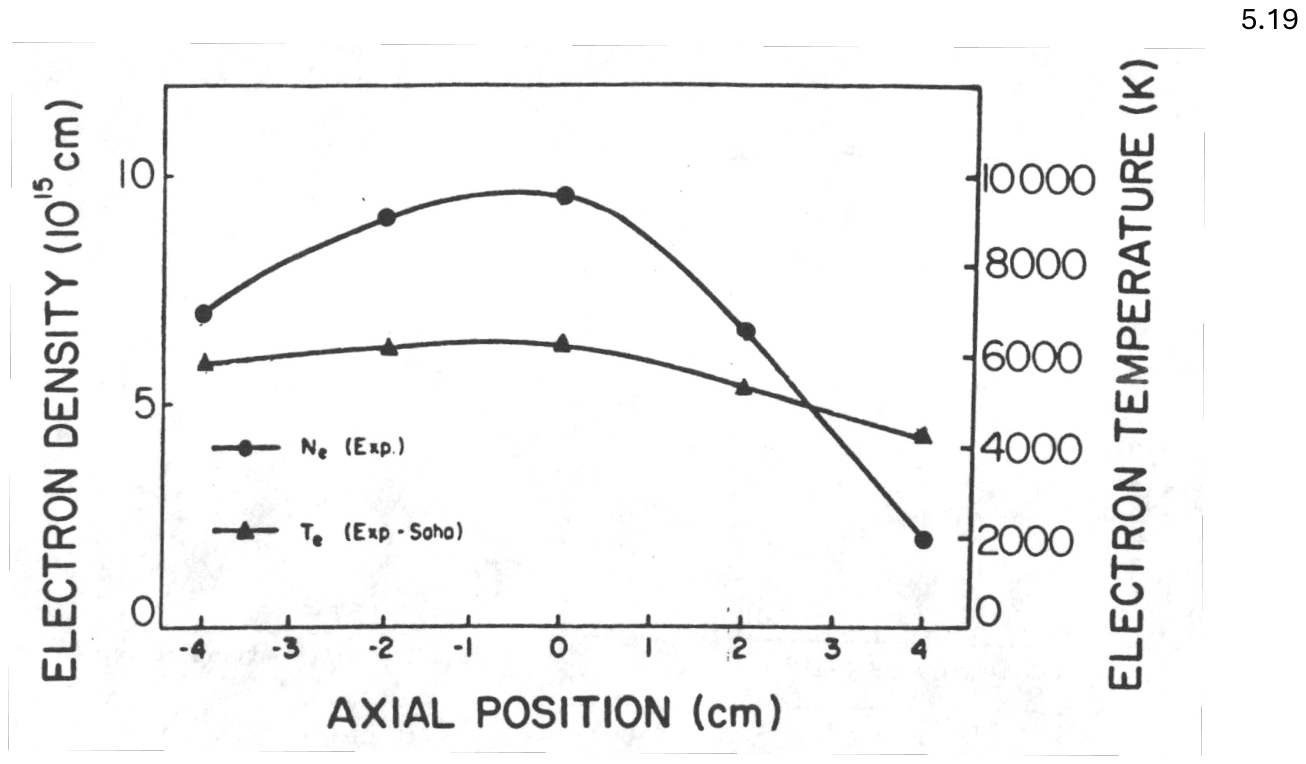}{Variation of $N_e$ with z at the five axial locations,
together with the variation of the electron temperature obtained from the
assumption of Saha equilibrium.}

A second experiment was performed at a higher sodium atom density
(figure \ref{fig:5-20}).  For this experiment, the laser beam was apertured to allow
only the core, 1.6 mm in diameter, to enter the heat sandwich oven.  The
purpose of this was to define a volume of vapor that is to be excited by a
more uniform cross-section of laser radiation.  The incident laser energy
entering the oven was estimated to be 12 mJ, and the plasma was observed at
z = -2 cm where N $\sim 1.3\times10^{16}$ cm$^{-3}$.  Sampling of the $4^2D$-$3^2P$ emission was
again performed over 2 ns, 65 ns after the onset of the incident laser
pulse.  The reconstructed inverted spectra along with the best fit computed
Voigt profiles for a number of radial positions are depicted in figure
\ref{fig:5-21}.  The resulting radial variation in the free electron density is
illustrated in figure \ref{fig:5-22}.  It should be noted that aperturing of the
incident laser beam does not appreciably steepen the electron density radial
gradient.  This was expected, as the results of aperturing did not
appreciably change the overall shape of the laser beam cross section (recall
figure \ref{fig:4-9}) but rather, the diameter.

\figstub{fig:5-20}{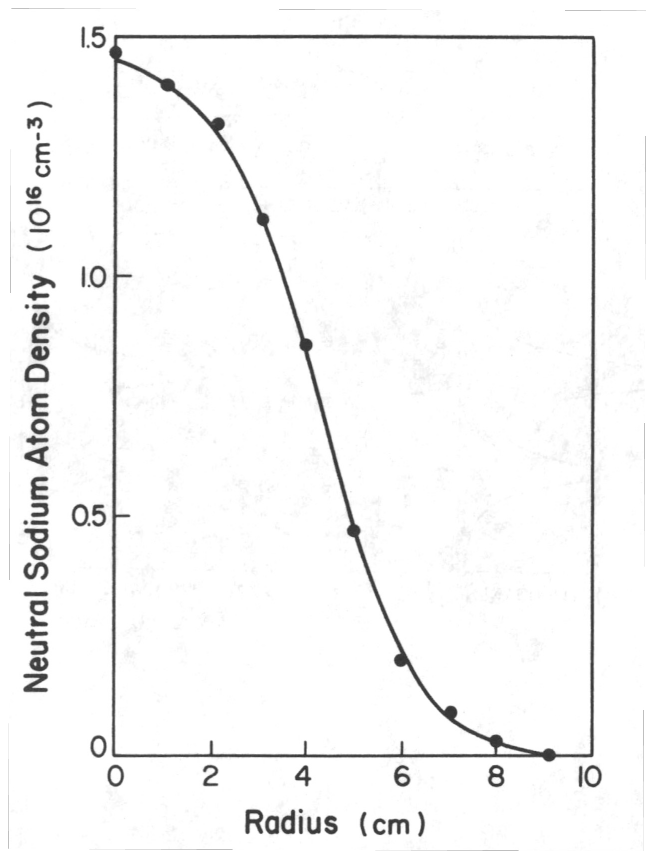}{Sodium atom density distribution for the second
experiment.}

\figstub{fig:5-21}{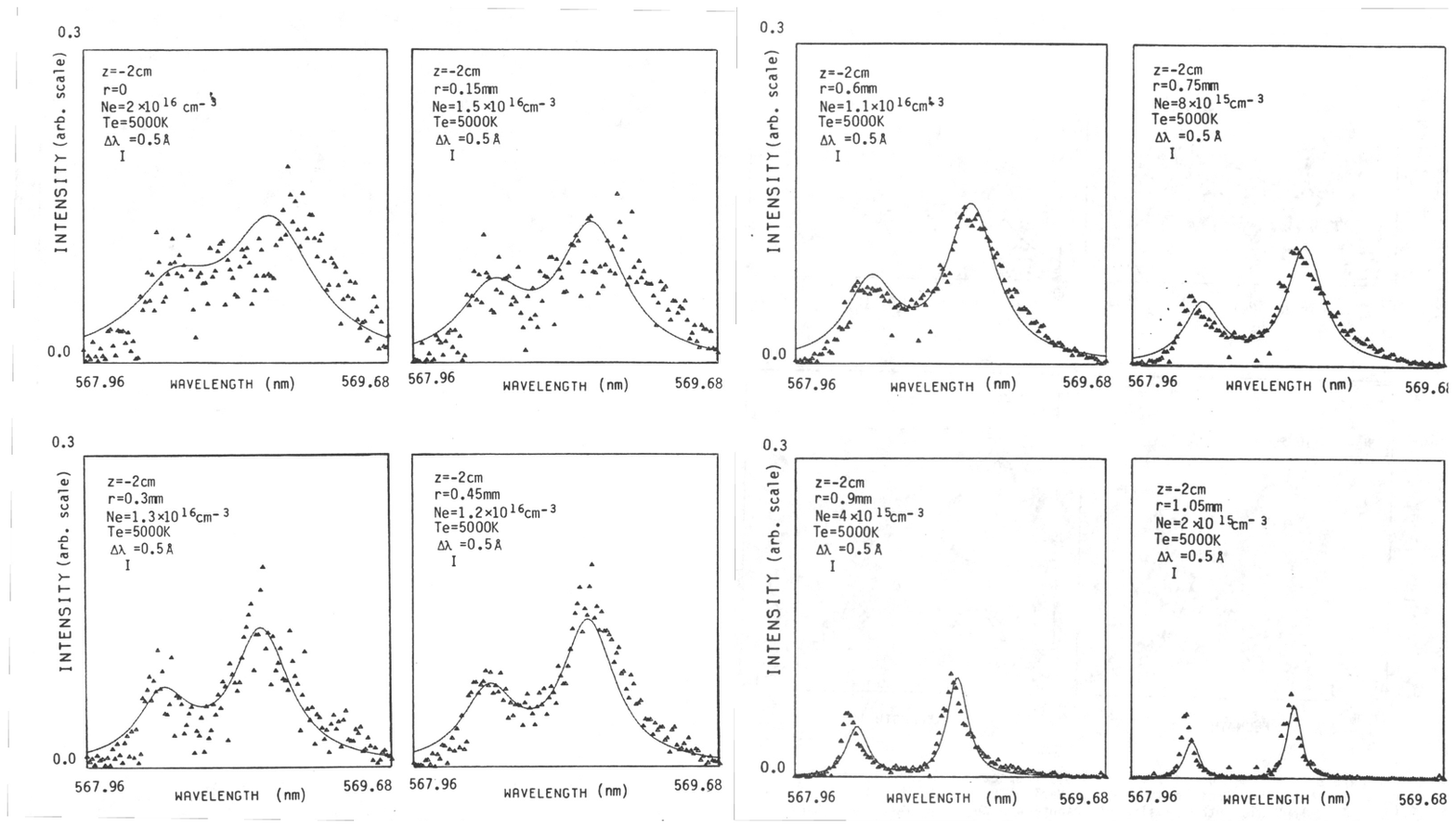}{Reconstructed inverted spectra along with the best fit
computed Voigt profiles for a number of radial positions.}

\figstub{fig:5-22}{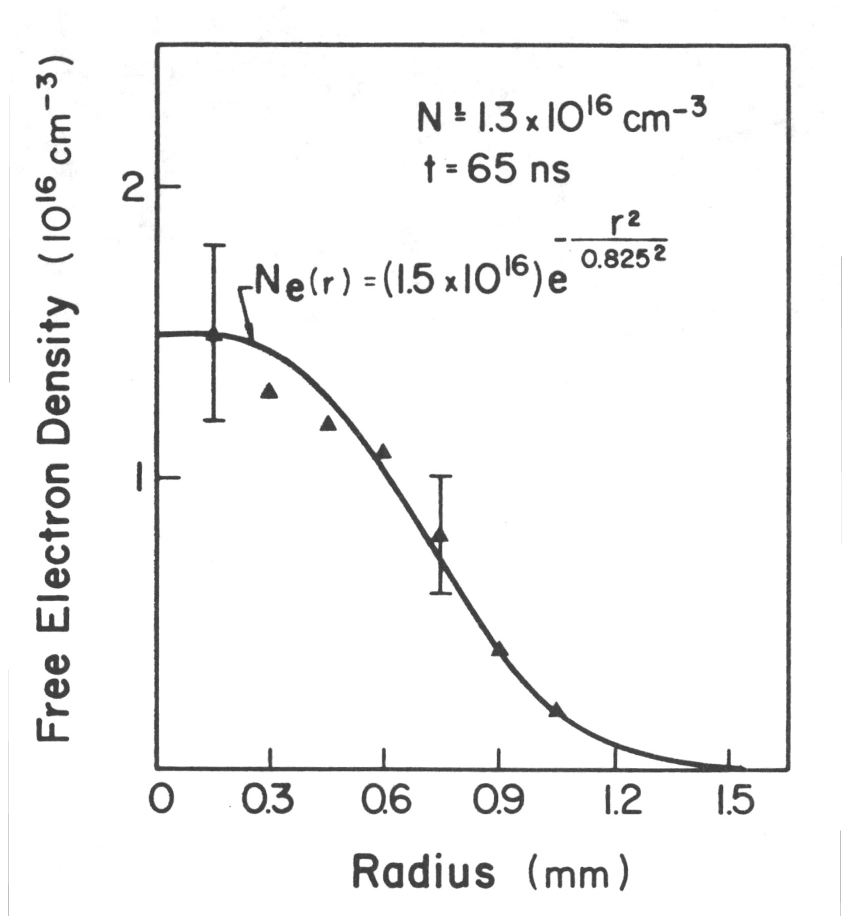}{Resulting radial variation in the free electron density.}

The lengthy acquisition time necessary for each wavelength scan (4-5
minutes per lateral position) does not permit us to perform detailed
measurements of the radial electron density distribution along many z
positions (the limitation is the stability of the neutral density within the
oven), however, we can again estimate the effective electron density and
plasma radius at other z positions by using the results displayed in figure
\ref{fig:5-22} along with the lateral emission spectra for y = 0, measured at a number
of z locations.  Figure \ref{fig:5-23} (like figure \ref{fig:5-18}) illustrates the results of
the measured lateral spectra for the five axial positions, along with the
spectra computed by the solution of the radiative transfer equation on the
basis of a uniform plasma.  Once again, the full radial electron density
measurement at z = -2 cm provided a means of scaling the plasma radius and
absolute intensity.

\figstub{fig:5-23}{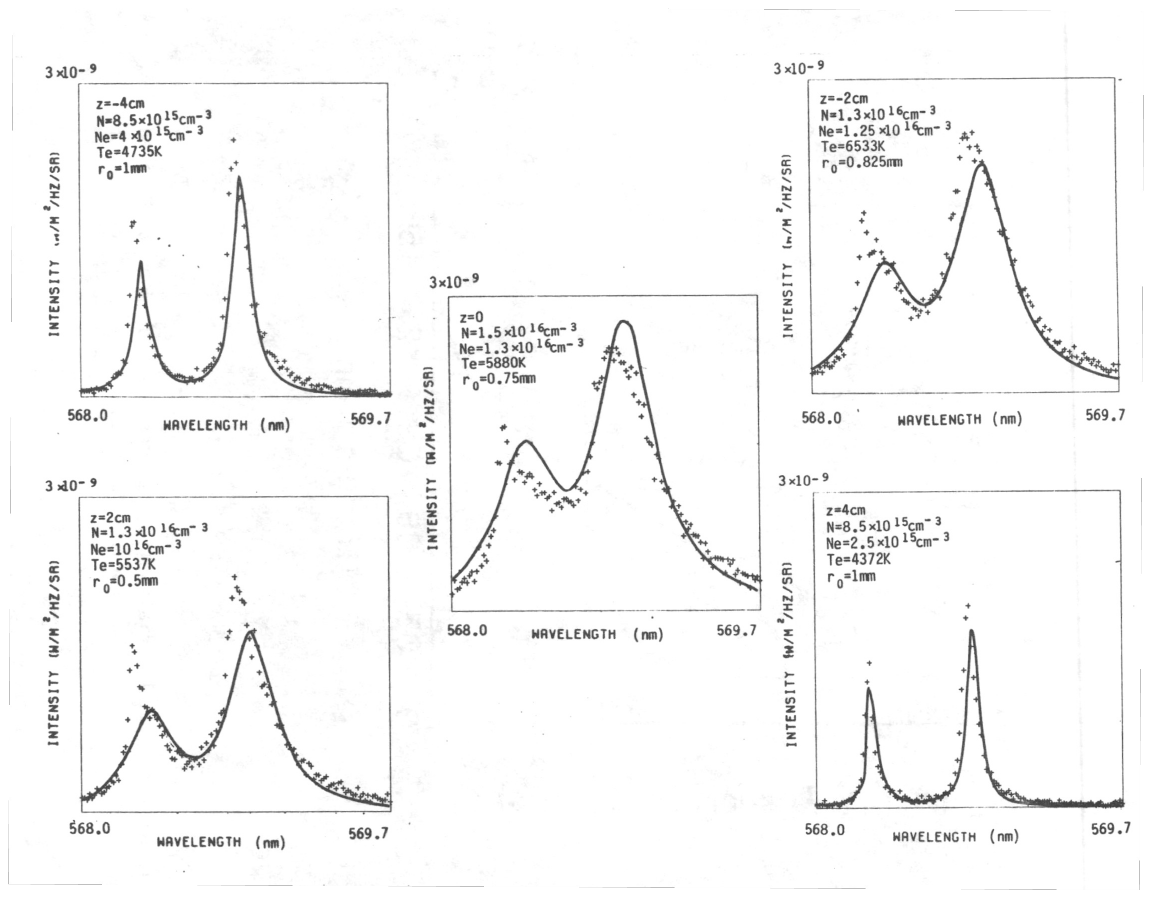}{Measured lateral spectra for the five axial positions,
along with the spectra computed on the basis of a uniform plasma (12 mJ).}

This same analysis was performed with the incident laser beam
attenuated such that the energy was approximately 6 mJ (half of the original
amount).  Figure \ref{fig:5-24} shows the results of this experiment, clearly
depicting (in comparison to figure \ref{fig:5-23}) a drop in the electron density
created along z.  The variation of the free electron density, electron
temperature and plasma radius derived using this analysis are depicted in
figures \ref{fig:5-25}, \ref{fig:5-26} and \ref{fig:5-27}.  Although the difference between the 12 mJ and
6 mJ run is small near the front of the vapor column where the laser enters
the oven, the plasma channel created at z $\geqslant$ 2 cm drops in electron density
by as much as 50 \% clearly indicating the sensitivity in the degree of
ionization along z, with incident laser energy.  The increase in the plasma
radius that is found to be necessary to match the profiles at z $\geqslant$ 2 cm, can
be attributed to the simplified ``single uniform cylinder'' analysis.  It may
be more appropriate to superimpose spectra similar to that obtained at z = 4
cm onto the spectra tabulated for z $<$ 4 cm, which would represent a
contribution arising from a low ionized plasma halo that could result from
radiation trapping.  The theoretical curves in figures \ref{fig:5-23} and \ref{fig:5-24} were
selected to fit the bulk of the data in the spectral wings and the
discrepancy near the unshifted line centre could also be accounted for as a
contribution to the observed spectra from this colder halo.

\figstub{fig:5-24}{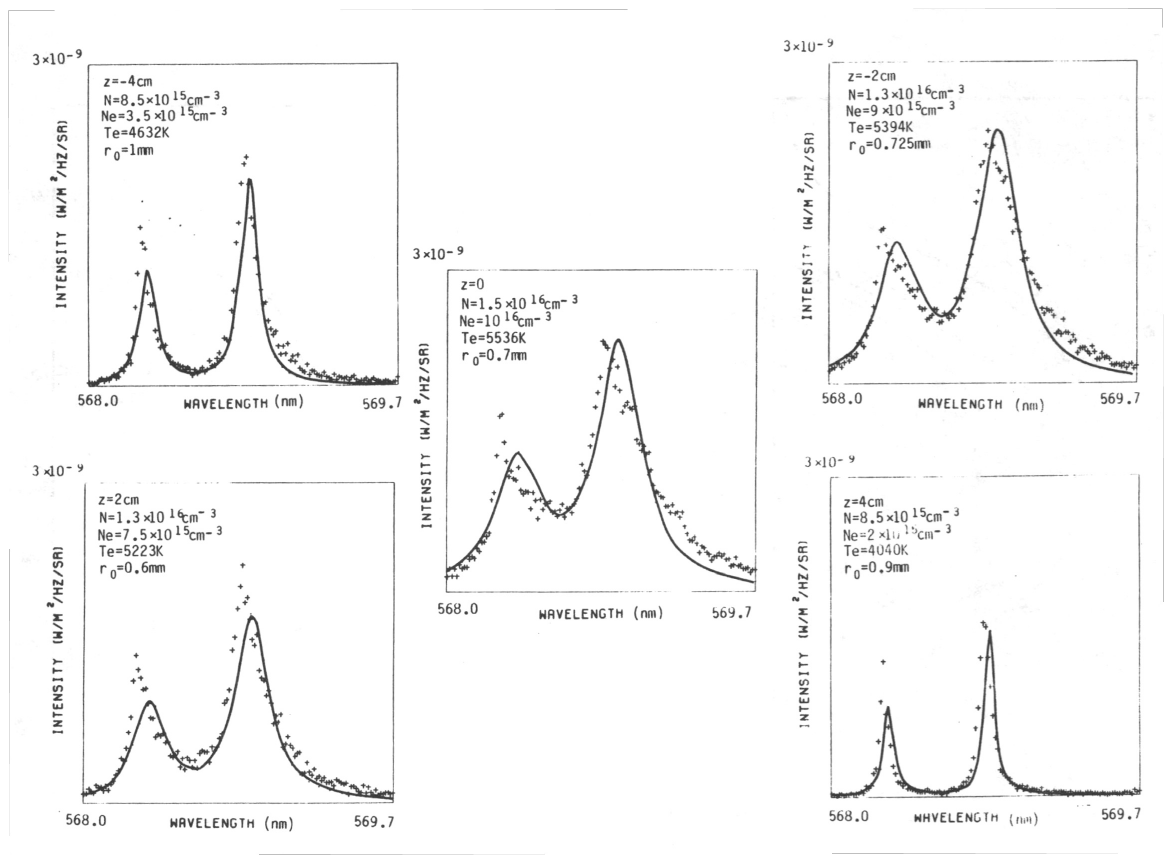}{Results with the incident laser beam attenuated to
approximately 6 mJ.}

\figstub{fig:5-25}{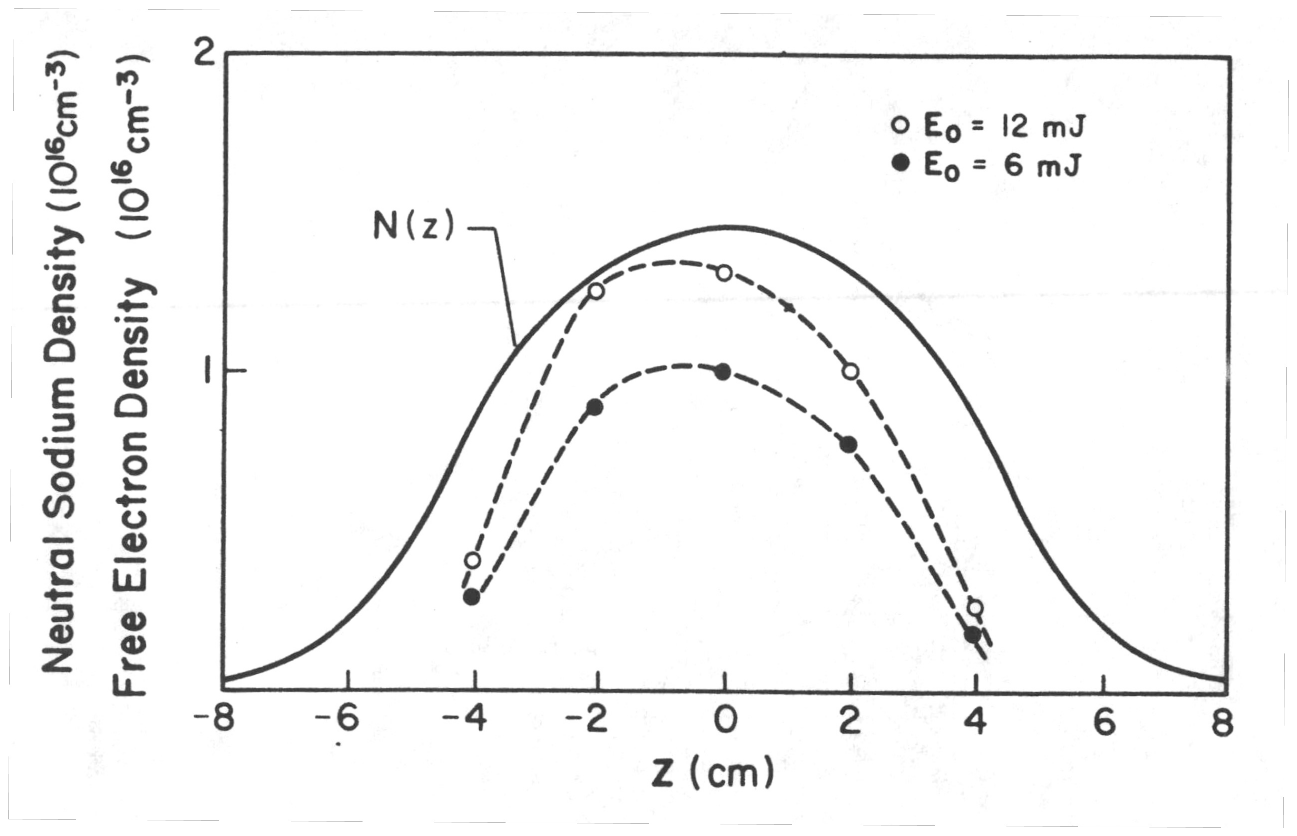}{Variation of the free electron density derived using this
analysis.}

\figstub{fig:5-26}{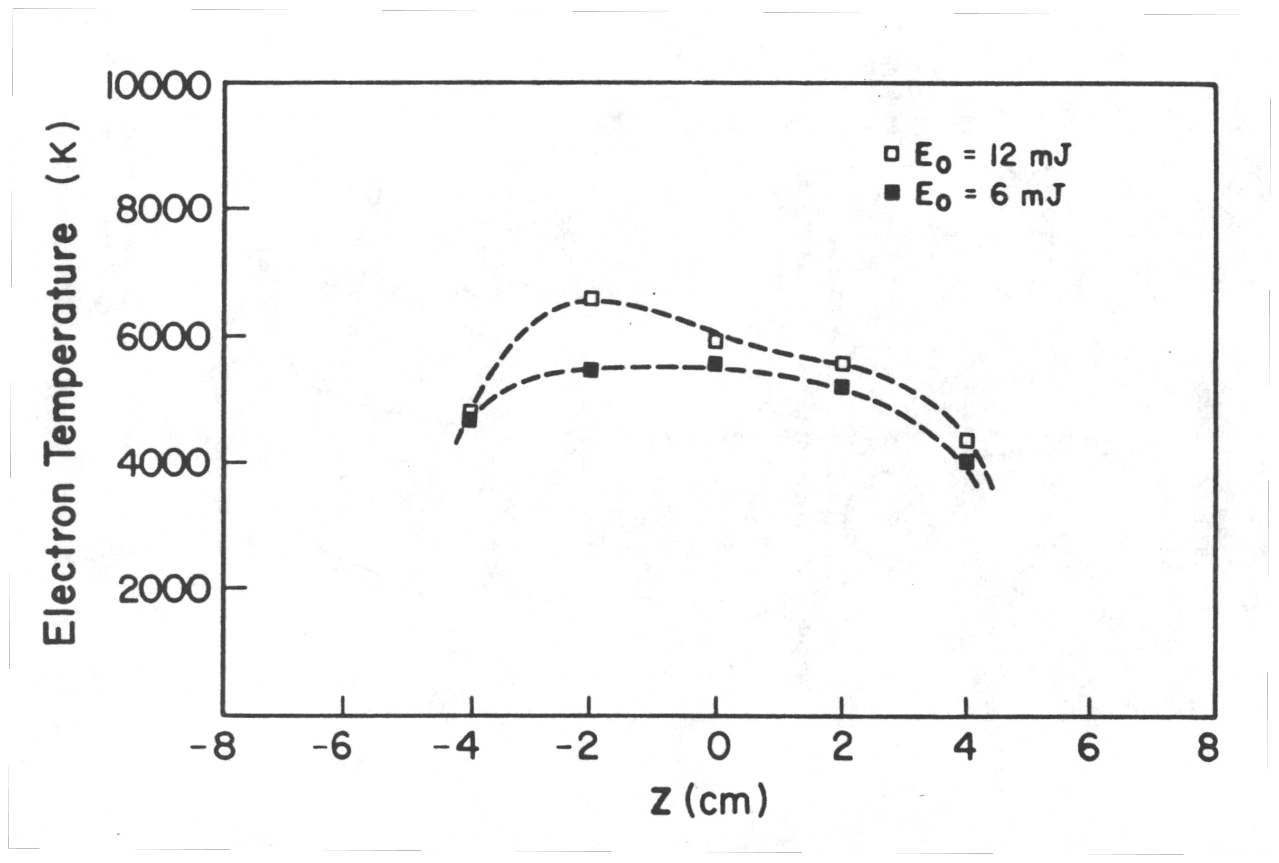}{Variation of the electron temperature derived using this
analysis.}

\figstub{fig:5-27}{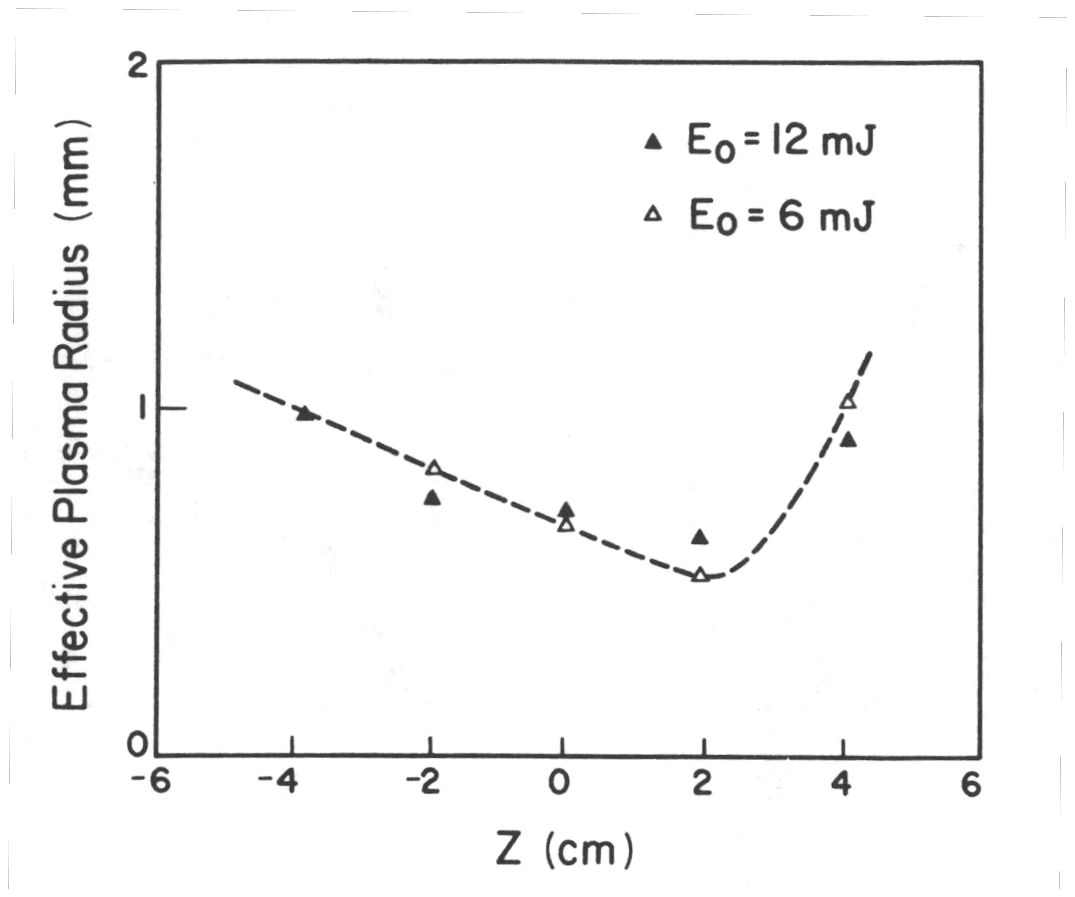}{Variation of the plasma radius derived using this
analysis.}

Although this analysis gives some indication of the effect of a change
in the incident laser energy on the plasma channel created, a measurement of
the change in the radial variation of the free electron density with laser
energy may be more meaningful and is the subject of the following section.

\subsection{Variation of $N_e(r)$ with Incident Laser Energy Fluence}
\label{sec:5-1-2}

A rigorous test of any LIBORS model, would be its ability to predict
the variation in the radial distribution of the free electron density, with
incident laser energy fluence.  An experiment was performed with the laser
beam apertured (1.6 mm diameter) and attenuated to provide us with three
incident laser energies.  Unattenuated, the total laser energy incident onto
the sodium vapor was estimated to be 6 mJ.  The thermopile admittance cone
has a diameter much greater than the aperture diameter so that the incident
laser energy can be described as

\begin{equation}
E_0 \;=\; \int_{0}^{\infty} 2\pi r\,\varepsilon_0(r)\,dr\;,
\label{eq:5-1}
\end{equation}

where $\varepsilon_0(r)$ represents the incident laser energy fluence distribution which
is not necessarily rectangular (as one might initially assume since it is
apertured), as diffraction as well as scattering from the front oven window
can contribute to radial structure.  Measurements were taken at z = -2 cm
where the sodium atom density was estimatd at $2\times10^{16}$ cm$^{-3}$ (figure \ref{fig:5-28}).
The electronics was triggered and gated to sample 2 ns, 65 ns after the
onset of the incident laser pulse.  The laser for this particular experiment
was tuned to the 589.0 resonance transition.  Energies of 1.5 mJ and 0.7 mJ
were obtained by attenuating the incident laser beam with calibrated neutral
density filters.

\figstub{fig:5-28}{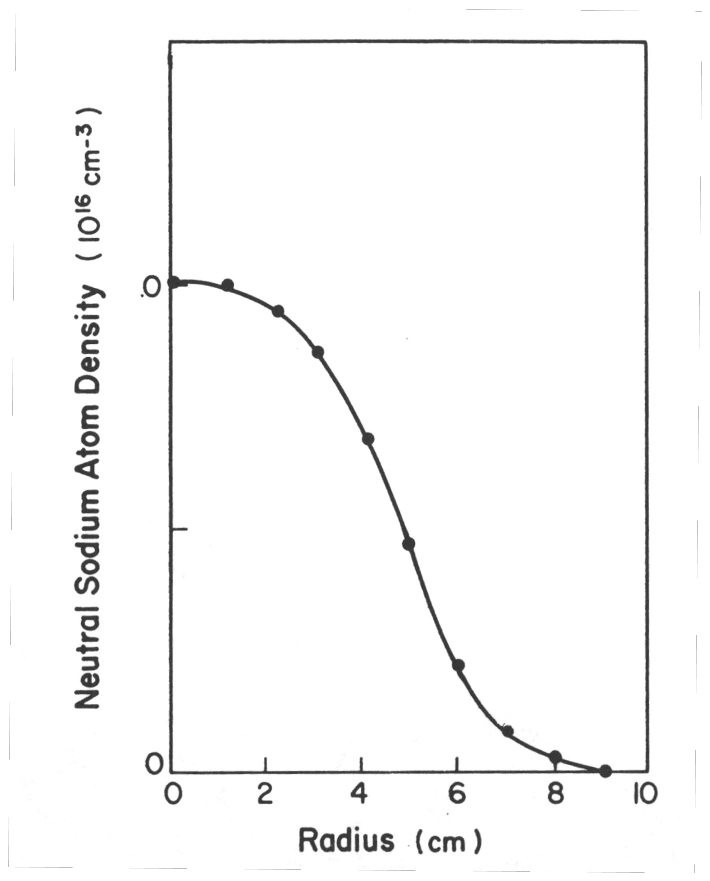}{Sodium atom density distribution for the incident laser
energy fluence experiment.}

Figures \ref{fig:5-29}, \ref{fig:5-30} and \ref{fig:5-31} give examples of the radially inverted
spectra obtained for the three values of incident laser energies.  The
electron density radial distributions derived from these spectra, are
illustrated in figure \ref{fig:5-32}.  Inspection of the results, reveal that at
higher laser energies, the plasma extends well beyond the volume defined by
the aperture.  This could arise from either the incident laser beam
extending beyond this diameter as a result of diffraction from the aperture,
or possibly, electron thermal conduction, extending the plasma in the radial
domain.  The effects of thermal conduction on the radial variation in the
free electron density are currently being investigated (Kissack 1987).
Figure \ref{fig:5-32} clearly shows that the core electron density and plasma radius
decrease with decreasing laser energy.  From the form of the empirical
functions used to represent the measured data (solid lines in figure \ref{fig:5-32}),
one can see that the radial electron density gradient increases and that the
effective 1/e radius decreases with decreasing incident laser energy.

\figstub{fig:5-29}{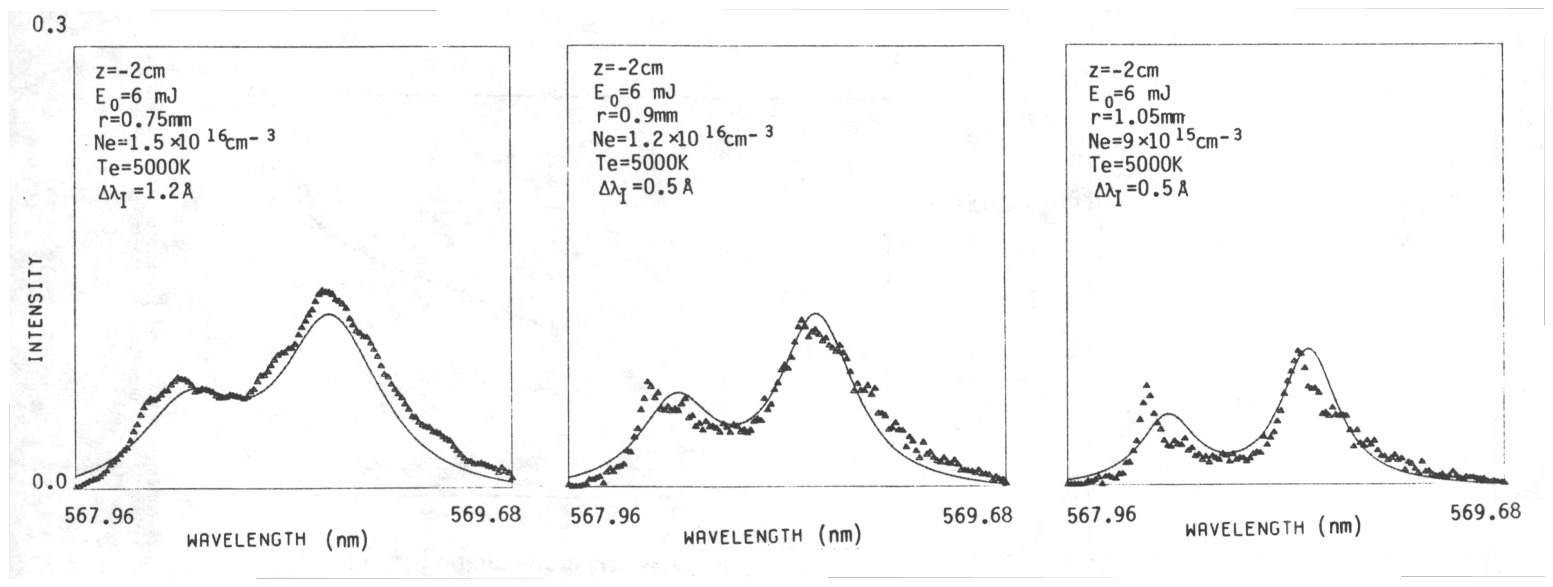}{Radially inverted spectra obtained for an incident laser
energy of 6 mJ.}

\figstub{fig:5-30}{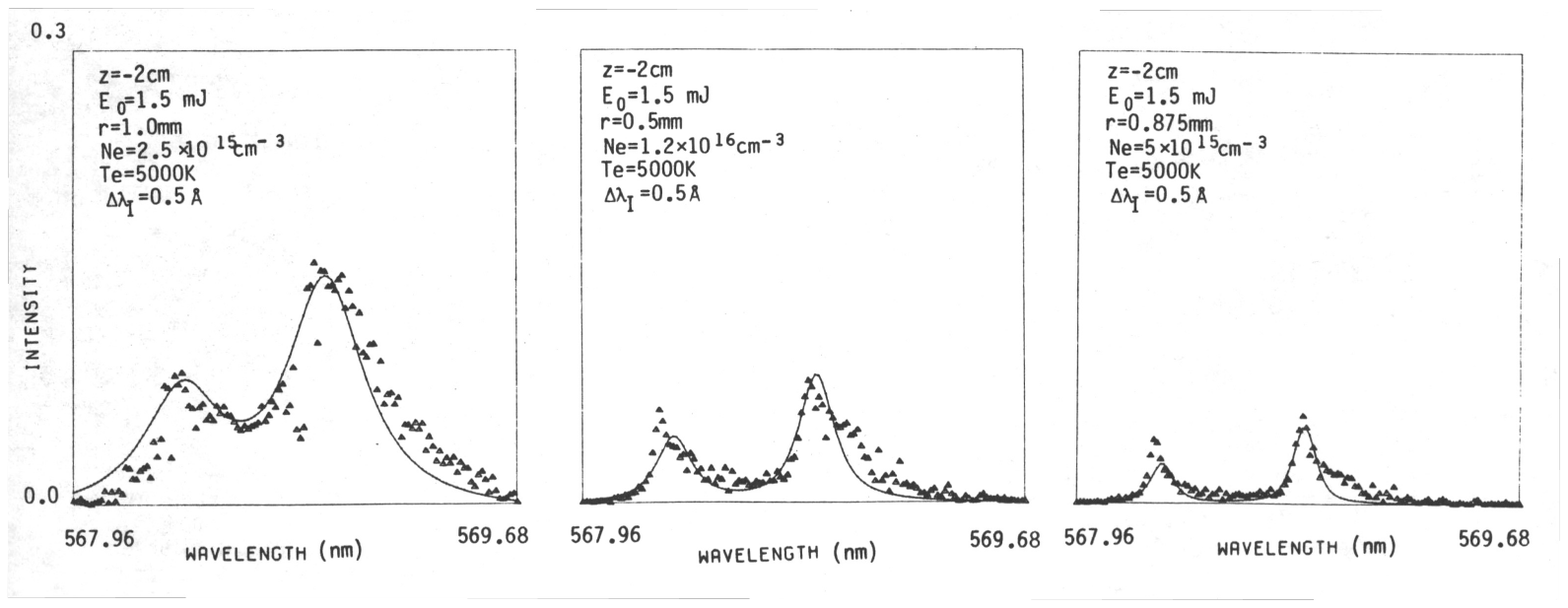}{Radially inverted spectra obtained for an incident laser
energy of 1.5 mJ.}

\figstub{fig:5-31}{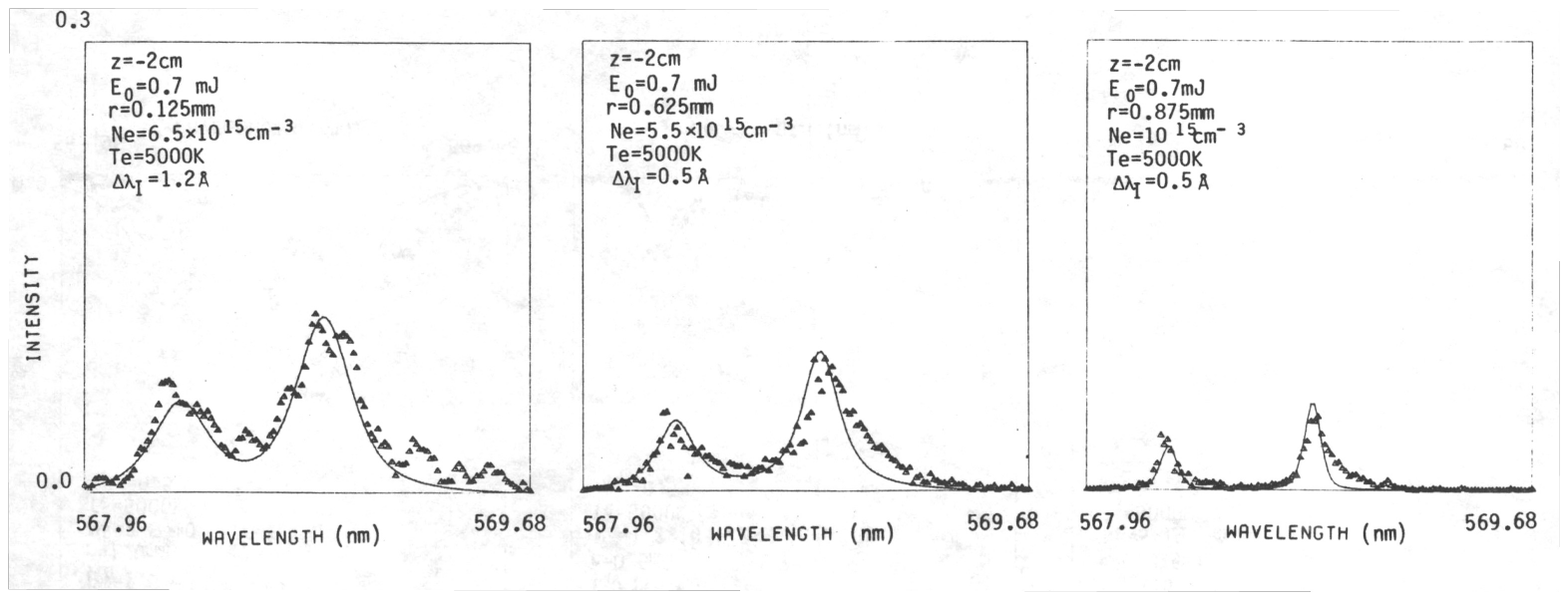}{Radially inverted spectra obtained for an incident laser
energy of 0.7 mJ.}

\figstub{fig:5-32}{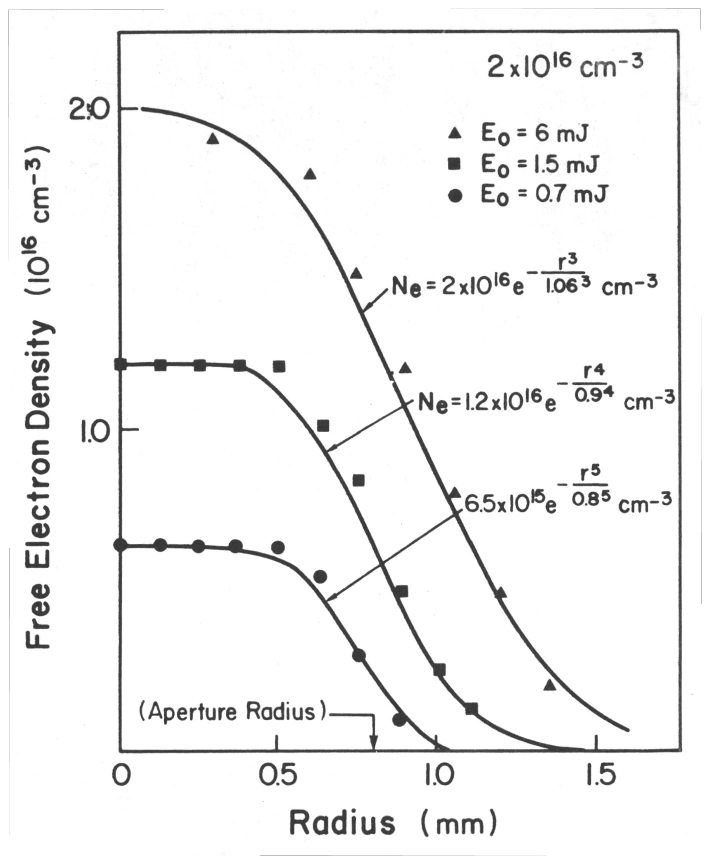}{Electron density radial distributions derived from the
spectra of figures \ref{fig:5-29}--\ref{fig:5-31}.}

One could work backwards from the results presented in figure \ref{fig:5-32}, to
reconstruct the radial variation in incident laser energy fluence (providing
that radial thermal conduction can be ignored).  For example, the energy
fluence in the core $\varepsilon_0(0)$ for $E_0$ = 1.5 mJ produces the same free electron
density as that at r = 0.9 mm for $E_0$ = 6 mJ.  If we represent the incident
laser energy fluence for the 6 mJ run as $\varepsilon_0(r,6\text{mJ})$, then

\begin{equation}
\varepsilon_0(0.9,6\text{mJ}) \;=\; (1.5/6.0)\,\varepsilon_0(0,6\text{mJ})\;,
\label{eq:5-2}
\end{equation}

and in a similar fashion for all radii, using the three curves in figure
\ref{fig:5-32}, we can construct the variation $\varepsilon_0(r,6\text{mJ})/\varepsilon_0(0,6\text{mJ})$.  We have found
that from the data given in figure \ref{fig:5-32}, a reasonable representation can be
described by

\begin{align}
\varepsilon_0(r)/\varepsilon_0(0) &= 1\;, &&\text{for}\qquad r < 0.5\ \text{mm}\;,\nonumber\\
&= 1.9 - 1.8r, &&\text{for}\quad 0.5 \leqslant r \leqslant 1.05\ \text{mm.}
\label{eq:5-3}
\end{align}

The distribution given by equation (\ref{eq:5-3}) is the radial distribution
of incident laser energy fluence that should be used as an input parameter
for the three-dimensional theoretical model that would be used to predict
the experimental results observed in this experiment.  The core value of the
incident laser energy fluence can be obtained from equation (\ref{eq:5-1}), that is

\begin{equation}
\int_{0}^{\infty}\varepsilon_0(r,6\text{mJ})\,2\pi r\,dr \;=\; 6\ \text{mJ}\;,
\label{eq:5-4}
\end{equation}

and substituting equation (\ref{eq:5-3}) into equation (\ref{eq:5-4}) above, we arrive at

\begin{equation}
\varepsilon_0(0) \;=\; 227\ \text{mJ/cm}^2\;.
\label{eq:5-5}
\end{equation}

In accordance with the assumption that thermal conduction has a
negligible influence on the electron density radial profiles, then the
radial domain can be mapped into incident laser energy fluence using
equation (\ref{eq:5-3}).  The results shown in figure \ref{fig:5-32} can be replotted to
illustrate the variation of the free electron density with incident laser
energy fluence (see figure \ref{fig:5-33}).  The trend in the data presented in figure
\ref{fig:5-33}, suggests that a limiting value of the free electron density is being
approached for $\varepsilon_0(0) >$ 227 mJ/cm$^2$.  In fact, it is not surprising that this
limiting value is very near the neutral sodium density of $2\times10^{16}$ cm$^{-3}$.
The variation of $r_{\text{eff}}$ (1/e plasma radius) with incident laser energy fluence
shows a similar behaviour (figure \ref{fig:5-34}), suggesting that a great fraction of
the volume defined by the laser beam cross-section is highly ionized at z =
-2 cm for the case where the incident laser energy is as high as 6 mJ.

\figstub{fig:5-33}{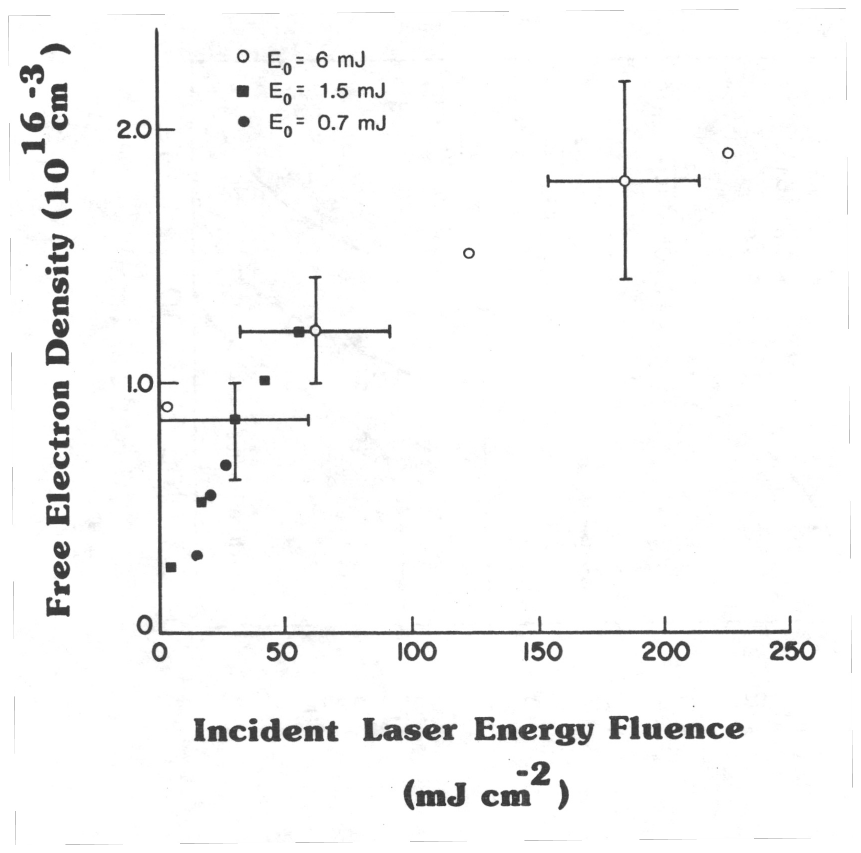}{Variation of the free electron density with incident laser
energy fluence.}

\figstub{fig:5-34}{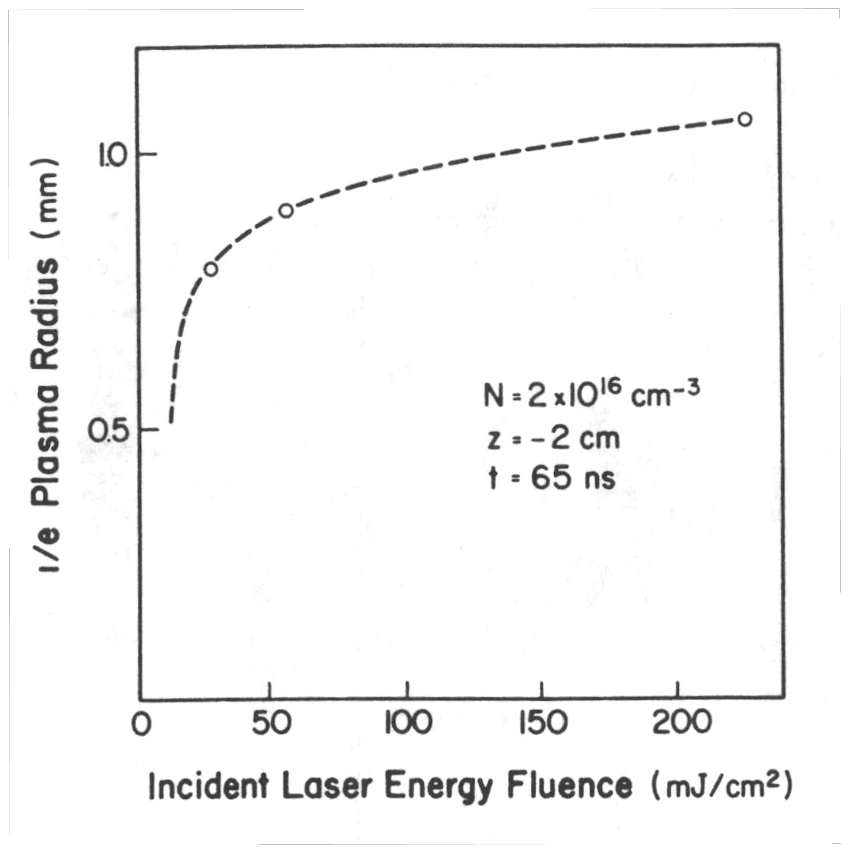}{Variation of $r_{\text{eff}}$ (1/e plasma radius) with incident laser
energy fluence.}

\subsection{Variation of $N_e(r)$ with Incident Laser Wavelength}
\label{sec:5-1-3}

It seems somewhat in place to conclude the results section on the free
electron density measurements, with the results of an experiment which
clearly illustrates the resonance behaviour in the plasma channel creation.
Measurements of the radial distribution in the free electron density were
made at z = -3 cm, for incident laser wavelengths of 588.3, 588.85 and 588.9
nm.  The laser beam was again apertured to a diameter of 1.6 mm.  The laser
energy for this experimental run was estimated at 10 mJ.  The neutral sodium
atom density at z = -3 cm was approximately $1.1\times10^{16}$ cm$^{-3}$ (see figure
\ref{fig:5-35}).  In this particular experiment, the electronics was triggered and
gated to sample over 2 ns, 95 ns after the onset of the incident laser
pulse.

\figstub{fig:5-35}{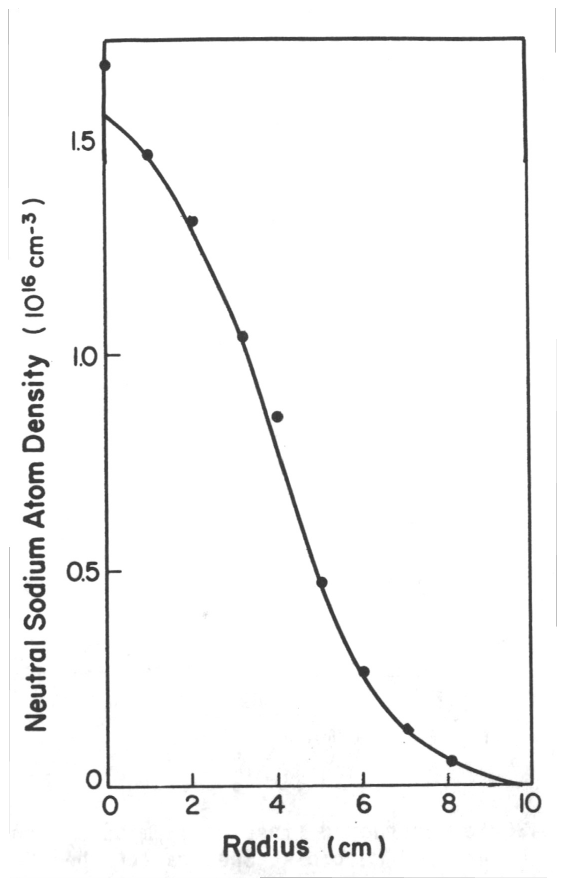}{Neutral sodium atom density distribution for the incident
laser wavelength experiment.}

Examples of the inverted spectra at three radial positions for the
three laser wavelengths are shown in figures \ref{fig:5-36}, \ref{fig:5-37} and \ref{fig:5-38}
respectively.  A compilation of these results clearly illustrate the
noticeable drop in the free electron density near the plasma core for the
case of 0.7 nm detuning, in comparison to the near resonance cases (see
figure \ref{fig:5-39}).  More surprising however, is the relatively insignificant
difference between the 588.85 and 588.9 nm cases.  Closer inspection and
comparison of figures \ref{fig:5-37} and \ref{fig:5-38} however, reveals a decrease in the
spectrally integrated emission suggesting that the excited state population
density is not quite as high for the 588.85 nm case as it is for the case
where the laser is closely tuned to resonance (589.0 nm).

\figstub{fig:5-36}{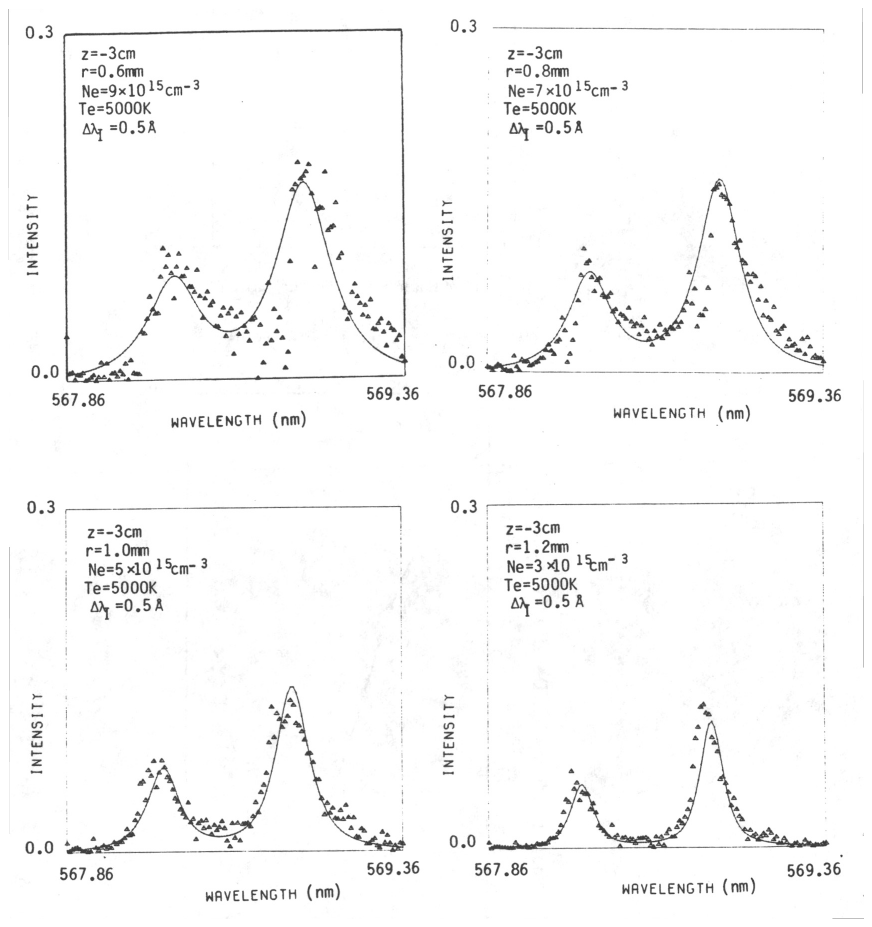}{Inverted spectra at three radial positions for an incident
laser wavelength of 588.3 nm.}

\figstub{fig:5-37}{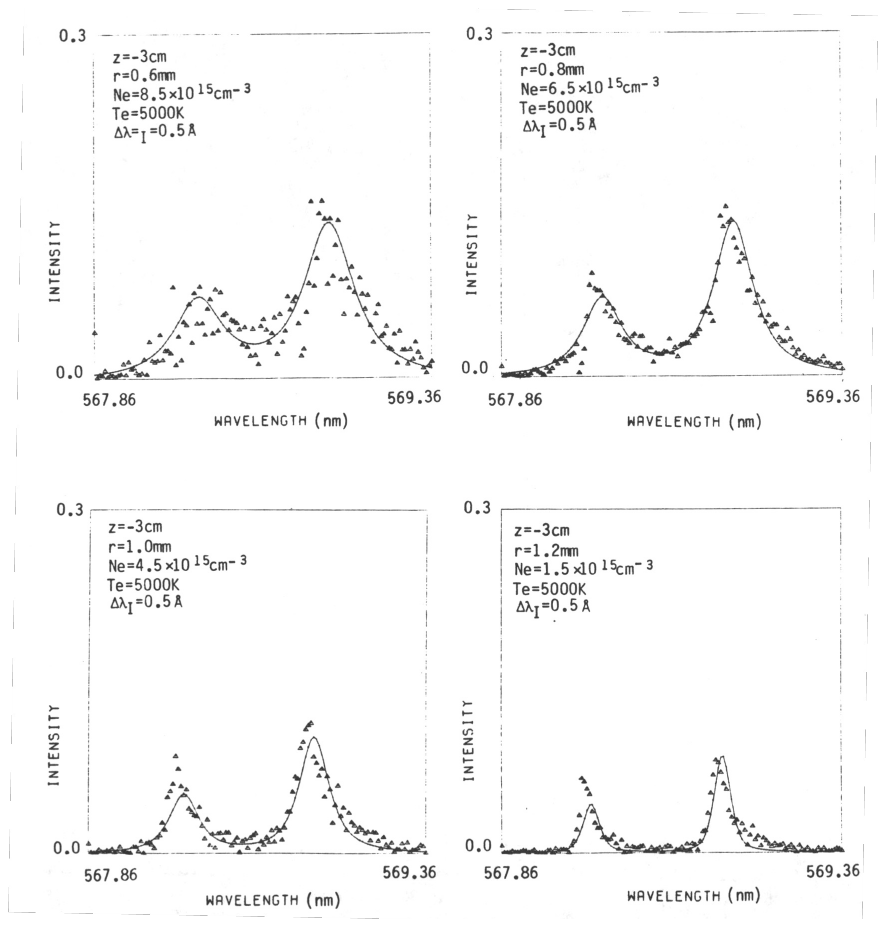}{Inverted spectra at three radial positions for an incident
laser wavelength of 588.85 nm.}

\figstub{fig:5-38}{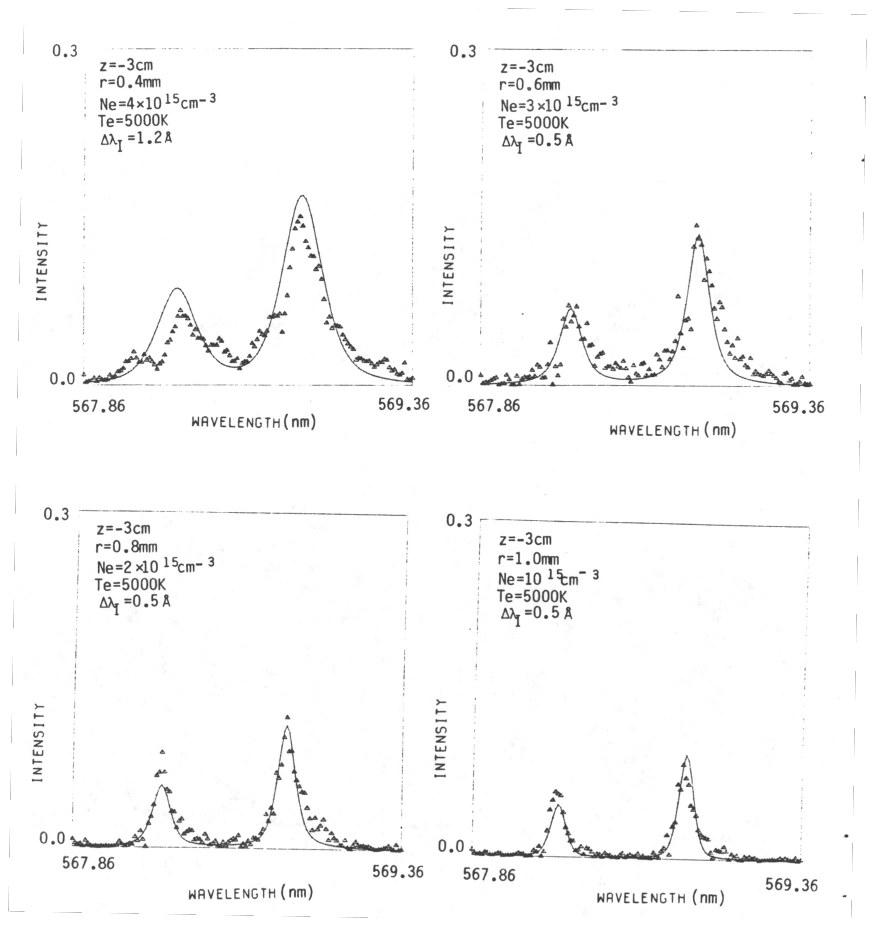}{Inverted spectra at three radial positions for an incident
laser wavelength of 588.9 nm.}

\figstub{fig:5-39}{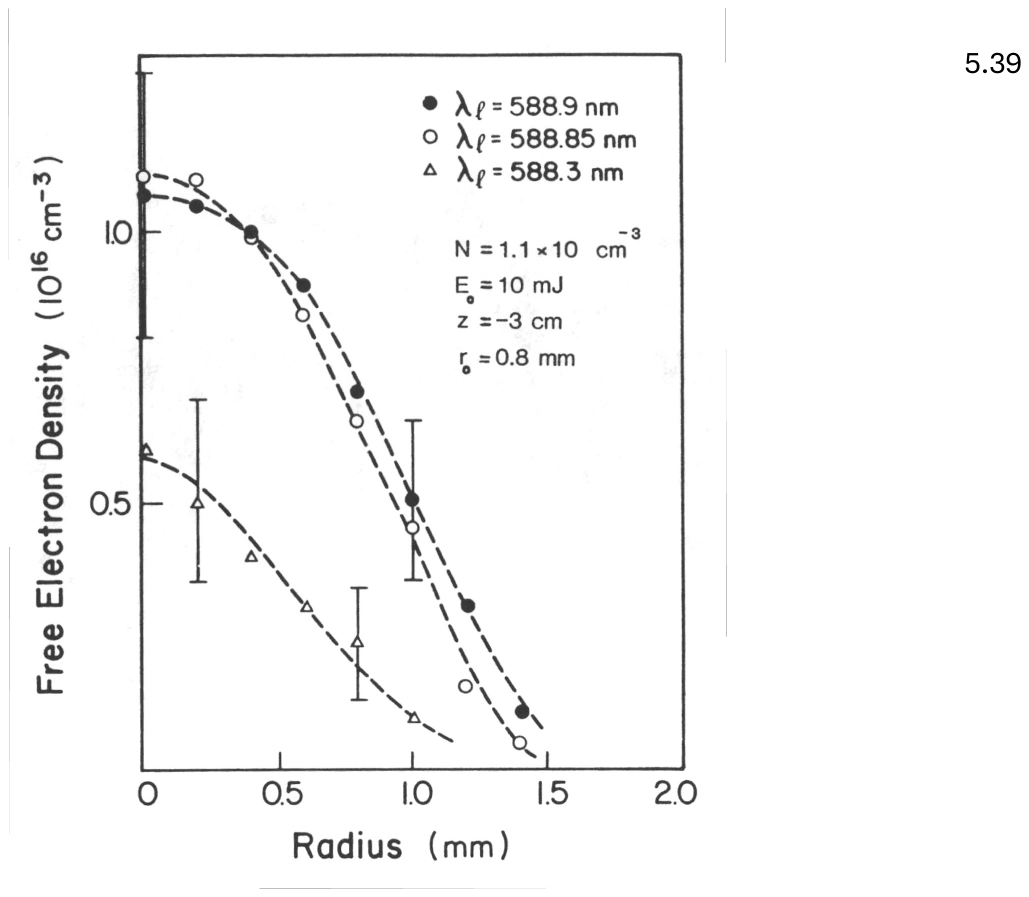}{Compilation of the free electron density radial profiles
for the three incident laser wavelengths.}

When the laser is tuned to or close to resonance (within an
Angstrom), the free electron density gradients are found to be comparable to
those obtained for the other experiments.  Strong evidence of
self-focussing and defocussing was observed for laser wavelengths between 588.2
and 588.7 nm (see section \ref{sec:5-3}).  This could have a pronounced effect on the
quality of the plasma channel created, and its complex nature makes it
difficult to include in a theoretical model.  As we shall see in section
\ref{sec:5-3}, these results lead us to put more emphasis on the fact that the laser
must be as near to resonance as possible, to create a highly ionized,
uniform and stable plasma channel.

\section{Electron Temperature Measurements}
\label{sec:5-2}

A detailed analysis of electron temperature was performed on the
results of the experiment described in the opening paragraphs of section
\ref{sec:5-1}.  In addition to the $4^2D$-$3^2P$ spectra measured at various lateral
positions (to reconstruct the radial variation in the free electron
density), the spectrally integrated $4^2D$-$3^2P$ and $6^2D$-$3^2P$ emission was
recorded at eleven positions across the plasma column.  In accordance with
the theoretical development of section \ref{sec:3-3-2}, we have assumed that the exit
slits are of sufficient width such that $\Theta_{nm}\approx 1$.  The implications of this
will be discussed in more detail later in this section.

Measurements of the spectrally integrated emission from the $n^2D$-$3^2P$ (n
= 4,5,6) and $n^2S$-$3^2P$ (n = 5,6,7) spectral series at y = 0 mm were also
recorded, to reconstruct a Boltzmann plot, the slope of which gives an
indication of the electron temperature (section \ref{sec:3-1}).

To conclude this section, a comparison will be made between the
results of the above two analyses and the electron temperature derived from
the Saha relationship with the measured radial distribution of the free
electron density and neutral sodium atom density.

\subsection{Electron Temperature Assuming a Uniform Optically Thin Plasma:
The Boltzmann Plot}
\label{sec:5-2-1}

The intensity of the above mentioned multiplets versus excited to
resonance state energy difference $E_{n2}$, is illustrated in figure \ref{fig:5-40}.  The
intensities have been corrected for the relative system response, with

\begin{equation}
K_n \;=\; \bigl\{(S_{nm}/S_{qp})\,\nu_{nm}A_{nm}g_n\bigr\}^{-1}
\label{eq:5-6}
\end{equation}

and we have assumed that the exit slit encompasses the majority of the
emission line.  A linear regression analysis on these data gives a slope of
0.4736 eV$^{-1}$ which translates to an electron temperature of 5496 K.  The
scatter of these data from a straight line is assumed to arise from plasma
inhomogeneity, optical depth and perhaps more importantly, deviations of
$\Theta_{nm}$ from unity (see Appendix G).  The temperature arrived at is
significantly greater than the temperatures reported by Krebs and Schearer
(1982) and by Landen et al.\ (1985).

\figstub{fig:5-40}{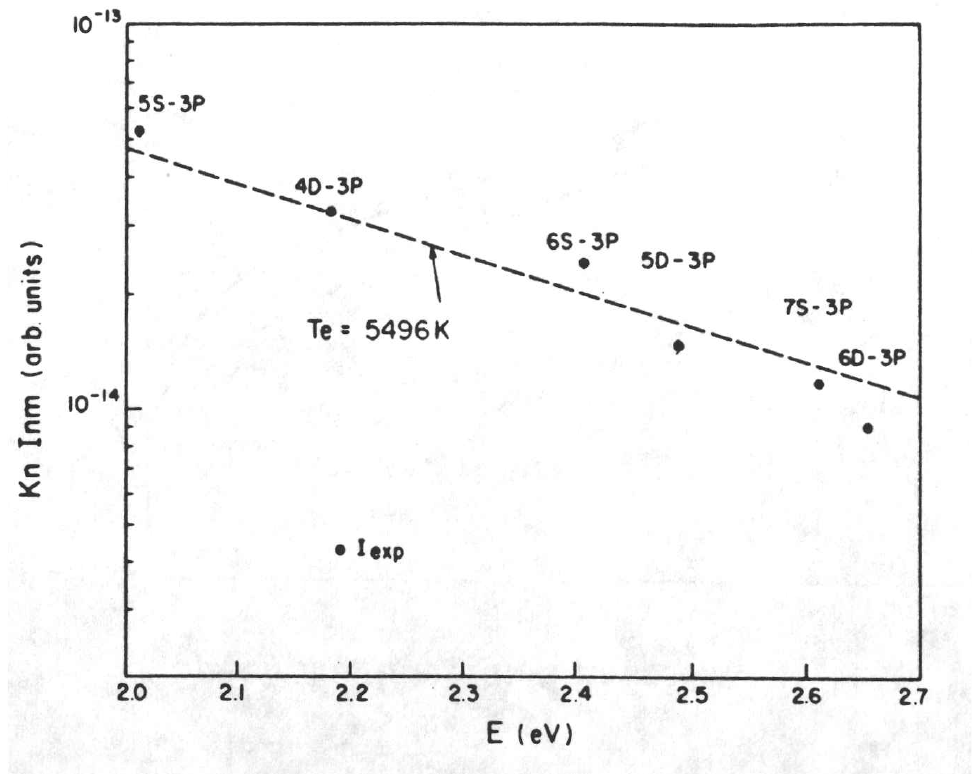}{Intensity of the multiplets versus excited to resonance
state energy difference $E_{n2}$ (the Boltzmann plot).}

\subsection{Radial Variation in the Electron Temperature from the Relative
Intensity of Emission Lines}
\label{sec:5-2-2}

The distribution of emission signals for both the $6^2D$-$3^2P$ and $4^2D$-$3^2P$
multiplets across the plasma column (z = 0) are presented in figure \ref{fig:5-41},
for the same experiment referred to in the above section.  The curves drawn
through the data points represent a least squares fit using a polynomial of
the form given by equation (\ref{eq:3-24}).  The Abel inversion of these polynomials
are readily attained by the procedure described in section \ref{sec:2-6}.  The
inverted radial emission profiles for these multiplets are displayed in
figure \ref{fig:5-42}.  With corrections for the system relative spectral response,
these curves generally represent the radial distribution of the excited
state population densities, providing that the plasma is optically thin and
once again, $\Theta_{nm}\approx 1$.  The condition that $\Theta_{nm} = 1$ would tend to break down
near y = 0, where the observation is along a path passing through the highly
ionized plasma core, and the emission line would tend to be more diffuse as
a result of the electron Stark broadening.  This in effect, would tend to
suppress the inverted emission near r = 0 for the more diffuse transition
(in this case, the $6^2D$-$3^2P$ transition) and would be interpreted as a lower
$6^2D$ state population density.  In all, it would lead to an underestimate of
the electron temperature in the plasma core.  In fact, the ratio of the
radial emission profiles lead to a radial electron temperature distribution
that has an unrealistic drop towards r = 0 (also in figure \ref{fig:5-42}).  The error
bars on the free electron temperature data are strictly a result of the
random scatter in the experimental data points.  They do not include the
uncertainty in the ratio of the Einstein spontaneous emission probabilities
for the two transitions, nor do they take into account the error arising
from the assumption that $\Theta_{nm} = 1$.  In the region where maximum emission
evolves (r $\leqslant$ 1 mm), the electron temperature is in reasonable agreement
with that obtained from the Boltzmann plot using the four additional
transitions and the assumption of plasma uniformity.

\figstub{fig:5-41}{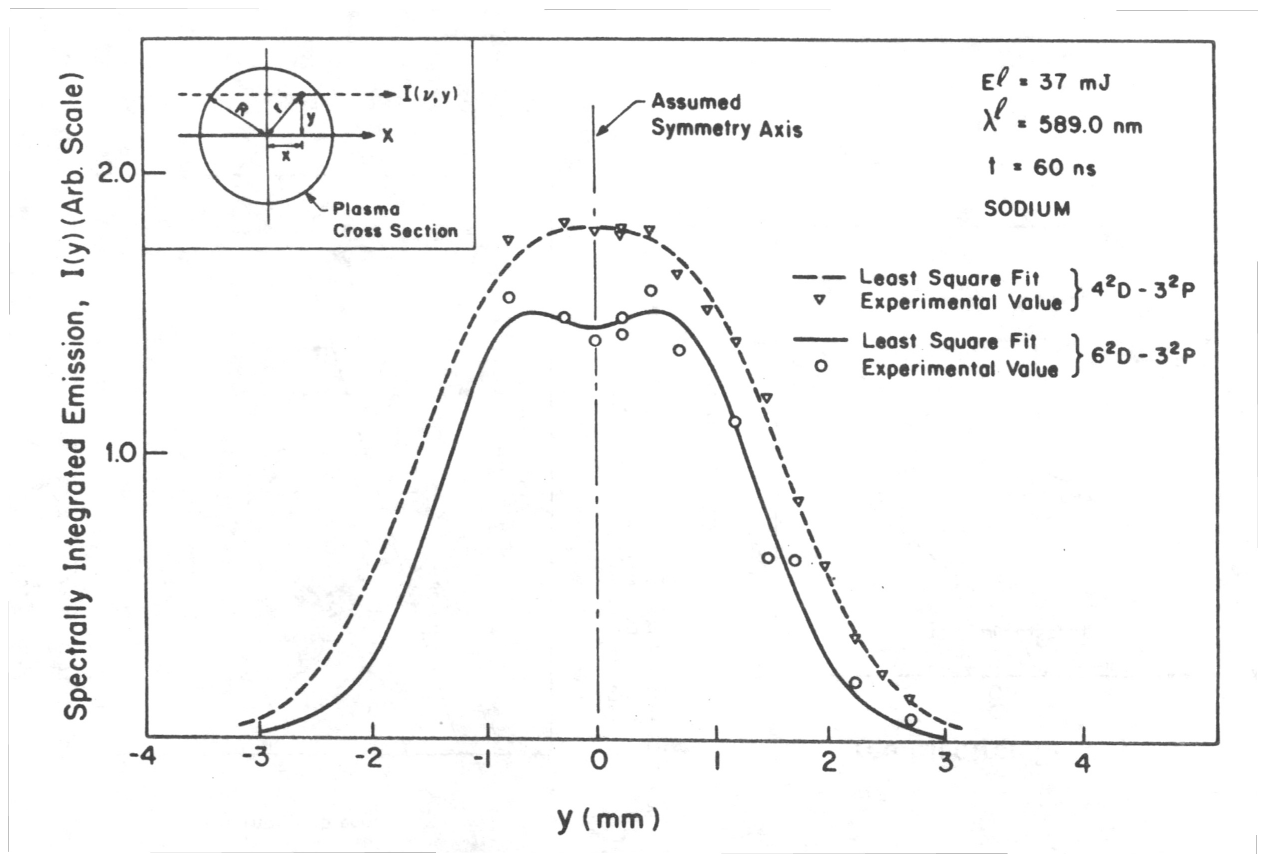}{Distribution of emission signals for both the $6^2D$-$3^2P$ and
$4^2D$-$3^2P$ multiplets across the plasma column (z = 0).}

\figstub{fig:5-42}{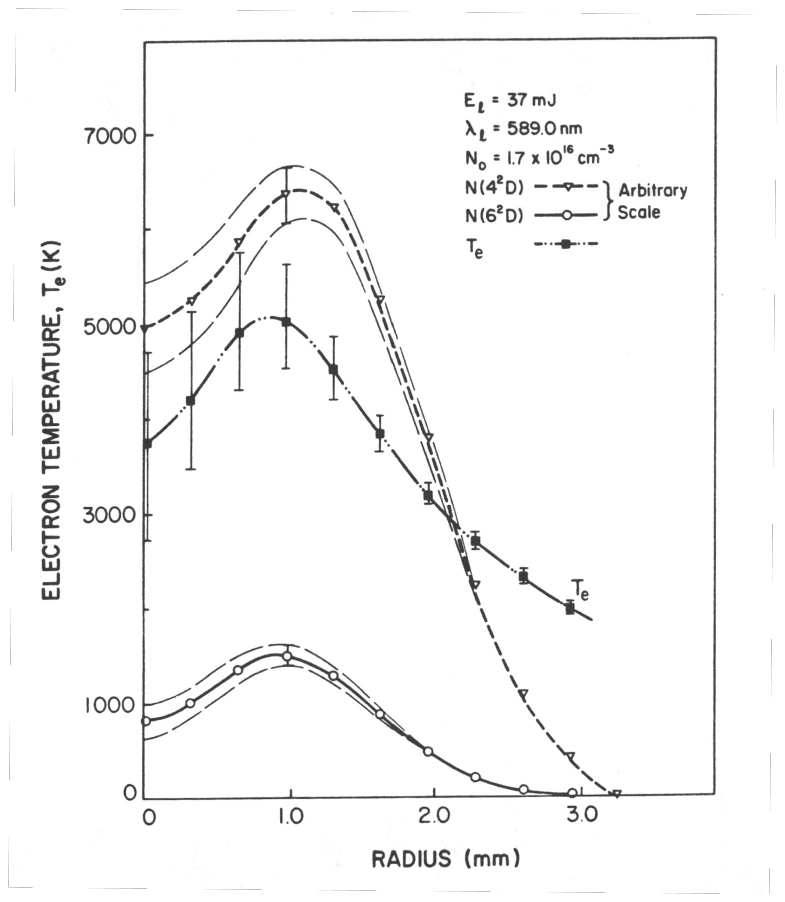}{Inverted radial emission profiles for these multiplets,
together with the derived radial electron temperature distribution.}

To illustrate that optical depth may play a minimal role in distorting
the spectral emission, the data were subjected to the iterative Abel
inversion procedure discussed in section \ref{sec:3-3-2-2}.  The plasma is assumed to
be in LTE down to the ground state and convergence is found to occur within
two iterative steps (see figure \ref{fig:5-43}).  As one can see from figure \ref{fig:5-43}, the
difference between the $0^{\text{th}}$ and final iteration is generally less than 10 \%
and within experimental uncertainty.  Of course, this iterative scheme
relies on the accuracy of the initial temperature distribution, and the
results should be used for qualitative discussion only.

\figstub{fig:5-43}{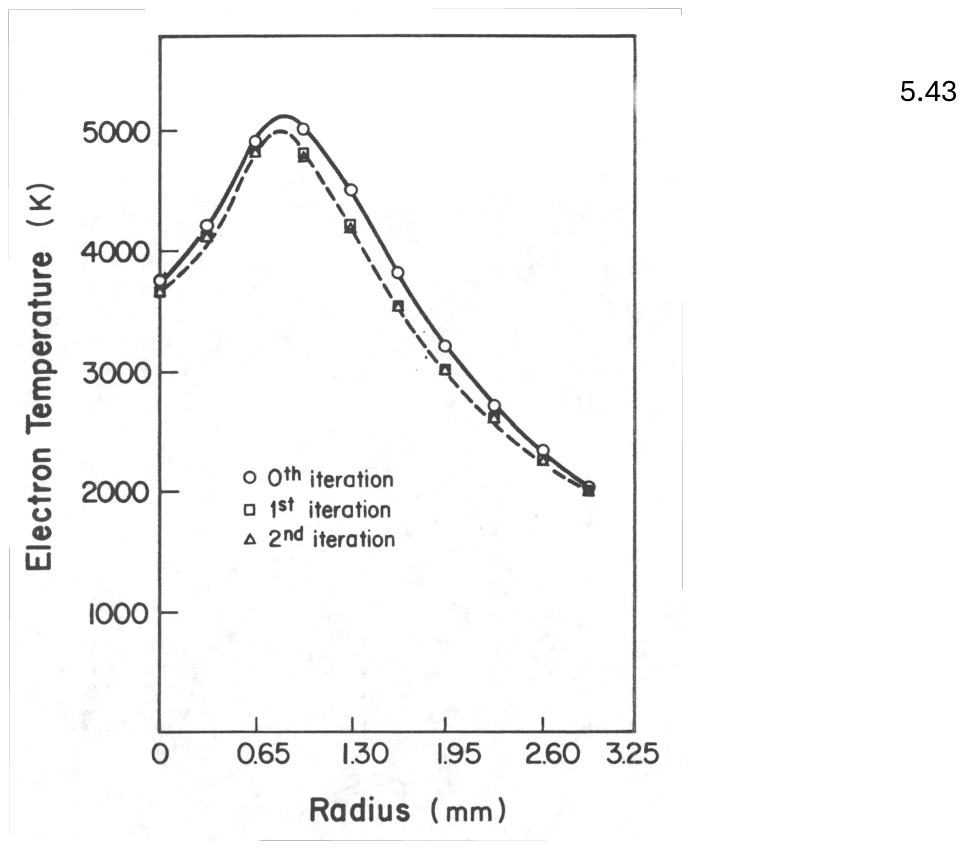}{Results of the iterative Abel inversion procedure,
showing convergence within two iterative steps.}

\subsection{Electron Temperature from Saha Equilibrium}
\label{sec:5-2-3}

An independent check on the above results can be made by using the
measured radial distribution of the free electron density (figure \ref{fig:5-9}) and
the estimated neutral sodium atom density with the assumption of Saha
equilibrium to generate the radial variation in $T_e$.  The result of this
computation is shown in figure \ref{fig:5-44} together with the electron temperature
derived from the two line intensity ratio technique.  It is apparent that
the Saha temperature is significantly higher for r $>$ 1 mm.

\figstub{fig:5-44}{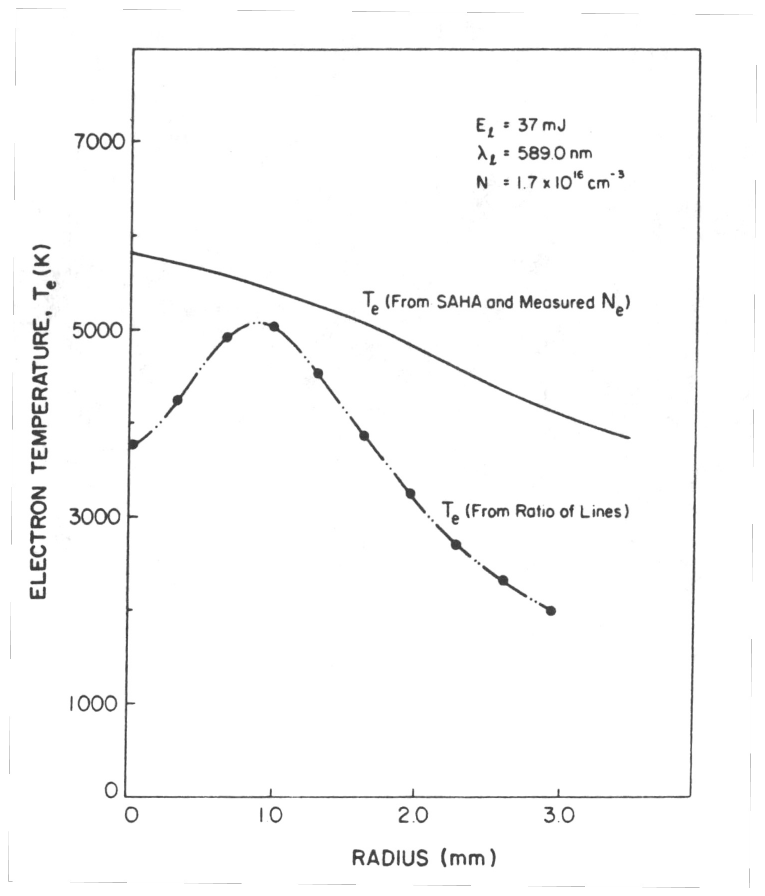}{Radial variation in $T_e$ generated from the assumption of
Saha equilibrium, together with the electron temperature derived from the
two line intensity ratio technique.}

In light of these results, we have compared the computer simulated
spectrally integrated intensity from the several high lying multiplets,
generated using this Saha temperature, to the experimentally measured data
shown in figure \ref{fig:5-40}.  This comparison is shown in figure \ref{fig:5-45}, with the
numerical computations normalized to the $4^2D$-$3^2P$ member of the series.  It
should be pointed out that the computation takes into account the finite
bandwidth of the spectrometer (defined by the slit widths used in this
particular experiment), that is, $\Theta_{nm}$ is not necessarily unity, and the full
solution (as opposed to the optically thin solution) to the radiative
transfer equation is used.  It can be seen that the agreement between the
two sets of data is fairly good and the computational results based on the
Saha temperature reproduces the scatter of the experimental data pattern
reasonably well, confirming that a Boltzmann plot can only be used to
provide an approximate indication of the electron temperature of a
cylindrical plasma having a strong radial variation.  This reasonable
agreement also suggests that the two-line ratio technique based on the
$4^2D$-$3^2P$ and $6^2D$-$3^2P$ transition intensities alone may have been influenced by
uncertainties in the relative oscillator strengths and detector spectral
response.

\figstub{fig:5-45}{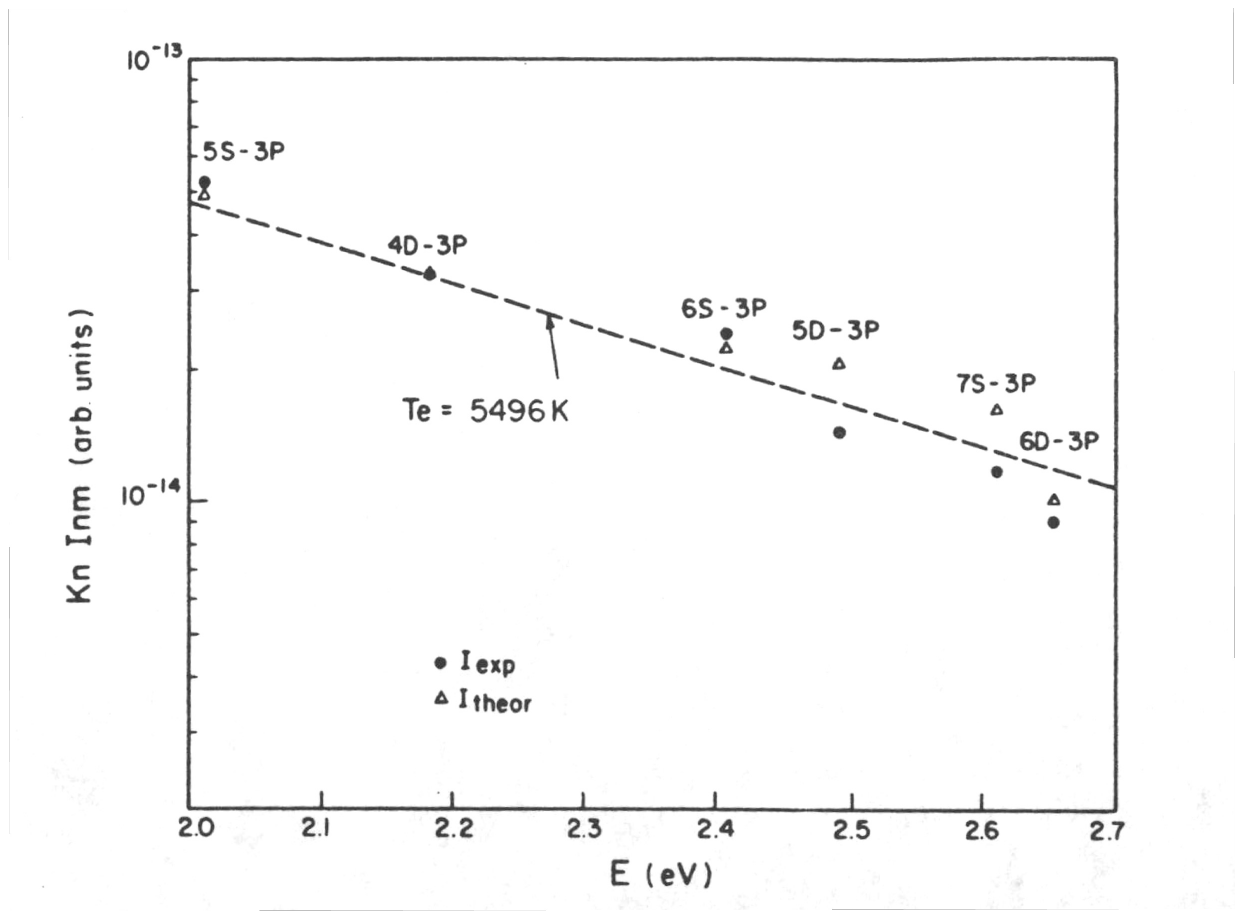}{Comparison of the computer simulated spectrally integrated
intensity, generated using the Saha temperature, with the experimentally
measured data of figure \ref{fig:5-40}.}

One further check on the self consistency of the Saha temperature
generated from the experimentally measured free electron density
distribution is to compute the $4^2D$-$3^2P$ multiplet spectral profiles for a
selected number of lateral positions and compare these with the
experimentally measured spectra.  Four such reconstituted spectra are
displayed with the corresponding experimental spectra in figure \ref{fig:5-46}.  The
absolute scaling of intensity was to the spectrally integrated emission at
y = 0.  The agreement is seen to be very good and provides us with
additional confidence in the measured free electron density distribution.
It should however be mentioned that ion broadening (section \ref{sec:2-2}) was not
included in the simulation, and the instrument broadening was not
deconvoluted from the experimental spectra.  These effects are of second
order and neither of these should appreciably detract from the excellent
agreement displayed in figure \ref{fig:5-46}.

\figstub{fig:5-46}{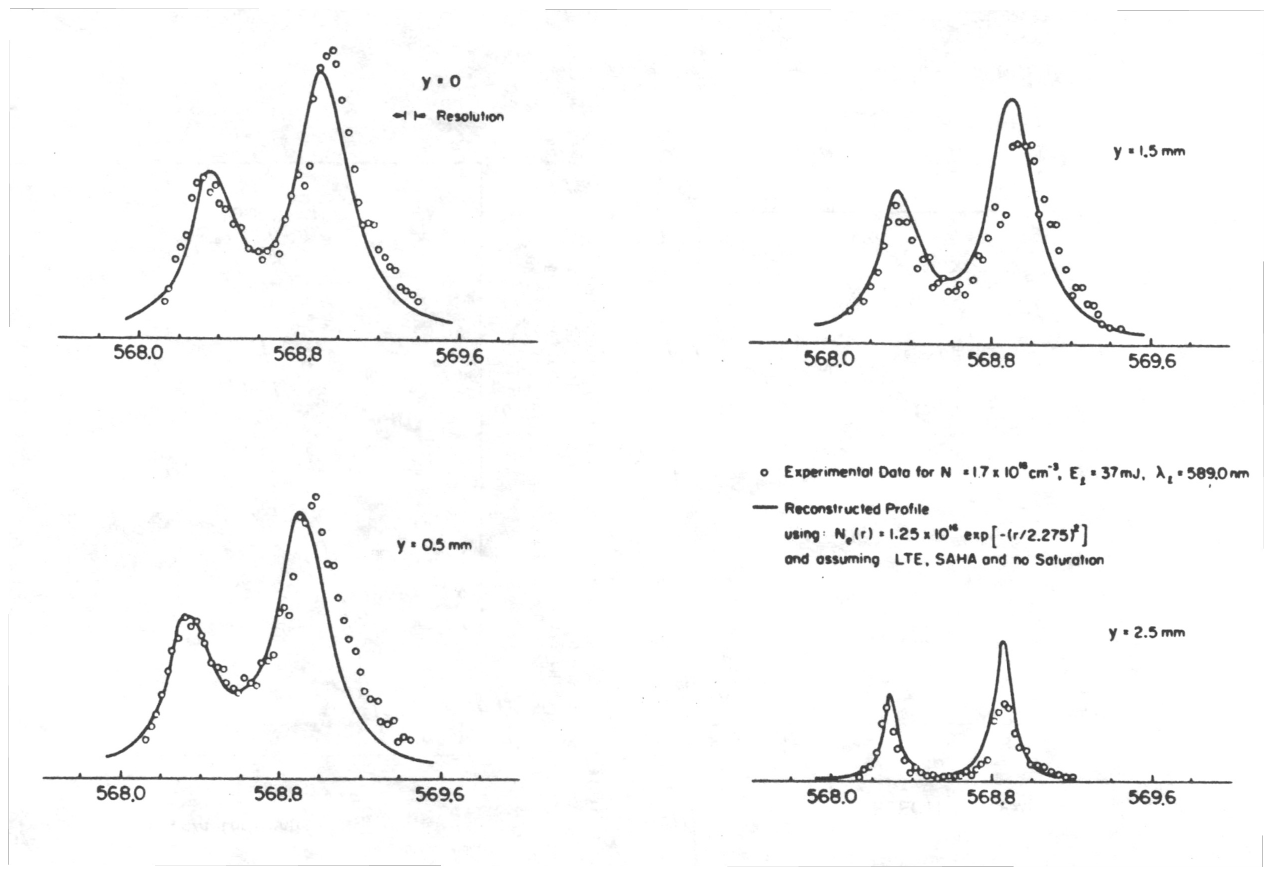}{Four reconstituted $4^2D$-$3^2P$ multiplet spectra displayed
with the corresponding experimental spectra.}

As a final comparison, we have spectrally integrated the computed
spectra displayed in figure \ref{fig:5-46} and compared it with the observed lateral
variation in the spectrally integrated emission for the $4^2D$-$3^2P$ transition
(see figure \ref{fig:5-47}).  The experimental points are normalized to the
theoretical radiance at y = 0 and as expected by inspection of figure \ref{fig:5-46},
the agreement is once again seen to be satisfactory.

\figstub{fig:5-47}{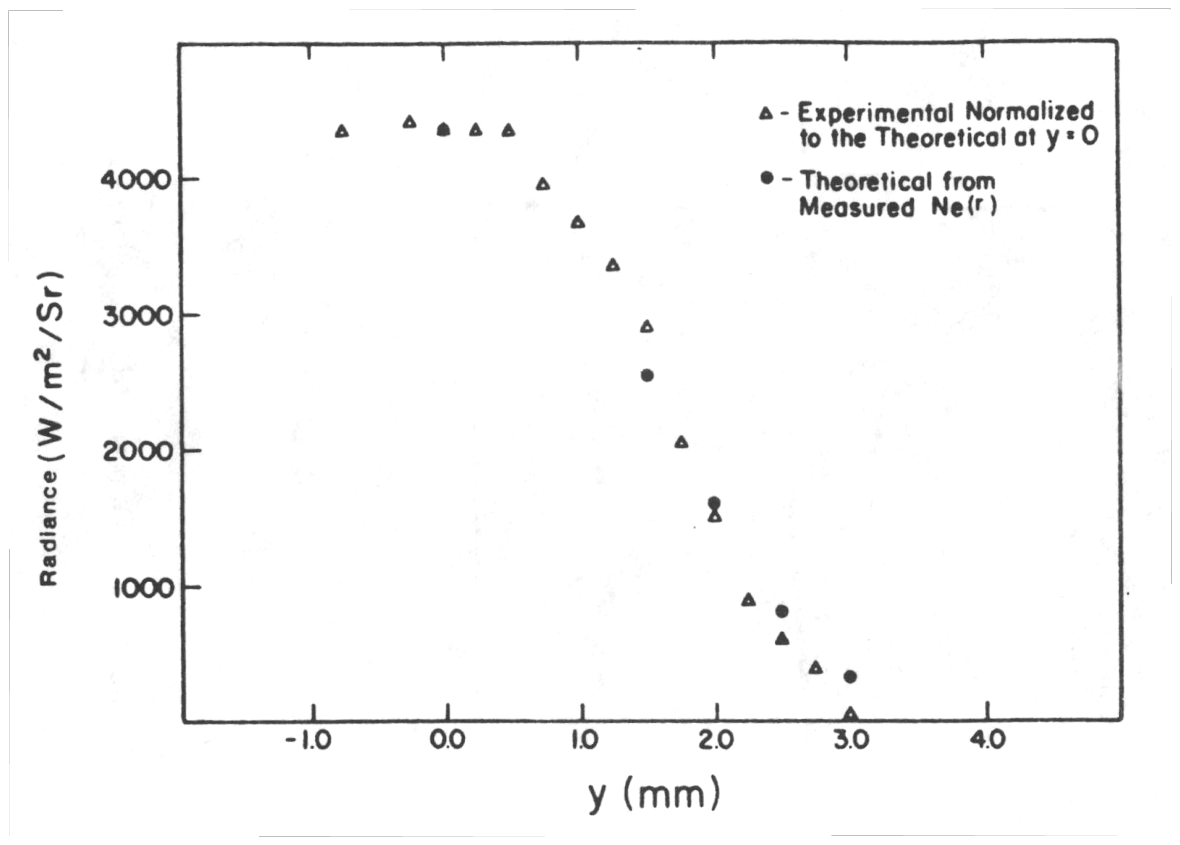}{Spectrally integrated computed spectra compared with the
observed lateral variation in the spectrally integrated emission for the
$4^2D$-$3^2P$ transition.}

\section{Laser Beam Attenuation}
\label{sec:5-3}

Laser resonance saturation leads to the formation of an enormous pool
of resonance state atoms.  Initially, these atoms can be ionized through
various seed ionization processes (Cardinal 1986).  An electron created by
this means can aquire the equivalent of one quanta of laser energy by
superelastically quenching one of the abundant laser excited atoms, this
rapidly heats the free electrons enabling them to further excite and
subsequently ionize the sodium vapor.  Direct evidence of this runaway
collisional ionization is the relatively high degree of ionization observed
in the plasma core (section \ref{sec:5-1}) and the corresponding electron temperatures
measured using both the emission line intensity ratios as well as from Saha
equilibrium using the measured neutral sodium atom density (section \ref{sec:5-2}).

The high efficiency associated with this type of laser ionization
leads to substantial absorption of the laser energy.  Early measurements of
the transmitted laser energy versus incident laser wavelength suggested that
even more energy is extracted from the laser pulse if the laser is slightly
detuned from resonance.  This observation was originally attributed to the
fact that off resonance excitation results in a longer ionization time
thereby allowing the $3^2P$ state to radiatively decay and the atom to be
cycled several times through its resonance transition by the laser field
before it is ionized.  On average then, more laser photons would be
extracted from the laser field when it is detuned and the large pool of $3^2P$
state atoms is not rapidly ionized.  In essence, laser absorption following
superelastic quenching and spontaneous emission, determines how many laser
photons are extracted.  When superelastic quenching dominates over radiative
decay (spontaneous emission) and the stimulated rates are such that the $3^2P$
state population density is rapidly replenished to its capacity (saturated),
then free electrons rapidly gain kinetic energy, leading to electron
collisional excitation and runaway ionization.

In the presence of an intense laser field, we have already seen
(section \ref{sec:2-3}) that the absorption spectra can be greatly modified as a
result of the AC Stark effect.  The combination of AC Stark splitting and
four-wave mixing (Harter and Boyd 1984) introduces further mechanisms (not
included in the LIBORS codes) which can compete with superelastic
collisions, and can be responsible for the removal of laser photons from the
path of the laser beam.  These include such processes as four-wave
parametric amplification in conjunction with self-focussing (Harter et al.\
1981).  This process can be thought of as a three-photon effect in a
strongly driven two level atom, leaving the atom in an excited state,
creating a gain medium for co-propagating coupled third and fourth waves.
These third and fourth waves are at the Rabi sidebands and are either
trapped and self-focused (third wave) or leave the volume (fourth wave)
defined by the saturated laser beam as a result of anomalous dispersion.
Filamentation (self-trapping) and conical emission have been observed over
the course of our experiments.  Rapid ionization (within the duration of the
laser pulse) can reduce the number of atoms available to participate in
these interactions.

Saturation of the volume defined by the laser beam is dependent on the
absorption cross-section of the atom at the laser wavelength.  In the weak
field limit ($\Delta\nu_L \gg \Omega$), the shape of the absorption line profile is
primarily determined by the broadening collisions discussed in chapter 2.
In the presence of a strong laser field ($\Delta\nu_L \approx \Omega$), one must consider the
interaction as leading to the creation of ``dressed states'' (Cohen-Tannoudji
and Reynaud 1977) separated by the generalized Rabi frequency $\Omega'$.
Significant conversion of the laser beam (of frequency $\nu_L$, detuned from
resonance by an amount $\Delta\nu_L = \nu_L - \nu_0$) into the third and fourth waves at $\nu_3$
$= \nu_L + \Omega'$ and $\nu_4 = \nu_L - \Omega'$, has been observed in sodium over a wide range of
sodium densities (Meyer 1980, Harter et al.\ 1981, Harter and Boyd 1984).
It is this consequence which may be of concern to those interested in
creating long plasma channels using LIBORS with resonance or near resonance
laser pumping.

This section describes the results of the first detailed investigation
of the dependence of laser penetration through the sodium vapor on laser
wavelength and incident laser energy.  The attenuation of the beam (ie, the
removal of photons from the volume defined by the incident laser beam cross
section) is found to be greater when the laser is slightly detuned from
either of the sodium D resonance lines, providing the energy is sufficient
to saturate and substantially ionize the vapor.  Saturation terminates
self-focussing (which has been found to further enhance the parametric
amplification - Harter and Boyd 1984).  Collisional excitation should
substantially modify the positive non-linear absorption coefficients
responsible for parametric gain, thereby preventing amplification of the
third and fourth wave.

Both the incident and transmitted laser pulses were monitored using
the photodiodes described in the previous chapter.  Careful aperturing of
the laser beam (2 mm diameter) was undertaken both before and after
transmission through the oven.  The purpose of these apertures, is to define
a volume from which photons are either absorbed, scattered, or scattered
into as a result of the two main mechanisms described above.

A representative laser absorption spectrum is presented in figure
\ref{fig:5-48}.  The incident laser energy for this scan was 21.6 mJ and the peak
sodium atom density was about $2\times10^{16}$ cm$^{-3}$ with a half peak density
diameter of the sodium vapor disc of about 8 cm.  Maximum attenuation of
laser energy can be seen to arise at a laser detuning of about 0.25 nm to
the blue of the 589.0 nm line and 0.2 nm to the red of the 589.6 nm line.
It is also quite apparent that the laser suffers increasing attenuation as
it is detuned to the red of the 589.0 nm line and to the blue of the 589.6
nm line.  This laser transmission spectrum is very different from that
observed with low intensity radiation (passing through a chord near the rim
of the vapor disk) from a broadband Xenon lamp source as depicted in figure
\ref{fig:5-49}.  Clearly, in the low intensity limit, where one observes distinct
minima near the sodium D line centres, figure \ref{fig:5-48} exhibits strong maxima.
Far field photographs of the spatial distribution of the transmitted laser
pulse (without aperturing the transmitted beam) provides some information as
to the structure observed in figure \ref{fig:5-48} (see figure \ref{fig:5-50}).  When detuned to
the blue side of the $D_2$ line centre, one observes an enlarged central spot
and a diffuse halo quite similar to that described by Meyer (1980) and
Harter et al.\ (1981).  Closer inspection reveals evidence of filamentation
characteristic of self-focussing (Meyer 1980).  These photographs suggest
that this emission observed at relatively large detunings can be responsible
for the large laser attenuation well away from line centre frequency.  As
the laser is detuned towards the red side of the $D_1$ line, a sharp ring
surrounding the core is observed to develop.  Unlike the ring reported by
Meyer (1980), the ring observed for our experimental conditions was present
over a much larger spectral range and disappeared when the detuning
approached 0.1 to 0.2 nm to the red of either of the resonance lines.  No
analysis of the spectral composition of these rings was performed in our
experiments.

\figstub{fig:5-48}{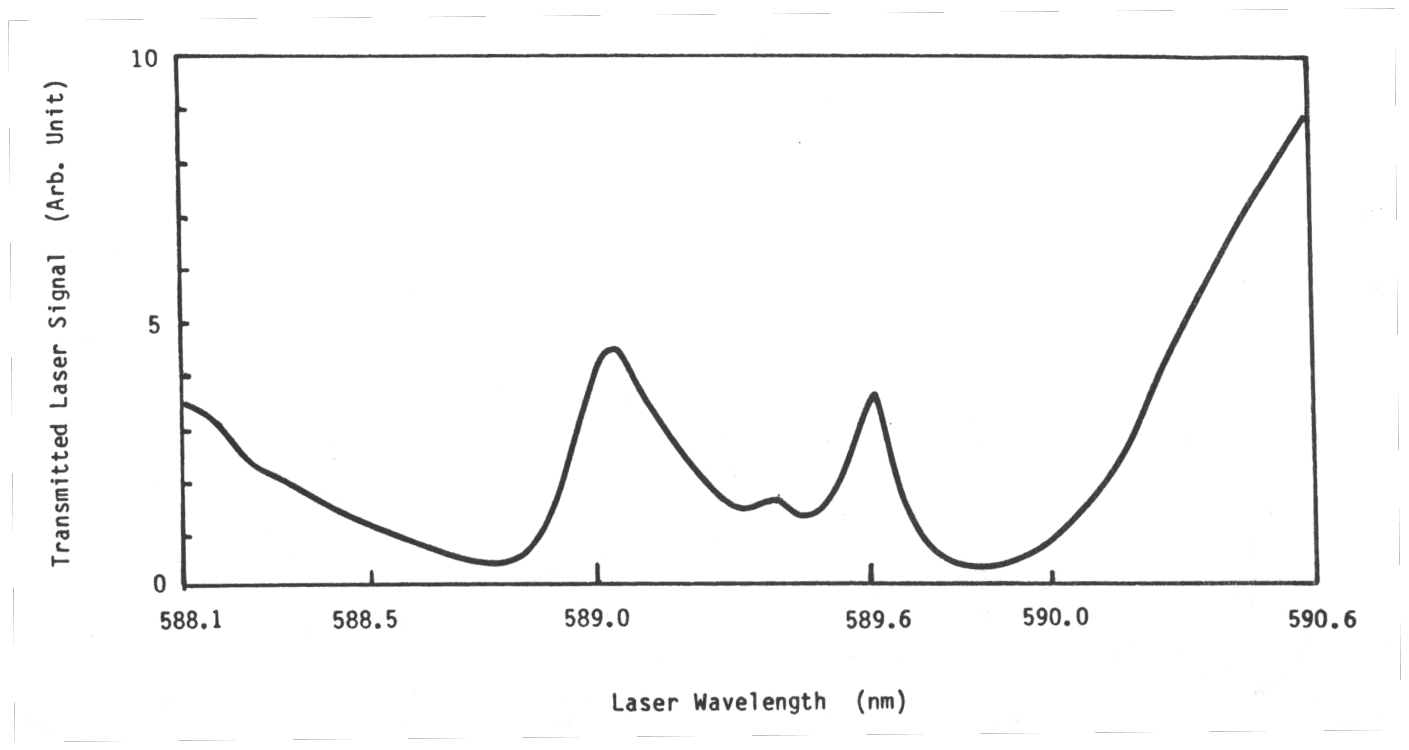}{Representative laser absorption spectrum for an incident
laser energy of 21.6 mJ.}

\figstub{fig:5-49}{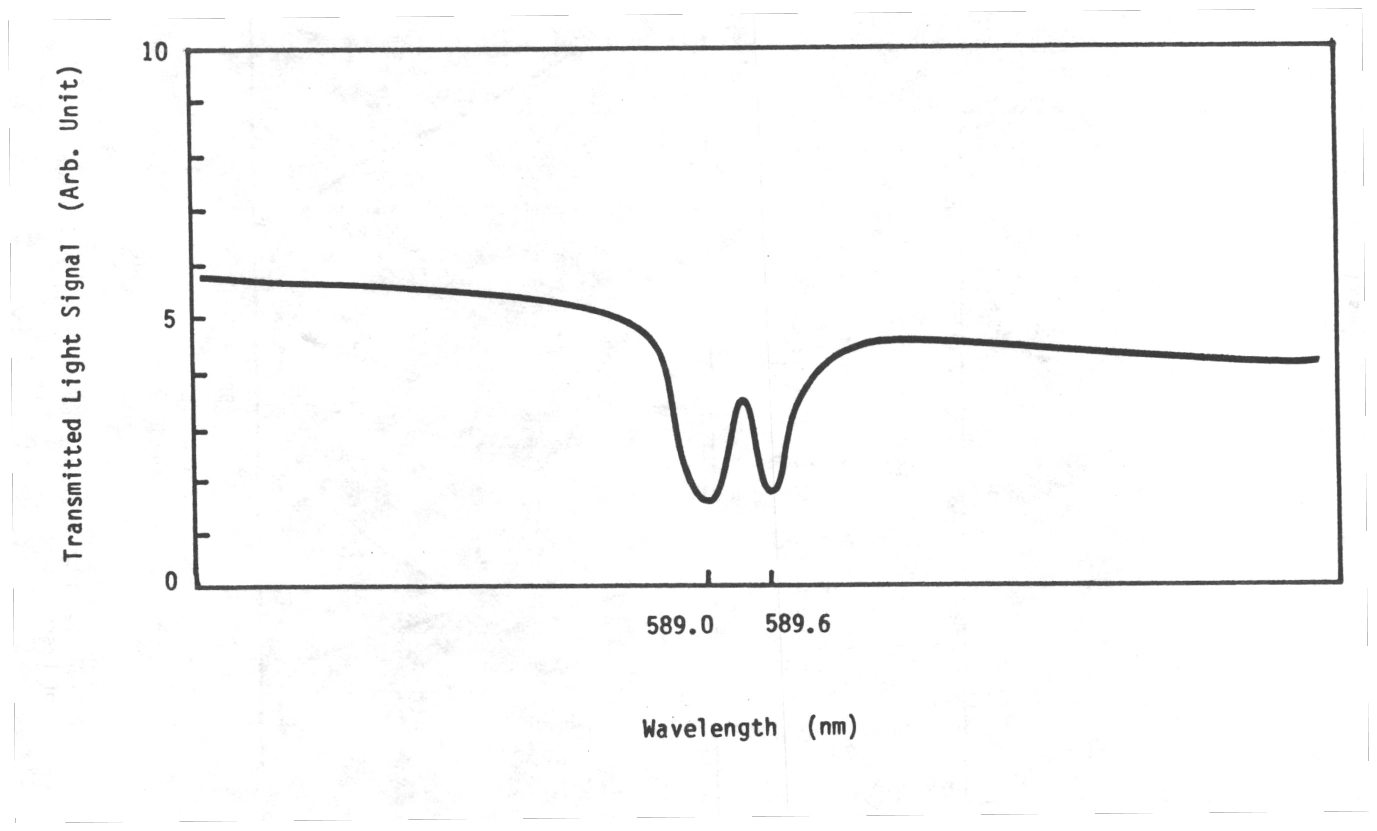}{Transmission observed with low intensity radiation from a
broadband Xenon lamp source.}

\figstub{fig:5-50}{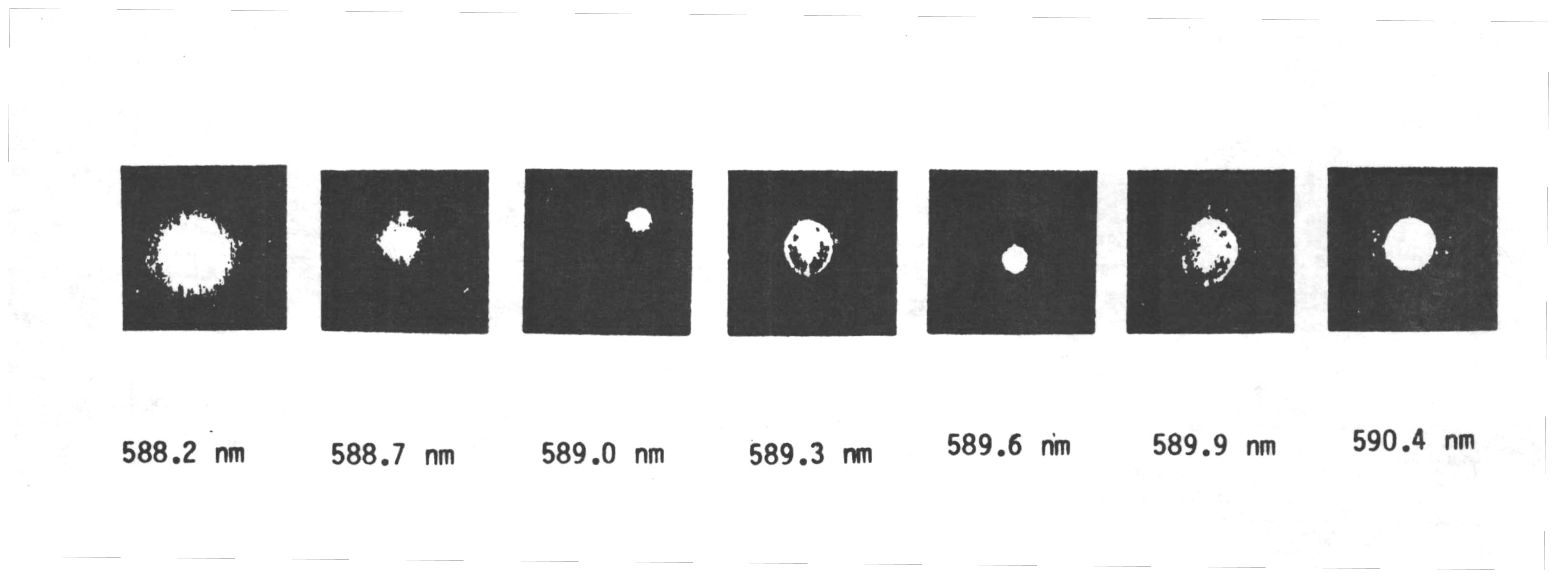}{Far field photographs of the spatial distribution of the
transmitted laser pulse.}

Our major concerns are the implications that these non-linear
processes have on the ionization of the vapor.  When the laser is detuned
far enough from the resonance lines, the optical coupling may be weak enough
to prevent both the laser resonance ionization as well as the parametric
process described.  When the laser is tuned to a wavelength slightly off
resonance, superelastic collisions and collisional ionization begin to
compete with these other processes in depopulating the resonance level.  On
resonance, providing conditions of vapor density and laser intensities are
met, the excited volume is saturated and efficiently ionized, with little
evidence of these non-linear processes.

An estimate of the degree of ionization at an axial position z = -2
cm along the path of the laser beam was made by measuring the Stark
broadened emission of the $4^2D$-$3^2P$ transition.  With an average incident
laser energy fluence of approximately $\overline{\varepsilon}_0$ = 127 mJ/cm$^2$, and the laser tuned
to the $D_1$ line, Stark broadening of the spectral emission indicated nearly
full ionization of the vapor.  The degree of ionization at this position
along z was found to decrease as the laser energy fluence is sequentially
halved (see figure \ref{fig:5-51}).  Figure \ref{fig:5-52} shows results of the transmitted
laser energy as a function of the incident laser wavelength for the same set
of laser energy fluences.  At the lowest energy fluence (corresponding to
peak laser irradiances of approximately 0.5 MW cm$^{-2}$), which may still be
sufficient to saturate the front portion of the sodium vapor column (but not
necessarily the complete volume), there is little evidence of the off
resonace maximum attenuation.  In fact, only when appreciable ionization is
observed ($\geqslant$ 50\%) does one begin to see appreciable transmission of energy
near line centre with a significant and rapid drop in transmission over a
small range of detuning.  By gating the boxcar averager to sample a 2 ns
window at selected times within the temporal profile of the output
photodiode trace, we can reconstruct the laser transmission versus
wavelength at various times within the duration of the laser pulse.  This
was performed in another experiment with a slightly higher neutral sodium
density (N(0) = $3\times10^{16}$ cm$^{-3}$).  Figure \ref{fig:5-53} clearly illustrates the
variation in the temporal history of the laser pulse after passing through
the sodium vapor with incident laser wavelength.  The absence of structure
at early times can be attributed to the absence of saturation (even when
tuned to resonance) in the laser excited volume and the spectrum is similar
to that obtained from the passage of low intensity radiation through the
vapor (Figure \ref{fig:5-49}).  By 14 ns, the medium becomes ``transparent'' to the
laser pulse when tuned to either of the sodium resonance lines.  As time
progresses further, the relative ratio of the peak transmission (on
resonance) to the farther off resonance transmission decreases as a result
of even more energy depletion resulting from ionization.


\figstub[0.6\textwidth]{fig:5-51}{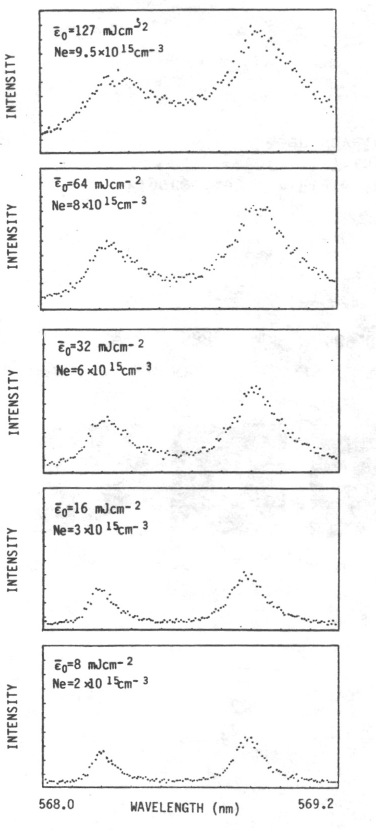}{Degree of ionization at z = -2 cm as the laser energy fluence is sequentially halved.}

\figstub[0.8\textwidth]{fig:5-52}{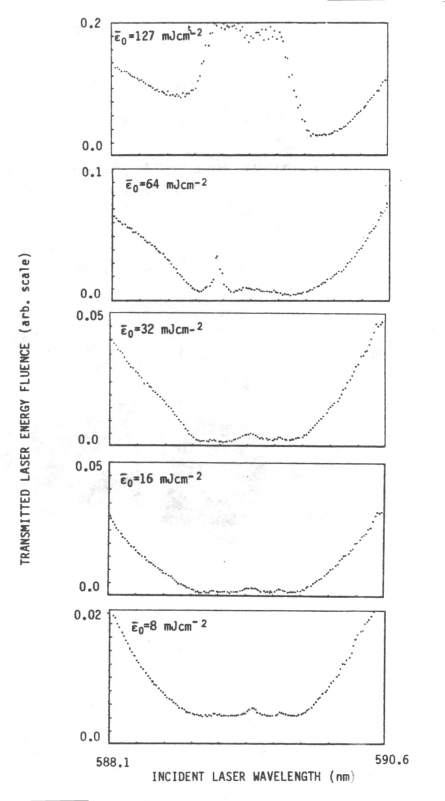}{Transmitted laser energy as a function of the incident
laser wavelength for the same set of laser energy fluences.}

\figstub{fig:5-53}{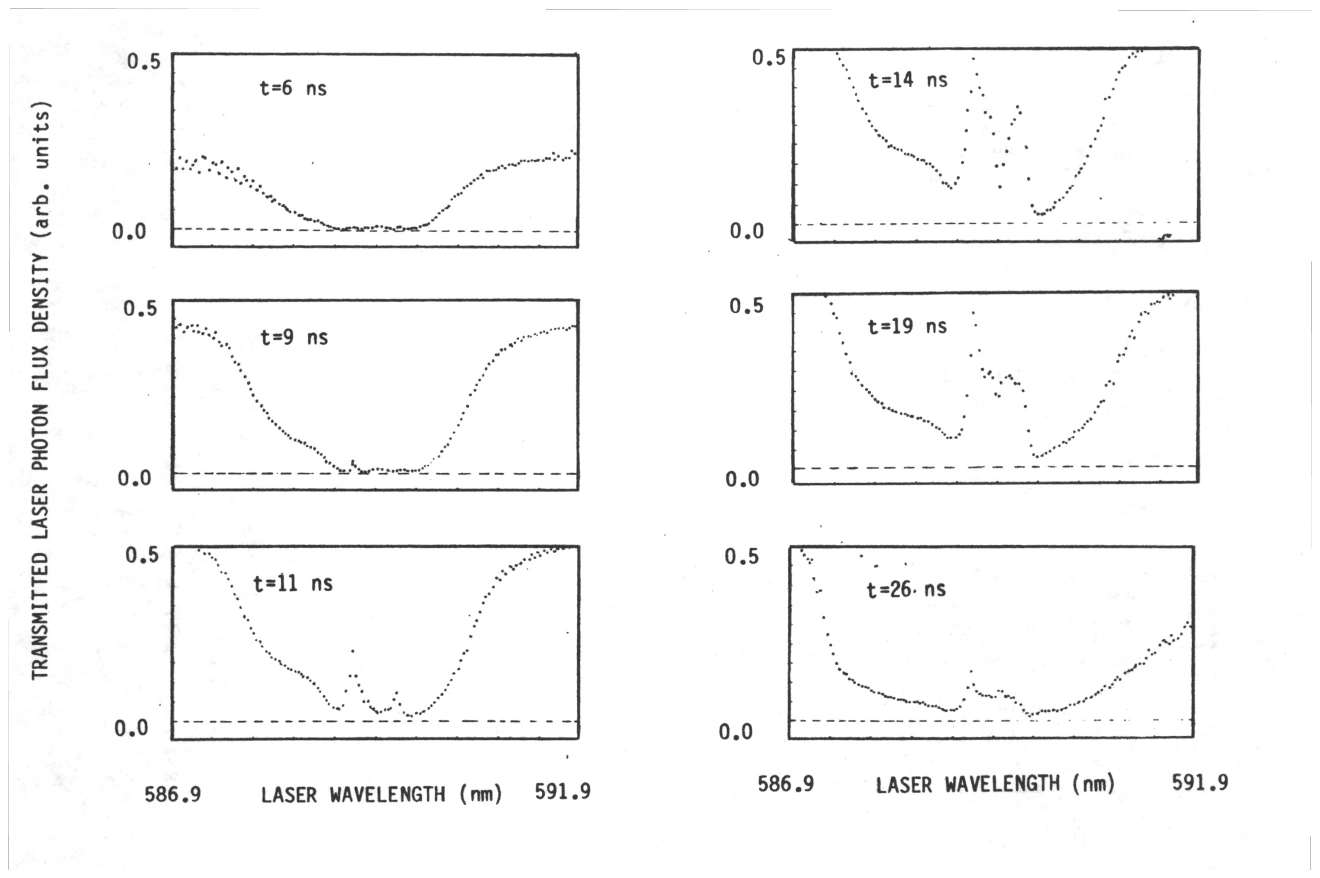}{Variation in the temporal history of the laser pulse after
passing through the sodium vapor with incident laser wavelength.}

The observed structure with time strongly supports the argument that
the sudden increase in transparancy of the sodium vapor to the laser beam
when tuned to the D lines is evidence of rapid and maintained saturation (as
a result of extended optical pumping).  This phenomenon was not reported by
Harter and Boyd (1984) as their laser pulse duration (2-7ns) was much less
than the radiative lifetime of the resonance state (16ns).

These observations suggest that the current LIBORS models are only
applicable over incident wavelength regions where superelastic collisions
dominate over amplification of the third and fourth waves in depopulating
the resonance state.  Clearly, at wavelengths 2-7 \AA\ from resonance, this may
not be the case.  The experimental observations reported here indicate that
at the sodium densities and laser intensities of interest (N $\geqslant 10^{16}$ cm$^{-3}$, $I^{\ell}$
$\geqslant 10^{7}$ W cm$^{-2}$) and with extended optical pumping ($\tau_L \geqslant 1/A_{21}$), this region
may lie within 1\,\AA\ from either line centre.  The LIBORS formulation may also
be applicable when the AC Stark effect plays a negligible role ($\Delta\nu_L \leqslant \Omega'$).
Figure \ref{fig:5-54} displays the first direct comparison of the measured transmitted
laser pulse with that predicted by the 5-level LIBORS code.  The
experimental pulse (for $\lambda_L$ = 5880\,\AA) has been normalized to that computed
using the incident laser pulse shown.  The measured transmitted laser pulse
very near resonance ($\lambda_L$ = 5890\,\AA) compares favourably to that computed ($\lambda_L$ =
5889\,\AA- the LIBORS code is too costly to run at 5890\,\AA).  The pulses exhibit
similar characteristics in their ability to predict the rapid onset at
10-13 ns, and the gradual depletion of energy from the laser tail (t $\geqslant$ 17 ns).
This agreement, although not excellent, is very encouraging in light of the
fact that the incident laser photon flux is only known to at best a factor
of two and there is in addition, limited accuracy in the measurement of the
neutral sodium atom density that is used as imput for the numerical model.

\figstub{fig:5-54}{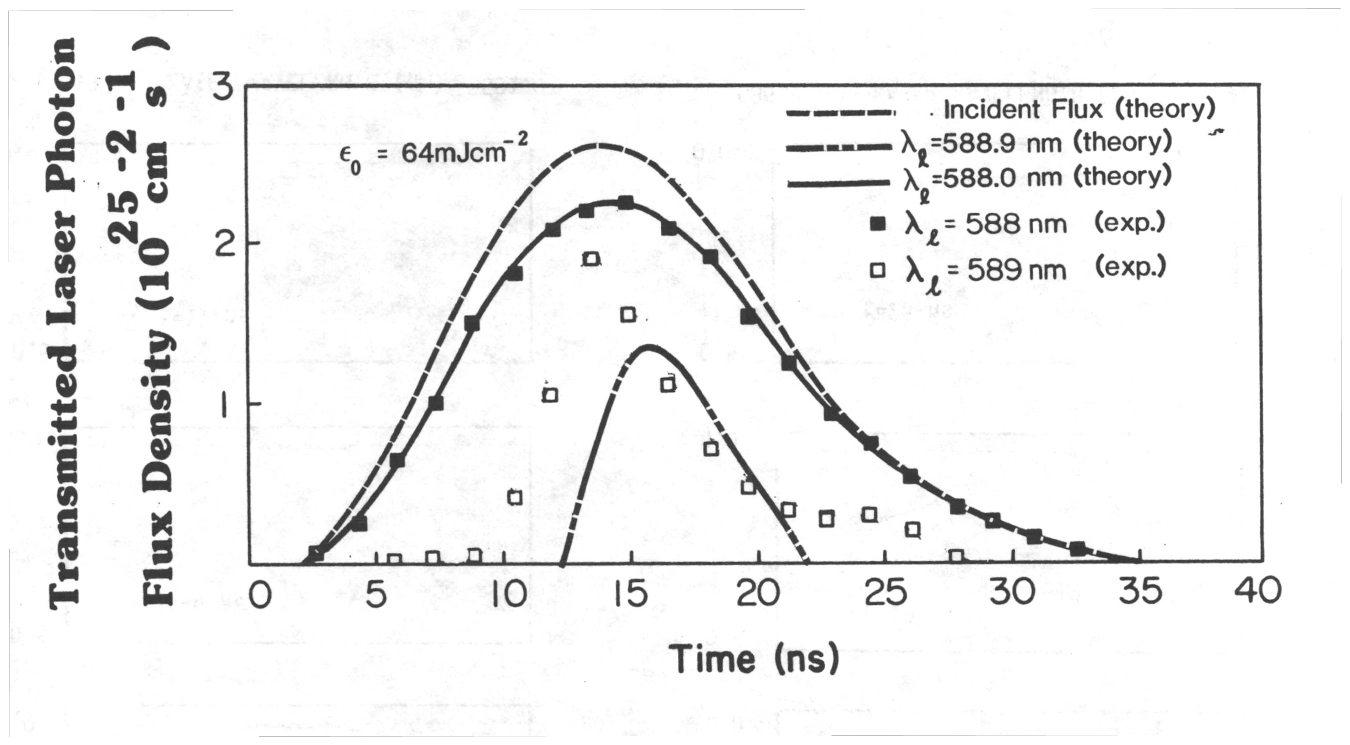}{First direct comparison of the measured transmitted laser
pulse with that predicted by the 5-level LIBORS code.}

\chapter{Comparison of Experimental Results to Theory}
\label{ch:comparison}

In order to test the most up to date theoretical computer models (Wong
1985, Kissack 1987, see also Appendix H), the author and co-worker R.S.\
Kissack have attempted to simulate, as closely as possible, the experiments
discussed in sections \ref{sec:5-1-1} and \ref{sec:5-1-2}.

The present form of the LIBORS computer code predicts the state of the
plasma along the direction of laser propagation for a given sodium atom
density distribution and a temporally prescribed laser field.  In reality,
the laser field also has a radial variation and this has to be taken into
account if we wish to model the three dimensional nature of the interaction.
To accomplish this, we have assumed that the incident laser pulse has a
Gaussian radial distribution (this assumption is in keeping with the
photodiode array measurements - for example, see figure \ref{fig:4-8}) and the same
temporal history (given by equation (\ref{eq:4-2})) at each radial position.

A series of computer runs for a range of incident laser energy
fluences was then undertaken.  The state of the plasma at any radial
positions can then be predicted from these computer runs by assigning the
appropriate laser energy fluence to each radial position.  By way of
example, displayed in figure \ref{fig:6-1}, is the predicted axial variation in the
free electron density $N_e(z)$ corresponding to three radial positions (r = 0,
1.25, and 2.55 mm) 65 ns after the start of the incident laser pulse for the
experimentally based sodium atom density distribution N(z) also shown.  The
incident laser pulse used to generate these results was assumed to have a
Gaussian energy fluence distribution described by $\varepsilon_0(r)$ and a 1/e radius of
2.5 mm.  The total energy of the laser pulse was measured to be
approximately 25 mJ.  In accordance with equation (\ref{eq:5-1}), the three radial
positions correspond to incident laser energy fluences of approximately 127,
95, and 45 mJ cm$^{-2}$.

\figstub{fig:6-1}{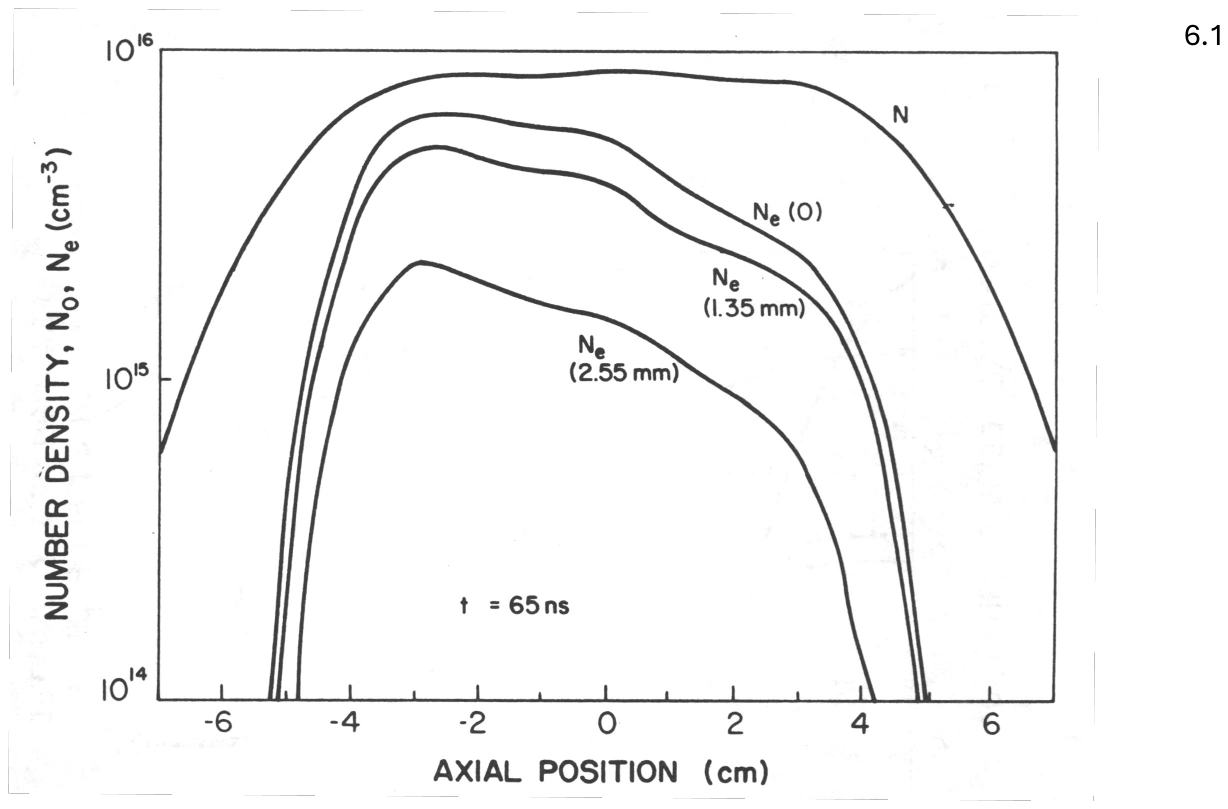}{Predicted axial variation in the free electron density
$N_e(z)$ corresponding to three radial positions (r = 0, 1.25, and 2.55 mm)
65 ns after the start of the incident laser pulse, with the experimentally
based sodium atom density distribution N(z).}

A comparison of the free electron densities achieved at equidensity
axial locations in figure \ref{fig:6-1} reveals that both the electron density and the
radius of the plasma column core diminishes as the laser pulse penetrates
farther into the sodium vapor.  This reduction of the core plasma density is
a direct consequence of the absorption suffered by the laser pulse as it
propagates through the sodium vapor (from -ve to +ve z values).  In figure
\ref{fig:6-2}, we present the axial variation of the fraction of the transmitted laser
energy fluence for several incident values of the laser energy fluence, for
the sodium atom density distribution shown in figure \ref{fig:6-1}.  It can be seen
that the smaller the incident laser energy fluence, the greater its
percentage attenuation in propagating throught the sodium vapor.  It follows
that the radial profile will tend to steepen as it propagates since its high
intensity core will be proportionately less reduced than its weaker outer
region.

\figstub{fig:6-2}{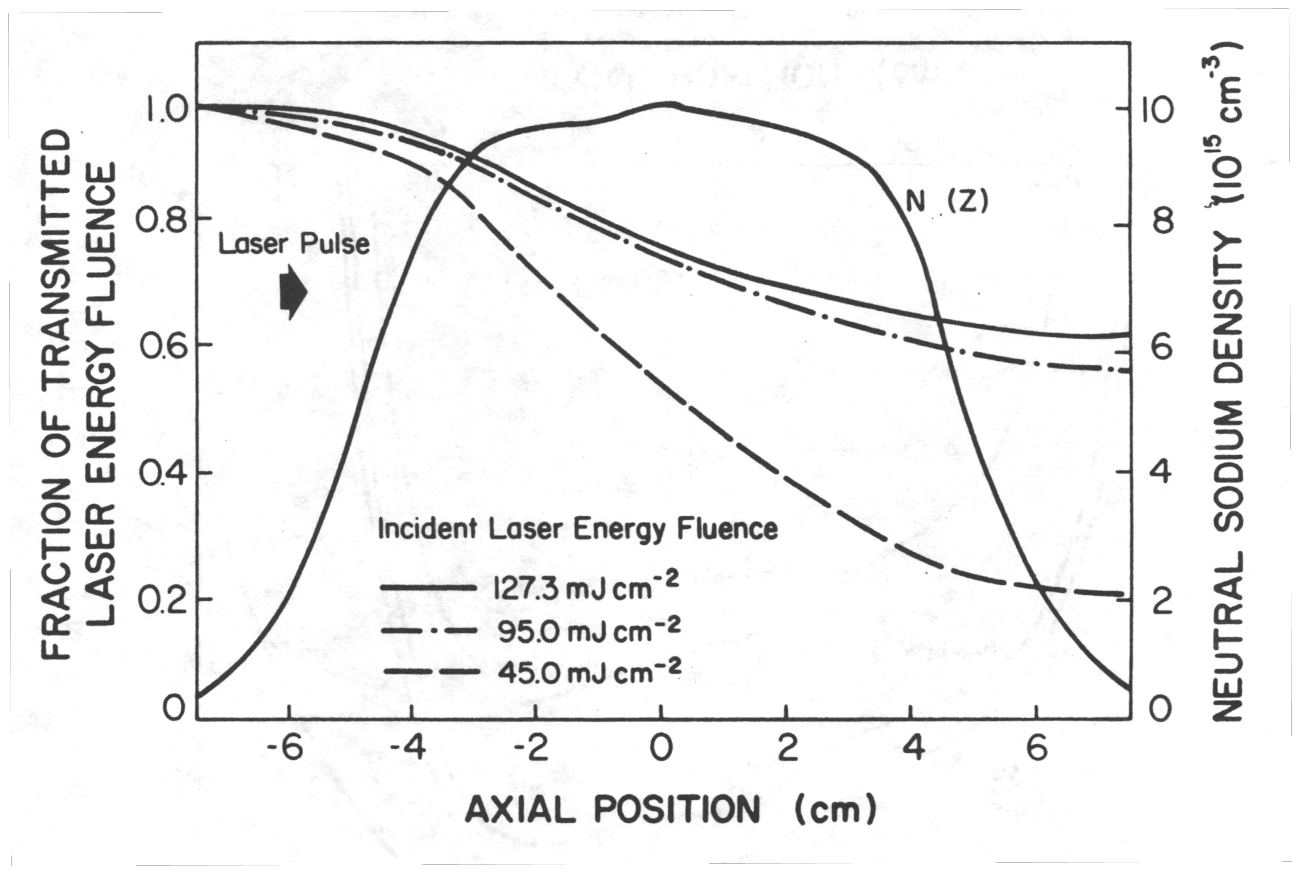}{Axial variation of the fraction of the transmitted laser
energy fluence for several incident values of the laser energy fluence.}

The reduction in electron density and plasma radius will tend to give
rise to a region of ionization that is somewhat conical in shape, in the
direction of laser propagation.  This can be inferred from the illustration
in figure \ref{fig:6-3}, where the variation in the free electron density at 65 ns and
z = -4, 0 and 4 cm are plotted as a function of the incident laser energy
fluence.

\figstub{fig:6-3}{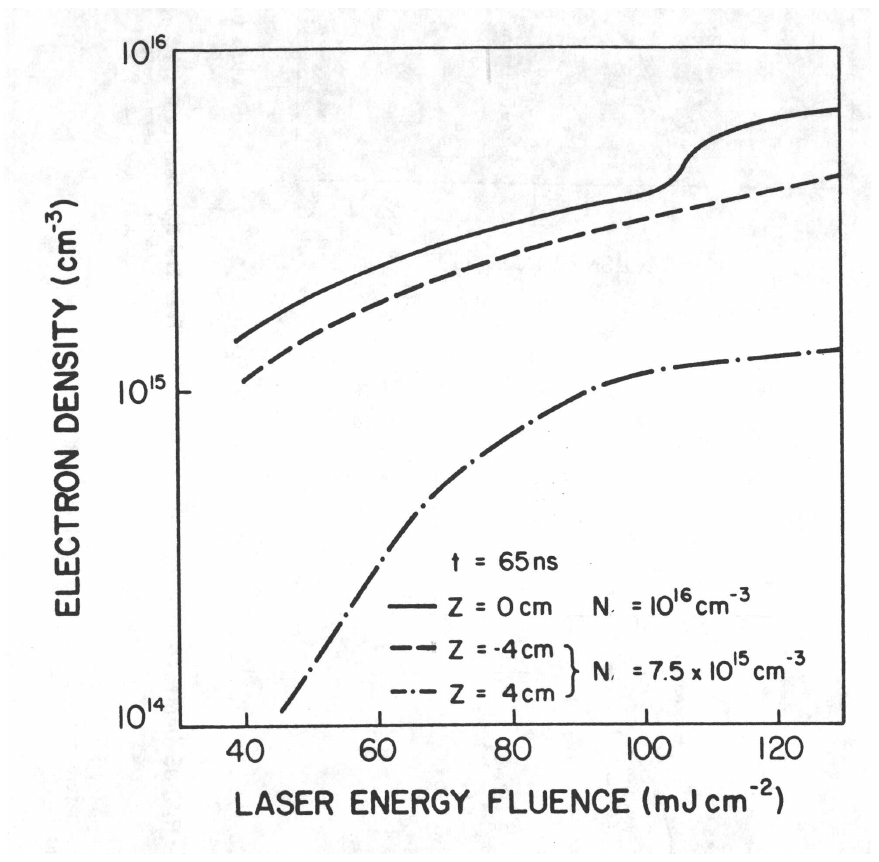}{Variation in the free electron density at 65 ns and
z = -4, 0 and 4 cm plotted as a function of the incident laser energy
fluence.}

In figure \ref{fig:6-4}a) we present both the experimental free electron density
radial profile (at z = -2 cm and t = 65 ns - recall figure \ref{fig:5-7}) and the
computed profile, while in figure \ref{fig:6-4}b) we present the corresponding radial
profiles at z = 2 cm.  While reasonable agreement is obtained at z=2 cm, the
code predictions appear to overpredict the electron density near the core
for z=-2 cm.  It should be noted that both our sodium atom density
measurements as well as the estimate of the laser beam radius have an
uncertainty of approximately 20\%.  Since the attenuation, along with the
temporal and spatial distortion suffered by the laser pulse depend upon
these variables, the predictions of the computer code would be of limited
accuracy.  To further complicate matters, the sensitivity of the predictions
to uncertainty depend upon the laser energy fluence so no single figure can
be quoted.  Nevertheless, if we were to assume that the uncertainty in the
code predictions are at least comparable to that in the neutral density,
then we can see with reference to figure \ref{fig:6-4}, that there would be agreement
between theory and experiment, within their respective uncertainties.

\figstub{fig:6-4}{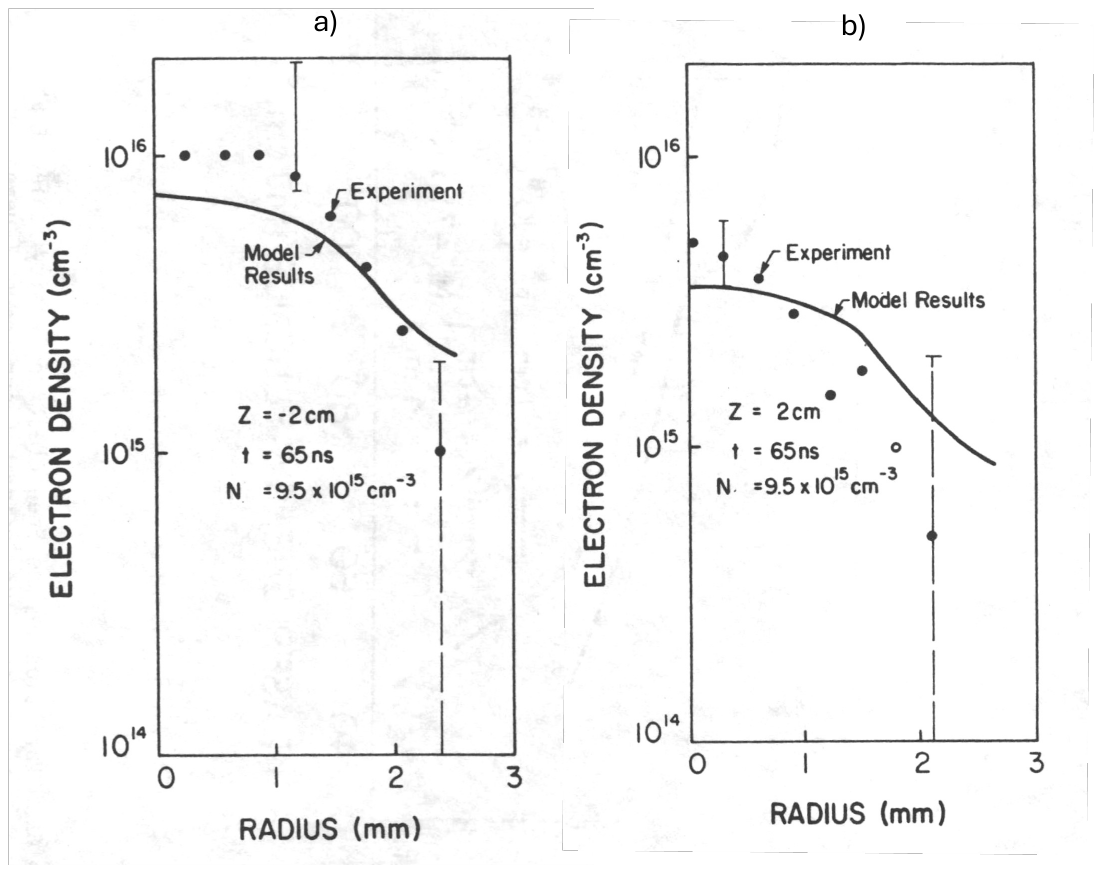}{a) Experimental free electron density radial profile (at
z = -2 cm and t = 65 ns) and the computed profile.  b) Corresponding radial
profiles at z = 2 cm.}

If we assume that the plasma is in LTE, then the experimentally
measured radial electron density distributions at z = -2 and z = 2 cm can be
used with the Saha equation and the inital atom density to determine the
corresponding radial profile in the electron temperature.  These temperature
distributions are displayed along with the corresponding profiles computed
by the LIBORS model in figure \ref{fig:6-5}a) and figure \ref{fig:6-5}b).  In both cases, the
experimental data are within 20\% deviation of the theoretical predictions.

\figstub{fig:6-5}{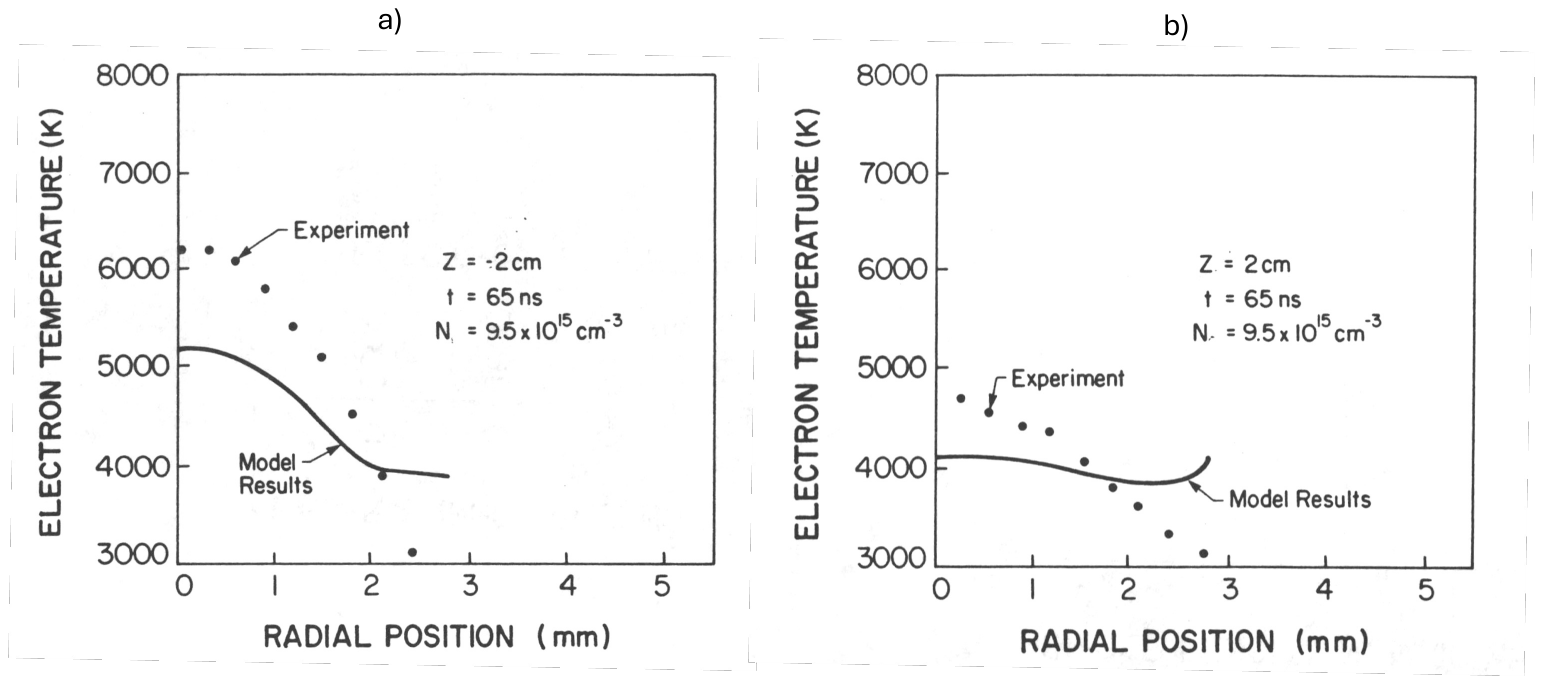}{Experimentally derived electron temperature distributions
displayed along with the corresponding profiles computed by the LIBORS
model: a) z = -2 cm; b) z = 2 cm.}

Both the experimental and predicted electron temperatures indicated in
figure \ref{fig:6-5}a) and \ref{fig:6-5}b) are somewhat lower than those predicted by earlier
LIBORS codes (Measures, Wong and Cardinal 1982; Wong 1985).  However, they
complement other experimental observations of low electron temperatures in
LIBORS type plasmas of the alkaline earths (Skinner 1980, Jahreiss and Huber
1983, Bachor and Kock 1981) and in sodium (Krebs and Schearer 1982, Landen
et al.\ 1985).  This is particularily so for z = 2 cm, where the mean
experimental temperature across the plasma column is approximately 4000K.
This low electron temperature, is also seen to be predicted by the current
LIBORS model after penetration of several centimetres of sodium vapor and
represents another manifestation of reduced laser energy fluence resulting
from absorption of the laser pulse.  This is clearly seen in figure \ref{fig:6-6}
where the LIBORS model predictions for the axial variation (at r=0) in the
free electron density and temperature are presented, along with the
variation in the laser energy fluence $\varepsilon_{\ell}$.  Also shown in the figure is the
measured axial variation of the neutral sodium density N that was used in
the computer code for these predictions.  Figure \ref{fig:6-7} presents a comparison
of the model predicted axial variation (at r=0) with that of the
experimentally derived values based on an assumed radially uniform plasma
and the measured $4^2D$-$3^2P$ spectra along y = 0 mm (recall figure \ref{fig:5-18}).  The
qualitative (and in some instances quantitative) agreement obtained between
the experimental values and theoretical predictions is viewed as acceptable
in light of the limitations imposed on both by the uncertainty in both the
neutral sodium atom density distribution along z, and the spatial
distribution of the incident laser beam.

\figstub{fig:6-6}{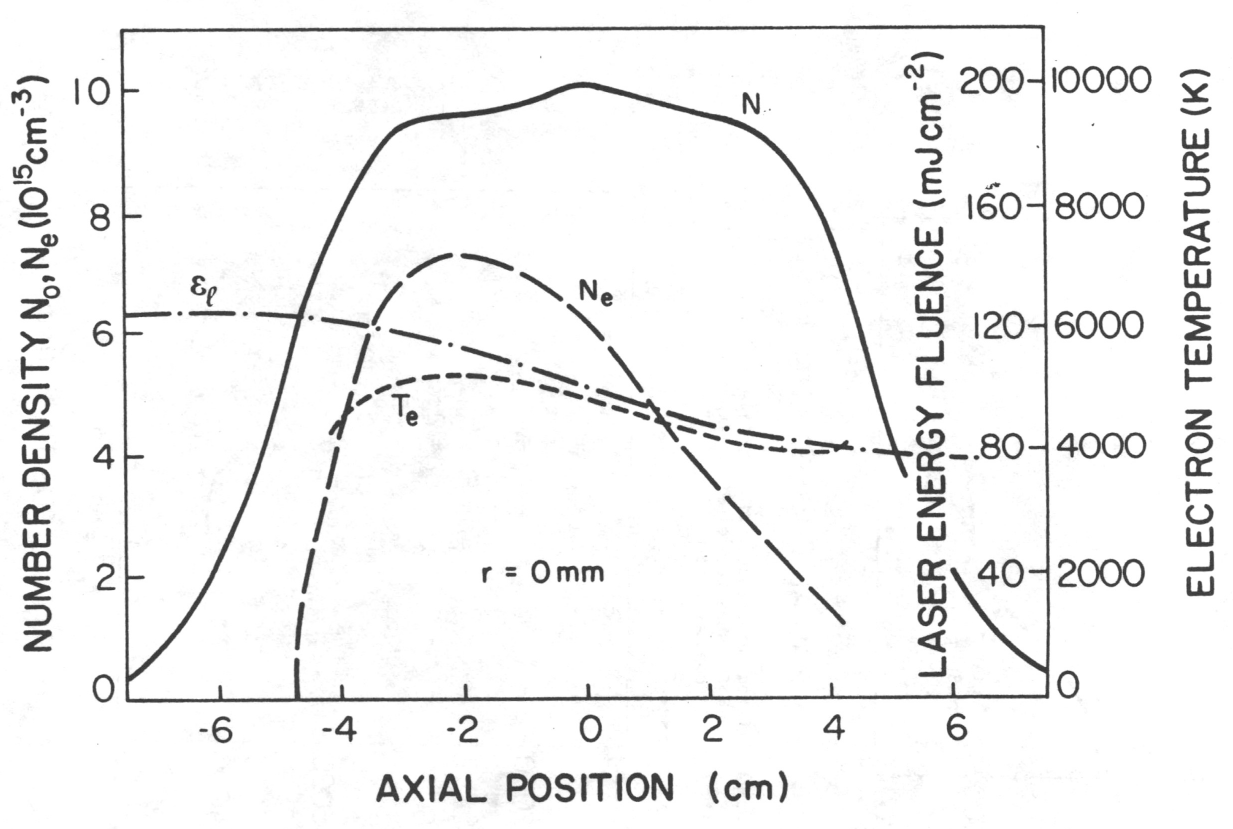}{LIBORS model predictions for the axial variation (at r=0)
in the free electron density and temperature, along with the variation in
the laser energy fluence $\varepsilon_{\ell}$ and the measured axial variation of the neutral
sodium density N.}

\figstub{fig:6-7}{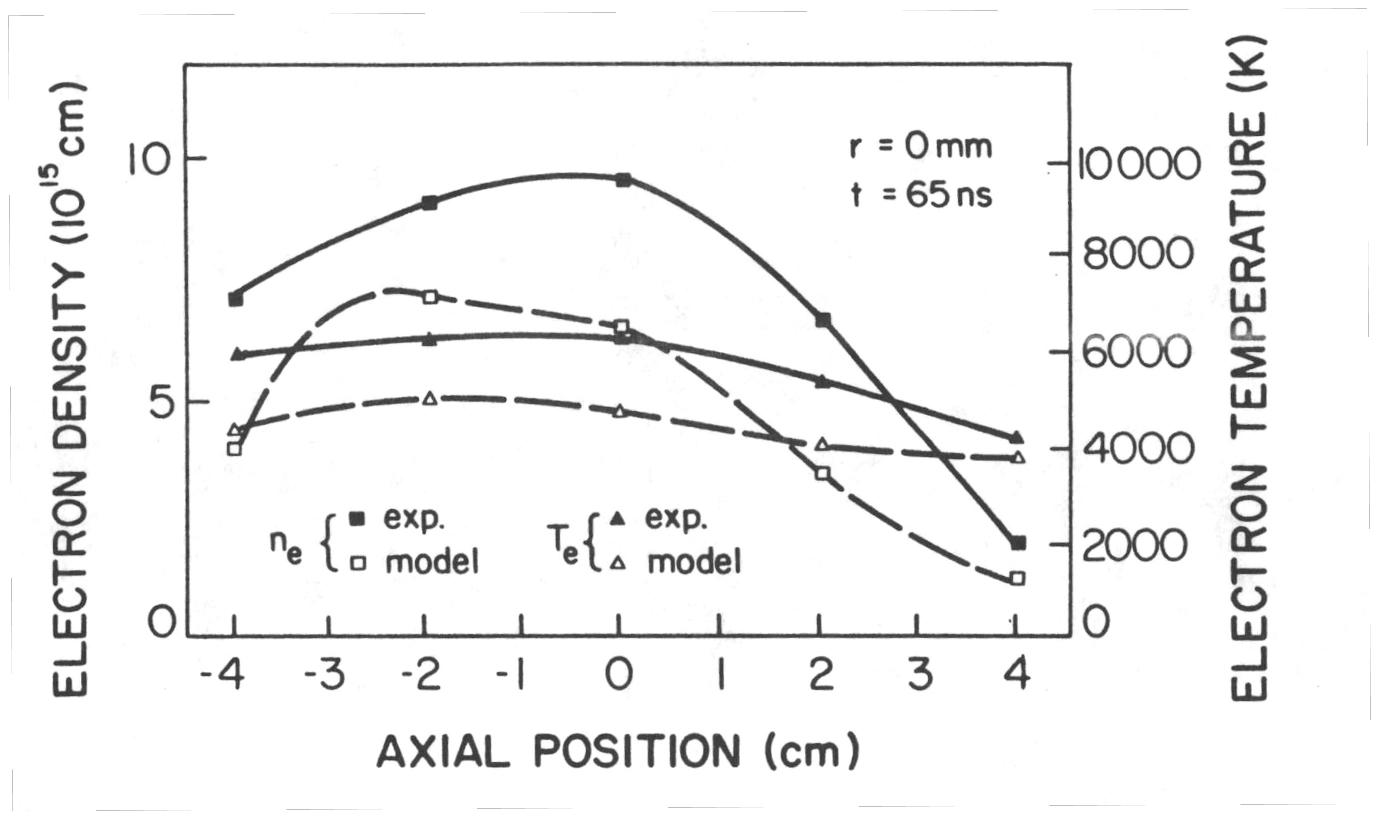}{Comparison of the model predicted axial variation (at r=0)
with that of the experimentally derived values.}

We can use the theoretical LIBORS code to predict the variation in the
free electron density at z=-2 cm with the incident laser energy fluence for
the experimental parameters described in section \ref{sec:5-1-2}.  A comparison
between these computations and the experimental results (recall figure \ref{fig:5-33})
is presented in figure \ref{fig:6-8}.  Once again, the agreement between theory and
experiment is within the relative uncertainty.  These encouraging results
have strongly supported the conclusions drawn as to the usefullness of the
current LIBORS computer code (see Chapter seven).

\figstub{fig:6-8}{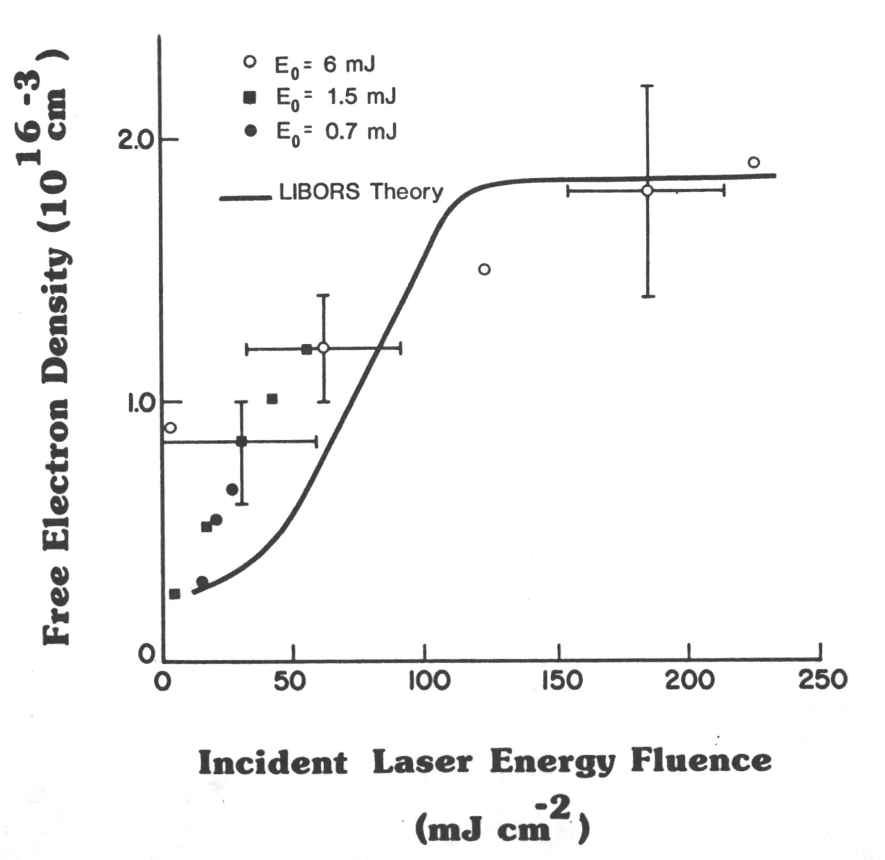}{Comparison between the LIBORS computations and the
experimental results for the variation in the free electron density at
z = -2 cm with incident laser energy fluence.}

\chapter{Summary}
\label{ch:summary}

Laser ionization based on resonance saturation (LIBORS) of a sodium
vapor has been studied experimentally.  The experimental parameters and
results have been catagorized in a form amenable for comparison with the
results of elaborate theoretical models used to describe the LIBORS
interaction.

A valid comparison of experiment to theory requires reasonable
knowledge of the incident laser pulse characteristics, as well as the sodium
vapor density variation along the path of laser propagation.  The sodium
vapor density distribution has been measured by a technique utilizing the
absorption of broadband radiation in the vicinity of the sodium D-lines.
The accuracy of this technique is limited by the uncertainties in the
current understanding of the dominant mechanisms responsible for the
spectral broadening of the atomic absorption profile, particularily in the
quasi-static line wings.

The incident laser energy fluence distribution and temporal pulse shape
have been measured.  In many cases, the energy fluence distribution for
these particular experimental runs can be considered to be Gaussian in
shape, with the total energy and 1/e radius specified.  Accurate collimation
of the beam prior to entering the heat sandwich oven is undertaken and
allows simplification in the theoretical modelling.

The spatial distribution (both radial and axial) of the free electron
density and neutral sodium vapor density are required to characterize the
plasma channel.  A technique has been developed, to measure the free
electron density radial profiles.  The technique is based on the spatial
inversion of the lateral $4^2D$-$3^2P$ multiplet emission.  The electron Stark
broadening of the inverted radial volume emission permits the radial
variation of the free electron density to be evaluated.  A survey of
competing broadening mechanisms is necessary to understand the range over
which the electron Stark broadening dominates.  The technique is found to be
particularily well suited for measuring the free electron density radial
gradients however, its accuracy is somewhat limited in determining the core
electron density when the core degree of ionization is very high ($>$90\%).

The free electron temperature is estimated from a Boltzmann analysis
of line intensities and is compared to the electron temperature obtained
from the assumption of Saha equilibrium between the free electrons and the
neutral species.  A critical evaluation of this Boltzmann analysis has been
performed, particular for cases where spectral lines suffer significant
broadening and self-absorption.

An analysis of the variation in the transmitted laser energy fluence
with incident laser energy and wavelength has been made.  A strong resonance
behavior has been observed in the vicinity of the sodium D-lines.  Although
the physical mechanisms responsible for these resonances have not been
definitively isolated, the results have allowed us to draw some conclusions
as to the importance of laser tuning to the creation of long stable
reasonably ionized plasma channels.

The results of a recently developed 3-dimensional LIBORS model have
been compared to results of a set of experiments.  The comparison required
experimentally derived input parameters such as the laser energy fluence
distribution and sodium density variation, and for this reason, the
comparison is of limited accuracy.  It has been shown that in some cases,
the dynamics of the plasma channel creation can be highly sensitive to these
variables.  Nevertheless, there is reasonable agreement between the
experimental results and the predictions of the current LIBORS computer
code.

\section{Conclusions}
\label{sec:7-1}

Electron Stark broadening of the $4^2D$-$3^2P$ multiplet transition in a
sodium plasma produced by laser resonance saturation has provided a means of
undertaking the first spatial measurements of the free electron density
across and along the plasma channel (section \ref{sec:3-2}).  From these measurements
and measurements of the neutral sodium density within the heat sandwich oven
(section \ref{sec:3-1}), we have been able to deduce the corresponding electron
temperature (section \ref{sec:5-2-3}).  These temperatures compare favourably with the
electron temperature estimated from a Boltzmann analysis of line intensities
(sections \ref{sec:5-2-1} and \ref{sec:5-2-2}) and suggests that a highly ionized ($>$10\%) plasma
of electron temperature $<$6000K can be produced within 100 ns of laser
excitation.  Although these temperatures are not necessarily in keeping with
earlier theoretical computations (Measures et al.\ 1981), the results are in
reasonable agreement (in general, within a factor of two over the plasma
cross-section) with a recently developed 3-dimensional model (Chapter 6).
The decrease in electron density and temperature along the path of the laser
beam has been attributed to significant depletion of laser energy.  These
results demonstrate that this LIBORS code is capable of predicting the
3-dimensional nature of this new mode of laser ionization with reasonable
accuracy, considering the range of experimental uncertainties and the
limited accuracy of the cross-sections used by the theoretical model
(Kissack 1987).  These results may also explain the low electron
temperatures and free electron densities observed by others (Landen et al.\
1985).

Detailed measurements of the variation of the radial electron density
profiles with laser detuning from resonance (section \ref{sec:5-1-3}) indicates that
when the laser is detuned by approximately 1.0 nm, the free electron density
is 50\% (in the core) of that achieved for the case where the laser is tuned
to resonance, having penetrated only a few centemetres of dense ($10^{16}$ cm$^{-3}$)
vapor.  Perhaps more importantly however, is the discovery that there is an
off resonance maximum attenuation which can be in part described by a
competition between superelastic electron heating, spontaneous emission and
parametric amplification followed by anomalous dispersion, when the laser is
moderately detuned (0.2-0.7 nm) from resonance.  These three-photon and
parametric phenomena become less pronounced as one approaches resonance
(0-0.2 nm) suggesting that rapid ionization of the saturated volume depletes
the source of atoms that are able to participate in these other processes.

Clearly, the experimental findings suggest that if LIBORS is to be
used as a means of generating fairly long uniform plasma channels, then the
laser energy fluence has to be sufficient to achieve close to full
ionization along the laser path with laser tuning critical to avoid
attenuation of the beam arising from these other processes.

It has been confirmed that at some positions along the laser path, the
interaction (free electron density achieved) is highly non-linear in
response to the incident laser energy fluence (section \ref{sec:5-1-2}).  The
implications of this with respect to multi-shot averaging have been
investigated (section \ref{sec:3-4}).  For the scatter observed in our incident laser
energy, the non-linearity of this interaction has been computed to have a
negligible effect on our experimental measurements of the free electron
density.

As an outcome of the work presented in this thesis, the following
contributions have been made to this field of research:

\begin{list}{}{%
  \setlength{\leftmargin}{3.5em}
  \setlength{\labelwidth}{2.5em}
  \setlength{\labelsep}{1em}
  \setlength{\itemsep}{1.2em}}

\item[(i)] A comprehensive facility has been developed to investigate
laser ionization based on resonance saturation.  An important
aspect of this facility is the characterization and
stabilization of neutral sodium vapor within a specially
designed heat sandwich oven.

\item[(ii)] The first emission based measurements of both the radial and
axial variation in the free electron density and temperatures
in an axially symmetric pulsed sodium plasma produced as a
result of laser resonance saturation have been reported.

\item[(iii)] The first observations of strong off-resonance attenuation have
been made which reemphasizes the importance of laser tuning
in the creation of long plasma channels in dense alkali metal
vapors with tuned lasers.

\item[(iv)] The first quantitative comparison between the 3-dimensional
measurements of electron density and temperature and the
predictions of a recently developed 3-dimensional computational
model (Wong 1985, Kissack 1987) of the resonance interaction
is presented.

\end{list}

\chapter*{References}
\addcontentsline{toc}{chapter}{References}
\markboth{REFERENCES}{REFERENCES}

\begingroup
\setlength{\parindent}{0pt}
\newcommand{\refitem}[1]{\par\hangindent=2.5em\hangafter=1 #1\par\vspace{0.45em}}

\refitem{Agnew, L., and Reichelt, W.H., 1968, J. Appl. Phys. \underline{39}, 3149.}

\refitem{Allen, C.W., \underline{Astrophysical Quantities}, London U.P.(Athlone), London, 1963.}

\refitem{Bachor, H.A., and Kock, M., 1980, J. Phys. B: Atom. Molec. Phys. \underline{13}, L369.}

\refitem{Bachor, H.A., and Kock. M., 1981, J. Phys. B: Atom. Molec. Phys. \underline{14}, 2793.}

\refitem{Baranger, M., 1958a, Phys. Rev. \underline{111}, 481. 1958b, Phys. Rev. \underline{111}, 494.\\
\hspace*{2.5em}1958c, Phys. Rev. \underline{112}, 855.}

\refitem{Bauer, J.F., and Cooper, J., 1977, J. Quant. Spectrosc. Radiat. Transfer \underline{17}, 311.}

\refitem{Bjorkholm and Liao, 1974, \underline{Laser Spectroscopy}, Proc. 2nd Int. Laser Spect. Conf., pg. 176.}

\refitem{Bober L., and Tankin R.S., 1979, J. Quant. Spectrosc. Radiat. Transfer \underline{9}, 855.}

\refitem{Boyd, R.W., Dodd, J.G., Krasinski, J., and Stroud, C.R. Jr., 1980, Opt. Lett. \underline{5}, 117.}

\refitem{Boyd, R.W., and Harter, D.J., 1980, Appl. Opt. \underline{9}, 2660.}

\refitem{Brehignac, C. and Cahuzac, Ph., 1982, Optics Commun. \underline{43}, 270.}

\refitem{Burgess, D.D., and Cooper, J., 1965, J. Sci. Instrum. \underline{42}, 829.}

\refitem{Cappelli, M.A., 1983, M.A.Sc. Thesis, University of Toronto (unpublished).}

\refitem{Cappelli, M.A., and Measures, R.M., 1984, Appl. Optics \underline{23}, 2107.}

\refitem{Cappelli, M.A., Cardinal, P.G., Herchen, H., and Measures, R.M., 1985, Rev. Sci. Instrum. \underline{56}, 2030.}

\refitem{Cappelli, M.A., and Measures, R.M., 1987a, accepted for publication in J. Opt. Sensors. 1987b, accepted for publication in Appl. Optics.}

\refitem{Cappelli, M.A., Wong, S.K., Kissack, R.S., and Measures, R.M., 1987, submitted for publication in Phys. Rev. A.}

\refitem{Cardinal, P.G., Wizinowich, P.L., and Measures, R.M., 1981, J. Quant. Spectrosc. Radiat. Transfer \underline{25}, 537.}

\refitem{Cardinal, P.G., 1986, UTIAS Report No. 299, University of Toronto.}

\refitem{Carre, B., Roussel, F., Breger, P., and Spiess, G., 1981a, J. Phys. B: Atom. Molec. Phys. \underline{14}, 4271. 1981b, J. Phys. B: Atom. Molec. Phys. \underline{14}, 4289.}

\refitem{Carrington, C.G., Stacey, D.N., and Cooper, J., 1973, J. Phys. B: Atom. Molec. Phys. \underline{6}, 417.}

\refitem{Chen, S.Y., and Takeo, M., 1957, Rev. Mod. Phys. \underline{29}, 20.}

\refitem{Chiang, W.T., Murphy, D.P., Chen, Y.G., and Griem, H.R., 1977, Z. Naturforsch. \underline{32a}, 818.}

\refitem{Choi, B.S. and Kim, H., 1982, Appl. Spectrosc. \underline{36}, 71.}

\refitem{Clough, A.V., and Barrett, H.H., 1983, J. Opt. Soc. Am. \underline{73}, 1590.}

\refitem{Cohen-Tanoudji, C., 1974, \underline{Laser Spectroscopy}, Proc. 2nd Int. Laser Conf., pg 324.}

\refitem{Cohen-Tanoudji, C., and Reynaud, S., 1977, J. Phys. B: Atom Molec. Phys. \underline{10}, 345.}

\refitem{Datla, R.U., and Griem, H.R., 1978, Phys. Fluids \underline{21}, 505.}

\refitem{Datla, R.U., and Griem, H.R., 1979, Phys. Fluids \underline{22}, 1415.}

\refitem{Deutsch, M., 1983, Appl. Phys. Lett. \underline{42}, 237.}

\refitem{Dimitrijevic, M.S., and Sahal-Brechot, S., 1985, J. Quant. Spect. Radiat. Transfer \underline{34}, 149.}

\refitem{Drawin, H.W., and Felenbok, P., 1965, \underline{Data for Plasmas in LTE}, Ganthier-Villars, Paris.}

\refitem{Dreike, P.L., and Tisone, G.C., 1986, J. Appl. Phys. \underline{59}, 371.}

\refitem{Drewell, N., 1979, UTIAS Report No. 279, University of Toronto.}

\refitem{Elder, P., Jerrick, T., and Birkeland, J.W., 1965, Appl. Optics \underline{4}, 589.}

\refitem{Goldbach, C., Nollez, G., Plomdeur, P., and Zimmermann, J.P., 1982, Phys. Rev. A \underline{25}, 2596.}

\refitem{Griem, H.R., 1964, \underline{Plasma Spectroscopy}, McGraw Hill, New York.}

\refitem{Griem, H.R., 1974, \underline{Spectral Line Broadening by Plasmas}, Academic, New York.}

\refitem{Grumberg, J., Coulaud, G., and Nguyen-Hoe, 1976, Phys. Lett. \underline{57A}, 227.}

\refitem{Harter, D.J., Narum, P., Raymer, M.G., and Boyd, R.W., 1981, Phys. Rev. Lett. \underline{46}, 1192.}

\refitem{Harter, D.J., and Boyd, R.W., 1984, Phys. Rev. A \underline{29}, 739.}

\refitem{Hashimoto, S., and Yamaguchi, N., 1983, Phys. Lett. \underline{95A}, 299.}

\refitem{Helbig, V., Kelleher, D.E., and Wiese, W.L., 1976, Phys. Rev. A \underline{14}, 1082.}

\refitem{Herchen, H., 1982, M.A.Sc. Thesis, University of Toronto.}

\refitem{Hohimer, J.P., 1984, Phys. Rev. A \underline{30}, 1449.}

\refitem{Hohimer, J.P., 1985, Phys. Rev. A \underline{32}, 676.}

\refitem{Holstein, T., 1947, Phys. Rev. \underline{72}, 1212.}

\refitem{Holstein, T., 1951, Phys. Rev. \underline{83}, 1159.}

\refitem{Huennekens, J., and Gallagher, A., 1983, Phys. Rev. A \underline{27}, 1851.}

\refitem{Jahreiss, L., and Huber, M.C.E., 1983, Phys. Rev. A \underline{28}, 3382.}

\refitem{Kaminsky, M.E., 1977, J. Chem. Phys. \underline{66}, 4951.}

\refitem{Kaminsky, M.E., 1980, J. Chem. Phys. \underline{73}, 3520.}

\refitem{Keikopf, J.F., 1974, J. CHem. Phys. \underline{61}, 4733.}

\refitem{Kelleher, D.E., 1981, J. Quant. Spectrosc. Radiat, Transfer \underline{25}, 191.}

\refitem{Kissack, R.S., 1984, private communication.}

\refitem{Kissack, R.S., 1987, UTIAS Report No. 305, University of Toronto.}

\refitem{Kolb, A.C., and Griem, H.R., 1958, Phys. Rev. \underline{111}, 514.}

\refitem{Konjevic, N., 1985, Phys. Rev. A \underline{32}, 673.}

\refitem{Krebs, D.J., and Schearer, L.D., 1982, J. Chem. Phys. \underline{76}, 2925.}

\refitem{Landen, O.L., Winfield, R.J., Burgess, D.D., Kilkenny, J.D., and Lee, R.W., 1985, Phys. Rev. A \underline{32}, 2963.}

\refitem{Lee, R.W., Kilkenny, J.D., Kauffman, R.L., and Matthews, D.L., 1984, J. Quant. Spectrosc. Radiat. Transfer \underline{31}, 83.}

\refitem{Lucatorto, T.B., and McIlrath, T.J., 1976, Phys. Rev. Lett \underline{37}, 428.}

\refitem{McCall, G.H., 1972, Rev. Sci. Instrum. \underline{43}, 865.}

\refitem{McIlrath, T.J., and Lucatorto, T.B., 1977, Phys. Rev. Lett. \underline{38}, 1390.}

\refitem{Measures, R.M., 1970, J. Quant. Spectrosc. Radiat. Transfer \underline{10}, 107.}

\refitem{Measures, R.M., Drewell, N., and Cardinal, P.G., 1979, J. Appl. Phys. \underline{50}, 2662.}

\refitem{Measures, R.M., and Cardinal, P.G., 1981, Phys. Rev. A \underline{23}, 804.}

\refitem{Measures, R.M., Cardinal, P.G., and Schinn, G.W., 1981, J. Appl. Phys. \underline{52}, 1269.}

\refitem{Measures, R.M., Wong, S.K., and Cardinal, P.G., 1982, J. Appl. Phys. \underline{53}, 5541.}

\refitem{Meyer, Y.H., 1980, Optics Commun. \underline{34}, 439.}

\refitem{Neiger, M., and Griem, H.R., 1976, Phys. Rev. A \underline{14}, 291.}

\refitem{Nesmeyanov, An.N., 1963, in \underline{Vapor Pressure of the Elements}, Academic Press, New York, pg. 443.}

\refitem{Niemax, K., and Pichler, G., 1974a, J. Phys. B: Atom. Molec. Phys. \underline{7}, 1204, 1974b, J. Phys. B: Atom. Molec. Phys. \underline{7}, 2355.}

\refitem{Niemax, K., and Pichler, G., 1975, J. Phys. B: Atom. Molec. Phys. \underline{8}, 179.}

\refitem{Oettinger, P.E., and Cooper, J., 1969, J. Quant. Spectrosc. Radiat. Transfer \underline{9}, 591.}

\refitem{Olsen, J.N., and Leeper, R.J., 1982, J. Appl. Phys. \underline{53}, 3397.}

\refitem{Roussel, F., Breger, P., Spiess, G., Manus, C., and Geltman, S., 1980, J. Phys. B: Atom. Molec. Phys. \underline{13}, L631.}

\refitem{Sahal-Brechot, S., 1969, Astron. Astrophys. \underline{1}, 91.}

\refitem{Salter, J.M., Burgess, D.D., and Ebrahim, N.A., 1979, J. Phys. B: Atom. Molec. Phys. \underline{12}, L759.}

\refitem{Salter, J.M., 1979, J. Phys. B: Atom. Molec. Phys. \underline{12}, L763.}

\refitem{Seaton, M.J., 1962, in \underline{Atomic and Molecular Processes} (Bates, D.R. ed.), Academic Press, New York.}

\refitem{Skinner, C.H., 1980, J. Phys. B: Atom. Molec. Phys. \underline{13}, 55.}

\refitem{Sobelman, I.I., Vainshtein, L.A., and Yukov, E.A., 1981, \underline{Excitation of Atoms and Broadening of Spectral Lines}, Springer, New York.}

\refitem{Srivastava, R.P., and Zaidi, H.R., 1975, Can. J. Phys. \underline{53}, 84.}

\refitem{Stacewicz, T., and Krasinski, J., 1981, Optics Commun. \underline{39}, 35.}

\refitem{Vaessen, P.H.M., Van Engelen, J.M.L.,and Bleize, J.J., 1985, J. Quant. Spectrosc. Radiat. Transfer \underline{33}, 51.}

\refitem{Vdovin, Yu.A., and Dobrodeev, N.A., 1969, Sov. Phys. JETP \underline{28}, 554.}

\refitem{Vidal, C.R., and Cooper, J., 1969, J. Appl. Phys. \underline{40}, 3370.}

\refitem{Waszink, J.H., and Flinsenberg, H.J., 1978, J. Appl. Phys. \underline{49}, 3792.}

\endgroup

\end{document}